\documentclass[useAMS,usenatbib,usegraphicx]{mn2e}
\usepackage{nicefrac,xfrac}
\usepackage{times}
\usepackage{float}
\usepackage{breqn}
\usepackage{graphicx}
\usepackage{pdfpages}
\usepackage{adjustbox}
\usepackage{amsmath}
\usepackage{amssymb}
\usepackage{booktabs}
\usepackage[utf8]{inputenc}
\usepackage{subfig}
\usepackage{array}
\usepackage{physics}
\usepackage{tabularx}
\usepackage{hyperref}
\usepackage{algorithm2e}

\DeclareGraphicsExtensions{.pdf,.png}

\newcommand{\apjl}{ApJ}
\newcommand{\apj}{ApJ}

\newcommand{\aap}{A\&A}

\newcommand{\mnras}{MNRAS}
\newcommand{\nat}{Nature}

\newcommand{\aapr}{The Astronomy and Astrophysics Review}
\newcommand{\apjs}{ApJ Supplement Series}

\newcommand{\rsun}{R$_\odot$}

\newcommand{\octo}{{\sc Octo-Tiger}}
\newcommand{\flash}{{\sc Flash}}
\newcommand{\flower}{{\sc Flower}}
\renewcommand{\vec}[1]{\mathbf{#1}}
\newcommand{\half}{\nicefrac{1}{2}}
\newcommand{\ddt}[1]{\frac{\partial}{\partial t} #1} 
\newcommand{\vect}[1]{{\bf #1}}

\newcommand{\revision}[1]{{#1}}

\title[\octo: a new, HPX-powered, hydro code]{Octo-Tiger: a new, 3D hydrodynamic code for stellar mergers that uses HPX parallelisation}
\author[Marcello et al.]
{Dominic C. Marcello$^{1,2,*}$, Sagiv Shiber$^{1,*}$, Orsola De Marco$^{3, 4}$, Juhan Frank$^1$,  \newauthor Geoffrey C. Clayton$^1$, Patrick M. Motl$^5$, Patrick Diehl$^{2}$, and Hartmut Kaiser$^{2}$ \\ \\
$^{1}$Department of Physics and Astronomy, Louisiana State University, Baton Rouge, LA, 70803 USA \\
$^{2}$Center for Computational Technologies, Louisiana State University, Baton Rouge, LA, 70803 USA\\
$^{3}$Department of Physics and Astronomy, Macquarie University, Sydney, NSW 2109, Australia \\
$^{4}$Astronomy, Astrophysics and Astrophotonics Research Centre, Macquarie University, Sydney, NSW 2109, Australia \\
$^5$The School of Sciences, Indiana University Kokomo, Kokomo, Indiana 46904, USA\\
$^*$The first two authors made equal contributions to this paper.\\
}

\date{Accepted XXX. Received 25 December 2020; in original form ZZZ}

\begin{document}
\maketitle
\label{firstpage}

\begin{abstract}
\octo\ is an astrophysics code 
to simulate the evolution of self-gravitating and rotating systems of arbitrary geometry based on the fast multipole method, using adaptive mesh refinement. \octo\ is currently optimised to simulate the merger of well-resolved stars that can be approximated by barotropic structures, such as white dwarfs or main sequence stars. The gravity solver conserves angular momentum to machine precision, thanks to a ``correction" algorithm.
This code uses HPX parallelization, allowing the overlap of work and communication and leading to excellent scaling properties, allowing for the computation of large problems in reasonable wall-clock times.
In this paper, we investigate the code performance and precision by running benchmarking tests. These include simple problems, such as the Sod shock tube, as well as sophisticated, full, white-dwarf binary simulations. Results are compared to analytic solutions, when known, and to other grid based codes such as {\sc flash}. We also compute the interaction between two white dwarfs from the early mass transfer through to the merger and compare with past simulations of similar systems. We measure \octo's scaling properties up to a core count of $\sim$80\,000, showing excellent performance for large problems. Finally, we outline the current and planned areas of development aimed at tackling a number of physical phenomena connected to observations of transients.

\end{abstract}


\section{Introduction}
\label{sec:introduction}

Approximately two thirds of observed stars are members of binary or multiple stellar systems \citep{duquennoy}. 
While the evolution of single stars is predictable given the initial mass and composition, the evolution of 
the components of a binary or multiple system can be dramatically altered if they are close enough to become 
interacting through mass transfer and mutual irradiation. The most consequential interactions in binary systems 
occur when the components are so close that one (or both) of the stars fill their 
Roche lobes and mass is transferred from one star to the other. In some cases, the transfer proceeds steadily for a long time; in others mass and 
energy are exchanged in a common envelope; while in the most dynamic and interesting cases the interaction leads 
to a merger giving birth to a single star of unusual properties that cannot be produced by the evolution of a single 
star. We know that all of these processes do occur in nature yielding observable outcomes such as contact binaries \citep{1984QJRAS..25..405S,2010AIPC.1314...29R}, helium white dwarfs (WD) \citep{HeWD1,HeWD2}, R Coronae Borealis 
stars \citep{clayton_rcb}, Type Iax supernovae \citep{typeiax,amcvn}, and dwarf 
novae \citep{warner03,2002apa..book.....F}. \revision{At least a fraction of all Type Ia supernovae, which are some of the most energetic explosion in the Universe, are likely caused} by the Roche-Lobe-overflow induced 
merger of two WDs \citep{typeia}. This particular type of supernova is used as a ``standard candle'' for measuring 
distances to other galaxies. Because of this, understanding Type Ia supernovae is important for understanding 
the origins of the universe and its eventual fate. 

Common envelope binary interactions \citep{Ivanovaetal2013} play a crucial role in the formation of close binaries of all types including X-ray binaries \citep{Tauris2014}, cataclysmic variables \citep{MeyerMHofmeister1979, Webbink1992}, close binaries in planetary nebulae \citep{DeMarco2015} and possibly even massive stellar double black hole binaries \citep{ricker2019}. Clearly, common envelope interactions are close cousins of mergers, the main differences being the final outcomes, which in turn depend on the types of stars involved, their mass ratios, and the energetics of mass loss through unbinding and ejection of the envelope.

Important observational developments in the past decade or so have lent further credence to the theoretical ideas
involving binary mergers and given further impetus to the development of accurate and efficient numerical tools for 
investigating binary mergers and common envelope evolution. For example, in September 2008, the contact binary, 
V1309 Sco, merged to form a luminous red nova \citep{tylenda2011}. When the merger 
occurred, the system increased in brightness by a factor of approximately $10^4$. \cite{Masonetal2010} observed the outburst spectroscopically, confirming it as a luminous red nova. Because the Optical Gravitational Lensing Experiment (OGLE) observed V1309 Sco prior to its merger for six years, this provided the rare opportunity to observe a stellar merger both before and after the event, prompting Tylenda to refer to V1309 Sco as the ``Rosetta Stone'' of contact-binary mergers. More recently, the Zwicky Transient Factory discovery of an eclipsing double white dwarf binary with an orbital period of 8.8 minutes which is destined to merge and become a hot subdwarf or an R Coronae Borealis star, makes it clear that 
such progenitor binaries do exist and provides further motivation for investigating stellar mergers and their outcomes \citep{Burdge20}. 
Discoveries of close-binary systems have multiplied in number due to surveys such as the Zwicky Transient Factory, and are due to increase exponentially with the arrival on the Vera Rubin Observatory in 2022 \citep{Ivezic2007}. 

Ultimately, one would want a high resolution 3D magneto- and radiation-hydrodynamic simulation with nuclear energy production, full chemistry, and resolution down to the dynamical timescale while also being able to follow the system over thermal timescales, but those capabilities are well beyond the horizon at this time. In lieu of such simulations, studies have used all manner of hybrid approaches. For example, 3D smooth particle hydrodynamics (SPH) simulations of merging WD binaries with 1.8 million particles are then mapped into grid codes to model SN detonations \citep{Pakmor2012} and the output is then used to study the light and polarisation signals \citep{Bulla2016}. In a different simulation, two core-hydrogen burning massive stars are merged using the moving mesh code, {\sc arepo}, with a resolution of 400\,000 to 4 million cells and the merged stars were then mapped into a 1D code to study the evolution of the remnant \citep[][]{Schneider2019}. A simulation of the V1309~Sco merger was carried out using 100\,000 SPH particles \citep{Nandez2014}, and then used as a starting point for a detailed discussion aided by analytical physics. A number of 3D hydrodynamic simulations were also used to model other binary interactions such as the mass transfer preceding  coalescence \citep[e.g.,][]{Pejchaetal2016b,MacLeod2020}. Also, \citet{Kashyap2018} carried out 200\,000 particle SPH simulations of WD mergers that then were mapped into a grid code to investigate the detonation properties.

Much of the research in 3D hydrodynamic simulations of WD mergers is aimed at understanding the dynamics of detonation with a goal to understand Type Ia supernovae. In their detailed review of 3D, WD merger simulations, \citet{Katz2016} described several efforts, most of which concentrate on the details of the properties at the time of merger but do not necessarily model the early mass-transfer and merger phase. This review cites the early pioneering efforts to model the entire merger by \citet{Motl_2002}, \citet{D_Souza_2006} and \citet{Motl2007}. In a recent counterpart to those papers, \citet{Motl2017}, carried out simulations of merging WD stars, using a finite difference technique code with up to 4 million cells and an SPH code with up to 1 million particles. \revision{Aside from the importance of code comparison, that publication explicitly shows the resolution and wall clock time constraints of this type of simulations.} 

In this paper, we present \octo, a code that aims at improving 3D simulations of interactions  using a number of computational techniques that increase accuracy and scalability. This will allow us to calculate full mergers with reasonably high resolution, and reasonable wall clock times and has the capacity to include a greater amount of physics without moving the computation into the realm of impossibility. \octo\ also conserves energy and angular momentum to excellent precision.
\octo's main application currently is simulating the merger of well resolved stars that can be represented via polytropes, such as main sequence stars or WDs.  \citet{Marcello2016} presented \octo, with a  description of the governing equations along with some preliminary tests. \citet{Staff2018} and \citet{Kadam2018} used \octo\ in parallel with a suite of other codes to simulate WD binary mergers leading to R Coronae Borealis stars and contact binaries. These studies provided a test of sorts for \octo, but the scope of those papers was such that a systematic verification and validation of \octo\ was not carried out, and neither was a scaling test aimed at measuring its speed. In the meantime, several improvements have been implemented in the code. It is therefore appropriate and timely to test and document \octo\ in a systematic way, by carrying out a suite of standard benchmark simulations, a comparison to other codes, complete with scaling tests of the latest code version.

This paper is structured in the following way. In Section~\ref{sec:octo} we describe \octo's underlying equations. In Section~\ref{sec:benchmarking_octotiger} we present the benchmarks, starting with the shock tube (Section~\ref{ssec:shocktube}), the Sedov blastwave (Section~\ref{ssec:sedov_blastwave}), a uniform static sphere (Section~\ref{ssec:static_sphere}) and continuing with a static pulsating polytrope (Section~\ref{ssec:polytrope}), a translating polytrope (Section~\ref{ssec:polytrope_translation}) and a rotating polytrope (Section~\ref{ssec:rotating_star}). We conclude with a binary simulation with a mass ratio of 0.5 in Section~\ref{sec:binarymerger}, and an assessment of the scaling properties of \octo\ in Section~\ref{sec:scaling}. We conclude in Section~\ref{sec:concl}.

\section{The AMR code Octo-Tiger}
\label{sec:octo}

\octo\ is an Eulerian AMR code, optimised for the simulation of inviscid, compressible fluids. Below we fully describe its governing equations and numerical methods.

\subsection{Hydrodynamic Evolution Equations}
\label{subsec:evolution_eq}

Octo-Tiger evolves Euler's inviscid equations of motion for a self-gravitating fluid on a rotating mesh. 
The evolution equations are:
\begin{equation}
\ddt{\rho_m}+\divergence{\rho_m\vect{v}}=0,
\label{rho}
\end{equation}
\begin{equation}
\ddt{\vect{s}}+\divergence{\vect{v}\vect{s} + \gradient{p}}=\rho \vect{g} - \Omega \times \vect{s},
\label{rhou}
\end{equation}
\begin{dmath}
\ddt{\left(E+\tfrac{1}{2}\rho\phi\right)}+\divergence{\vect{v}\left(E+\rho\phi\right)}+\divergence{\vect{u} p }=\tfrac{1}{2}\left(\phi\ddt{\rho}-\rho\ddt{\phi}\right) + \rho \Omega \cdot \left( \vect{x} \times  \vect{g}\right),
\label{energy_equation}
\end{dmath}
and 
\begin{equation}
\nabla^2 \phi = 4 \pi G \rho,
\label{poisson_equation}
\end{equation}
where $\rho_m$ is the mass density of the $m^{\mathrm{th}}$ species, $\vect{v}$ is the velocity in the rotating frame, $\vect{s}$ is the inertial frame momentum density, $p$ is the gas pressure, $\vect{u}$ is the inertial frame velocity,
$\vect{g}$ is the gravitational acceleration, $\Omega$ is the rotational frequency of the grid,
$E$ is the gas internal plus bulk kinetic energy density in the inertial frame, $\rho=\sum \rho_m$ is the total mass density, $\vect{x}$ is the position vector on the grid, $\phi$ is the gravitational potential and $G$ is the gravitational constant. The rotating frame velocity is related to the inertial frame momentum density by
\begin{equation}
\vect{s} = \rho \left( \vect{v} + \Omega \times \vect{x} \right).
\label{eq:momentum_density}
\end{equation}
The gravitational potential is related to the gravitational acceleration by $\vect{g}  = -\gradient{\phi}$. The time derivatives are taken in the rotating frame. These are related to time derivatives in the inertial frame by
\begin{equation}
\ddt{} = \left(\frac{\partial}{\partial t}\right)_{\mathrm{inertial}} + \left(\vect{x} \times \Omega\right) \cdot \vect{\nabla}
\end{equation}

\revision{
A previous version of \octo\ used a hydro-solver that conserved linear and angular momentum to machine precision. This was based on the work by \cite{DESPRES201528}. We successfully adapted this method to \octo\ for use when velocities are reconstructed in the inertial frame. However, our methodologies could not be applied to when velocities are reconstructed in the rotating frame. The rotating frame is a better choice when dealing with binary interactions, because equilibrium rotating stars retain their initial profiles much better. We thus abandoned the development of the hydrodynamics angular momentum conservation feature. A legacy of this development work is a separately evolved angular momentum field}, $\vect{l}$, described by the following equation:
\begin{equation}
\ddt{\vect{l}} + \divergence{\vect{v}\vect{l} + \vect{\nabla}} \times \vect{x} p = \rho \vect{x} \times \vect{g} - \Omega \times \vect{l}.
\label{eq:ang_mom_ev}
\end{equation}
In the current version, this quantity is evolved passively, meaning the evolution of the other variables does not depend on the value of $\vect{l}$. It is initialized to  $\vect{x} \times \vect{s}$ at $t=0$. Although our gravity solver exactly conserves angular momentum to machine precision, because of numerical viscosity in the hydrodynamics solver, angular momentum is not exactly conserved. The difference between $\vect{x} \times \vect{s}$ and $\vect{l}$ after $t=0$ gives a measure of the error in angular momentum conservation. 

The source terms on the righthand side (RHS) of Equation~\ref{rhou} include a gravitational and a rotational term. The rotational term accounts for the rotation of the momentum vector relative to the rotating
mesh. Because we evolve inertial frame quantities on a rotating mesh, as opposed to rotating frame quantities on a rotating mesh, this term is half the Coriolis force (only half because velocities are taken with respect to the rotating grid, but momenta are calculated in the inertial frame). Note that if $\int_V \vect{s} = 0$, 
this term does not violate momentum conservation.

Rather than solving Equation~\ref{poisson_equation} using an iterative approach, \octo\ uses the fast multipole method (FMM; described below).  To calculate the the first term on the RHS of Equation~\ref{energy_equation}, we solve for $\ddt{\phi}$
using the FMM with the numerically computed value of $\ddt{\rho}$ as the source term. This results
in the two parts of this term cancelling when summed over the entire grid.

Equation~\ref{energy_equation} is derived from the usual form of the energy equation:
\begin{equation}
\ddt{E}+\divergence{\vect{v} E}+\divergence{\vect{u} p}= -\rho \vect{u} \cdot \gradient{\phi},
\label{energyonly_equation}
\end{equation}
along with Equation \ref{rho}.
Equation~\ref{energy_equation} is written in a form that emphasizes that the conserved quantity is the gas energy, $E$ (internal plus bulk kinetic), plus the potential energy or $E + \frac{1}{2}\rho \phi$. The first term on the RHS of Equation~\ref{energy_equation} vanishes \revision{globally} because $\phi$ is linearly related to
$\rho$, while the second term vanishes \revision{globally} because the total change in angular momentum due to gravity over all space is zero. As discussed by \citet{Marcello2012}, evolving the energy equation in this
form prevents violation of energy conservation due to matter moving up or down a potential well because of numerical viscosity. Because our gravity solver conserves linear and angular momenta to machine precision, using this form of the energy equation conserves total energy to machine precision. In practice, we actually evolve $E$ instead of $E + \frac{1}{2}\rho \phi$, by taking the discretized evolution equation for Equation~\ref{energy_equation} and solving it for $E$. 

Following the dual energy formalism of \cite{Bryanetal1995}, \octo\ evolves a second variable for the energy. While they used the internal energy density as the second energy variable, we use the ``entropy tracer" \citep{Motl_2002}, defined as 
\begin{equation}
\tau = \left(\rho \varepsilon\right)^{\frac{1}{\gamma}},
\label{eq:entropy_tracer}
\end{equation}
where $\varepsilon$ is the specific internal gas energy and $\gamma = \frac{5}{3}$ is the ratio of specific heats.
When there are no shocks, $\tau$ is conserved and evolves as 
\begin{equation}
\ddt{\tau}+\divergence{\tau\vect{v}}=0.
\label{tau}
\end{equation}
The specific internal energy, $\varepsilon$, is computed according to
\begin{equation}
    \rho \varepsilon = 
\begin{cases}
    E - \tfrac{1}{2} \rho u^2, & \text{if } E - \tfrac{1}{2} \rho u^2  \geq \epsilon_1 E\\
    \tau^{\gamma},             & \text{otherwise}
\end{cases},
\label{eq:dual_energy1}
\end{equation}
where the default value of $\epsilon_1 =  0.001$. Once we have the internal energy density we can compute the pressure with the ideal gas equation:
\begin{equation}
p = (\gamma-1) \rho \varepsilon.
\end{equation}
At the end of every update of the evolution variables, the entropy tracer is then reset using Equation~\ref{eq:entropy_tracer} in computational cells which satisfy 
\begin{equation}
    E - \tfrac{1}{2} \rho u^2 > \epsilon_2 E
\label{eq:dual_energy2}
\end{equation}
for 
at least one of the adjacent cells or the cell itself, where the default value of $\epsilon_2 =  0.1$. Otherwise it is left alone.  
\revision{
    We use the values for $\epsilon_1$ and $\epsilon_2$ chosen by ~\cite{Bryanetal1995}
}.

This treatment allows for the proper evolution of shocks while simultaneously retaining numerical accuracy of the internal energy in high mach flows, where the kinetic energy dominates. It 
is roughly analogous to adding extra digits of precision to the internal energy. This method also guarantees positive values for the internal energy even when $E - \tfrac{1}{2} \rho u^2 < 0$.

\octo\ evolves the gas using an ideal gas equation of state (EoS), but by setting $\epsilon_1 = \epsilon_2 = 1$, shock heating is eliminated and effectively \octo\ evolves the gas with a polytropic EoS. If 
\begin{equation}
p_{\mathrm{poly}} = K \rho^{1+\frac{1}{n}},
\label{eq:poly}
\end{equation}
initially, where $K$ and $n$ are constants, this condition will continue to hold as the system evolves.

\subsection{Gravity Update and the Angular Momentum Correction}
\label{subsec:gravity_solver}

\octo\ uses a variant of the Cartesian FMM described by \cite{Dehnen2000}. This method conserves linear momentum to machine precision. The FMM adapts naturally to an oct-tree based adaptive mesh refinement (AMR) scheme such as that used by \octo. The multipoles of each cell are composed of the multipoles of its child cells, and the expansion in each child cell is derived from the expansion of its parent cell. Both multipole and expansions are relative to cell centres. 

\octo\ has an extension, detailed by \cite{Marcello2016}, which allows its gravity solver to conserve both angular and linear momentum to machine precision. Based on the FMM described by \cite{Dehnen2000}, which conserves linear momentum to machine precision, the \octo\ extension works by adding an additional higher order multipoles which is used to calculate a correction to the calculated force. This angular momentum correction cancels the angular momentum conservation violation while preserving linear momentum conservation of the original method. As a result of angular momentum conservation, the last term in the RHS of Equation ~\ref{energy_equation} also sums to zero over all space, enabling conservation of energy to machine precision in the rotating frame.

Conservation of angular momentum and energy is important for the self-consistent accurate modelling of near equilibrium astrophysical systems such as binary stars in the early phases of mass transfer. Violations in energy conservation can cause stars in equilibrium to ``evaporate", while the balance of angular momentum may affect the ultimate merger if one occurs. 

The FMM requires the specification of an opening criterion. Given two cells, $A$ and $B$, at grid locations $\vec{x}_A$ and $\vec{x}_B$, respectively, the opening angle is
    \begin{equation}
        \theta^\prime = \frac{\Delta x}{|\vec{x}_A - \vec{x}_B|},
    \end{equation}
where $\Delta x$ is the width of a grid cell. If $\theta^\prime < \theta$, where $\theta$ is a critical value less than unity, then two cells are well separated. Two well separated cells interact through the multipole interaction when they are well separated from each other, but their respective parents are not well separated. In the models presented in this paper, $\theta = 0.5$ or 0.34, the latter being used for the interacting binary models. 

We also define the quantity, $\theta_{\rm min}$, to refer to the lowest $\theta$ allowed by \octo. This is determined at compile time and its value depends on the size of the sub-grids, with larger sub-grids allowing for smaller values of  $\theta_{\rm min}$. For $8\times 8\times 8$ sub-grids, as used in this paper, $\theta_{\rm min}=0.34$.

It should be noted the current version of \octo\ does not conserve angular momentum in the hydrodynamics module. We have experimented with extending an angular momentum conserving hydrodynamics method described by \cite{DESPRES201528} for use in \octo. While we have had success when the grid is not rotating, the method fails in the rotating frame. The reconstruction of face values are computed in the rotating frame, eliminating the degree of freedom in the reconstruction required for \cite{DESPRES201528} to work.

\subsection{Hydrodynamic Update}
\label{ssec:hydro_update}

The hydrodynamic update begins by reconstructing cell averaged values on the surface of the cells. \octo\ reconstructs values at the geometric centres of the the $6$ cell faces, $12$ cell edges, and $8$ cell vertices, for a total of $26$ quadrature points. Note this results in each face having $9$ quadrature points, $1$ at the cell face centre, and $4$ each for the cell face's edges and vertices. This allows \octo\ to compute the flux through a face as the integral of the fluxes at $9$ points on the face. 

Rather than reconstructing the vector of conserved quantities, 
 \begin{align*}
 U &= \begin{bmatrix}
 	\rho_m \\
        \vect{s} \\
        E \\
        \tau \\
        \rho \phi \\
      \end{bmatrix},
\end{align*}
we reconstruct 
 \begin{align*}
 V &= \begin{bmatrix}
 	\rho_m \\
        \vect{v}\\
        E - \tfrac{1}{2} \rho u^2\\
        \tau \\
        \phi \\
       \end{bmatrix},
\end{align*}
and then transform back to $U$. The reconstruction of velocities is done in the rotating frame. \revision{In \octo\ the rotating frame is defined as rotating around the z-axis, and the frequency can be specified by the user. For the binary simulations in this paper the rotating frame frequency is the same as the initial orbital frequency of the binary.} Reconstructing the velocities in the inertial frame is also a valid choice. However, we have found that single stars in initial equilibrium retain their original density profiles better when reconstructing the rotating frame velocities. 

\revision{The the total energy, defined as the sum of the potential and gas energy, is a conserved quantity, but it is not a purely hydrodynamic quantity. The local potential plays virtually no role in the formation of discontinuities in the gas energy. When applying a central advection scheme we want the averaging of left and right states to occur over hydrodynamic variables. Furthermore, discontinuities in the potential energy are caused solely by the local mass density, the specific potential energy itself being smooth. For these reasons we treat the two quantities as separate when computing the reconstruction and fluxes.}

The lefthand side of Equation~\ref{energy_equation} is evolved as the sum of two parts, 
\begin{equation}
\ddt{E}+\divergence{E \vect{v} }+\divergence{E \vect{u} p }= 0,
\label{energy_equation1}
\end{equation}
and
\begin{equation}
\ddt{\tfrac{1}{2}\rho\phi}+\divergence{\vect{v}\rho\phi} = 0.
\label{energy_equation2}
\end{equation}
Below we show how we combine these two parts to form a single energy equation for the gas energy. 

There are left and right values for each interface, and here we denote those with $R$ and $L$ superscripts, respectively. For a quantity $u$ and for every $i$, $j$, and $k$ there are eight vertex values, 
\begin{dmath}
u_{i+\half{} j+\half{} k+\half{}}^{RRR},  
u_{i+\half{} j+\half{} k+\half{}}^{RRL}, 
u_{i+\half{} j+\half{} k+\half{}}^{RLR}, \\
u_{i+\half{} j+\half{} k+\half{}}^{RLL}, 
u_{i+\half{} j+\half{} k+\half{}}^{LRR}, 
u_{i+\half{} j+\half{} k+\half{}}^{LRL}, \\
u_{i+\half{} j+\half{} k+\half{}}^{LLR}, \ \mathrm{and}  \ 
u_{i+\half{} j+\half{} k+\half{}}^{LLL},
\end{dmath}
twelve edge values, 
\begin{dmath}
u_{i+\half{} j+\half{} k}^{RR}, 
u_{i+\half{} j+\half{} k}^{RL}, 
u_{i+\half{} j+\half{} k}^{LR}, 
u_{i+\half{} j+\half{} k}^{LL}, \\
u_{i+\half{} j k+\half{}}^{RR}, 
u_{i+\half{} j k+\half{}}^{RL}, 
u_{i+\half{} j k+\half{}}^{LR}, 
u_{i+\half{} j k+\half{}}^{LL}, \\
u_{i j+\half{} k+\half{}}^{RR}, 
u_{i j+\half{} k+\half{}}^{RL},  
u_{i j+\half{} k+\half{}}^{LR},  \ \mathrm{and} \
u_{i j+\half{} k+\half{}}^{LL}, 
\end{dmath}
and six face values,
\begin{dmath}
u_{i+\half{} j k}^{R}, 
u_{i+\half{} j k}^{L}, \\
u_{i j+\nicefrac{1}{2} k}^{R}, 
u_{i j+\nicefrac{1}{2} k}^{L}, \\
u_{i j k+\half{}}^{R},  \ \mathrm{and} \
u_{i j k+\half{}}^{L}.
\end{dmath}
Note that the $u$ used here is not the inertial frame velocity used elsewhere in this paper. We use the piecewise parabolic method (PPM) of \citet[][with contact discontinuity detection]{COLELLA1984174} to compute these values.
The five cell stencil required for PPM is
taken along the line from the cell centre to each face, edge, or vertex.  For example, the reconstructed values at $u_{i+\half{} j k}^L$ and  $u_{i-\half{} j k}^R$ are computed by applying PPM to the five cell stencil formed by $u_{i-2,j,k}$,$u_{i-1,j,k}$,$u_{i,j,k}$,$u_{i+1,j,k}$, and $u_{i+2,j,k}$, while the reconstructed values at $u_{i+\half{} j-\half{} k}^{LR}$ and  $u_{i-\half{} j+\nicefrac{1}{2} k}^{RL}$ are computed with the stencil formed by $u_{i-2,j+2,k}$,$u_{i-1,j+1,k}$,$u_{i,j,k}$,$u_{i+1,j-1,k}$,and $u_{i+2,j-2,k}$. 
\revision{ Reconstructing the primitive variables at $26$ points across the cell's surface, using PPM, results in multi-dimensional third order convergence. Without this feature, the atmospheres of equilibrium stars turn into box like structures.}

Once a cell's evolved quantities are reconstructed at each of the $26$ quadrature points on the cells' surfaces, \octo\ computes the fluxes at each of the $9$ quadrature points for each cell's face. We use the central-upwind scheme described by \cite{KURGANOV00}. \revision{This scheme was originally chosen because it was straightforward to adapt it for use with \octo's hydrodynamics angular momentum conservation feature, as described in Section~\ref{subsec:evolution_eq}.} It can be defined in terms of left and right face values, 
\begin{equation}
    H\left(U_L,U_R\right) = \frac{a^+ F(U_L) - 
    a^- F(U_R)}{a^+ - a^-} + \frac{a^+ a^-}{a^+ - a^-} \left[ U_R - U_L\right], 
\end{equation}
where $H$ is the numerical flux, $F$ is the physical flux, and $a^+$ and $a^-$ are the positive and negative signal speeds. \octo\ obtains $H$ for each dimension using the physical flux,
\begin{align}
F_x = \begin{bmatrix}
 	    v_x \rho \\
         v_x s_x + p \\
         v_x s_y \\
         v_x s_z \\
         v_x E + u_x p \\
         v_x \tau \\
         v_x\rho \phi
      \end{bmatrix} 
    &
F_y = \begin{bmatrix}
 	    v_y \rho \\
         v_y s_x \\
         v_y s_y  + p\\
         v_y s_z \\
         v_y E + u_y p \\
         v_y \tau \\
         v_y\rho \phi
      \end{bmatrix} 
    &
F_z = \begin{bmatrix}
 	    v_z \rho \\
         v_z s_x \\
         v_z s_y \\
         v_z s_z + p \\
         v_z E + u_z p \\
         v_z \tau \\
         v_z\rho \phi
      \end{bmatrix}, 
\end{align}
and the signal speeds
\begin{equation}
\begin{matrix}
a_x^+ = \max\left({c_{s,R}+v_{x,R},c_{s,L}+v_{x,L},0}\right), \\
a_x^- = \min\left({c_{s,R}-v_{x,R},c_{s,L}-v_{x,L},0}\right), \\
a_y^+ = \max\left({c_{s,R}+v_{y,R},c_{s,L}+v_{y,L},0}\right), \\
a_y^- = \min\left({c_{s,R}-v_{y,R},c_{s,L}-v_{y,L},0}\right), \\
a_z^+ = \max\left({c_{s,R}+v_{z,R},c_{s,L}+v_{z,L},0}\right),  \\
a_z^- = \min\left({c_{s,R}-v_{z,R},c_{s,L}-v_{z,L},0}\right),
\end{matrix}
\label{eq:signal}
\end{equation}
where the $c_s$ are the sound speeds, 
\begin{equation}
    c_s = \sqrt{\frac{\gamma p}{\rho}}.
\end{equation}

The total flux through a face is then obtained by summing the fluxes taken at each quadrature point on the face. For the x-face, for instance, we have
\begin{dmath}
   \mathcal{H}_{x,i+\half{} j k} = \\ \frac{16}{36} \left[H_x(U^L_{i+\half{} j k}, U^R_{i+\half{} j k} )\right] + \\
   \frac{4}{36} \left[ 
   H_x(U^{LL}_{i+\half{} j+\half{} k}, U^{RL}_{i+\half{} j+\half{} k} ) + \\
   H_x(U^{LR}_{i+\half{} j-\half{} k}, U^{RR}_{i+\half{} j-\half{} k} ) + \\
   H_x(U^{LL}_{i+\half{} j k+\half{}}, U^{RL}_{i-\half{} j k+\half{}} ) + \\
   H_x(U^{LR}_{i+\half{} j k-\half{}}, U^{RR}_{i-\half{} j k-\half{}} ) \right] + \\
   \frac{1}{36} \left[
   H_x(U^{LLL}_{i+\half{} j+\half{} k+\half{}}, U^{RLL}_{i+\half{} j+\half{} k+\half{}} ) + \\
   H_x(U^{LLR}_{i+\half{} j+\half{} k-\half{}}, U^{RLR}_{i+\half{} j+\half{} k-\half{}} ) + \\
   H_x(U^{LRL}_{i+\half{} j-\half{} k+\half{}}, U^{RRL}_{i+\half{} j-\half{} k+\half{}} ) + \\
   H_x(U^{LRR}_{i+\half{} j-\half{} k-\half{}}, U^{RRR}_{i+\half{} j-\half{} k-\half{}} )
   \right],
\end{dmath}
where $\mathcal{H}_{x,i+\half{} j k}$ is the total numerical flux through the x-face. 

The semi-discrete form of the evolution equations then becomes
\begin{dmath}
    \frac{d}{dt} U_{i j k} + \\
    \frac{H_{x,i+\half{} j k } - H_{x,i-\half{} j k }}{\Delta x}  + \\
    \frac{H_{y,i j+\half{} k } - H_{y,i j-\half{} k }}{\Delta x}  + \\
    \frac{H_{z,i j k+\half{} } - H_{z,i j k-\half{} }}{\Delta x} = S_{i j k},
\label{eq:flux}
\end{dmath}
where we have added the source term $S_{i j k}$ (defined below). 

We still need to re-combine the gas energy and potential energy parts of the flux to form a single equation for the gas energy. This is accomplished through the transformation
\begin{equation}
    \begin{matrix}
    H_E & \rightarrow & H_E + H_{\phi} \\
    H_\phi & \rightarrow & 0,
    \end{matrix}
\label{eq:trans}
\end{equation}
where $H_E$ refers to the gas energy flux and $H_\phi$ refers to the potential energy flux.

The source term, $S$, is equal to the RHS of the evolution Equations~\ref{rho}, \ref{rhou}, \ref{energy_equation}, \ref{poisson_equation} and \ref{tau},
\begin{equation}
    S = \begin{bmatrix}
    0 \\
    \rho \vec{g} + \Omega \times \vec{s} \\
     \tfrac{1}{2}\left(\phi\ddt{\rho}-\rho\ddt{\phi}\right) + \Omega \times \left(\rho \vect{x} \times  \vect{g}\right) \\
      0 \\ 
      0
      \end{bmatrix}
\end{equation}
Equations \ref{eq:flux} and \ref{eq:trans} account for all terms of Equations \ref{rho} through \ref{energy_equation} and Equation \ref{tau} except for the $\ddt{\tfrac{1}{2}\rho\phi}$ term on the LHS of Equation \ref{energy_equation}. This term is accounted for by updating the total energy every time the FMM solver is called to update the potential,
\begin{equation}
E_{\mathrm{after}} = E_{\mathrm{before}} + \tfrac{1}{2}\rho\left(\phi_{\mathrm{before}} - \phi_{\mathrm{after}}\right),
\end{equation}
where the ``before" and ``after" subscripts refer to before and after the FMM solver is called.

The $\frac{d}{dt}$ operator from Equation \ref{eq:flux} is discretized using the third order Total Variation Diminishing Runge Kutta integrator of \cite{SHU1989}. The time step size, $\Delta t$, is chosen by the Courant-Friedrichs-Lewy (CFL) condition 
\begin{equation}
    \Delta t = \eta_{\mathrm{CFL}} \  \mathrm{max}_{all}\left[\frac{\Delta x}{a^\pm_{x|y|z}}\right],
    \label{cfl}
\end{equation}
where the maximum is taken over all the computational cells in the domain and $\eta_{\mathrm{CFL}}$ is a positive dimensionless constant less than $\frac{1}{2}$. For the binary simulations presented below, $\eta_{\mathrm{CFL}}=\frac{4}{15}$.

A schematic representation of the execution of the steps is shown in Algorithm~\ref{fig:algorithm}.

\begin{algorithm}
	Set initial conditions\;
	Do initial output\;
  	\While{}{
  		\For{$i\gets 1$ \KwTo $3$}{
    			\If{$i$ == $1$}{
      				$U_0 \gets U$\;
      				Compute $\Delta t$\;
    			}
    			Compute $\ddt{U}$ due to advection only\;
    			Compute $\vect{g}$, $\phi$, and $\ddt{\phi}$\ using the FMM\;
    			Add the source terms to $\ddt{U}$\;
    			Update $U$ from $\ddt{U}$ and $U_0$ using the 3rd order R-K method\;
    			Update the entropy tracer using the dual energy formalism\;
    			\If{using floors}{
				Apply floor values for density and entropy\;
  			}    
  		}
		\If{time to check refinement}{
			refine and de-refine the AMR mesh as needed\;
			redistribute the work-load\;
		}
		\If{time to output}{
			output to SILO file\;
		}
	}
\caption{The algorithm Octo-Tiger uses for evolution}
\label{fig:algorithm}
\end{algorithm}

\subsection{Adaptive Mesh Refinement}
\label{ssec:amr}

\octo\ uses an oct-tree based AMR. Each node of the oct-tree has associated with it a single $N\times N\times N$ sub-grid and is either fully refined with eight child nodes (an interior node) or not refined at all (a leaf node). For the simulations in this paper, the sub-grids' interior size is $8 \times 8 \times 8$. PPM requires a 3 cell boundary for a total sub-grid size of $14 \times 14 \times 14$. \octo\ properly nests the sub-grids, meaning there can be no more than one jump in refinement levels across sub-grid boundaries. 

We define an ``AMR boundary" as a sub-grid boundary across which the refinement level changes. Because of proper nesting, this involves only two refinement levels, which we refer to here as the ``fine" and ``coarse" level. The fine sub-grid at an AMR boundary cannot get its boundary cells from a neighboring sub-grid of the same refinement level, so it must interpolate its boundary from the coarse sub-grid sharing the same boundary. To describe our AMR boundary interpolation scheme, we use an indexing system, which is aligned such that $\vect{x}_{i-\half j-\half k-\half}^C = 
 \vect{x}_{2 i-1, 2 j-1, 2 k-1}^F$, where the superscripts "C" and "F" refer to the coarse and fine levels, respectively. The coarse sub-grid has cell centres at $\vect{x}_{i j k}$ while the fine sub-grid has cell centres at half integer locations, $\vect{x}_{2 i\pm\half, 2 j\pm\half, 2 k\pm\half}^F$. The slopes for our  interpolation scheme for a variable $u$ (note, the $u$ variable used here is not the inertial frame velocity used everywhere else in this paper) are
 \begin{equation}
    \begin{matrix}
     u^C_{x,i j k} =
     \mathrm{minmod}[u^C_{i+1 j k},u^C_{i j k},u^C_{i-1 j k}] \\
     u^C_{y,i j k} =
     \mathrm{minmod}[u^C_{i j+1 k},u^C_{i j k},u^C_{i j-1 k}] \\
     u^C_{z,i j k} =
     \mathrm{minmod}[u^C_{i j k+1},u^C_{i j k},u^C_{i j k-1}] \\
     u^C_{xy,i j k} =
     \mathrm{minmod}[u^C_{i+1 j+1 k},u^C_{i j k},u^C_{i-1 j-1 k}] \\
     u^C_{xz,i j k} =
     \mathrm{minmod}[u^C_{i+1 j k+1},u^C_{i j k},u^C_{i-1 j k-1}] \\
     u^C_{yz,i j k} =
     \mathrm{minmod}[u^C_{i j+1 k+1},u^C_{i j k},u^C_{i j-1 k-1}] \\
     u^C_{xyz,i j k} =
     \mathrm{minmod}[u^C_{i+1 j+1 k+1},u^C_{i j k},u^C_{i-1 j-1 k-1}].
     \end{matrix}
\end{equation}
where {\it minmod} is zero if the signs are opposite, or the minimum absolute value is taken if the signs are the same. The interpolation scheme is then
\begin{dmath}
    u^F_{2 i+q,2 j+r,2 k+s} = 
    u^C_{i j k} + \frac{9}{64}\left(
    {\rm sgn}[q] u^C_{x,i,j,k} +
     \mathrm{sgn}[r] u^C_{y,i,j,k} + 
      \mathrm{sgn}[s] u^C_{z,i,j,k}\right) + 
      \frac{3}{64}\left(
    \mathrm{sgn}[q]\mathrm{sgn}[r] u^C_{xy,i,j,k} + 
     \mathrm{sgn}[q]\mathrm{sgn}[s] u^C_{yz,i,j,k} + 
      \mathrm{sgn}[r]\mathrm{sgn}[s] u^C_{xz,i,j,k}\right) + 
 \frac{1}{64}\mathrm{sgn}[q]\mathrm{sgn}[r]\mathrm{sgn}[s] u^C_{xyz,i,j,k},
\end{dmath}

\noindent where $q=\pm\half$, $r=\pm\half$, and $s=\pm\half$ and we emphasise that the variable $s$ does not mean the inertial frame momentum density of Equation~\ref{eq:momentum_density} and is used only here. The string ``sgn" means the sign of $q$, $r$ and $s$.

To ensure conservation across the AMR boundary, coarse fluxes on AMR boundaries are taken to be the sum of the fine fluxes through the cell face:
\begin{dmath}
    H^C_{x,i+\half j k} = 
    \frac{1}{4} H^F_{x,2 i + 1, 2 j+\half, 2 k+\half} +
    \frac{1}{4} H^F_{x,2 i + 1, 2 j+\half, 2 k-\half} +
    \frac{1}{4} H^F_{x,2 i + 1, 2 j-\half, 2 k+\half} +
    \frac{1}{4} H^F_{x,2 i + 1, 2 j-\half, 2 k-\half}.
\end{dmath}

The refinement criteria for a cell to be refined is either
\begin{equation}
     \frac{\Delta x}{\rho}\frac{d \rho}{ d x_i} > 0.1 ,
\end{equation}
or
\begin{equation}
     \frac{\Delta x}{\tau}\frac{d \tau}{ d x_i} > 0.1 ,
\end{equation}
for any $i$ direction, where $\rho$ is the density and $\tau$ is the entropy tracer of Equation~\ref{eq:entropy_tracer}.

\subsection{Boundary Conditions}
\label{ssec:boundary_conditions}

\octo\ has three boundary conditions available. With ``inflow" boundary conditions, the ghost cells at the edge of the physical domain are copied from the closest interior cell. ``Outflow" boundary conditions are the same as inflow except that the momentum is set to zero in the case where the momentum in the closest interior cell points inwards. This prevents the artificial creation of inflows. For our binary simulations we generally use the outflow boundary condition. These boundary conditions are meant to simulate isolated systems, and do not need to be accounted for by the FMM gravity solver. 

A reflecting boundary condition is available for pure hydrodynamics  runs with no gravity. Although it is certainly feasible to incorporate reflecting boundary conditions into the FMM and periodic boundary conditions into both the FMM and the hydrodynamics solver, to incorporate these boundary conditions into the FMM solver is a non-trivial task. Since \octo\ does not use these types of boundary conditions to simulate binary systems they were not implemented at this time.

\subsection{Minimum Values for Density and Entropy}
\label{ssec:floors}

Both the mass density and entropy tracers should be always positive. This condition is difficult to maintain numerically when there is a large range of densities present. For our the polytropic binary simulation detailed in Section~\ref{sec:binarymerger}, we fill the regions around the stars with a density equal to $10^{-10}$ times the maximum density on the grid. The density can drop below this value, because the star's gravity is constantly pulling matter from the near-vacuum region onto the stars. \octo\ also uses a third order Runge Kutta scheme, so even without the presence of gravity, it is not possible to know {\it a priori} the maximum time step size needed to guarantee positivity. For these reasons, \octo\ includes options to set minimum values for the mass density and entropy tracer. These floor values are denoted $\rho_f$ and $\tau_f$, respectively.

Imposing a density floor, results in imposing values on other variables that depend on density. We thus define the density floor scaling parameter,
    \begin{equation}
    f_\rho = 1-\mathrm{max}\left[1 - \frac{\mathrm{max}[\rho,0]}{\rho_f}, 0\right]
    \end{equation}
allowing us to transform the variables according to
    \begin{equation}
    \begin{matrix}
        \vect{s} & \rightarrow & \vect{s}  f_\rho \\
        E  & \rightarrow & E f_\rho + \tau_f^\gamma (1-f_\rho) \\
        \tau & \rightarrow & \tau f_\rho + \tau_f (1-f_\rho) \\
        \rho_m & \rightarrow & \mathrm{max}[\rho,\rho_f] \frac{\rho_m}{\rho} 
        \end{matrix}
    \end{equation}
    This deteriorates the strict machine precision conservation of the affected variables, however it only occurs in the near-vacuum regions.

\subsection{AMR Refinement Criteria}
\label{ssec:refinement_criteria}

\octo\ checks for refinement every $(2/\eta_{\mathrm{CFL}})$ time steps, where $\eta_{\mathrm{CFL}}$ is from Equation \ref{cfl}. A window of two cells is used for refinement. This condition prevents waves from propagating through an AMR boundary before the relevant cells are checked for refinement. If a sub-grid contains one or more cells flagged for refinement, the sub-grid is refined, converting its leaf node to an interior node with 8 children. Conversely, interior nodes whose 8 children are all leaf nodes are de-refined if none of their cells and none of the children's cells are flagged for refinement. Sub-grids are also flagged for refinement as needed to ensure the difference in refinement levels across grid boundaries is no greater than one.

For the binary simulations in this paper, we use a density based refinement criterion. 
If $l_{\mathrm{max}}$ is the maximum allowed refinement level and $\rho_r$ is the refinement density cut-off, a cell is flagged for refinement if the maximum level $l$ for which 
\begin{equation}
    \rho > 8^{l_{\mathrm{max}} - l}\rho_r
\end{equation}
holds true is greater than the cell's level of refinement. For the binary simulations $\rho_r$ is held constant initially. Once material ejected from the binary begins to fill the grid, $\rho_r$ is adjusted dynamically in a manner that attempts to keep the total number of sub-grids at a specified level. After every refinement, 
\begin{equation}
    \rho_r \rightarrow \left(\frac{N_{\mathrm{grids}}}{N_{\mathrm{target}}}\right)^2 \rho_r,
\end{equation}
where $N_{\rm grids}$ is the total number of sub-grids and $N_{\rm target}$ is the desired number.

The CFL factor (Equation~\ref{cfl}) typically used by simulations is $\sim$0.4. Technically there is a limit to the maximum CFL number one should use. In three dimensions with PPM, it is $1/7$. This is because the maximum ratio between the density reconstructed at the cell's face and the cell averaged density is $7/3$. As a result, the limit should be $3/7$ in one dimension or $3/(3 \times 7)$ in three dimensions. This limit is stringent, in that it prevents any sharp transition in density from emptying a cell of its content possibly resulting in zero or negative densities. In \octo\ negative densities are corrected by introducing a ``floor" density value in that cell (see Section~\ref{ssec:floors}), which in turn results in mass non-conservation. Preempting the results of Section~\ref{ssec:polytrope_translation}, this can result in mass growth at the level of 0.0001 percent from one time step to the next when highly supersonic motions are present. It is therefore important to critically appraise the value of the CFL factor to be used in each simulations, particularly if strong shocks are present. We discuss these choices and their impacts further in Section~\ref{ssec:polytrope_translation}.

\subsection{Code Units}
\label{ssec:code_units}

There are no physical constants present in the hydrodynamic equations, therefore the simulations without gravity enabled are unit-less. With gravity we have a single physical constant, $G$. In the code the value of this constant is set to unity. We convert between code units and physical units in the cgs system using three conversion factors, $m_{\mathrm{cgs}}$, $l_{\mathrm{cgs}}$, and $t_{\mathrm{cgs}}$ for mass, length, and time, respectively. Because the value of $G$ is fixed at unity, we may specify two of these conversion factors, with the third being determined by the relation
\begin{equation}
\frac{l_{\mathrm{cgs}}^3}{m_{\mathrm{cgs}} t_{\mathrm{cgs}}^2} = G_{\mathrm{cgs}},
\label{eq:codeunits}
\end{equation}
where $G_{\mathrm{cgs}}$ is the value of $G$ in cgs units. In \octo\ the user specifies $l_{\mathrm{cgs}}$ and $m_{\mathrm{cgs}}$ and \octo\ calculates $t_{\mathrm{cgs}}$ using Equation \ref{eq:codeunits}. When outputting the grid to file, \octo\ converts all quantities to their cgs equivalents using these unit conversion factors.

\subsection{The Temperature in OCTO-TIGER}
\label{ssec:temperature}

\octo\ evolves the density, $\rho$, total gas energy density (internal and bulk kinetic energy density), $E$, and velocities, and derives all the other physical quantities based on these values and the EoS. The evolution equations, Equations~\ref{rho} to \ref{poisson_equation}, do not require knowledge of the temperature. For post processing purposes, the temperature can be computed by assigning atomic mass and atomic numbers to each mass density species and assuming a fully ionized gas according to
\begin{equation}
    T = \frac{(\gamma-1) \rho \varepsilon}{k_B \sum \frac{\rho_m (1+N_{Z,m})}{m_H N_{A,m}}},
\end{equation}
where $N_{A,m}$ and $N_{Z,m}$ are the atomic mass and atomic numbers of the $m^{\mathrm{th}}$ species and the specific internal energy $\varepsilon$, comes from Equation~\ref{eq:dual_energy1}.

\subsection{The C++ Standard Library for Concurrency and Parallelism (HPX)}
\label{ssec:HPX}

\octo\ is parallelized for distributed systems using the C++ Standard Library for Concurrency and Parallelism \citep[HPX,][]{Kaiser2020,BILLIONS,daiss2019piz}. HPX is an open source C++ Standard Library for Concurrency and Parallelism and is within the class of the so-called asynchronous many-task AMT runtime systems. 

Another AMT utilized in astrophysics simulation, e.g.,\ ChaNGa \citep{jetley2008massively} or Enzo-P \citep{10.5555/2462077.2462081}, is Charm++ \citep{10.5555/871085}. We focus on the comparison of these two AMTs in this paper, for a more comprehensive review for various AMTs we refer to \cite{thoman2018taxonomy}. The commonality if HPX and Charm++ is the usage of the same concepts for ``parallelism" and ``concurrency". The distinctness of HPX is that it fully conforms to the \texttt{C++17} ISO standard \citep{cxx17_standard} and implements proposed features of the upcoming \texttt{C++20} ISO standard \citep{cxx20_standard}. This means that HPX's features that are available in the C++ standard can be replaced without changing the function arguments. From a programmer's perspective, HPX is more an abstraction of the C++ language while Charm++ is more a standalone library. The requirement for \octo\ (distributed, task-based, asynchronous) are met by only few AMTs and HPX has the highest technology readiness level according to this review \citep{thoman2018taxonomy}.

\octo\ takes advantage of four main features of HPX: (i) fine grained, (ii) task based parallelism through light weight user space threads, (iii) the use of C++ futures to encapsulate both local and remote work, and (iv) an active global address space (AGAS), whereby global objects are remotely and locally accessible \citep{amini2019assessing,kaiser2014hpx}. These global objects reside in the memory on a given node but can be accessed remotely from any node. 

Through HPX futures, \octo\ is able to overlap work with communication in a straightforward and efficient manner.  For a given sub-grid, the hydrodynamic and gravity computations are  performed by an HPX thread. This thread spawns threads sending boundary data to sibling sub-grids. HPX threads \citep{kaiser2009parallex} are lightweight and \octo\ may spawn hundreds or even thousands of threads per system thread. The sub-grid thread also creates a set of futures encapsulating the boundary data it expects from its siblings.  This allows the sub-grid thread to sleep while waiting for its boundary data. When this data becomes available, HPX automatically wakes the thread, allowing computation to begin. The use of HPX futures in this manner allows \octo\ to overlap work with communication in a natural way.  AGAS allows each node of the \octo\ oct-tree to be distributed across the system in a relatively simple manner, and each oct-tree node can access its children or its siblings using the same constructs regardless of whether a particular child or sibling resides locally or on a remote processor. Another benefit of HPX is that a unified application programmer interface (API) for local and remote functionality is provided. Thus, there is some simplification for the application programmer, since there is no need to deal with two different interfaces, like combining the two hybrid parallel approaches such as MPI and OpenMP. Note that HPX utilises MPI for the communication, but provides an abstraction to the application programmer to hide the direct interaction with the MPI API.

To integrate acceleration cards, like GPUs, HPX provides two approaches: \texttt{hpx::compute} \citep{copik2017using} and \texttt{hpx::cl} \citep{8638479} to overlap GPU kernel execution with CPU work and networking. The GPU kernel execution returns a future allowing asynchronous integration of the GPU work into the overall asynchronous execution flow. \texttt{hpx::cl} provides features to integrate existing CUDA kernels and texttt{hpx::compute} automatically generates CUDA kernels from C++ code. \octo\ extends texttt{hpx::compute} to launch hand-written CUDA kernels. For more implementation details, we refer to \citep{daiss2019piz}.

\begin{figure*}
\centering     
\subfloat[Shock aligned to axes, density]{\includegraphics[width=7.5cm]{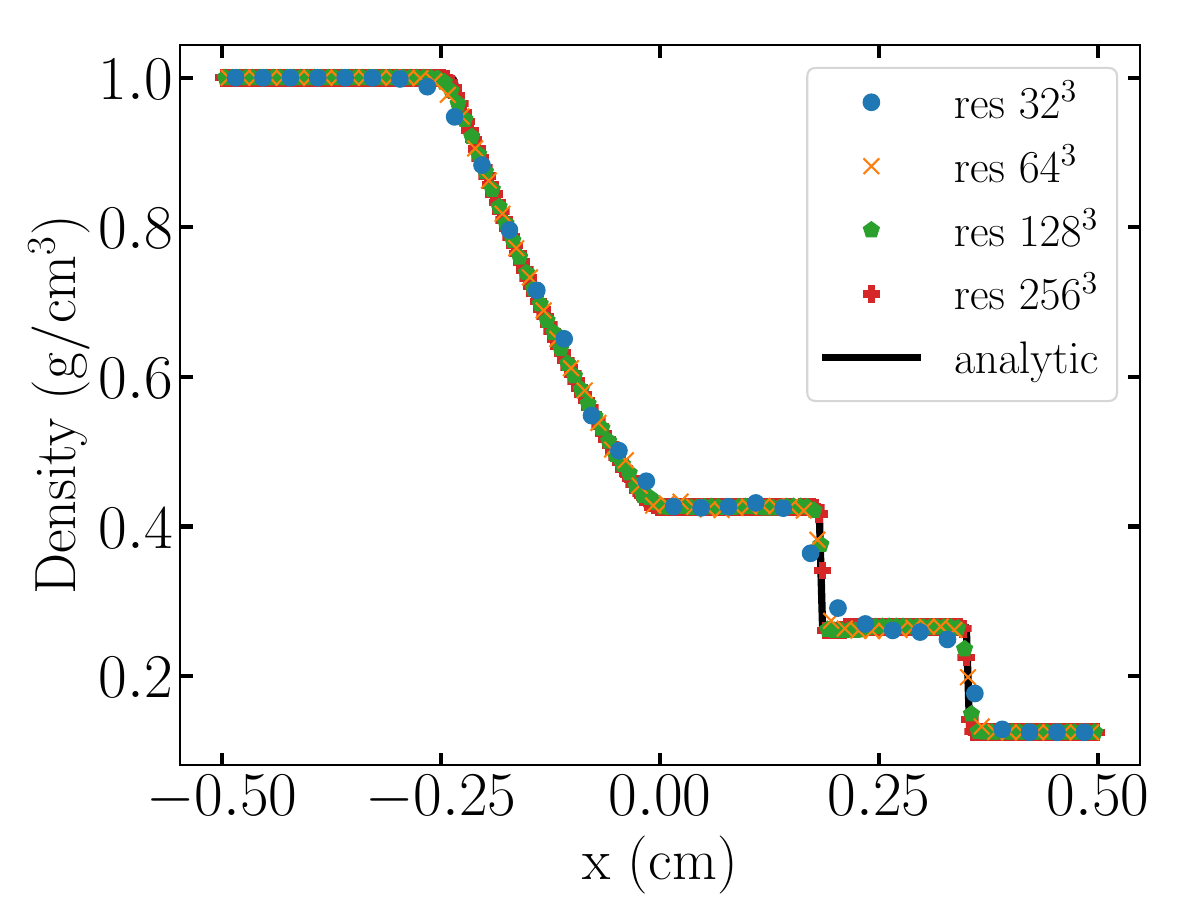}}
\subfloat[Shock aligned to axes, velocity]{\includegraphics[width=7.5cm]{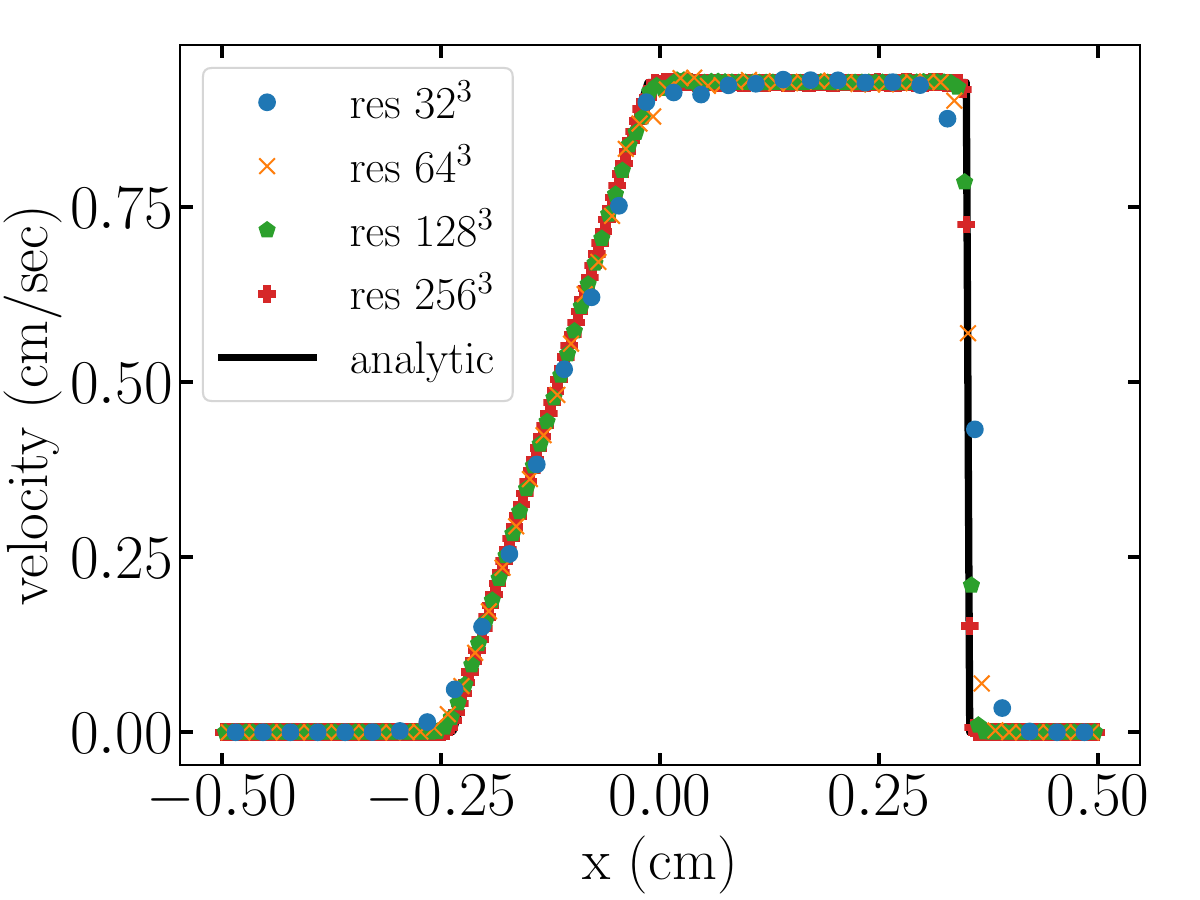}}
\qquad
\subfloat[Shock diagonal to axes, density]{\includegraphics[width=7.5cm]{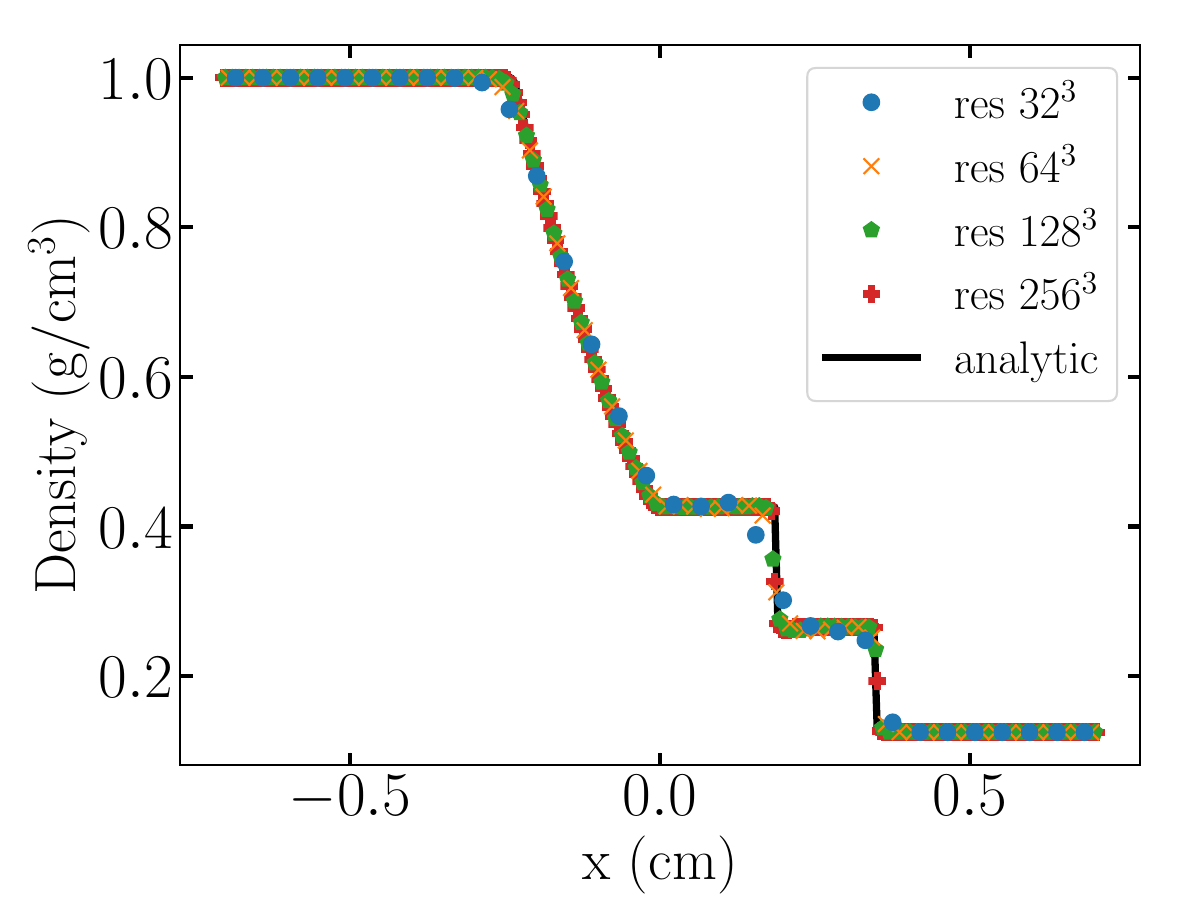}}
\subfloat[Shock diagonal to axes, velocity]{\includegraphics[width=7.5cm]{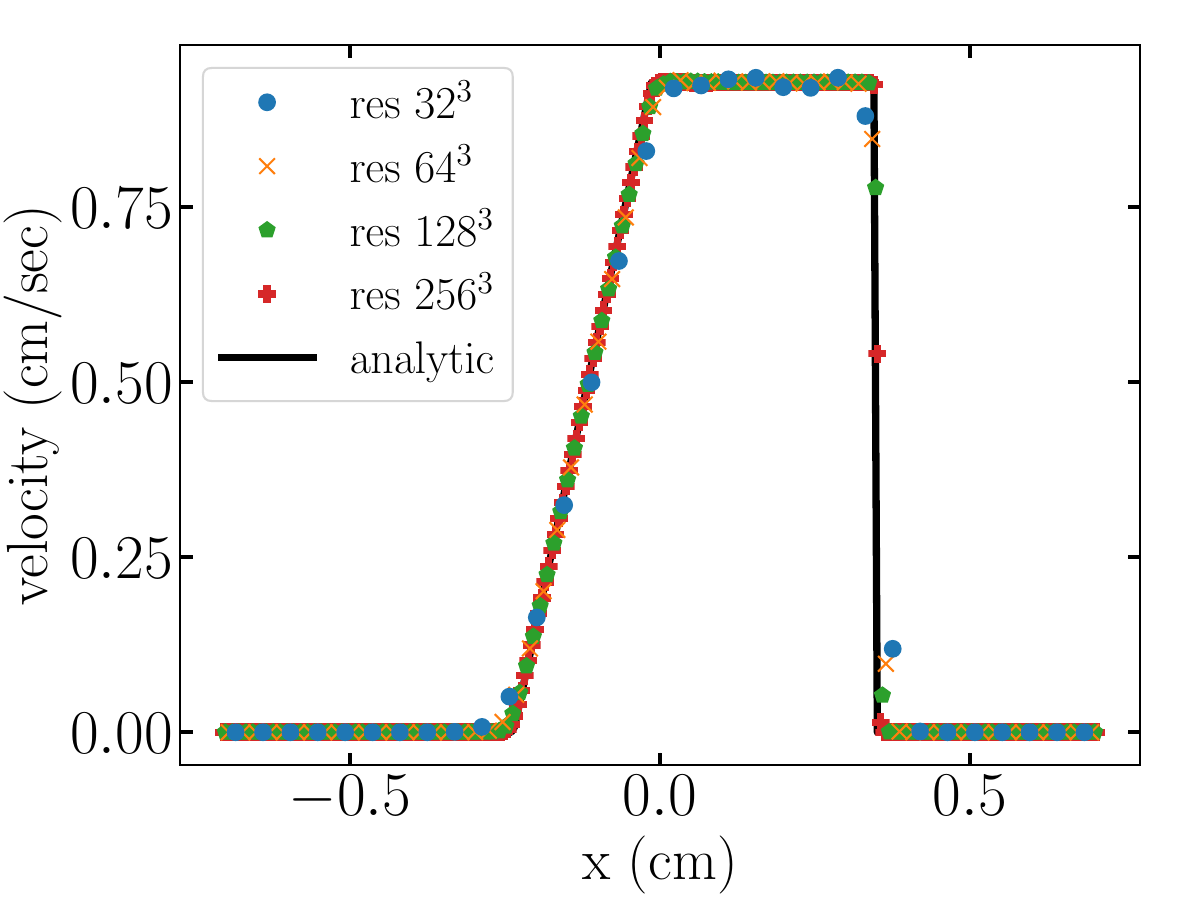}}
\caption{\protect\footnotesize{Comparison in density and velocity between the numerical results with higher resolution of \octo\ simulations and the exact solution for the shock tube at $t = 0.2$. Top row: The density and velocity are plotted in the x direction, respectively, at the end of the simulation, along a line perpendicular to the discontinuity and through the centre of the grid. Bottom row: The density and the parallel component of the velocity are plotted along the line that starts from the bottom left corner and ends at the upper right corner (in the xy plane) }}
\label{fig:sod_x_with_res}
\end{figure*}  
    
\section{Benchmarking OCTO-TIGER}
\label{sec:benchmarking_octotiger}

Below we carry out a number of simulations aimed at verifying and validating \octo. In doing so, we sometimes compare our benchmark test results with a similar test performed with \flash\ \citep{Fryxel2000} a well used hydrodynamic code often used for astronomical applications including binary interactions.

\subsection{Benchmark Design}
\label{ssec:benchmark_design}

We start testing the code by running pure hydrodynamic problems, namely, the shock tube problem (Section~\ref{ssec:shocktube}), and the blastwave problem (Section~\ref{ssec:sedov_blastwave}). The purpose is to validate and examine the performance of the hydrodynamic solver of \octo\ in isolation. This is an important task as this current version of \octo\ has an extended, more accurate (and more computationally demanding) hydrodynamic solver, which includes reconstruction of the cell averaged values not only on the cell faces and edges but also on the cell vertices (see Section~\ref{ssec:hydro_update}).

To test our gravity solver in isolation we utilize the grid with the mass distribution of a uniform density sphere and let the code compute the gravitational potential without evolving it dynamically. The sphere is surrounded by a negligible amount of gas (see Section~\ref{ssec:static_sphere}). \octo's gravity solver uses an opening angle parameter to modulate the accuracy of the gravity solution (see Section~\ref{subsec:gravity_solver}). We examine the accuracy of the computed potential with different values of this parameter as well with different resolutions.

The next test is a simulation of a polytropic structure (which could model certain types of stars) evolved over a number of dynamical timescales. We check the structure stability over several dynamical timescales, for different resolutions, values of the gravity solver opening angle and EoS (see Section~\ref{ssec:polytrope}).

Next, we test the polytrope in a wind tunnel to check how the star behaves as it moves through a hot, low density medium (see Section~\ref{ssec:polytrope_translation}). We also test a rotating polytrope and we check for diffusion of the stellar rotation profile over time (see Section~\ref{ssec:rotating_star}).

Finally, in Section~\ref{sec:binarymerger}, we simulate two polytropic stars orbiting one another in a detached configuration and then merging, with a mass ratio of 0.5. We check stability of the two structures over a number of orbits with a number of resolutions. 

We compare  some of these tests with equivalent ones using the code \flash\ \citep{Fryxel2000}. We also carry out scaling tests in Section~\ref{sec:scaling}.

\subsection{Shock Tube}
\label{ssec:shocktube}

To test the hydrodynamic solver we run the Sod shock tube problem \citep{Sod1978}.
We use the conventional initial configuration of this problem: $\rho_l=1,\;p_l=1,\;v_l =0;\;\rho_r=0.125,\;p_r=0.1,\;v_r=0$, where $l$ denotes the left side of the discontinuity, and $r$ the right side. The variables $\rho$, $p$, and $v$ are the gas mass density, pressure and velocity, respectively. The gas is taken to be an ideal gas with an adiabatic index of $\gamma = 7/5$ (this value is historical as it pertained to molecular gas typical of air). Although the problem is one-dimensional in nature, we run it in three dimensions for testing. We simulated two configurations: in the first the discontinuity plane is $x=0$, and in the second the discontinuity plane is $x+y=0$. From the problem symmetry, planes parallel to the xy plane are identical. We discuss the results of the simulations in the next two subsections.

\subsubsection{Shock front aligned along the x-axis}
\label{ssec:sod_xaxis}

In this simulation we set the initial discontinuity plane to be $x=0$. As the simulation evolves in time, a shock wave propagates to the right of the box, along the positive x-axis, while a rarefaction wave propagates at the sound speed of the unperturbed denser gas to the left, along the negative x-axis. Between them the density discontinuity moves to the right. To show the convergence of the numeric solution to the analytic one, we run four simulations in which the grid is uniform and have a growing resolution of $64^3$, $128^3$, $256^3$, and $512^3$ cells. We stop the simulation at time $t=0.2$, when the shocks fronts have not yet reached the grid boundary. 
In Figure~\ref{fig:sod_x_with_res}, top row, we plot the density and velocity in the x direction, respectively, at the end of the simulation, along a line perpendicular to the discontinuity and through the centre of the grid and compare it to the analytic solution. 

The simulations, even with low resolution, nicely fit the rarefaction wave.
Around the discontinuity, there are cells that underestimate the density behind the shock and overestimate the density in front of the shock. This discrepancy diminishes with higher resolution. We ran, in addition, two simulations in which the shock is aligned to the y-axis and the z-axis. We find the same behaviour along the direction of the shock. The velocities perpendicular to the normal direction of the plane of discontinuity vanish everywhere as they should.

Overall, the Sod problem with a shock aligned to an axis shows agreement between the \octo\ simulation and the analytical solution. As expected, the numerical solution approaches the analytic one with higher resolution.

\subsubsection{Shock front aligned diagonally}
\label{ssec:sod_diagonal}

To check the effect of a discontinuity that is not aligned along as axis we fixed the discontinuity plane at an angle of 45 degrees from the x-axis (the $x+y=0$ plane). A strong shock front propagates towards the higher $x$ and $y$ values, to the less dense upper right corner on the xy plane. A discontinuity propagates in this direction as well but with lower speed. A rarefaction wave propagates toward the bottom-left corner at the sound speed. In this configuration the wavefronts do cross the grid boundary. As the simulation evolves, an increasingly larger part of the waves encounters the grid boundary. In principle, the analytic solution is only valid in the regions where the wavefronts did not reach the boundaries, e.g., diagonally from the bottom-left corner towards the upper-right corner. 

In Figure~\ref{fig:sod_x_with_res}, bottom row, we plot the density and the parallel component of the velocity along the line that starts from the bottom left corner and ends at the upper right corner (on the xy plane) of four \octo\ simulations that differ only in their resolutions.
Across this line, we notice similar behaviour to when the shock was aligned along the x axis. The simulations reproduce accurately the rarefaction wave. Around the discontinuity, some cells underestimate the density behind the shock and overestimate the density in front of it, a discrepancy that diminishes with higher resolution.

It is, however, interesting to understand the features that appear at the opposite corners as a function of particular boundary conditions. For this test, we run two \octo\ simulations with $256^3$ cell resolution that differ only in their boundary conditions. We also compare the \octo\ simulations to an identical \flash\ (version 4.6) simulation. 
We run \flash\ with the split hydro solver that uses the PPM method. The first boundary condition termed ``outflow" in \flash\ and ``inflow" in \octo\ (see Section \ref{ssec:boundary_conditions}), means gas can inflow back to the grid. The second boundary condition, called ``diode" in \flash\ and ``outflow" in \octo\, means no inflow is allowed. 

\begin{figure*}
\centering   
    \subfloat[\octo, outflow bc]{\includegraphics[trim=110 100 20 50,clip,width=5.5cm]{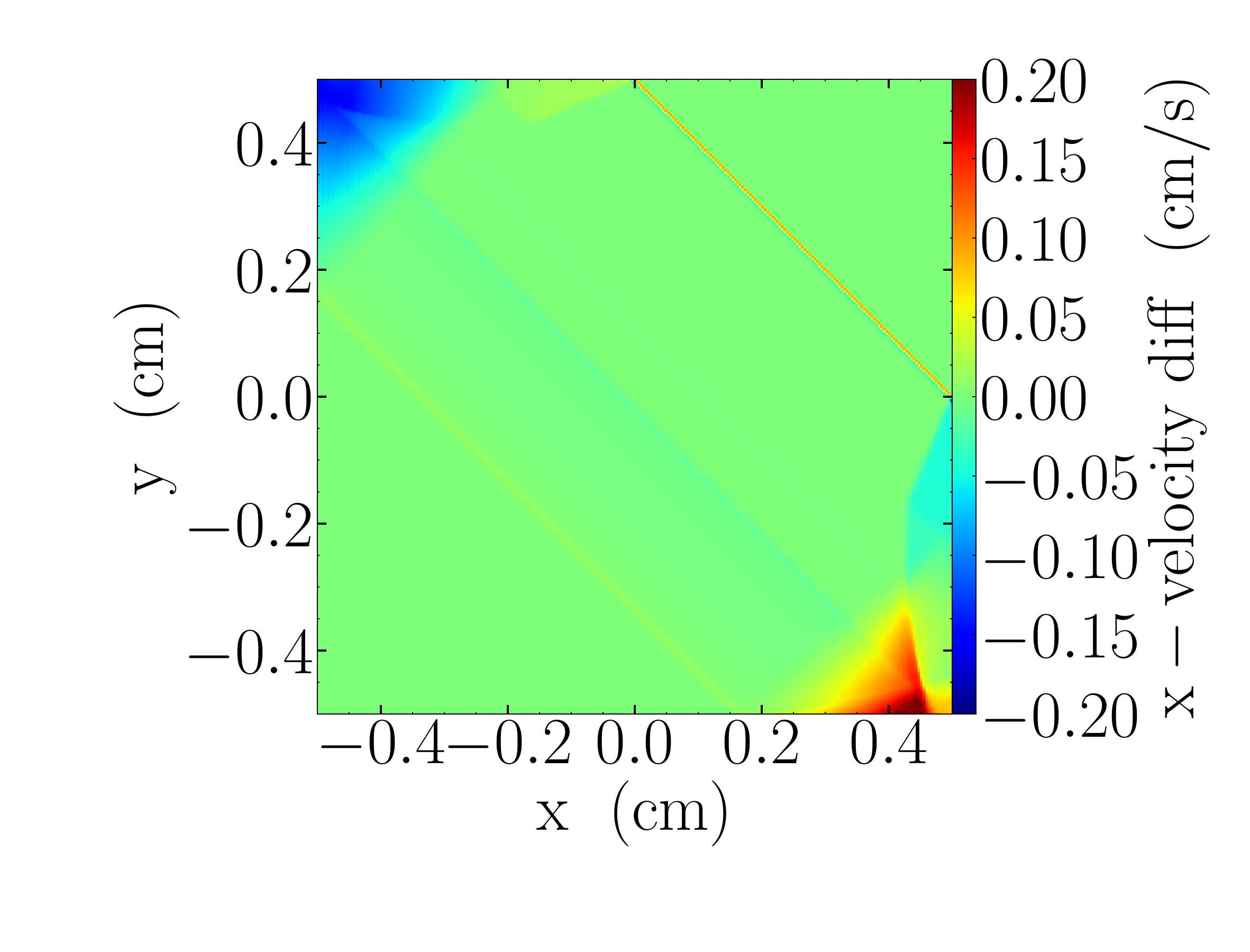}}%
    \subfloat[\flash, outflow bc]{\includegraphics[trim=110 100 20 50,clip,width=5.5cm]{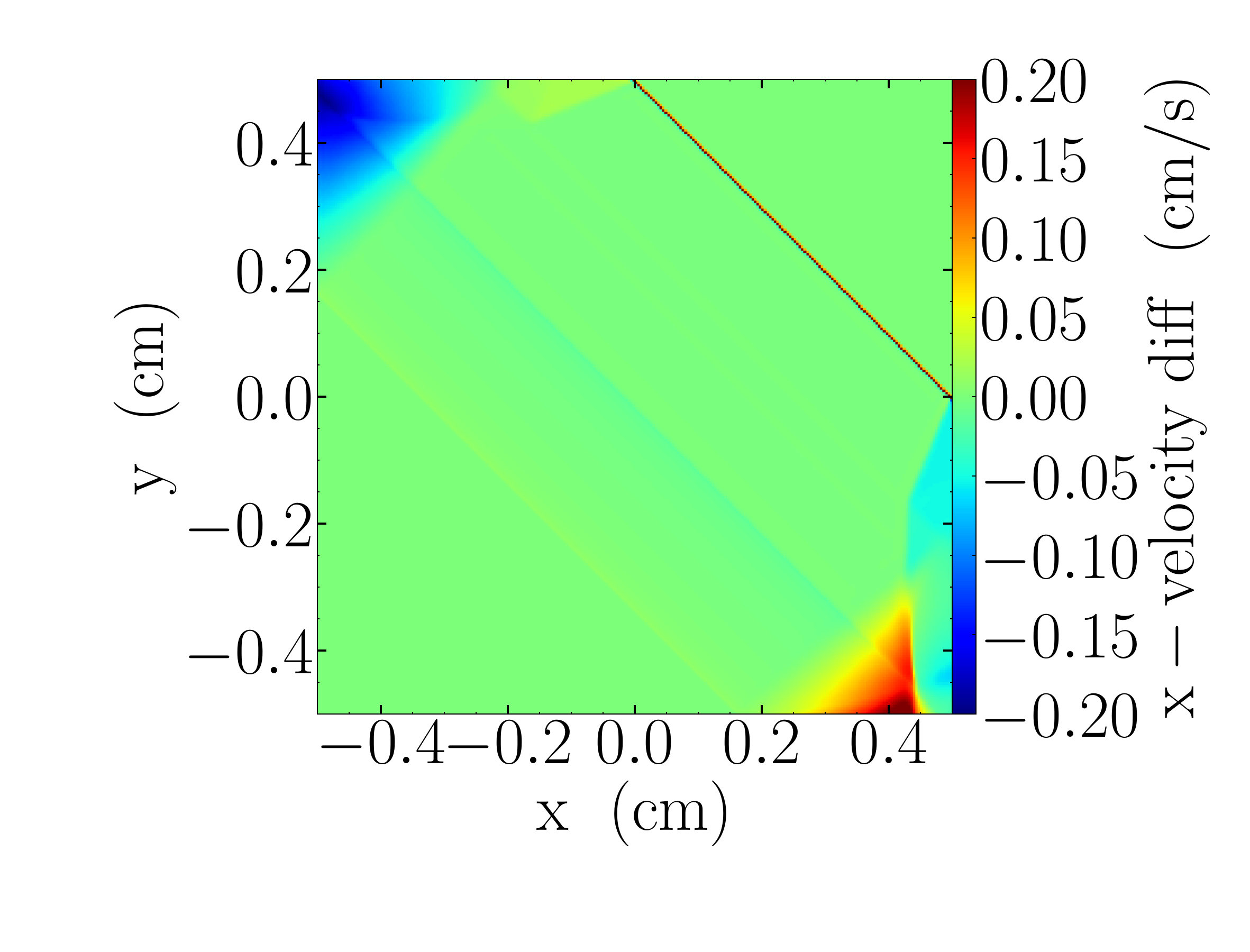} }
    \qquad
    \subfloat[\octo, diode bc]{\includegraphics[trim=110 100 20 50,clip,width=5.5cm]{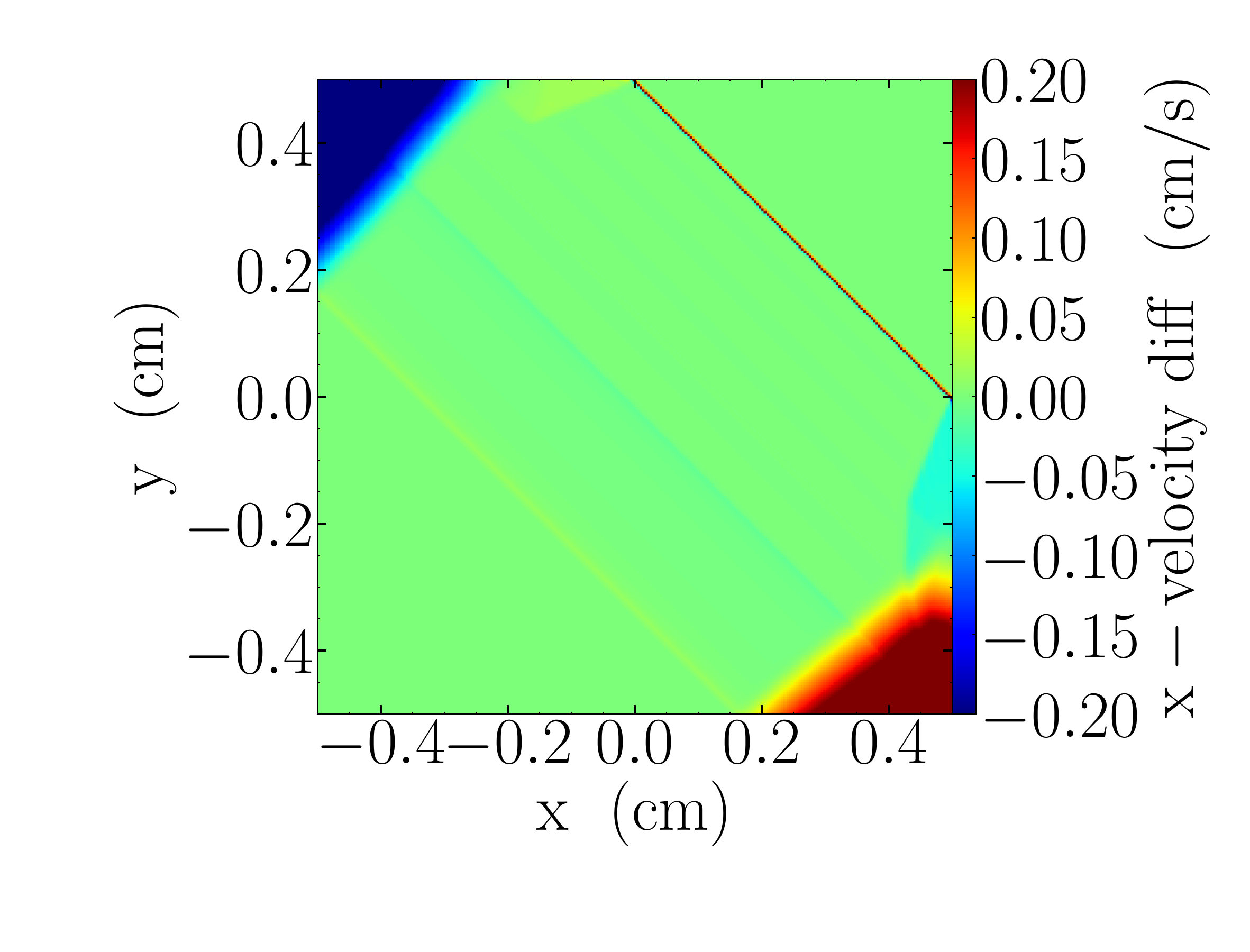} }%
    \subfloat[\flash, diode bc]{\includegraphics[trim=110 100 20 50,clip,width=5.5cm]{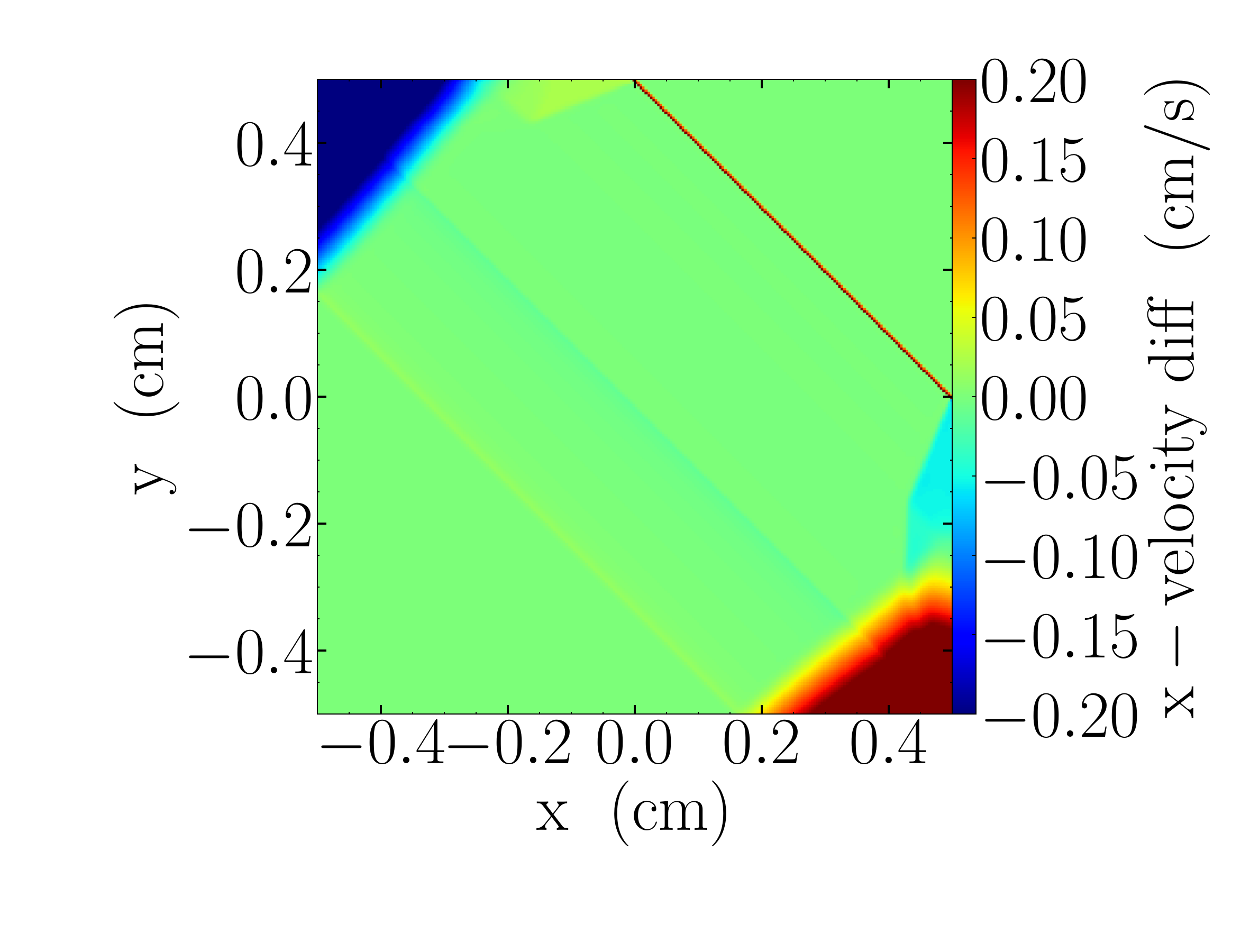} }    
\caption{\protect\footnotesize{ \octo\ vs. \flash\ for a resolution of $256^3$ and an angle of 45~deg for the shock tube. Difference in the x-velocities between simulations and analytic solution at the end of the simulation, time $t=0.2$. The boundary condition in the top row is outflow without material inflow (diode), while in the lower row it is an outflow condition that allows material to inflow back to the simulation domain (outflow) }}
\label{fig:sod_256_diff_velocity_x}
\end{figure*}

In Figure~\ref{fig:sod_256_diff_velocity_x}, we present slices along the $xy$ plane ($z=0$) of the difference in the x components of the velocity between the simulations and the analytic solution, which assumes that there is no boundary. 
Despite the excellent agreement with the analytical solutions (\revision{convergence is first order near the shock front and second order at the rarefaction wave}), both \flash\ and \octo\ simulations show similar boundary features at the upper left and bottom right corners. At those corners the density is smaller than the analytical solution due to gas that escapes from the grid. The gas in the upper left corner escapes easily through the $y=0.5$ boundary and accelerates in the negative x-direction, but it also encounters the gas to its right and decelerates in the positive x-direction. The exact opposite happens at the bottom right corner. For the same reason, the differences in the y-velocities have exactly the same mirror picture. The diode boundary, by virtue of setting to zero the momentum parallel to the boundary component, does not allow velocities in the boundary direction to evolve. The rarefaction wavefront diverts near the boundary and propagates in a perpendicular direction to the boundary. This creates a square pattern at the corners.

\subsection{Sedov-Taylor Blast Wave}
\label{ssec:sedov_blastwave}

The Sedov-Taylor blast wave test \citep{Sedov1946} 
 has an  analytic solution in three dimensions. The shock wave it produces is much stronger than that produced by the usual Sod shock tube and its spherical geometry provides a stringent test of the hydrodynamics. The initial conditions are a constant density medium with a value of 1 code unit. The internal energy density is setup so as to be described by a delta function: we approximate this in \octo, in a manner similar to \flash, namely by setting the internal energy density to $E_0 / (160 \Delta x^3$) for the $160$ computational cells satisfying $\left|\vect{x}\right| < 3.5$, and setting its value to $1\times10^{-20}$ for all other cells. We ran this model using AMR with 2, 3, 4, 5, 6, and 7 levels of refinement and the usual 8$^3$ base grid. The finest grid cell size of the model with 2 levels is $3.1\times 10^{-2}$ code units and the finest grid cell size of the model with 7 levels is $9.8 \times 10^{-4}$  code units.  

\begin{figure*}
\centering   
\subfloat[Density slice, 6 levels of refinement]{
\includegraphics[width=7.5cm]{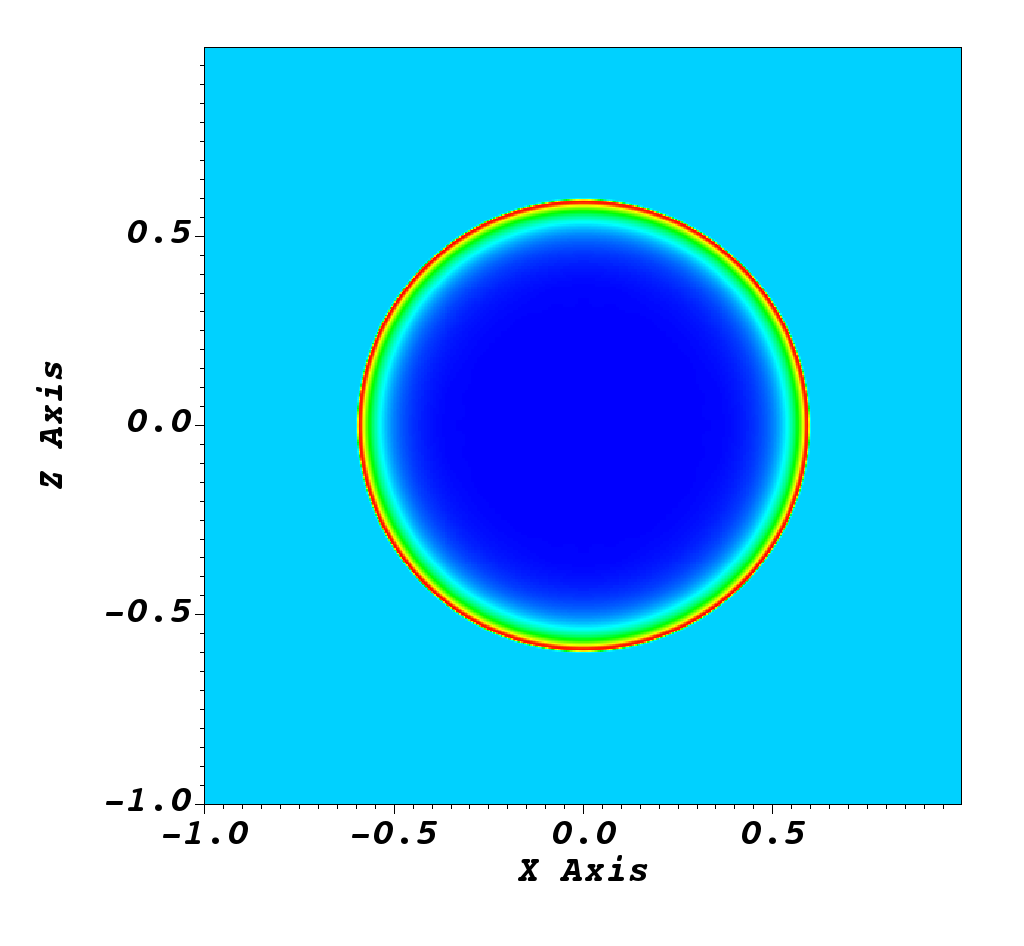}} %
\subfloat[Density slice, 7 levels of refinement]{
\includegraphics[width=7.5cm]{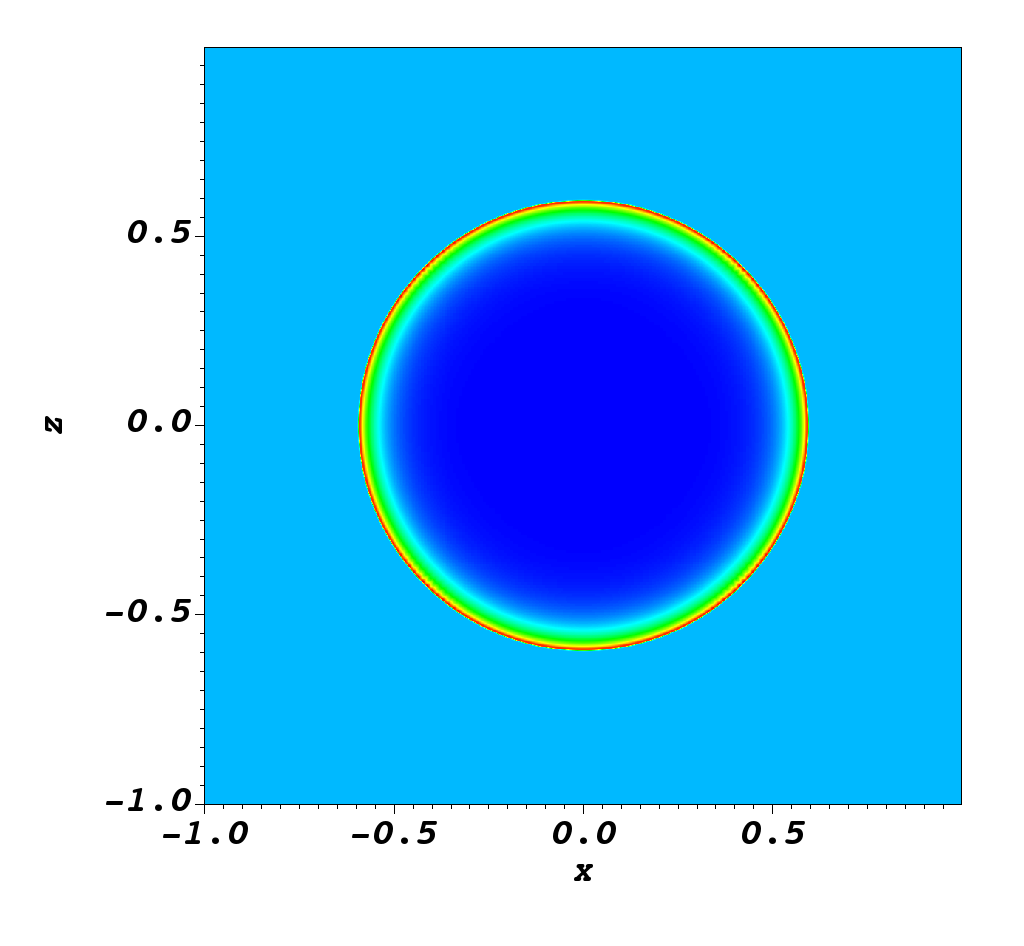} }
\qquad
\subfloat[Grid structure with 7 levels of refinement]{
\includegraphics[width=7.5cm]{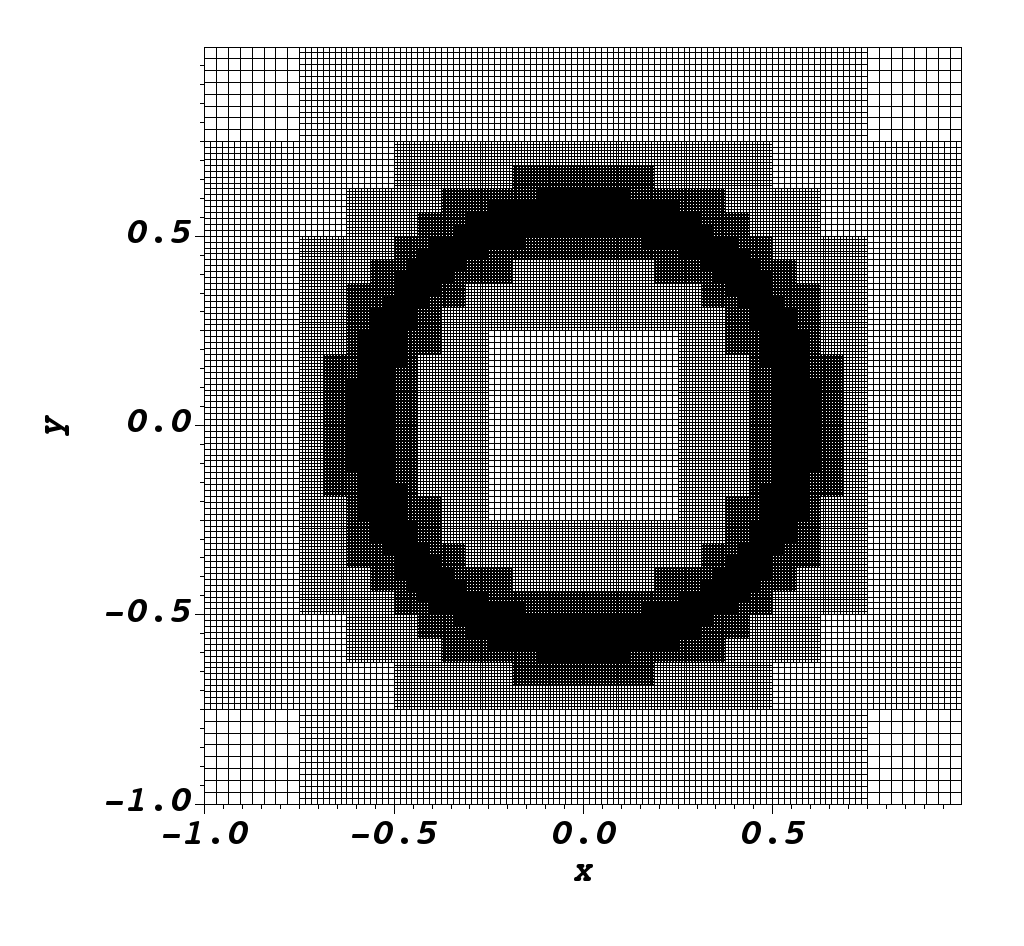}} %
\subfloat[Density residuals]{
\includegraphics[width=7.5cm]{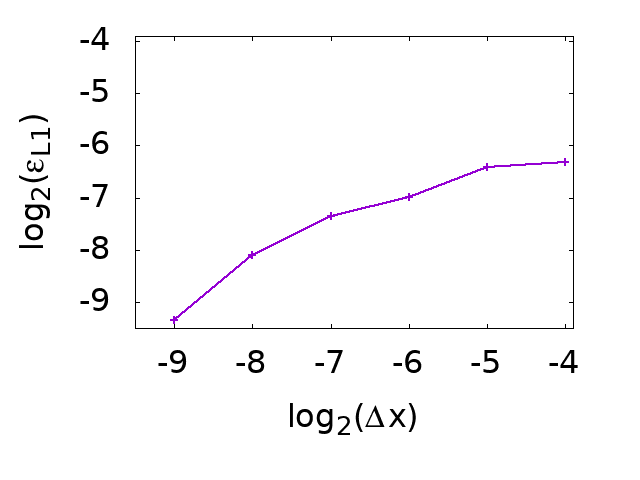}} 
\caption{\protect\footnotesize{The Sedov-Taylor blast wave test. Density slices using \octo\ 
with 6 (a) and 7 (b) levels of refinement at $t=0.25$ code units. Panel (c): the AMR grid structure of the simulation with 7 levels; Panel (d): the absolute deviation of the density from the analytic solution as a function of grid cell size \revision{for 6 simulations carried out with a maximum of 7 levels of refinement to a minimum of 2}}}
\label{fig:sedov_octo19}
\end{figure*}

Density slices at $t=0.25$ code units are shown for the models with 6 and 7 levels in Figure~\ref{fig:sedov_octo19}. We also show the AMR grid structure for the run with 7 levels. In Figure~\ref{fig:sedov_octo19}(d)  we show the difference between the computed density and the analytic solution as a function of fine grid cell size. The blast wave is a difficult problem to obtain convergence as it requires high resolution to resolve the shock front. \octo\ obtains slightly better than 1st order convergence from the run with 6 level of refinement to that with 7 levels.

\subsection{Uniform Static Sphere}
\label{ssec:static_sphere}

To test the performance of the gravity solver that solves the Poisson equation we initiate a problem with a static uniform density sphere. This problem has a simple analytic solution and is widely used to test solvers for self-gravitating fluids \citep{Motl_2002}.

\begin{figure*}
\centering   
    \subfloat[\octo, $\theta=0.5$, Res $128^3$]{\includegraphics[trim=110 100 80 40,clip,width=7.5cm]{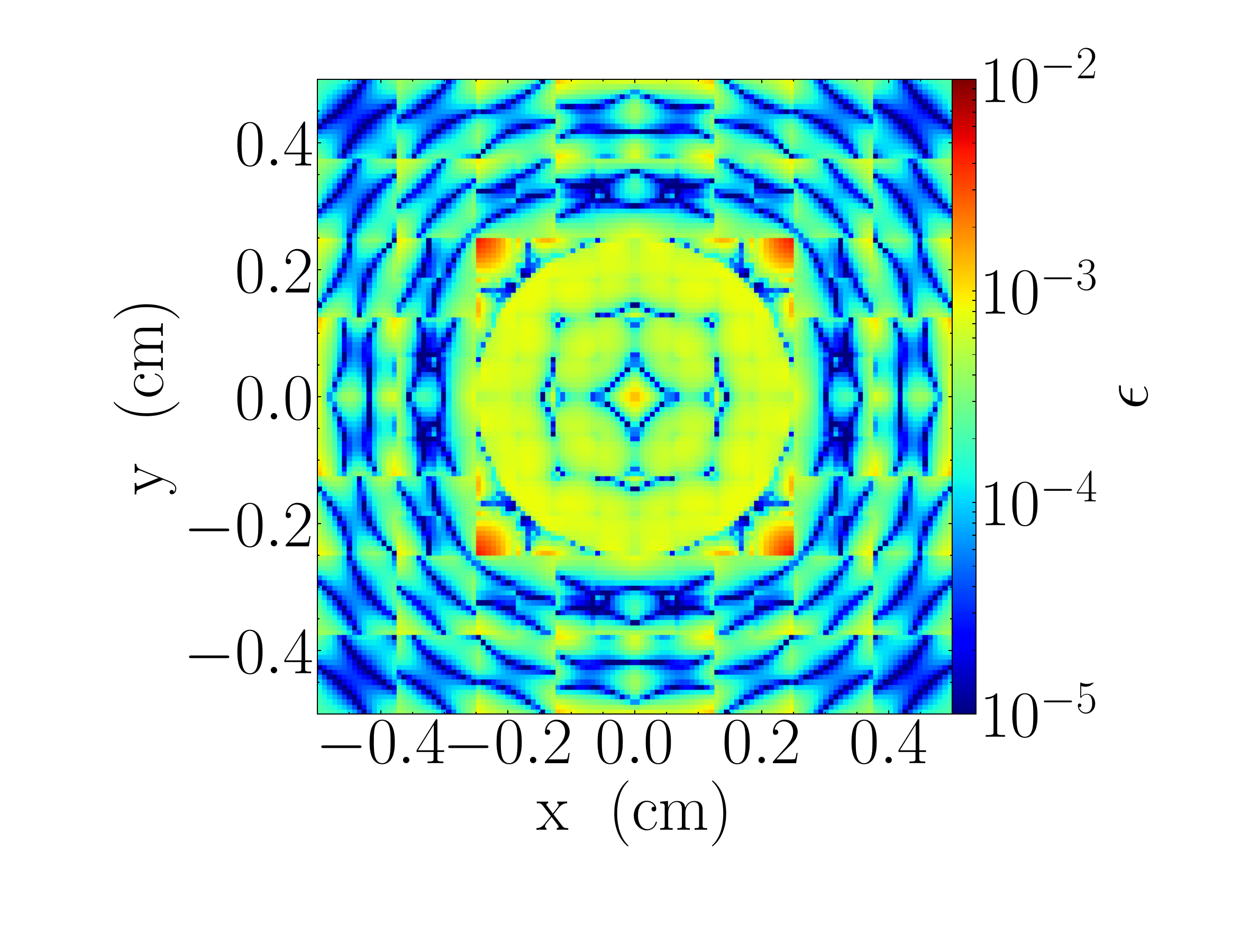} }
    \subfloat[\octo, $\theta=0.5$, Res $256^3$]{\includegraphics[trim=110 100 80 40,clip,width=7.5cm]{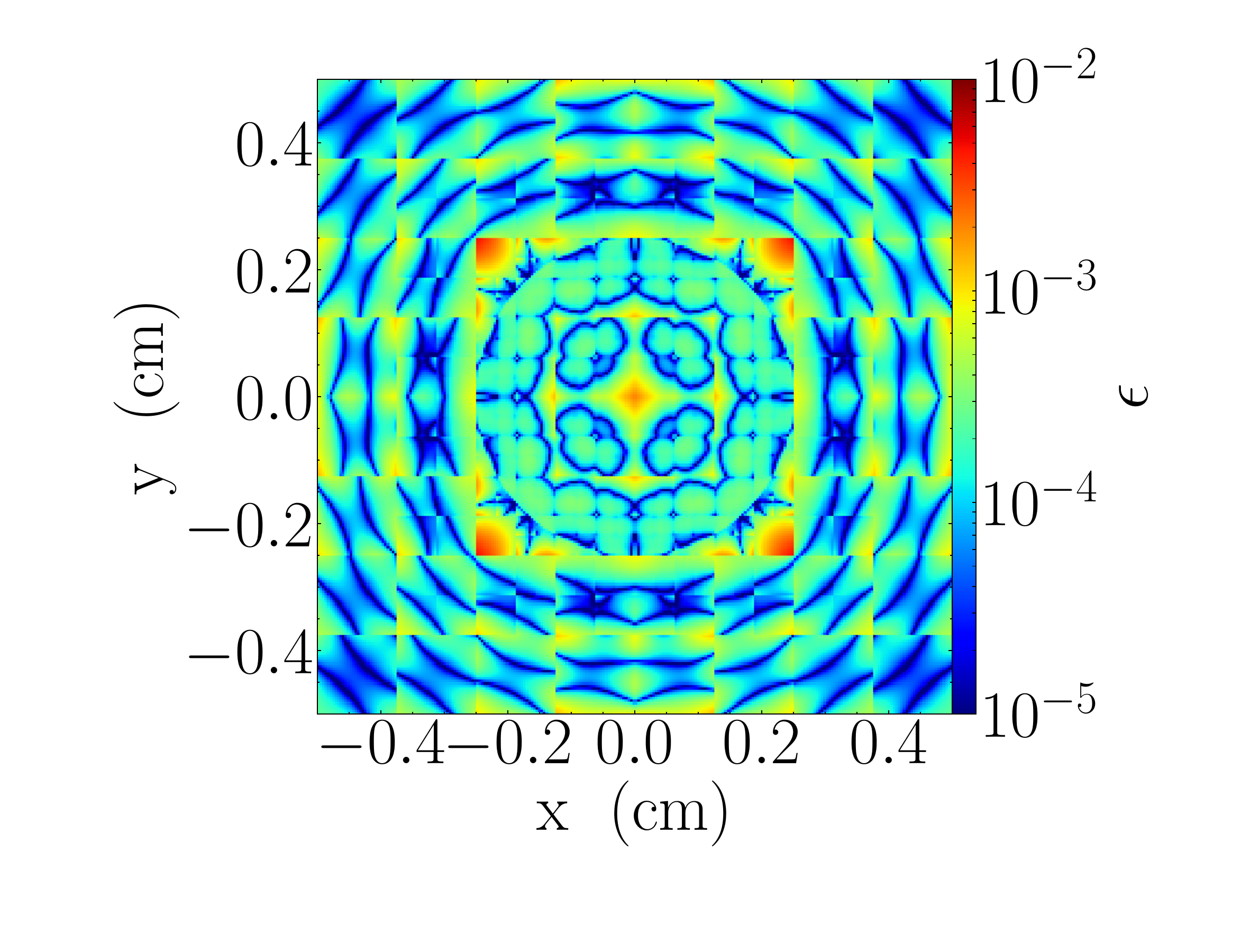} }%
    \qquad    
    \subfloat[\octo, $\theta=0.35$, Res $128^3$]{\includegraphics[trim=110 100 80 40,clip,width=7.5cm]{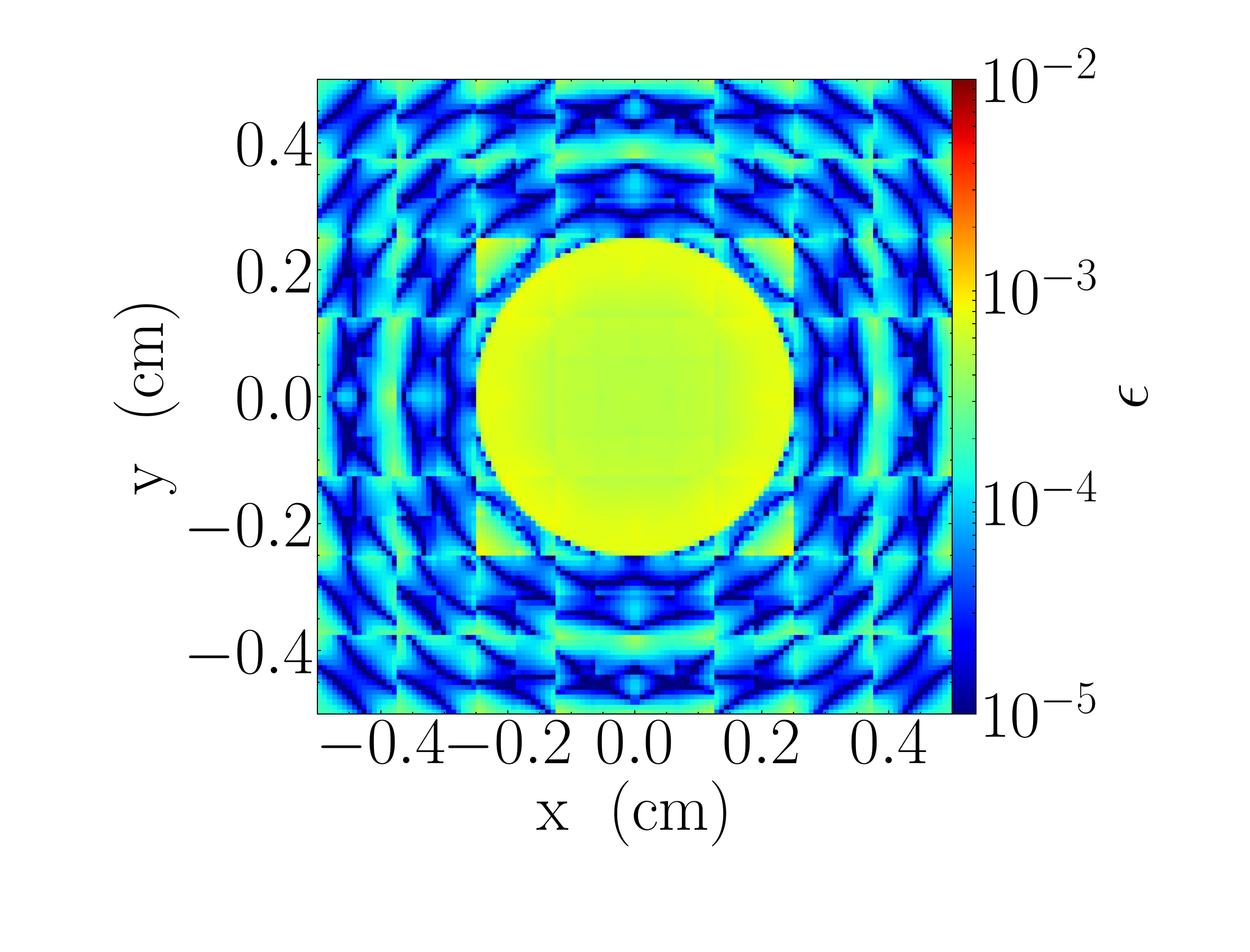}}%
    \subfloat[\octo, $\theta=0.35$, Res $256^3$]{\includegraphics[trim=110 100 80 40,clip,width=7.5cm]{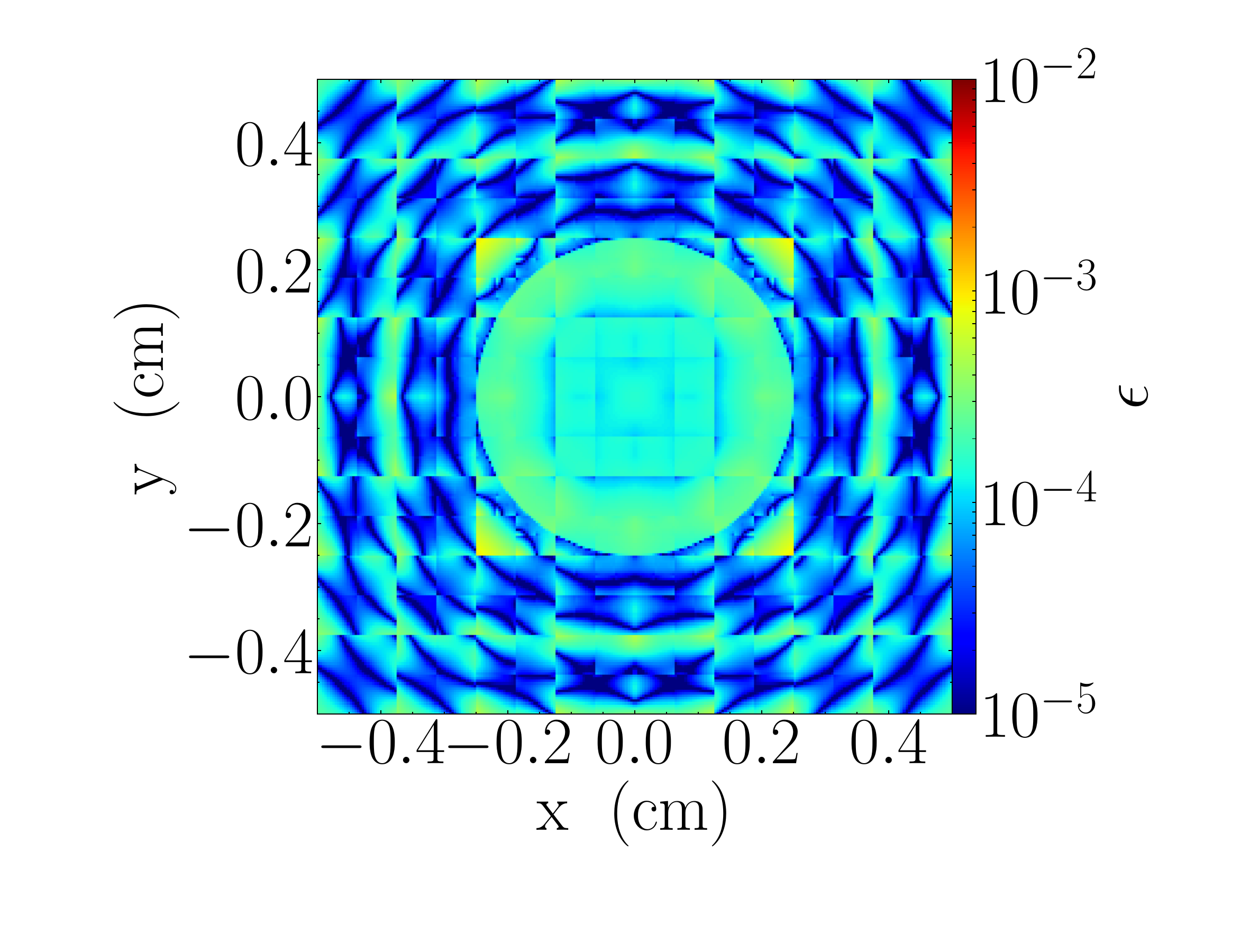}}       
    \qquad
    \subfloat[\flash, Multigrid, Res 256]{\includegraphics[trim=110 100 80 40,clip,width=7.5cm]{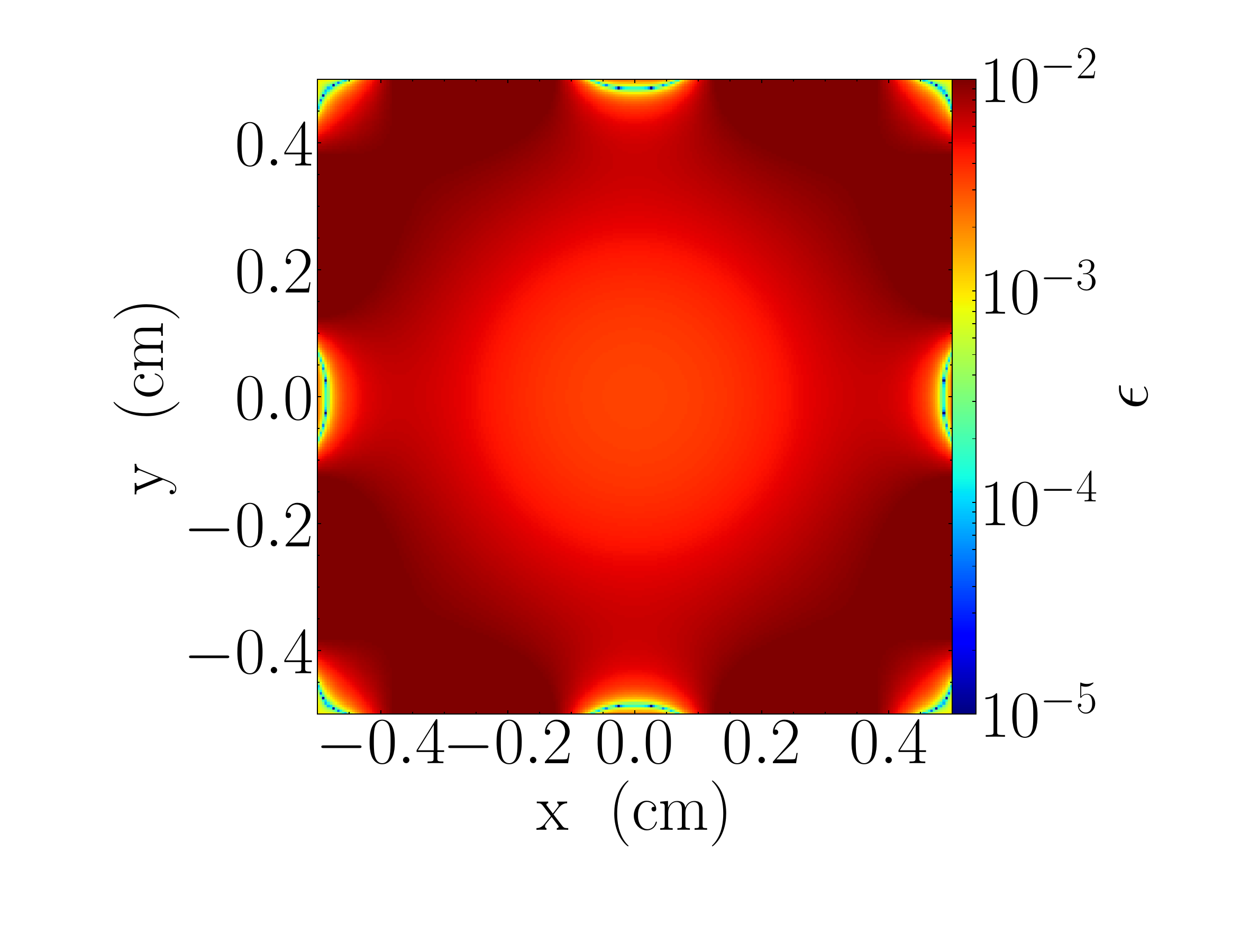} }
    \subfloat[\flash, BHTree, res 256]{\includegraphics[trim=110 100 80 40,clip,width=7.5cm]{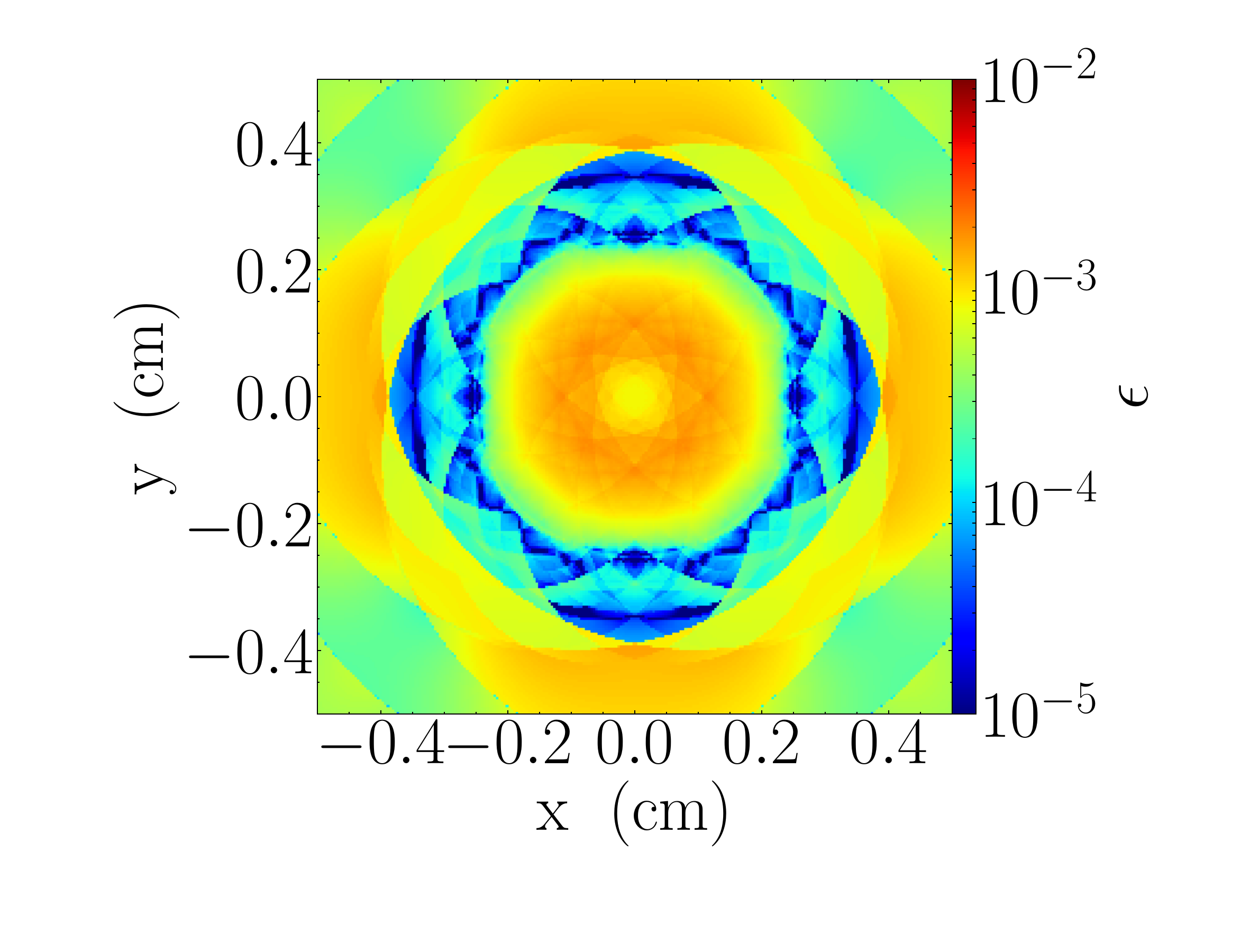} }
\caption{\protect\footnotesize{Uniform static sphere test. Maps of the residuals of the gravitational potential,  $\epsilon$, at the equatorial plane $z=0$ as a function of resolution, modelling code and opening angle $\theta$ }}
\label{fig:sphere_residual_maps_resolution}
\end{figure*}

In this test, we place a sphere with a radius of 0.25 code units at the centre of a cubic grid with a length of 1 code units. The total mass inside this sphere is 1 code unit. We fill the grid outside the sphere with a low density of $1\times10^{-10}$ code units. We used a uniform grid with four resolutions: $64^3,\;128^3,\;256^3$,  and $512^3$. To demonstrate the effect of changing $\theta$, the opening-angle parameter (see Section~\ref{subsec:gravity_solver}), we carried out two additional simulations with resolution of $128^3$, and $256^3$, where $\theta=0.35$ instead of the default value of $0.5$. 
The \octo\ simulations reproduce very accurately the analytic gravitational potential. The residuals of the potential on the equatorial plane 
\begin{equation}
    \epsilon=\abs{\frac{\phi-\phi_{\rm analytic}}{\phi_{\rm analytic}}}.
\label{eq:residual_def}
\end{equation} 

\noindent are plotted in Figure~\ref{fig:sphere_residual_maps_resolution} for four of the \octo\ simulations. A square pattern is apparent for the residuals in the \octo\ simulations with the maximum residuals appearing at the corners of a square that encloses the sphere. These larger residuals do not decrease with higher resolution, something that is expected of the FMM method.  \revision{The FMM computes the multipole expansion of the potential between grid cells at the same refinement level as each other. Adding extra refinement to those levels only increases the quality of the solution between cells at the extra refinement level. The cells at the coarser levels still interact with each other in the same way, and the same expansions are passed to  the more refined levels. The only way to cause cells to interact with each other at finer levels of refinement  is to decrease the opening criterion.}

We tested whether a smoother transition of the sphere to the background density can lower the residuals by carrying out simulations of two other density distributions (which have known analytic solutions), one where the density of this sphere decreases to the ambient value by a parabolic dependence with radius (continuous but not differentiable), and a second where the density distribution obeys an $n=1$ polytropic profile (continuous and differentiable). We find that regardless of the smoothness of the density decrease, high residuals values remain at the corners. To decrease the high residuals at the corners, a lower value of $\theta$ needs to be set. We also tried to simulate an off-centre sphere and a range of additional sphere sizes. Changing the sphere location does not affect the residuals. However, increasing the sphere radius does result in lower maximum residual values. We conclude that the maximum values are related to how much mass is concentrated in a given volume. 

For comparison, in  Figure~\ref{fig:sphere_residual_maps_resolution}, we present equatorial maps of several $128^3$ and $256^3$ cell \flash\ simulations that used two gravity solvers: multigrid, panel (e) and BHTree, panel (f). Here we see that each solver has a different residual shape. 

\begin{figure*}
\centering   
    \subfloat[\octo, histogram]{\includegraphics[trim=0 5 0 14,clip,height=5.35cm]{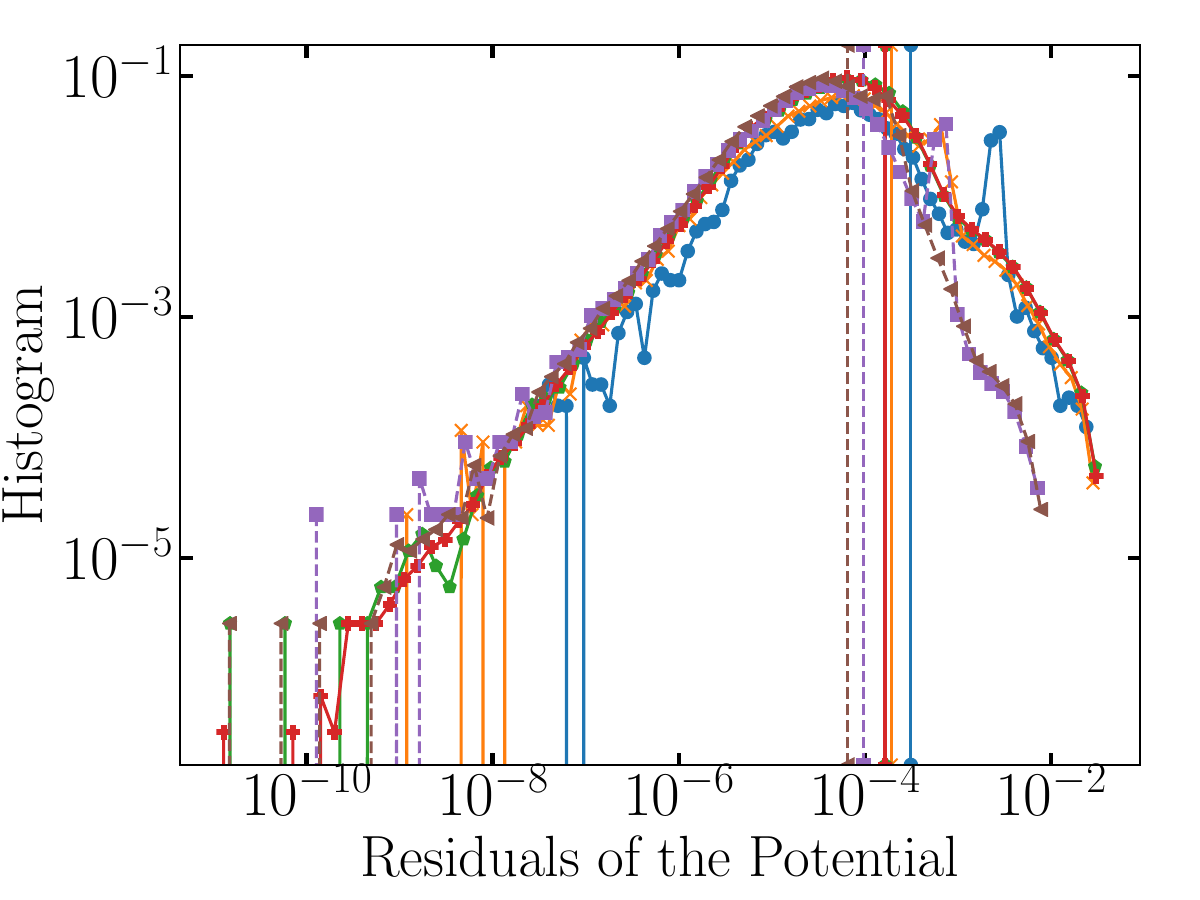}}
    \subfloat[\octo, mass distribution]{\includegraphics[height=5.35cm]{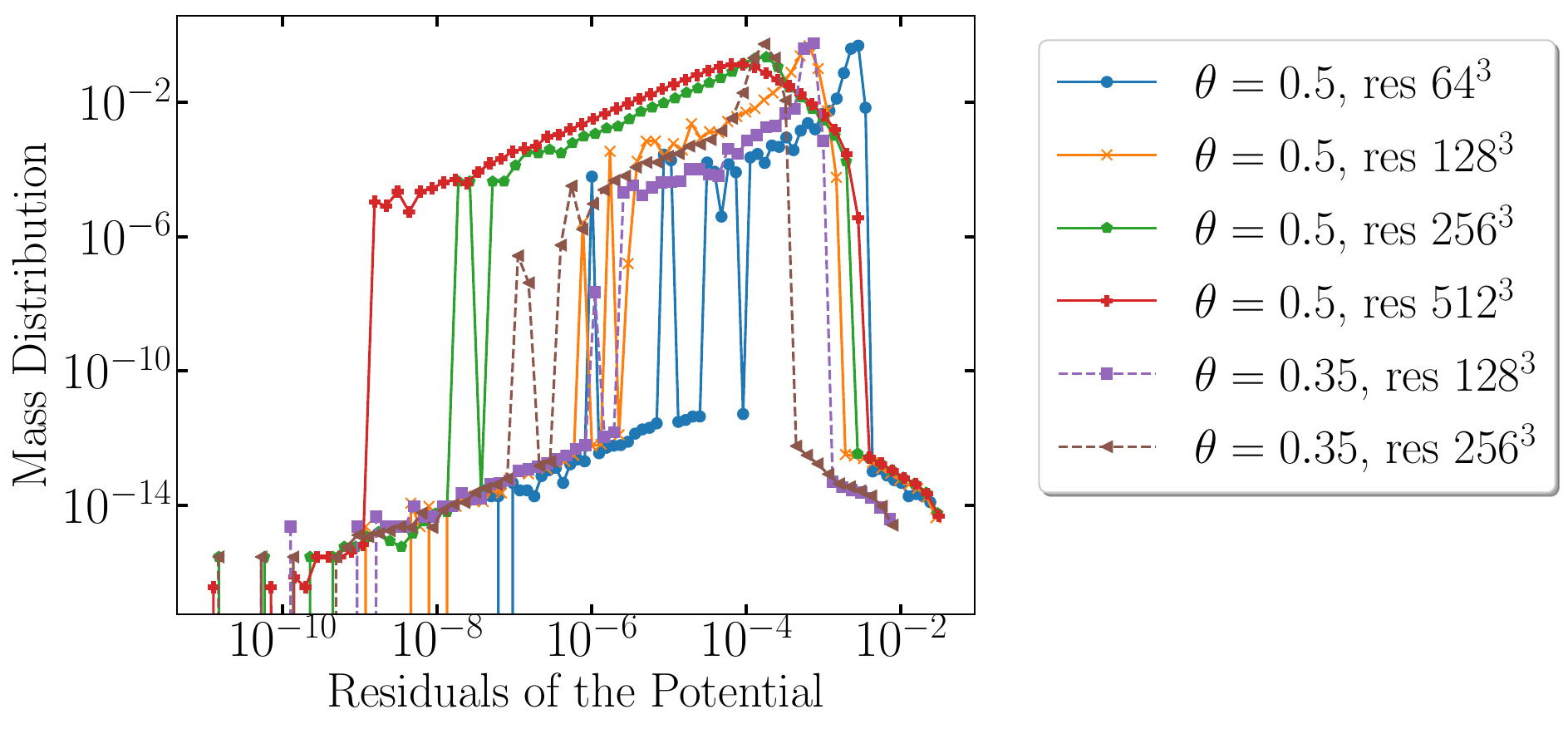}}
    \qquad
    \subfloat[\flash, histogram]{\includegraphics[trim=0 5 0 14,clip,height=5.35cm]{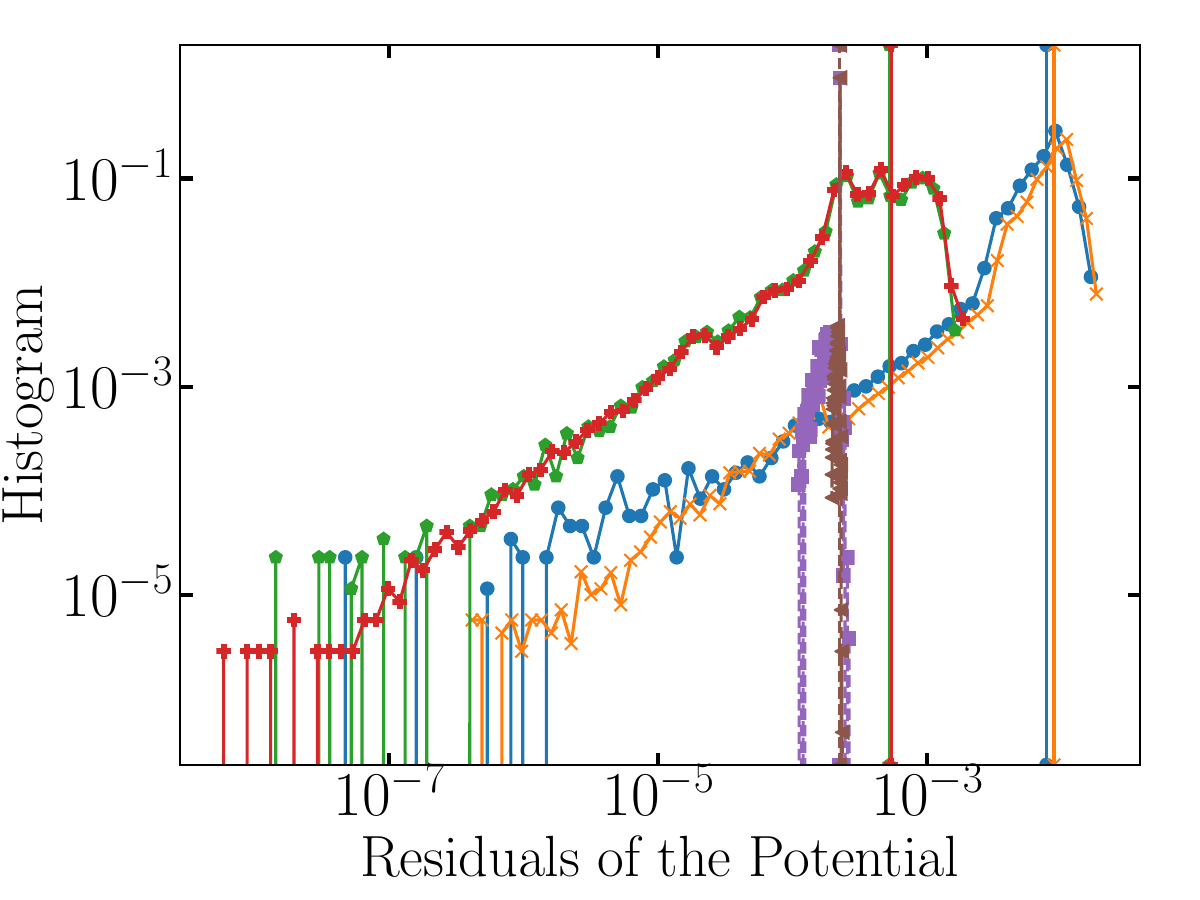}}
    \subfloat[\flash, mass distribution]{\includegraphics[height=5.35cm]{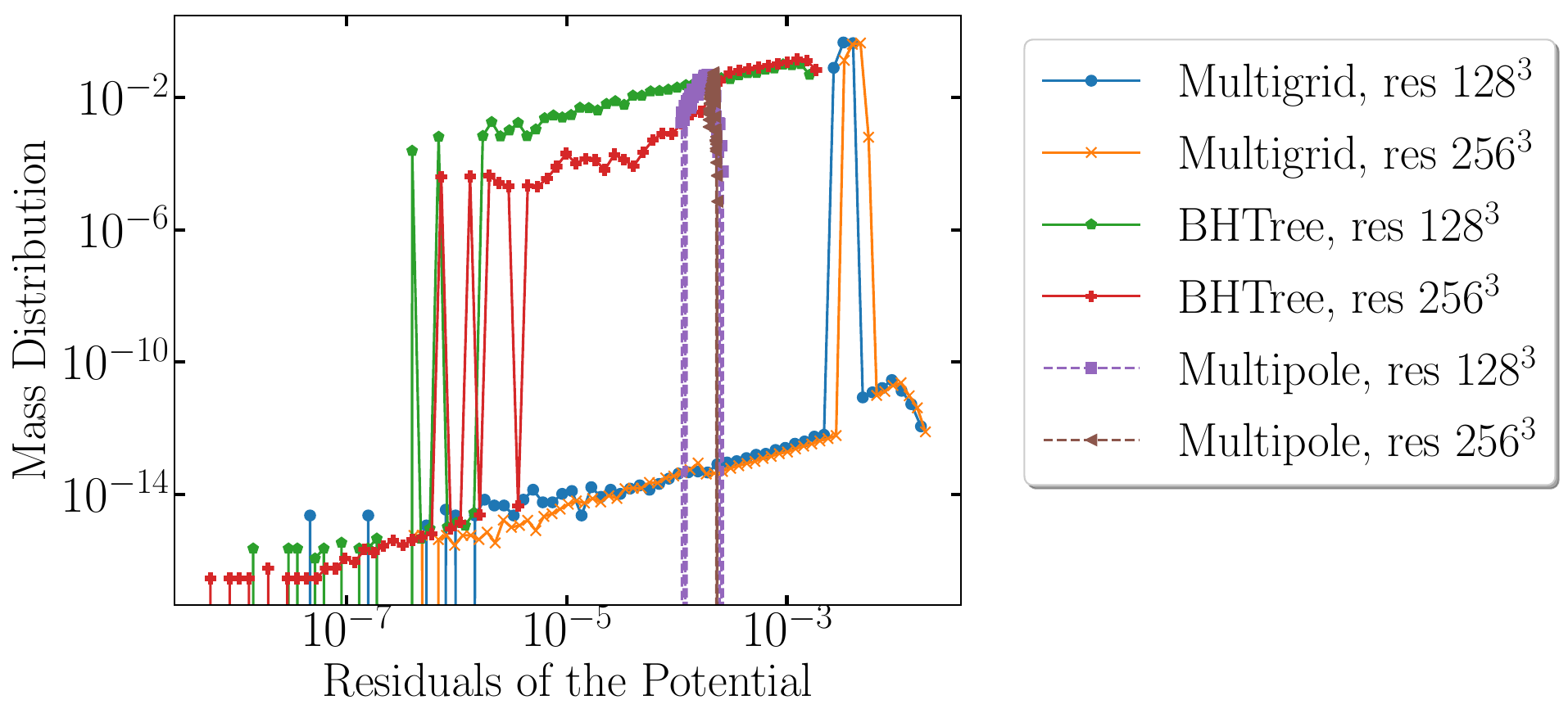}}
\caption{\protect\footnotesize{Uniform static sphere test. Histograms and mass distributions of the potential's 
residuals of \octo\ and \flash\ simulations. The vertical lines in panels (a) and (c) mark the residuals' mean value for each simulation}}
\label{fig:sphere_residual_histograms}
\end{figure*}

In Figure~\ref{fig:sphere_residual_histograms}, we plot histograms of the residuals by volume (panels (a) and (c)) and by mass (panels (b) and (d)). Figure~\ref{fig:sphere_residual_histograms}(a) shows that the residuals' maxima do not decrease by increasing the resolution; however, the residuals' mean and minimum values do. By increasing the resolution, the distribution is stretched to lower residual values. The notable second peak of the histogram for the lower resolution simulations practically disappears for the two higher resolutions. The most common residual value is consistently around $10^{-4}$ irrespective of resolution. 

\begin{table*}
    \centering
    \begin{tabular}{lcccccc}
    \hline
    $N_{\rm cells}$ &  $64^3$    & $128^3$   & $256^3$ & $512^3$   & $128^3$   & $256^3$ \\
    $\theta$ &  0.5   & 0.5  & 0.5 & 0.5   & 0.35   & 0.35 \\
    \hline
    max   & $2.66\times 10^{-2}$ & $3.23\times 10^{-2}$ & $3.49\times 10^{-2}$ & $3.61\times 10^{-2}$ & $0.82\times 10^{-2}$ & $0.90\times 10^{-2}$ \\
    mean  & $3.13\times 10^{-4}$ & $1.93\times 10^{-4}$ & $1.67\times 10^{-4}$ & $1.65\times 10^{-4}$ & $9.6\times 10^{-5}$ & $6.5\times 10^{-5}$  \\
    min   & $3\times10^{-8}$ & $1\times10^{-9}$ & $1\times10^{-11}$ & $1\times10^{-11}$ & $1\times10^{-10}$ & $1\times10^{-11}$ \\
    \hline
    \end{tabular}
    \caption{\revision{Gravitational potential residuals of the static uniform sphere in \octo. \octo\ uses the fast mutipole method to solve for the gravity, which has an opening angle parameter $\theta$. Full distributions appear in Figure \ref{fig:sphere_residual_histograms}}}
    \label{tab:sphere_residuals_octo}
\end{table*}

\begin{table*}
    \centering
    \begin{tabular}{lcccccccccccc}
    \hline
    Solver     & BHTree & BHTree & Multigrid & Multigrid & Multipole & Multipole \\
    $N_{\rm cells}$ & $128^3$   & $256^3$ & $128^3$   & $256^3$ & $128^3$   & $256^3$ \\
    \hline
    max   & $0.18\times 10^{-2}$&$0.20\times 10^{-2}$&$1.81\times 10^{-2}$&$1.96\times 10^{-2}$&$2.6\times 10^{-4}$&$2.3\times 10^{-4}$\\
    mean  &$5.21\times 10^{-4}$& $5.38\times 10^{-4}$&$7.691\times 10^{-3}$&$8.759\times 10^{-3}$&$2.23\times 10^{-4}$&$2.24\times 10^{-4}$ \\
    min   & $1 \times 10^{-8}$ & $5 \times 10^{-9}$ & $4 \times 10^{-8}$ & $4 \times 10^{-7}$ & $1.1 \times 10^{-4}$ & $1.9 \times 10^{-4}$  \\
    \hline
    \end{tabular}
    \caption{\revision{Gravitational potential residuals of the static uniform sphere in \flash. Full distributions appear in Figure \ref{fig:sphere_residual_histograms}}}
    \label{tab:sphere_residuals_flash}
\end{table*}



In Figure~\ref{fig:sphere_residual_histograms}(c) and (d), we present the residuals' histograms and mass distributions of the \flash\ simulations, where we used three commonly-used \flash\ gravity solvers: Multigrid, BHTree, and Multipole.  
The \flash\ Multipole gravity solver has the lowest maximum residuals. The $128^3$ simulation that used the multipole solver has a lower mean value, $\simeq 2.23\times 10^{-4}$, than every one of the other \flash\ simulations (including the high resolution simulations). Only the \octo\ simulations have lower mean residuals: $1.93\times 10^{-4}$, $1.67\times 10^{-4}$ and $1.65\times 10^{-4}$ for the $128^3$, $256^3$ and $512^3$ simulations, respectively. The \flash\ multipole solver easily produces an accurate solution to this problem because it has a spherical symmetry. Including additional terms, besides the monopole, does not contribute to the numeric solution in the \flash\ multipole solver. The maximum and mean values of the residuals slightly increase in the high resolution \flash\ simulations. The \octo\ simulations show overall low mean residuals and low most common residuals (the peak in the distributions both with respect to volume and mass). 

\revision{ We summarize the residuals' minimum, maximum, and mean values for the \octo\ and \flash\ simulations in Tables \ref{tab:sphere_residuals_octo} and \ref{tab:sphere_residuals_flash}, respectively. Table \ref{tab:sphere_residuals_octo} shows that in \octo, a higher resolution decreases the mean residual value, up to a certain resolution, where the mean residual value converges to some low value. To decrease the mean residual further, one must set a lower $\theta$ value.}

From the mass distributions (Figure~\ref{fig:sphere_residual_histograms}(b)), we see that the cells which have the highest residual values actually contain a negligible amount of mass. Namely, the cube corners which have high residuals reside outside the uniform density sphere, where the density is very low. In addition, increasing the resolution results in having most of the mass in lower residuals. The residual value, that has the maximum mass, shifts from  $3\times10^{-3}$ in the $64^3$ simulation to $10^{-4}$ in the $512^3$ simulation, and the mass distribution becomes flatter towards smaller residuals with higher resolution. We infer that these regions with high values of the potential residuals will have minimum effect on the evolution of a simulation, once we evolve it, because of their low total mass.

As can be seen from Figures~\ref{fig:sphere_residual_maps_resolution} and \ref{fig:sphere_residual_histograms}, by decreasing the opening-angle parameter, $\theta$, we can increase the level of accuracy in the \octo\ simulations even more. The residuals altogether are reduced, including the maximum residual values at the corners. With such low value of $\theta$, we get the lowest mean residual values among all the simulations even at the lower resolution of the $128^3$ cells. As with the simulations with $\theta=0.5$, improving the resolution does not reduce the residual maximum, but does improve the minimum and mean values. By increasing the resolution, the main peak is shifted to lower residual values of around $4 \times 10^{-5}$. The mean residual value in the high resolution simulation indicates a deviation of 0.0065 percent from the analytical value. We note, though, that the peak in the mass distribution of the residuals remains approximately the same for the $\theta=0.35$ simulations and for the $\theta=0.5$ simulations (compare the peaks of the purple squares line and the yellow crosses line or the peaks of the green circles line with the brown triangles line in  Figure~\ref{fig:sphere_residual_histograms}(b)). Reducing $\theta$ has a higher computational cost (see Section~\ref{sec:scaling}). This compromise between the accuracy of the simulation and the running time should be taken into account when simulating a problem.

\subsection{Stationary Star}
\label{ssec:polytrope}

In this test, we set up a polytrope with an index of $n=3/2$. The stellar diameter is the same as we have used in the uniform static sphere test, equal to half of the domain size, and the polytrope's centre coincides with the centre of the domain. A polytrope can be scaled to model different types of stars. We show two such scalings in Table~\ref{tab:poly_models}, one for a low-mass, fully convective, main-sequence star and a second one for a low-mass white dwarf. Outside the star we fill the domain with gas with a density that is 10 orders of magnitudes smaller than the star's central density. While the star itself is in good hydrostatic equilibrium, the outside medium is not in equilibrium and it starts free falling onto the surface of the star as the simulation starts. The falling of the gas together with the diode boundary condition, which prevents inflow, create outer regions with very low densities. To avoid the creation of a vacuum, whenever cell densities become lower than the threshold floor density of $10^{-15}$ times the central density, we set the density to be that value. In addition, to reduce shocks due to supersonic in-falling gas we give the ambient medium a high internal energy (and hence temperature). We list the ambient medium's properties in Table~\ref{tab:poly_models}.
\begin{table*}
\begin{center}
\begin{adjustbox}{max width=\textwidth}
\begin{tabular}{lcccccccccccc}
\hline
model & $M$                  & $R$                   & $\rho_c$             & $p_c$                              & $T_{c}$ & $c_{s,c}$     & $T_{s}$     & $c_{s,s}$ & $T_{\mathrm{medium}}$ & $c_{s,\mathrm{medium}}$ & $T_{\rm f}$ & $K$ \\
 &(M$_{\rm \odot})$& &$({\rm g~cm^{-3}})$&$({\rm dyne/cm^2})$& (K) &$({\rm km/s})$& (K) &$({\rm km/s})$ & (K) &$({\rm km/s})$& (sec/min)& (erg~cm$^2$/g$^{5/3}$) \\
\hline
MS    & $0.27$               & $0.25 R_{\rm \odot}$  & $150$                & $1.7\times10^{17}$     &  $1.8 \times 10^7$       & $430$ & $6300$ &$9.3$&  $1.9\times10^8$                     &  $2300 $     & 24 min  &  $4\times10^{13}$        \\
WD    & $0.35 $              & $10^9$ cm             & $10^6$               & $2.5\times10^{22}$       & $6.1\times 10^8$        & $2000$ & $7.3\times10^4$  &$44$&                     $4.3\times10^9$  &  $11\,000 $   & 18 sec  & $2.5\times 10^{12}$          \\
\hline
\multicolumn{13}{l}{Legend: MS = main sequence; WD = white dwarf; $M$ = mass; $R$ = radius; $\rho_c$ = central density; $p_c$ = central pressure; $T_c$ = central temperature; $c_{s,c}$ = central sound speed; } \\
\multicolumn{13}{l}{$T_s$ = surface temperature;  $c_{\rm s,s}$ = surface sound speed; $T_{\rm medium}$ = medium temperature; $c_{\rm s,medium}$ = medium sound speed; $T_{\rm f}$ = fundamental period; $K$ = polytropic constant.}\\
\end{tabular}

\end{adjustbox}
\end{center}
 \begin{quote}
  \caption{\protect\footnotesize{Examples of how our polytope model of Section~\ref{ssec:polytrope} to \ref{ssec:rotating_star} can be scaled to represent stars of different types. We assume $\mu_{\rm He^{+2}}=4/3$, and $\mu_{\rm CO}=2$, $\mu_{\rm H}=1$, $\mu_{\rm H^{+}}=1/2$, and for calculating the temperature for the main sequence star center, WD centre, main sequence star surface, and WD surface, respectively. To calculate the medium's temperature we assume $\mu_{\rm H^{+}}=1/2$}. Note that $T_f$ is the fundamental period, while $T_s$, $T_c$ and $T_{\rm medium}$ are temperatures}
\label{tab:poly_models}
\end{quote}
\end{table*}

Any small perturbation from hydrostatic equilibrium will induce an oscillation with polytrope eigenfrequencies and modes. \citet{Hurley1966} computed numerically the fundamental modes, as well as the first and second harmonic modes of several polytropes with different adiabatic indices. They found that a pulsating $n=3/2$ polytrope has a fundamental frequency of
\begin{equation}
\omega_F^2(n=3/2;\;\gamma_{\rm ad}=5/3)=0.3764 \frac{8 \pi G \rho_c}{5}.
\label{eq:fundamental_frequency}
\end{equation}
The pulsation periods are of the order of the free-falling time of the star. For the low mass main sequence star model, for example, the fundamental period, $T_F$, is 24 minutes, while for the WD model the fundamental period is 18 seconds.

\begin{figure*}
\centering   
    \subfloat[AMR grid]{\includegraphics[trim=110 100 80 40,clip,width=7.5cm]{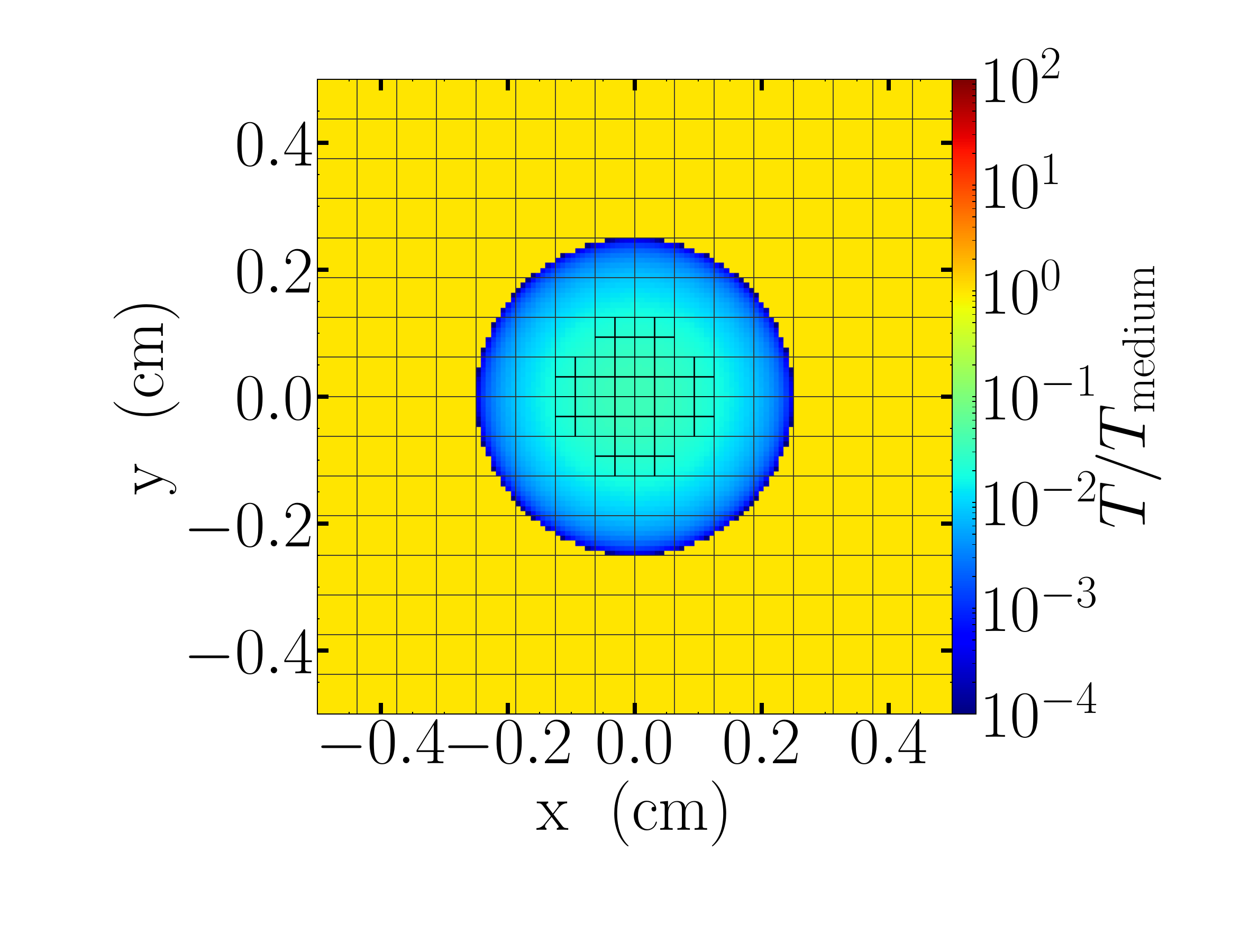}}%
    \subfloat[Density profile]{\includegraphics[width=7.5cm]{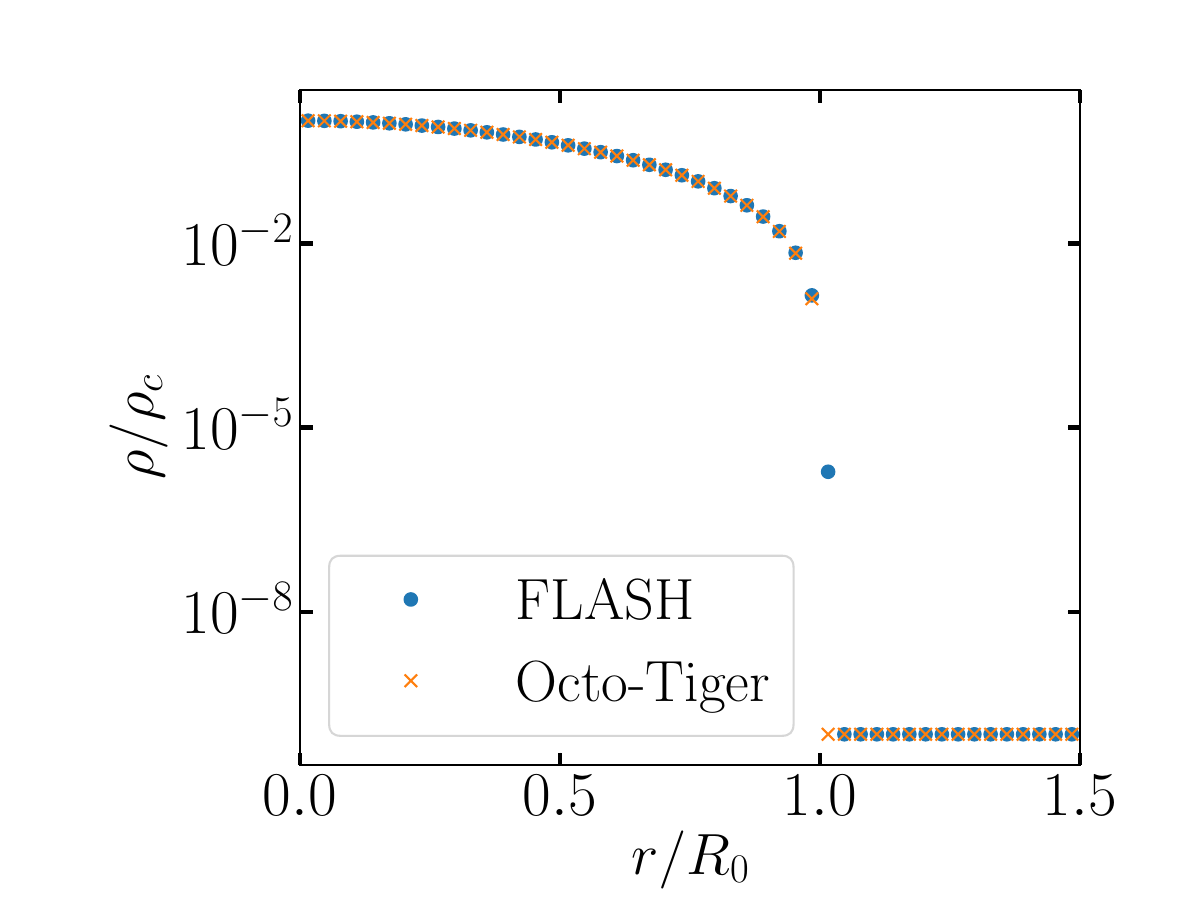}}
    \caption{\protect\footnotesize{Stationary star initial state. (a) AMR grid in a temperature slice at the equatorial plane of the AMR \octo\ simulation. Each cube represents a sub-grid containing $8^3$ equal volume cubic cells; (b) the initial density profile of the stars in \octo\ and \flash. Small differences are present due to different interpolation schemes}}
\label{fig:poly_n1.5_initial}
\end{figure*}

We compare the oscillations observed in the simulations with the analytical solutions. We also monitor the diffusion of the outer layers of the star and the behaviour of the low-density medium that surrounds the structure and we verify the conservation of some basic physical quantities.

We ran six simulations in total, two \octo\ uniform grid simulations with resolutions of $128^3$ and $256^3$ cells and one \octo\ AMR simulation with base resolution of $128^3$ and one level of refinement that increases linearly the resolution by a factor of 2. We refine based on a criterion of density (see Section~\ref{ssec:amr}). These three simulations use the \octo\ FMM gravity solver, with an opening angle parameter $\theta=0.5$. We also ran a fourth uniform grid \octo\ simulation with resolution $128^3$, and $\theta=0.35$. 

These four simulations used an ideal gas EoS. The fifth simulation, a uniform grid of $128^3$ resolution and $\theta=0.5$, used a polytropic EoS (see Equation~\ref{eq:poly}). The sixth simulation is carried-out with \flash\ on a uniform grid resolution of $128^3$, using the BH-tree gravity solver. 

In \octo\ we solve for each cell in the domain of the Lane-Emden equation to obtain the polytrope, while in \flash\ we interpolated a one-dimensional polytropic solution into the three-dimensional grid. This results in small differences between the initial states of the \octo\ and \flash\ simulations. In Figure~\ref{fig:poly_n1.5_initial}(a), we show the AMR grid in a temperature slice at the equatorial plane of the AMR \octo\ simulation. Each cube represents a subgrid containing $8^3$ equal volume cubic cells. As we refined by the density criterion, only the inner region of the star is refined to the maximum level. In Figure~\ref{fig:poly_n1.5_initial}(b), we plot the initial density profile of the stars in both \octo\ and \flash.

\begin{figure*}
\centering   
    \subfloat[Central density]{\includegraphics[width=7.5cm]{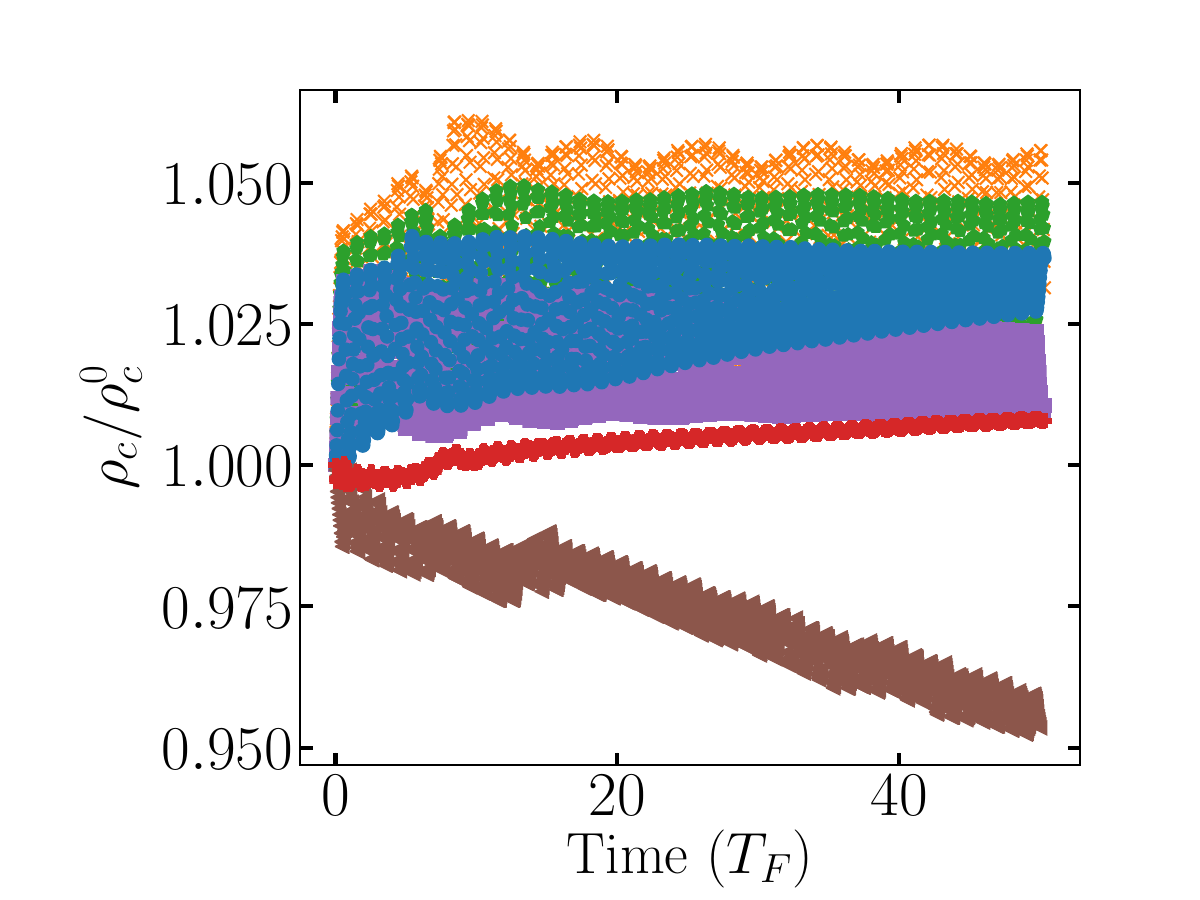}}%
    \subfloat[Fourier transform of the central density]{\includegraphics[width=7.5cm]{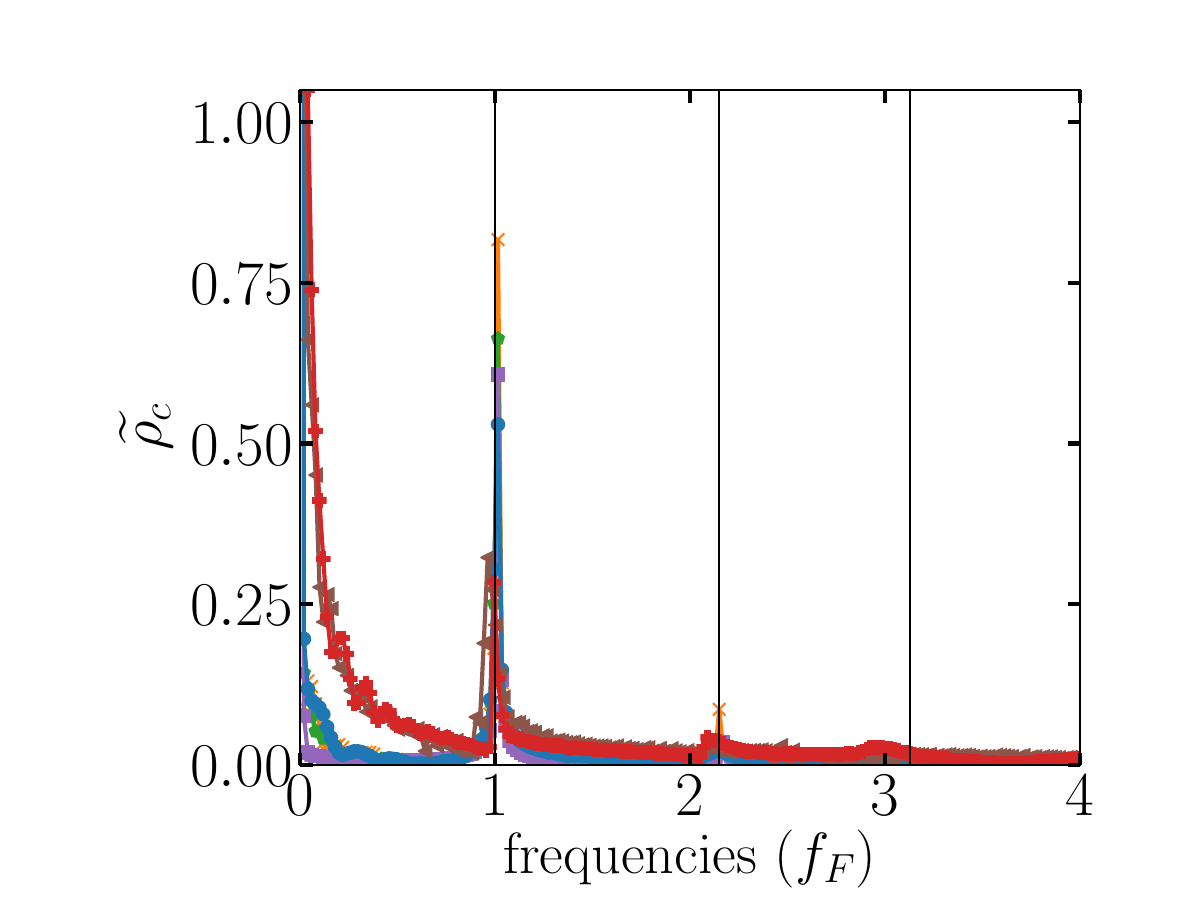}}
    \qquad
    \subfloat[Total mass conservation]{\includegraphics[width=7.5cm]{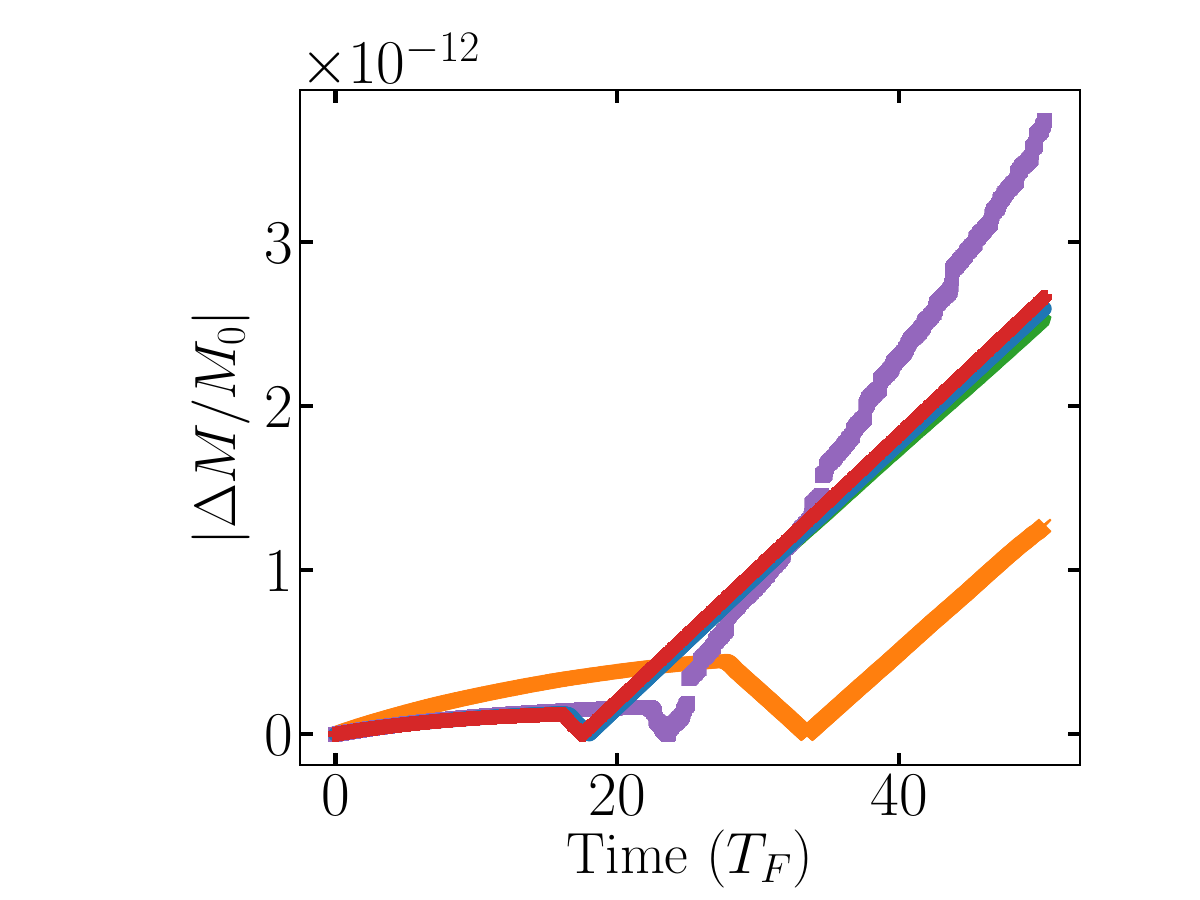}}
    \subfloat[Mass within the polytrope boundary]{\includegraphics[width=7.5cm]{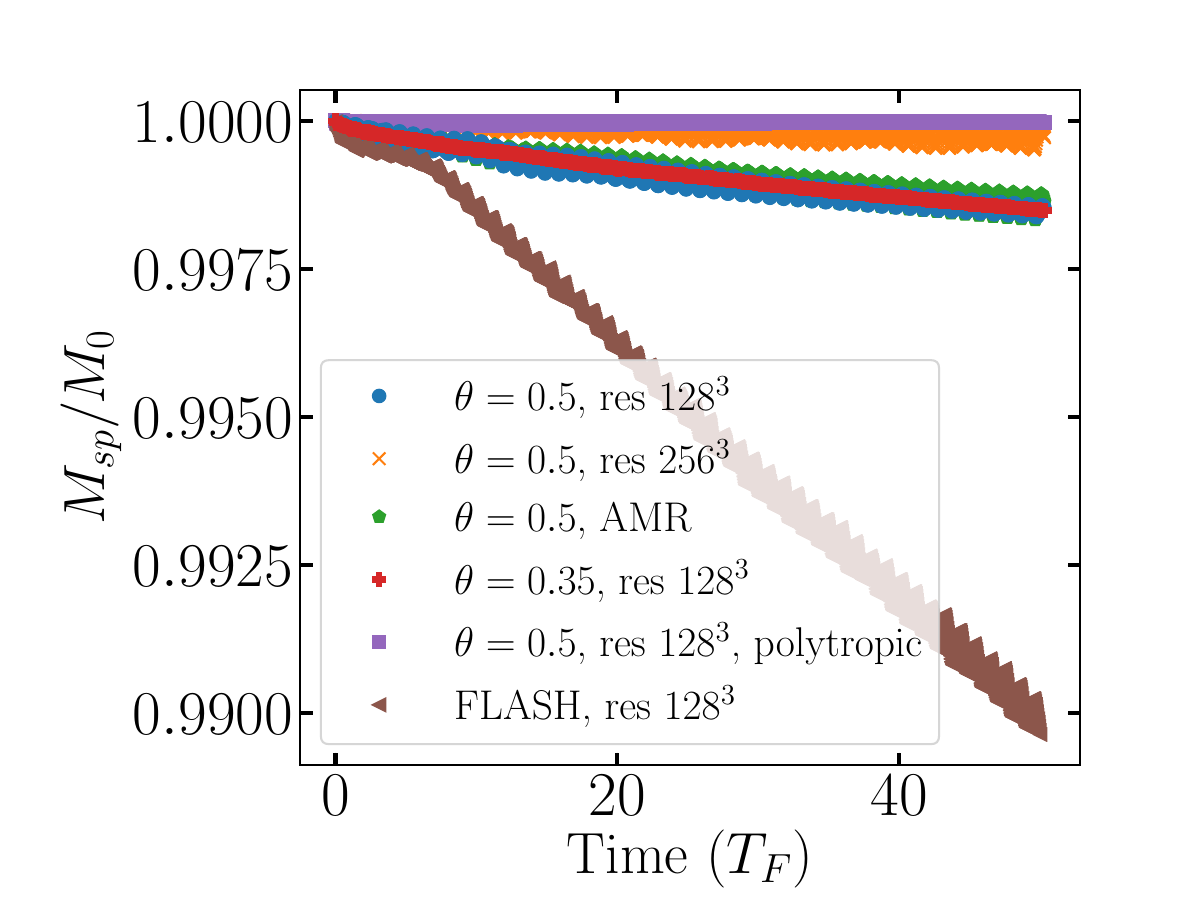}}%
    \qquad    
    \subfloat[Total energy conservation]{\includegraphics[width=7.5cm]{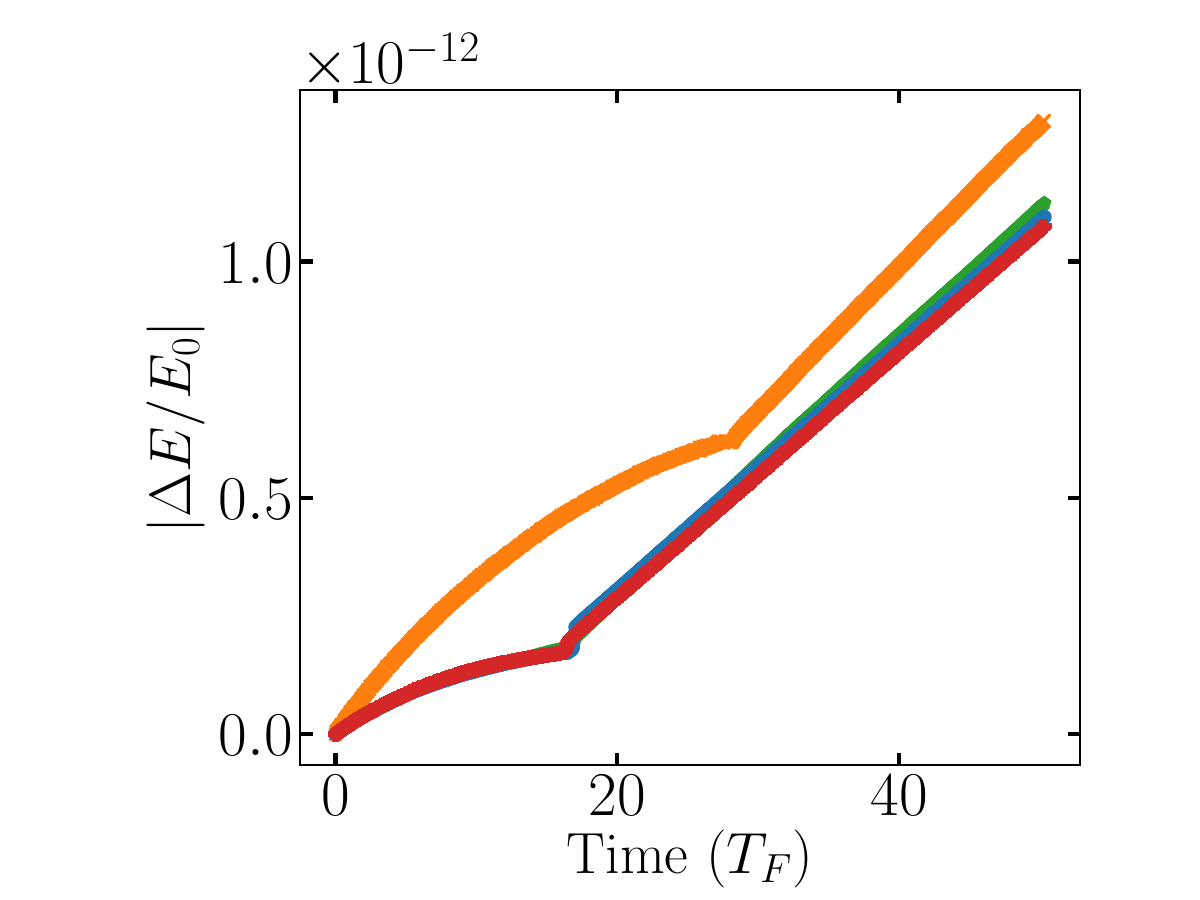}}%
    \subfloat[Total entropy conservation]{\includegraphics[width=7.5cm]{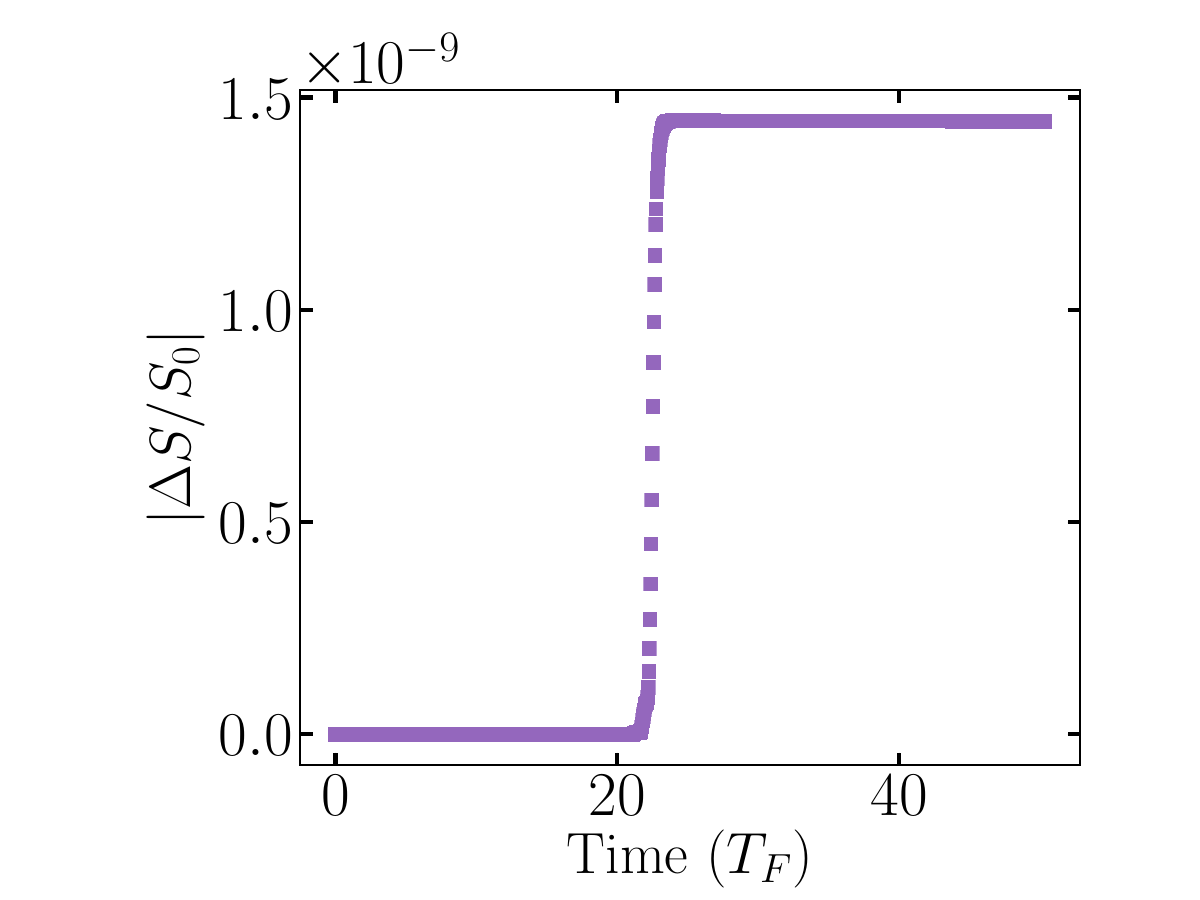}}%
        \caption{\protect\footnotesize{The stationary polytrope benchmark test. (a) Central density over time divided by the initial central density. (b) The Fourier transform of the central density pulsations normalised by the initial pulsation amplitude. The vertical lines show the fundamental frequency and the first and second harmonic frequencies. (c) Conservation of mass (only for \octo\ simulations). (d)  Total mass enclosed in a sphere with a radius of the initial polytropic radius. (e) Conservation of energy (only for \octo\ simulations with ideal gas EoS). (f) Conservation of entropy (only for the polytropic EoS \octo\ simulation). $T_F$ is the fundamental period of a polytrope with $n=3/2$ and $\gamma=5/3$}}
\label{fig:poly_n1.5_puls_octo1.9.3}
\end{figure*}

In Figure~\ref{fig:poly_n1.5_puls_octo1.9.3}(a), we plot the central density of the star, divided by the initial central density, over time (in units of $T_F$, the fundamental pulsation period of our polytope). As expected, in all simulations, the  central density of the star shows clear oscillations. In the \octo\ simulations, the central density oscillates around some converging value, while in \flash, there is a slow decrease with time overlaid on the central density oscillations. By setting a lower value for the opening angle parameter of the gravity solver ($\theta=0.35$, red crosses in Figure~\ref{fig:poly_n1.5_puls_octo1.9.3}), we can lower the amplitude for these oscillations. 

Although the oscillation amplitude might affect the amount of noise in the solution, the clear-cut test is whether the oscillation frequencies are aligned with the fundamental frequency predicted by theory. The assumption is that both the initial state and the numerical solving scheme inaccuracies will induce random noise which will oscillate at the fundamental frequency of the polytrope.
For that, we plot the Fourier transform of the stars' central density (Figure~\ref{fig:poly_n1.5_puls_octo1.9.3}(b)), normalised by the initial pulsation amplitude, in units of  the fundamental frequency, $f_F$. The vertical lines in Figure~\ref{fig:poly_n1.5_puls_octo1.9.3}(b) show the fundamental frequency and the first and second harmonic frequencies, $2.15 f_F$ and $3.13 f_F$, respectively. The Fourier transform of the densities of all the simulations notably peak close to the fundamental frequency. Again, the star, that is evolved with a $\theta=0.35$, reproduced most accurately the fundamental mode, with a frequency peak that deviates by only ~0.5~percent from the theoretical fundamental frequency. The other \octo\ simulations deviate by 1.5  percent, while \flash\ deviates by 3.8  percent. The star that is evolved with a polytropic EoS, naturally suffers least from low-frequency noise. 

We next plot (Figure~\ref{fig:poly_n1.5_puls_octo1.9.3}(c)), the deviation from conservation of mass, $\Delta M / M_0$, in the \octo\ simulations. $\Delta M = M\left( t \right) + M_{\rm out} - M_0$, where $M\left( t \right)=\int_V\rho dV$, $  M_{\rm out}$ and $M_0$ are the mass inside the simulation domain, the mass that outflowed from the grid and the initial mass, respectively. \octo\ calculates automatically, for every time step, the outflow of quantities such as mass, energy, and entropy, which simplify following of every small change of those quantities. Therefore, we plot only the conservation of mass in \octo\ simulations. We see conservation at the 10$^{-13}$ level up to a time when the density flooring adds mass to the grid. Even then the conservation, by the end of the simulation is still at the level of a few $\times 10^{-12}$. 

In these simulations, we set a density floor which causes a deviation from conservation at a machine precision level. The floor density is 5 orders of magnitude less than the initial density of the ambient medium, and usually will not affect mass conservation. However, at machine precision level, a deviation that grows approximately linearly with time appears. This can be explained by considering the following. As a consequence of ambient medium gas falling onto the star surface, the density outside the star decreases with time. Eventually, the density will decrease below the density floor threshold at the outer regions, which is then filled with density at the floor value. This happens after approximately 20, 25, and 35 fundamental periods in the low-resolution atmosphere with ideal gas EoS simulations, polytropic EoS simulation, and the high-resolution simulation, respectively. 
If the region that is being filled with a floor density is a fraction $\alpha$ of the domain volume, we will get a deviation from mass conservation of roughly $\alpha V \rho_{\rm floor} / M_0 =\alpha \rho_{\rm floor} / \overline{\rho}_0$, where $\overline{\rho}_0$ is the initial mean density in the simulation's domain. In our simulations $\rho_{\rm floor} / \overline{\rho}_0\simeq 9\times10^{-14}$. If only one layer of cells at the grid boundary is being filled with the floor density then $\alpha\simeq0.05$. If, in addition, the cells will be refilled every 5 time steps, after $\sim$20\,000 time steps a deviation of $2.2\times10^{-11}$ from conservation of mass will occur. In the simulations themselves, fewer cells are refilled less frequently so the deviations are smaller. Additionally, as shock heating is eliminated in the polytropic EoS simulation, the ambient gas continuously falls onto the star's surface without disturbance, which results in a bigger volume of flooring, and in a somewhat higher deviation from conservation of mass (purple line).

In Figure~\ref{fig:poly_n1.5_puls_octo1.9.3}(d), we show the total mass enclosed in a sphere with a radius of the initial star's radius. Here we see that \flash\ loses 1  percent of the stellar mass in 40 pulsations, while the \octo\ simulations lose at most 0.2 percent of the mass flowing out. The lower resolution, polytropic EoS as well as the higher resolution simulations perform best.

In Figure~\ref{fig:poly_n1.5_puls_octo1.9.3}(e), we show the deviation from conservation of energy $\Delta E / E_0 $, where $\Delta E = E\left( t \right) + E_{\rm out} - E_0$, and $E\left( t \right)=\int_V\left( E + \frac{1}{2}\rho \phi \right) dV$, $E_{\rm out}$ and $E_0$ are the energy inside the simulation domain, the energy that outflows from the grid and the initial energy, respectively. In Figure~\ref{fig:poly_n1.5_puls_octo1.9.3}(f), we show the deviation from conservation of entropy $\Delta S / S_0$, where $\Delta S = S\left( t \right) + S_{\rm out} - S_0$, and $S\left( t \right)=\int_V\tau dV$, $S_{\rm out}$, and $S_0$ are the entropy inside the simulation domain, the entropy that outflows from the grid and the initial entropy, respectively. Beside the small effect of the flooring, the ideal gas EoS simulations conserve energy at machine precision, while the polytropic EoS simulation conserves entropy at a machine precision.

\noindent Click on the link\footnote{\url{https://youtu.be/4-ra6fY982Q}} to view a movie of the \octo, $256^3$, $\theta = 0.5$ simulation.

\begin{figure*}
\centering   
    \subfloat[Centre of mass position accuracy]{\includegraphics[width=7.5cm]{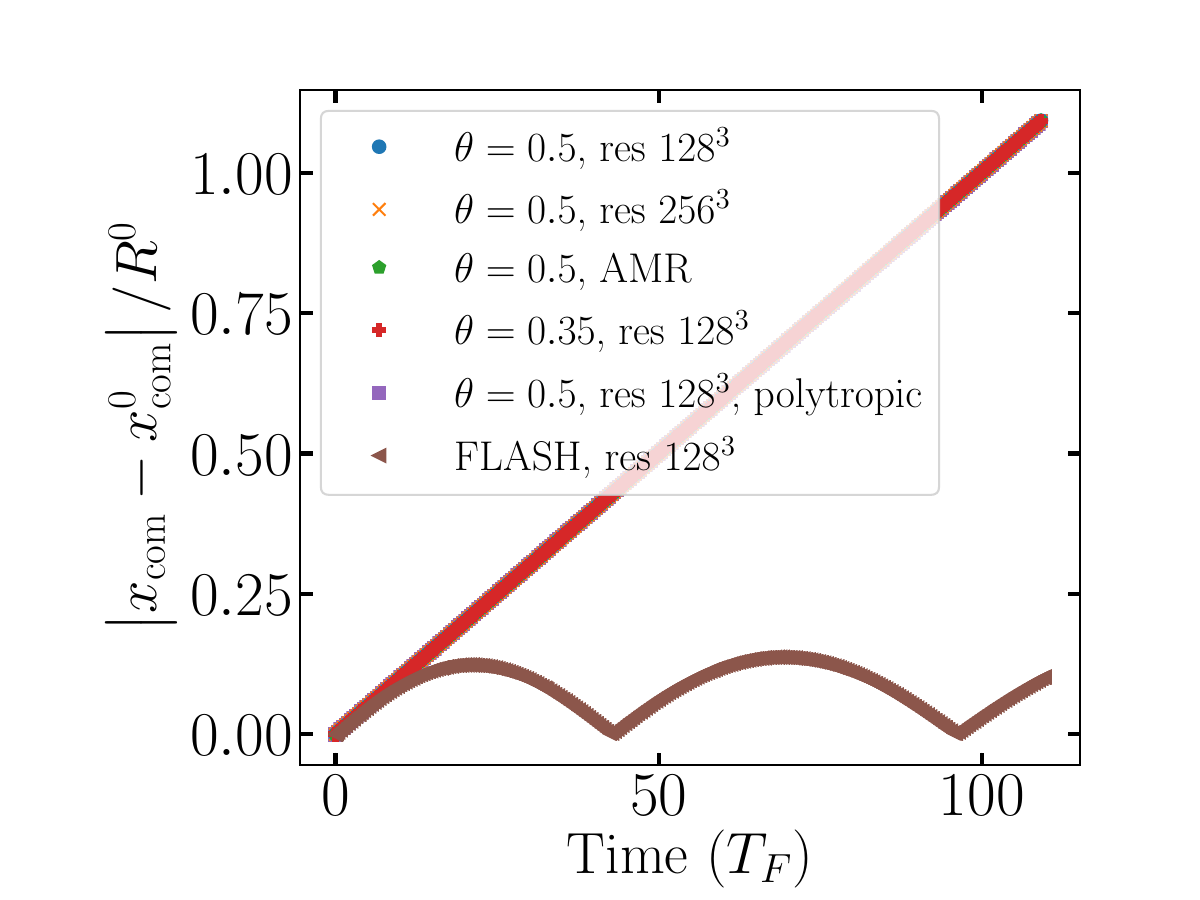}}    
    \subfloat[Centre of mass velocity accuracy]{\includegraphics[width=7.5cm]{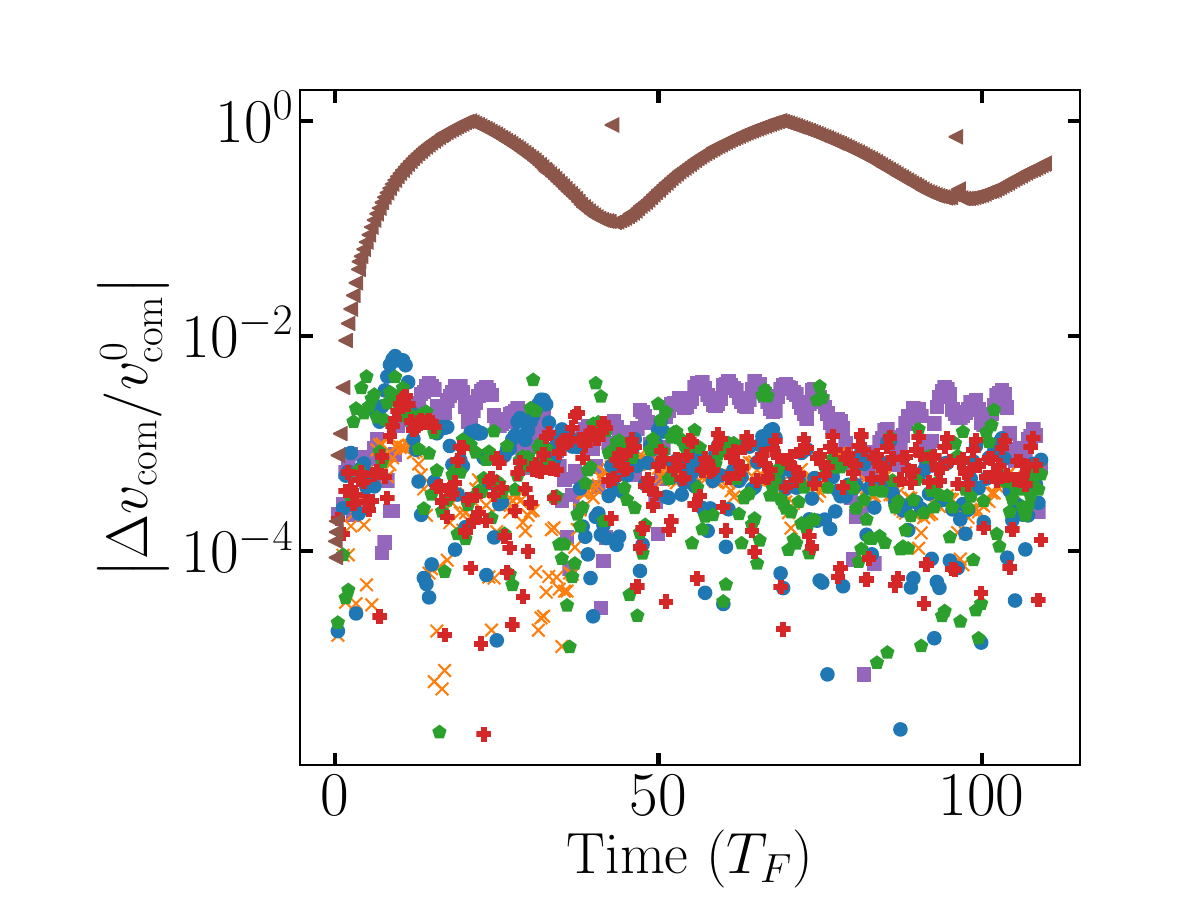}}    
    \qquad
    \subfloat[Ambient medium total x-linear momentum]{\includegraphics[width=7.5cm]{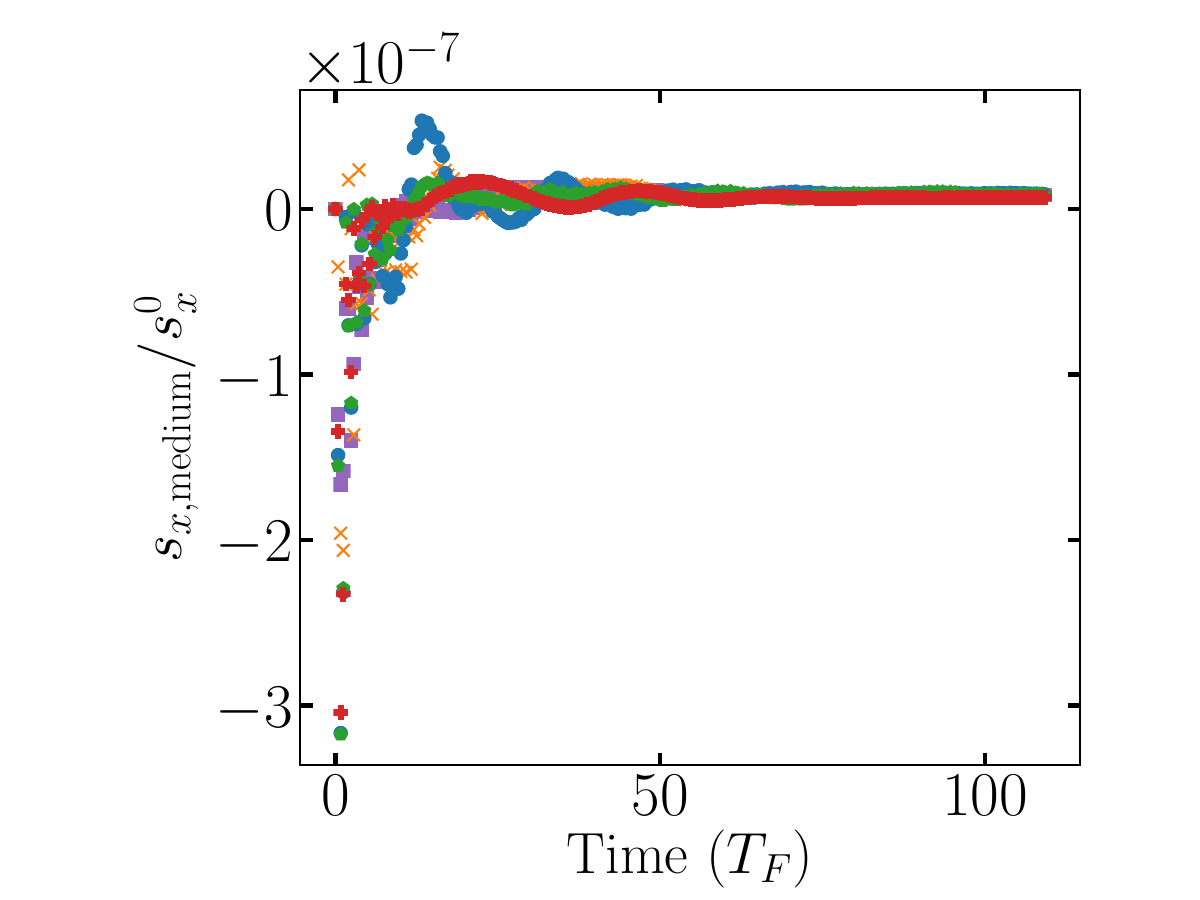}}%
    \subfloat[Total x-momentum conservation]{\includegraphics[width=7.5cm]{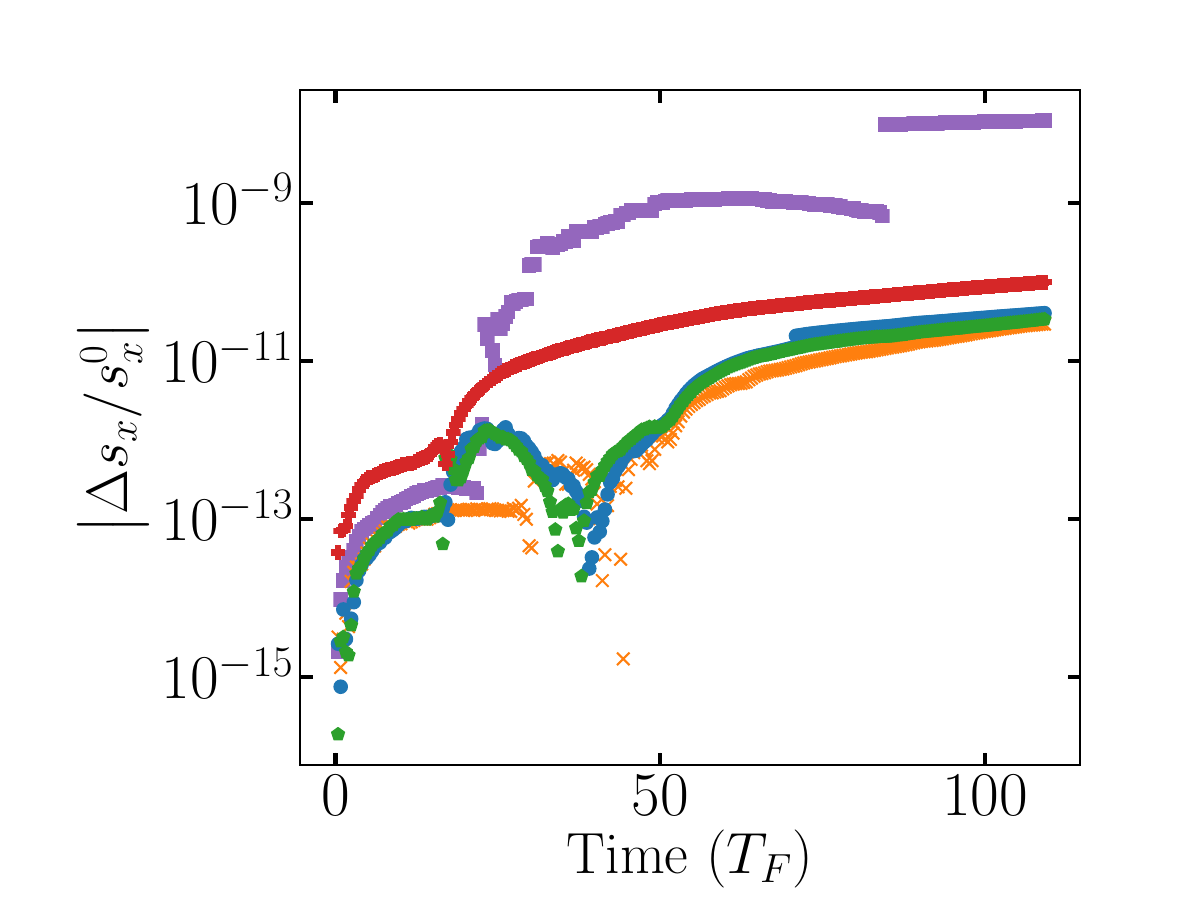}}
        \caption{\protect\footnotesize{The translating star benchmark test with the star moving at Mach $5.2\times10^{-4}$ (with respect to the diffuse medium). We plot various physical quantities of interest over time. Time is in units of $T_F$, the fundamental pulsation period}}
\label{fig:moving_sub_poly_n1.5_puls}
\end{figure*}


\subsection{Star Moving Linearly in the Grid}
\label{ssec:polytrope_translation}

We next simulate a star as in Section~\ref{ssec:polytrope} except we initialize the star with a non-zero bulk velocity, allowing it to translate at a given constant velocity through the grid. This exercise tests the degree to which the surface layers of our structure shear away due to interaction with the low density medium permeating the background. It also tests how well a spherical star moves through our cubical grid and how well our code conserves the total, non-zero linear momentum.

\begin{figure*}
\centering   
    \subfloat[Central density]{\includegraphics[width=7.5cm]{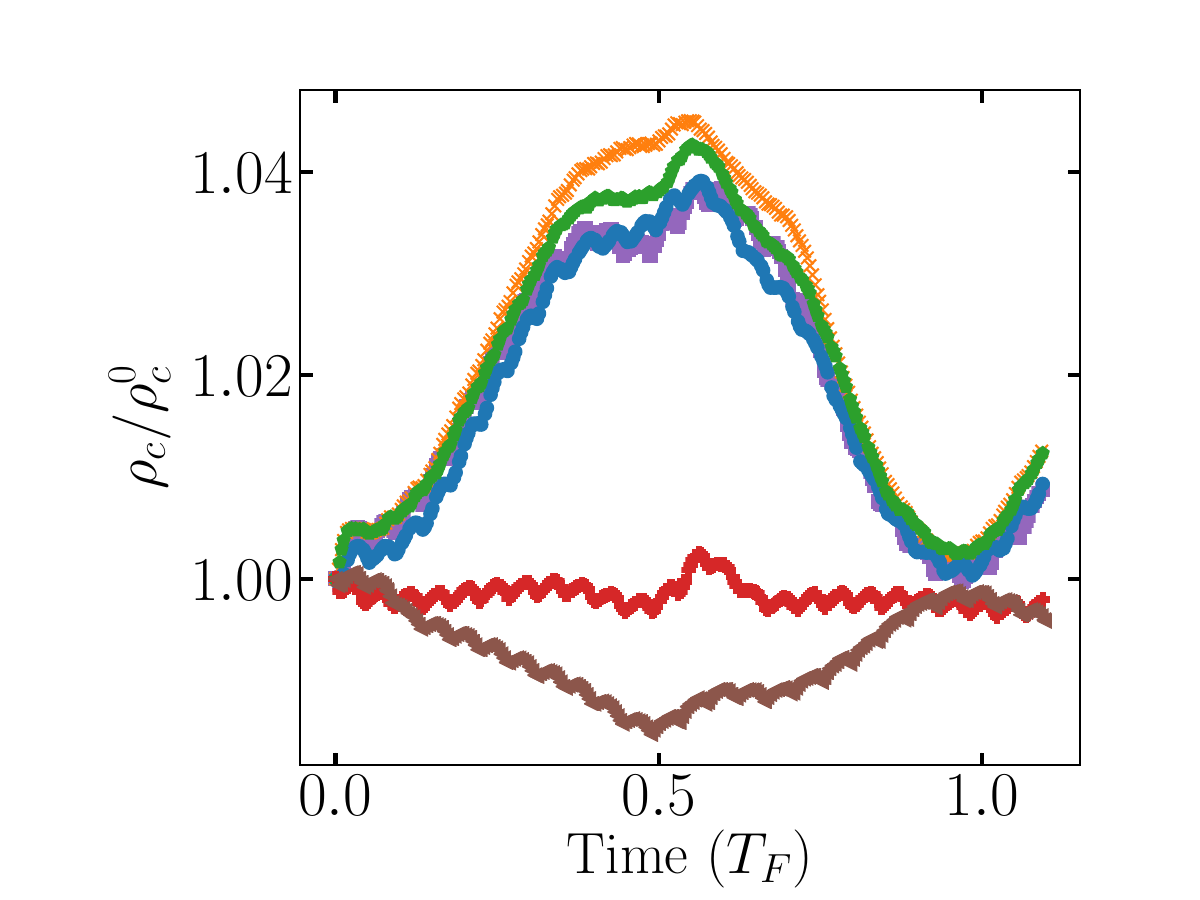}}
    \subfloat[Centre of mass velocity accuracy]{\includegraphics[width=7.5cm]{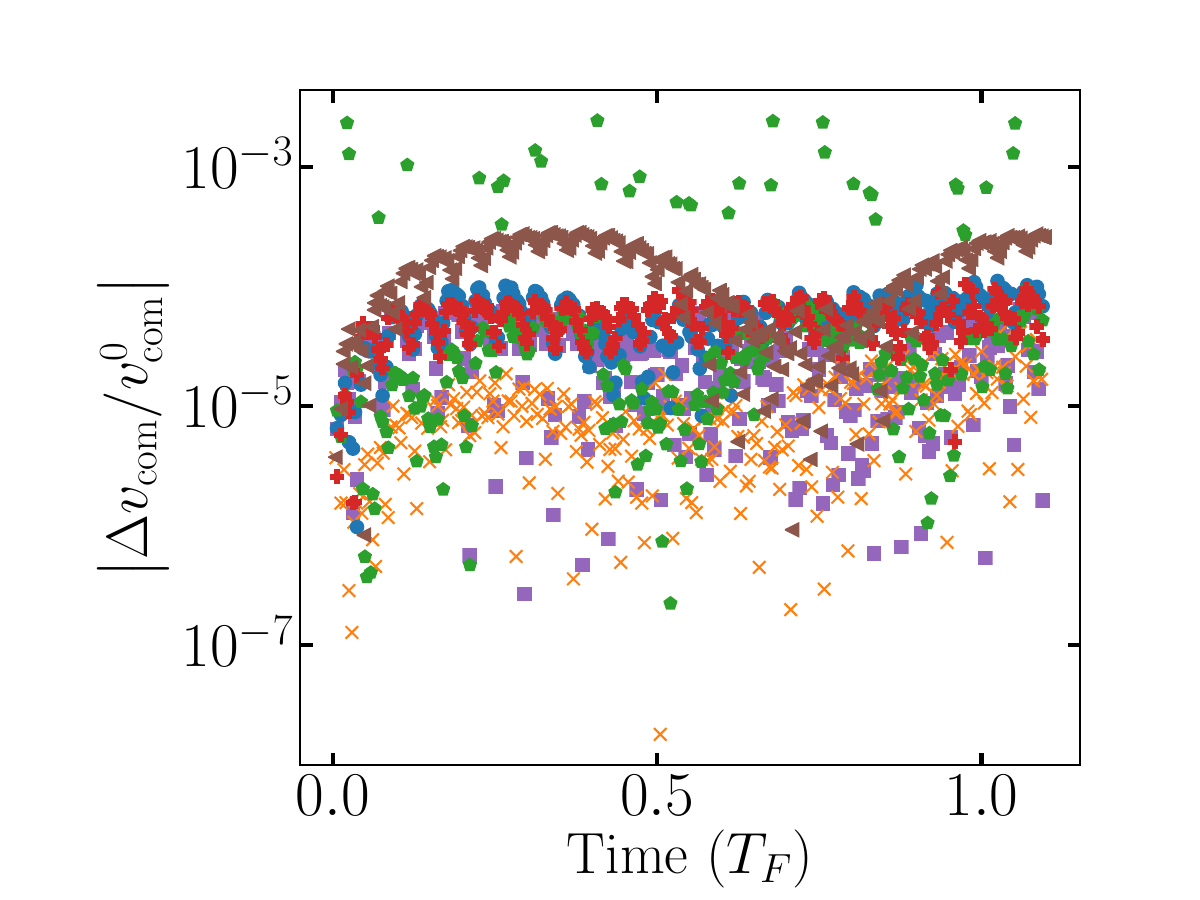}}
    \qquad    
    \subfloat[Ambient medium total x-linear momentum]{\includegraphics[width=7.5cm]{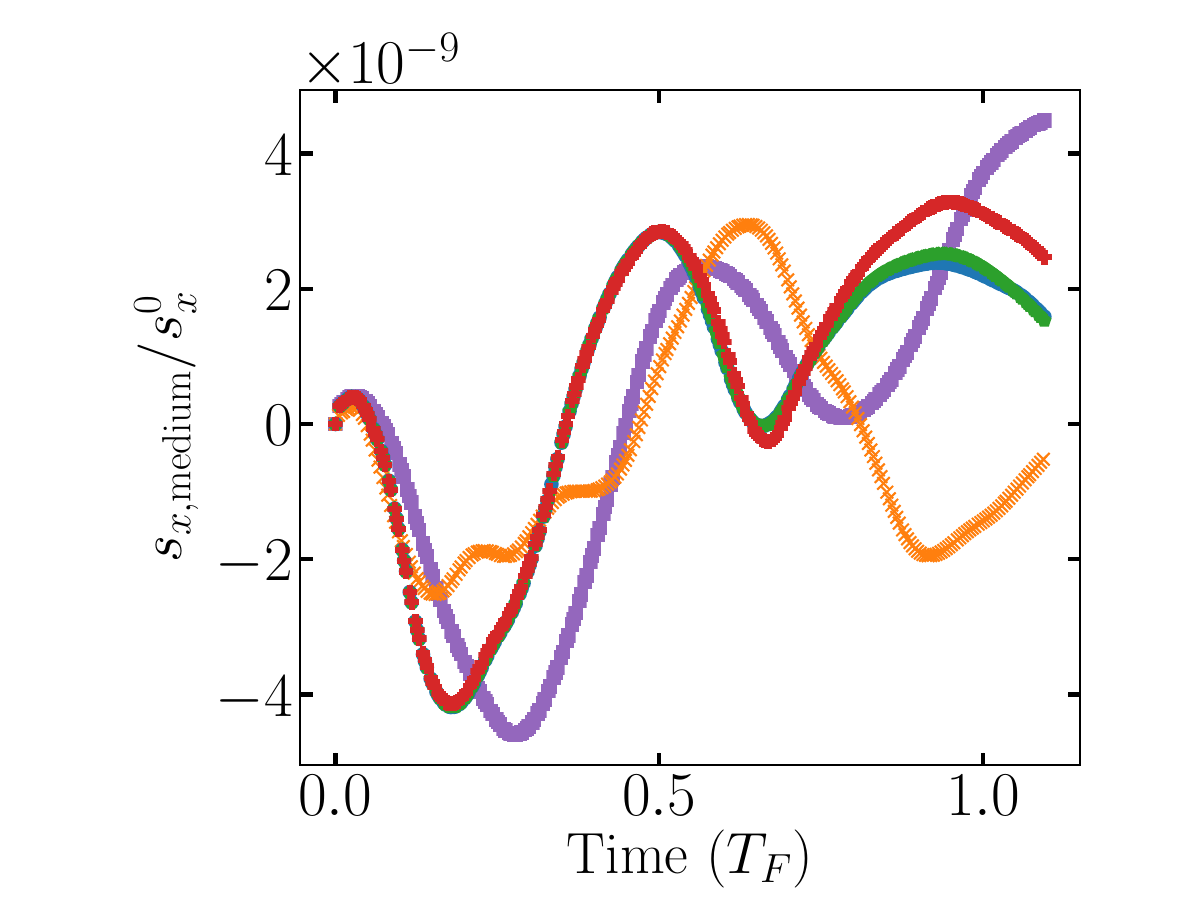}}%
    \subfloat[Mass within polytrope boundary]{\includegraphics[width=7.5cm]{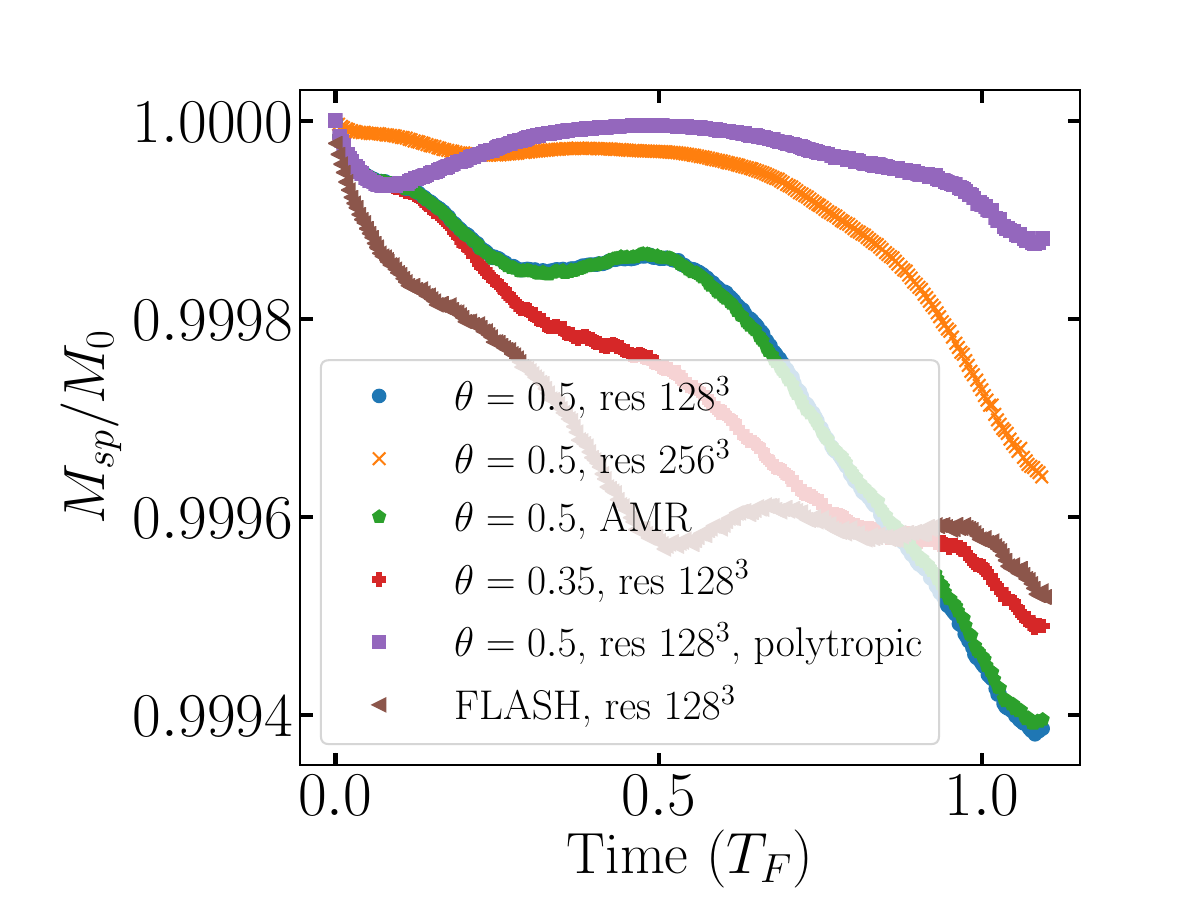}} 
    \caption{\protect\footnotesize{The translating star benchmark test with the star moving at Mach $5.2\times10^{-2}$ (with respect to the diffuse medium). We plot various physical quantities of interest over time. Time is in units of $T_F$, the fundamental pulsation period }}
\label{fig:moving_super_poly_n1.5_puls}
\end{figure*}

We test three velocities, (a) a subsonic velocity, $v_1=5.2\times10^{-4}c_{s,\mathrm{medium}}$, where $c_{s,\mathrm{medium}}$ is the speed of sound of the ambient hot medium (see Table~\ref{tab:poly_models}); (b) an intermediate velocity, $v_2=5.2\times10^{-2}c_{s,\mathrm{medium}}$; and (c) a high, supersonic velocity, $v_3=5.2c_{s,\mathrm{medium}}$. We list these three velocities scaled to a main sequence and WD models, as well as the simulation time (shorter for the faster stars) and the distance the star travels in Table~\ref{tab:moving_poly_models}. The distance the star travels equals to 1.1 times the star's initial radius for all simulations. We placed the star initially at (-0.125, -0.125, 0) and gave it a velocity in the direction towards grid position (1,1,0).

\begin{table*}
\begin{center}
\begin{adjustbox}{max width=\textwidth}
\begin{tabular}{lcccccccccc}
\hline
Model & $v_1$              & $v_2$             & $v_3$                & $t_{{\rm sim} 1}$ & $t_{{\rm sim} 2}$ & $t_{{\rm sim} 3}$ & $R_{\rm travel}$     \\
& (km~s$^{-1}$) & (km~s$^{-1}$) &  (km~s$^{-1}$) & \multicolumn{3}{c}{(sec/min/hr)} & (\rsun) \\
\hline
MS    & 1.20 & 120 & 12\,000 & 44 hr               & 27 min             & 16 sec            & 0.28 \\
WD    & 5.6  & 560 & 56\,000 & 33 min             & 20 sec             & 0.2 sec            & 0.016   \\
\hline
\end{tabular}
\end{adjustbox}
\end{center}
 \begin{quote}
  \caption{\protect\footnotesize{Translating polytrope benchmark test. Scaled quantities: motion velocities, total simulation times and displacement. For other scaled quantities refer to Table~\ref{tab:poly_models}}}
\label{tab:moving_poly_models}
\end{quote}
\end{table*}

Each velocity regime contains a set of six simulations (as we have done for the stationary star): two uniform grid, $128^3$ and $256^3$ cell simulations, one AMR simulation ($128^3$ cells and one level of refinement), with ideal gas EoS and $\theta=0.5$; a uniform grid $128^3$ simulation with ideal gas EoS and $\theta=0.35$; a uniform grid $128^3$ simulation with polytropic EoS and $\theta=0.5$; and a \flash\  uniform grid $128^3$ simulation. In Sections~\ref{sssec:subsonic_translation} -- \ref{sssec:supersonic_translation}, we describe the results of the simulations.

\begin{figure*}
\centering   
    \subfloat[Total mass conservation ]{\includegraphics[width=7.5cm]{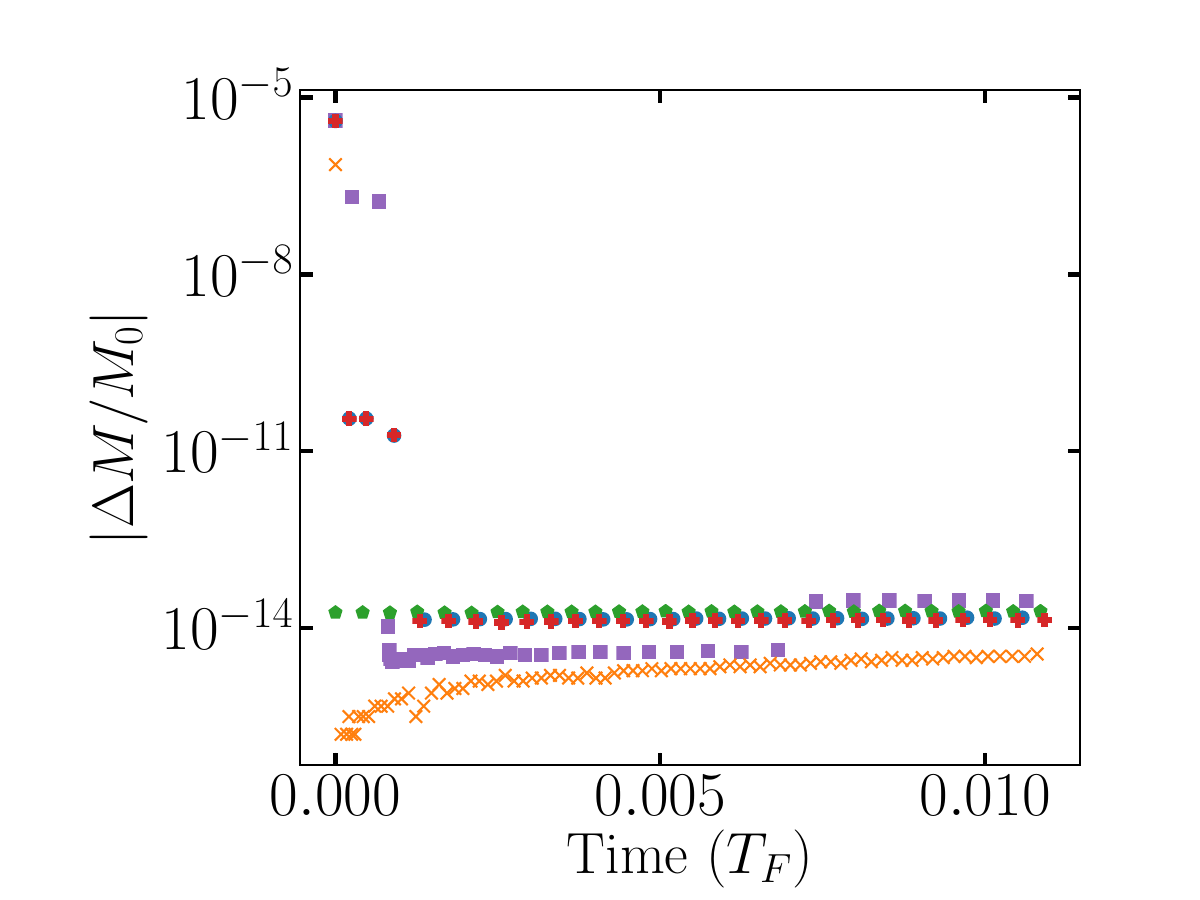}}%
    \subfloat[Centre of mass velocity accuracy]{\includegraphics[width=7.5cm]{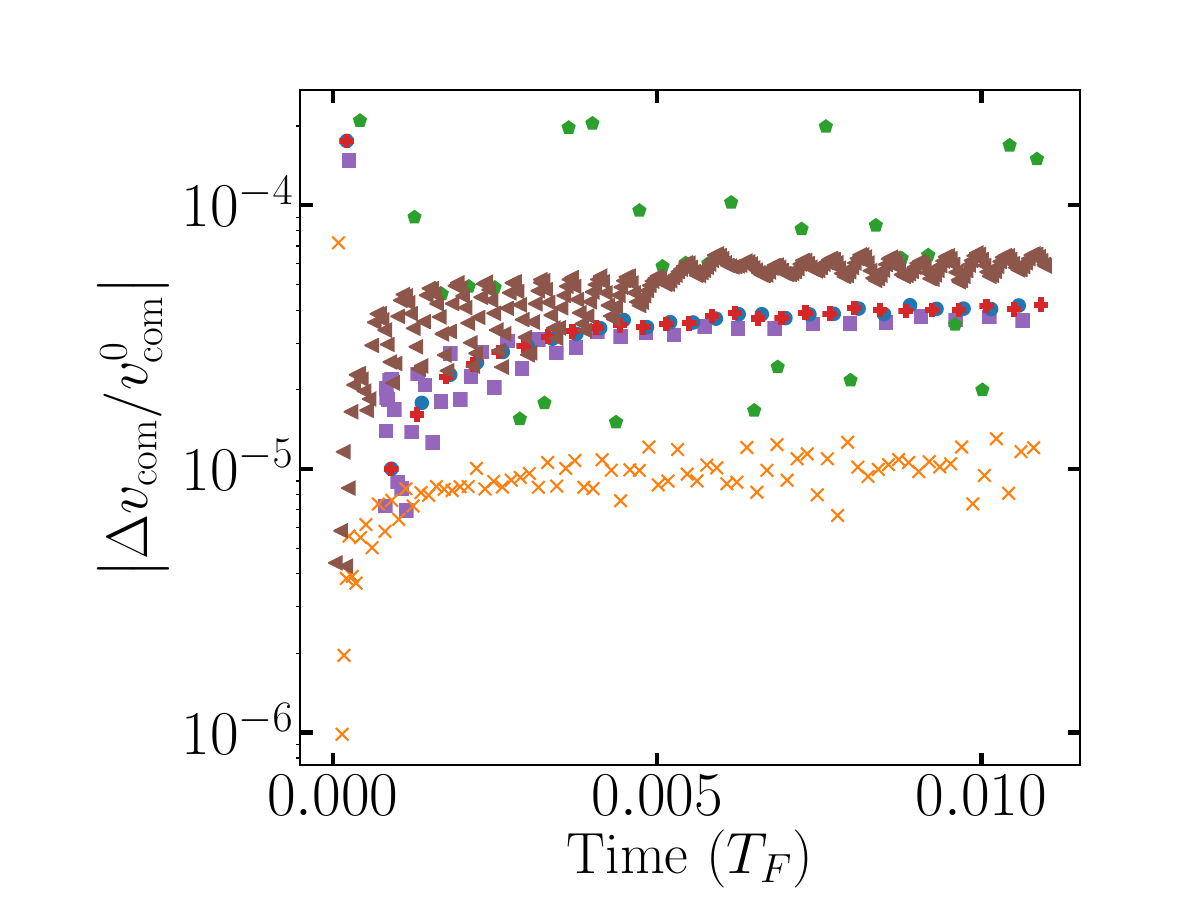}}    
    \qquad    
    \subfloat[Ambient medium total x-linear momentum]{\includegraphics[width=7.5cm]{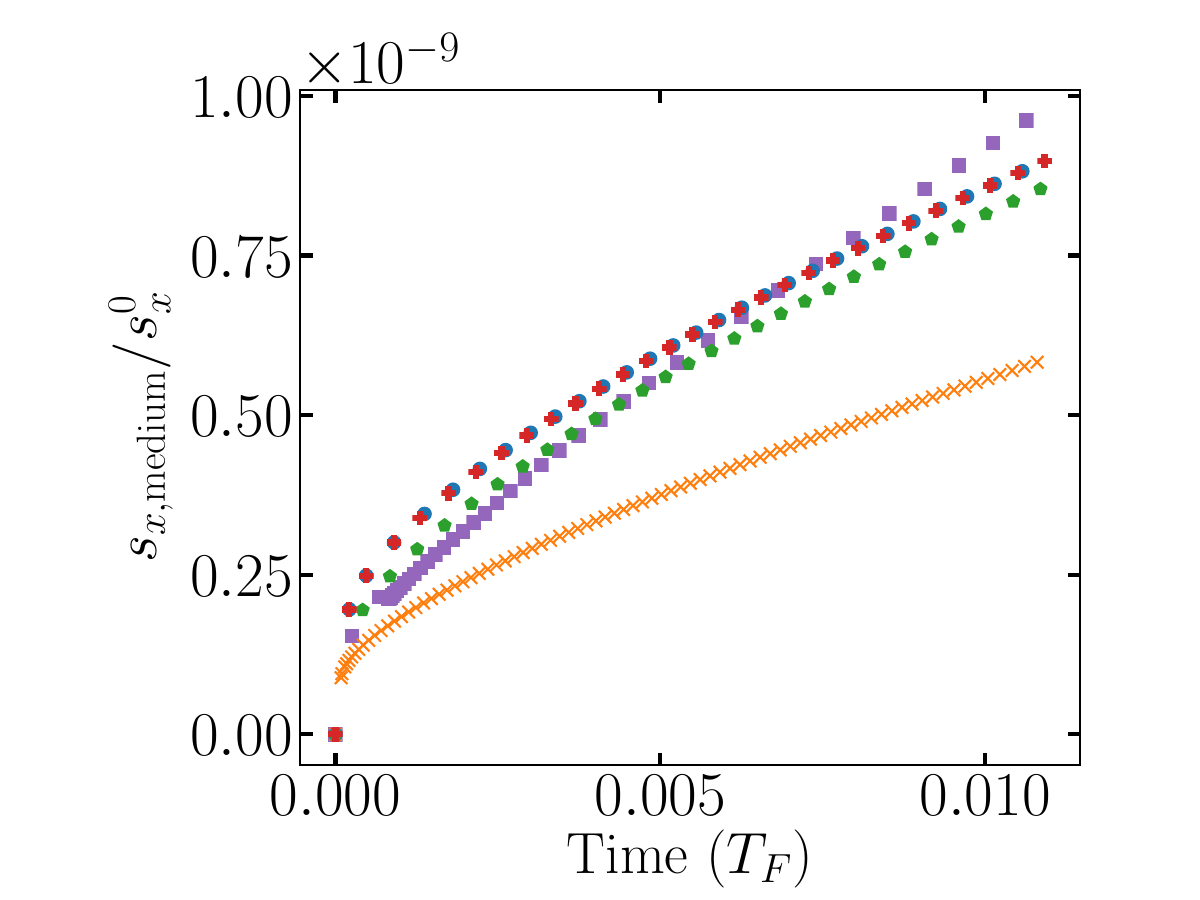}}%
    \subfloat[Mass in star]{\includegraphics[width=7.5cm]{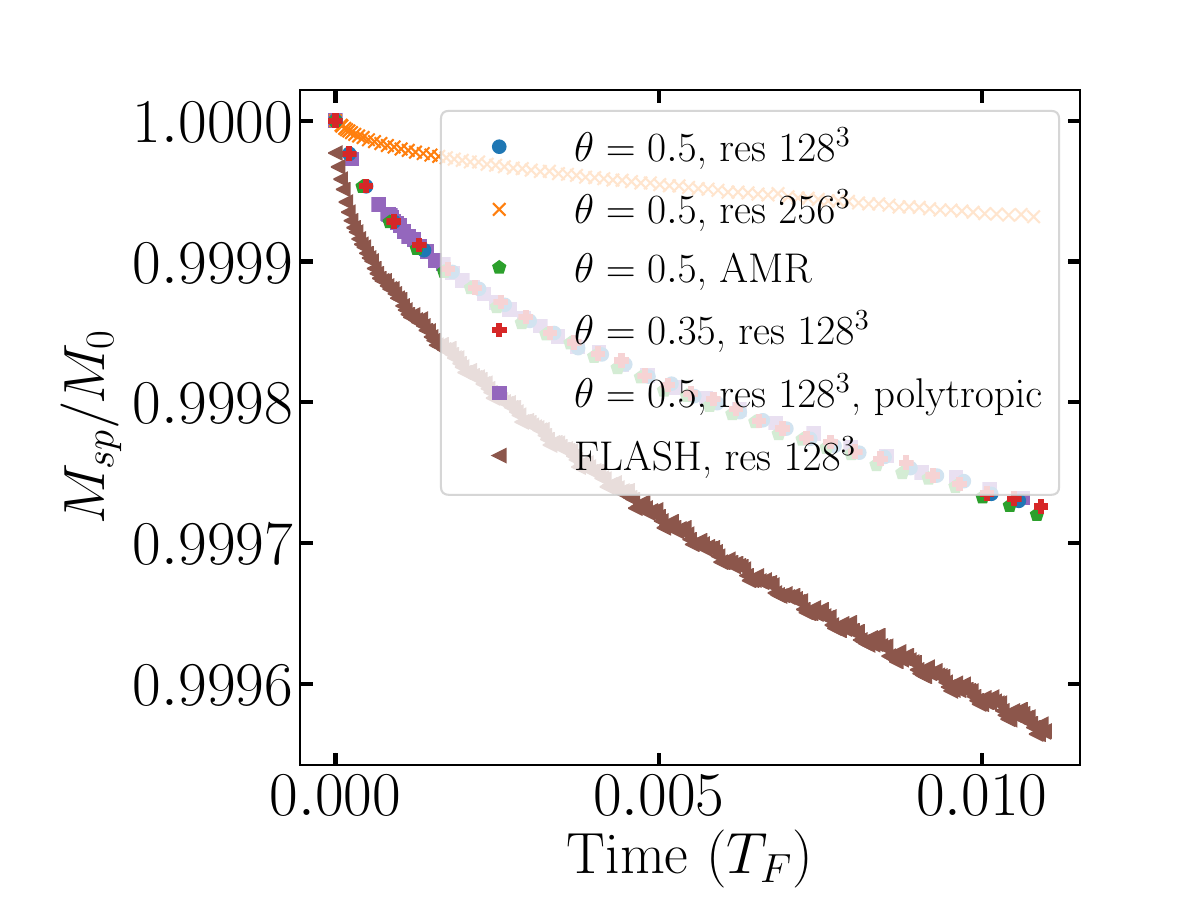}}
    \caption{\protect\footnotesize{The translating star benchmark test with the star moving at Mach $5.2$ (with respect to the diffuse medium). We plot various physical quantities of interest over time. Time is in units of $T_F$, the fundamental pulsation period }}
\label{fig:moving_super2_poly_n1.5_puls}
\end{figure*}

\subsubsection{Star Translating at Low Velocity}
\label{sssec:subsonic_translation}

In this regime, the star translates at a very low velocity, equivalent to a Mach number of $5.2\times 10^{-4}$, not only with respect to the ambient medium speed of sound, but also with respect to the speed of sound of the coldest regions at the star's surface. Such low bulk velocities are a challenge for hydrodynamic codes. Low level subsonic noise that develops can advect the stellar momentum to the surrounding low density gas, diminishing the star's initial low velocity. In addition, the star's low velocity allows us to run this test for over an hundred fundamental pulsations periods, the longest time we have run a polytrope simulation. During this time, the star should keep pulsating at its Eigenmodes as it translates through the grid. 

The star in the \octo\ simulations slowly moves from the bottom left corner to the upper right corner (in the xy plane). As in the stationary star simulations, the star remains in hydrostatic equilibrium, while the ambient medium is free falling onto the stellar surface. This slowly cools the ambient medium and creates shocks on the star's surface that heat the gas for ideal gas simulations. 
We performed the same analysis as we did for the stationary star. We find that the star's movement does not affect the star's pulsations. Almost identically to the stationary star in Figure~\ref{fig:poly_n1.5_puls_octo1.9.3}(a), the central density oscillates at the fundamental frequency in all \octo\ simulations, while the mean central density of the star simulated with \flash\ decreases with time. Additionally, the star's mass ($M_{\rm sp}$) remarkably decreases at the same rate as for the stationary star in Figure~\ref{fig:poly_n1.5_puls_octo1.9.3}(d), decreasing by 2 percent in the \flash\ simulation and by less than 0.4 percent with \octo, after 100 pulsations.        

In Figure~\ref{fig:moving_sub_poly_n1.5_puls}(a), we plot the centre of mass position as a function of time, while in panel (b) we show the deviation of the centre of mass velocity from its initial value. The star in the \octo\ simulations moves through the grid at a constant velocity, while in the \flash\ simulations, the star slows down, starts moving in the opposite direction and then oscillates around the star's initial position. The \octo\ simulations deviate from the initial velocity by less than 1 percent throughout the entire evolution.

We exploited \octo's capability to track the provenance of gas, to calculate the diffusion of linear momentum from the star to its environment. In  Figure~\ref{fig:moving_sub_poly_n1.5_puls}(c), we plot the x-momentum of the diffuse medium divided by the initial {\it total} x-momentum in the \octo\ simulation. The initial oscillation is due to the sharp density gradient at the star's surface. Soon after the value settles at values close to zero. We mention though, that the mass ratio between the ambient medium gas and the star is of the order of $10^{-8}$. No comparison with \flash\ is possible due to \flash\ not tagging gas provenance.

We find, similarly to the stationary star test, that mass, energy, and entropy are conserved in the \octo\ simulations to excellent precision, except for small deviations due to adding mass when a cell's density drops below the minimum floor value (see Sec.~\ref{ssec:floors}). We additionally follow the conservation of x-linear momentum (Figure~\ref{fig:moving_sub_poly_n1.5_puls}(d)), finding a machine precision level until flooring starts operating at which point we observe a slow linear growth. The deviation, though, is minimal, less than $10^{-10}$ after 100 pulsation periods for all simulations except the polytropic EoS. The polytropic simulations suffers more from flooring (because shock heating is absent) and the deviation settles on a value of $8\times10^{-9}$. 

\noindent Click on the link\footnote{\url{https://youtu.be/8ArZqP9F93Y}} to view a movie of the low velocity translating star.

\subsubsection{Star Translating at Intermediate Velocity}
\label{sssec:midvelocity_translation}

With a speed of Mach $5.2\times 10^{-2}$ with respect to the outside medium, the star translates at a subsonic velocity, but at a supersonic speed with respect to the star's surface. The simulation time is comparable to the star's fundamental period. This allows the star to pulsate only once while it translates at a higher speed than the previous test.

The star's gas diffuses out in the wake of the star, but then reverses motion and trails the star moving in a flow that resembles accretion. Shocks are apparent at this wake in the ideal EoS which increases the diffusing out of material from the star.

In Figure~\ref{fig:moving_super_poly_n1.5_puls}, we plot only the salient quantities that demonstrate the level of accuracy of the simulations. In panel (a) we demonstrate how the simulation time is approximately the time of the fundamental period of pulsation and the amplitude is similar to the stationary star simulations of Figure~\ref{fig:poly_n1.5_puls_octo1.9.3}(a). The error on the star's velocity is less than $10^{-3}$ for {\it all} simulations (Figure~\ref{fig:moving_super_poly_n1.5_puls}(b)), where \octo\ simulations have an error that is about 10 times smaller than the \octo\ simulations in the case of the slow moving star of Figure~\ref{fig:moving_sub_poly_n1.5_puls}(b) (Section~\ref{sssec:subsonic_translation}). In Figure~\ref{fig:moving_super_poly_n1.5_puls}(c) we see that the ambient medium momentum in \octo\ simulations remains negligible. After an oscillatory period, we expect it to converge to a value that is of the order of $10^{-9}$, which is over 10 times smaller than for the equivalent, slow moving test. 
Finally, in Figure~\ref{fig:moving_sub_poly_n1.5_puls}(d) the mass in the original polytrope is also retained with a precision that is 10 times that of the slow moving polytrope. An amount that is relatively high if we consider the short simulation time.

\noindent Click on the link\footnote{\url{https://youtu.be/b3MMyDCPY60}} to view a movie of the intermediate velocity translating star.

\subsubsection{Star Translating at a High Velocity}
\label{sssec:supersonic_translation}

For the highest velocity simulation, at Mach 5.2 with respect to the sound speed of the ambient medium, the running time is shorter than the dynamical time of the star so the star does not relax.  
There is a slight overall mass increase in the two unigrid simulations, but not for the AMR one (Figure~\ref{fig:moving_super2_poly_n1.5_puls}(a)) between the first and fifth time steps in the simulation. This is due to more mass leaving a cell than the mass in that cell at the sharp stellar edges of the polytrope. Since the mass in a cell is not allowed to be zero nor negative, mass is added to reach the density floor value, hence introducing mass in the grid. At subsequent time steps when numerical diffusion has softened all edges, this does not happen. This is not a large problem and can be resolved with a reduction of the Courant time to values $\sim 0.14$ (see justification in Section~\ref{ssec:refinement_criteria}).

In Figure~\ref{fig:moving_super2_poly_n1.5_puls}(b-d) we see that the velocity error, ambient x-momentum and mass in sphere retain their values to a precision that is comparable to that of the polytrope moving at intermediate speed (Section~\ref{sssec:midvelocity_translation}). The best simulation is the high resolution unigrid with $\theta = 0.5$ (yellow line), something that is also true but to a lesser extent for the other two velocity tests.

\noindent Click on the link\footnote{\url{https://youtu.be/rZiLMJ6Cxc4}} to view a movie of the high velocity translating star.

\begin{figure*}
\centering   
    \subfloat[Central density, inertial frame]{\includegraphics[width=6.5cm]{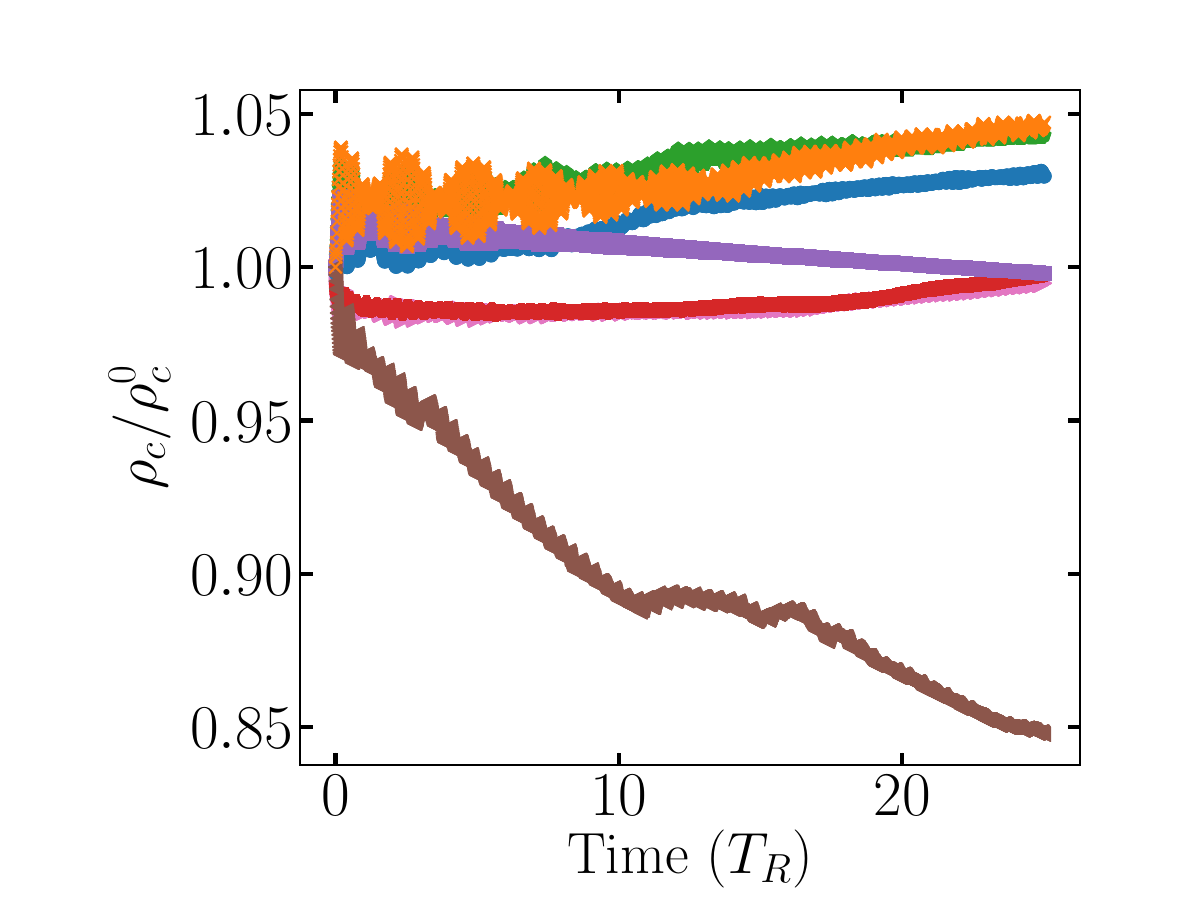}}%
    \subfloat[Central
    density, rotating frame]{\includegraphics[width=6.5cm]{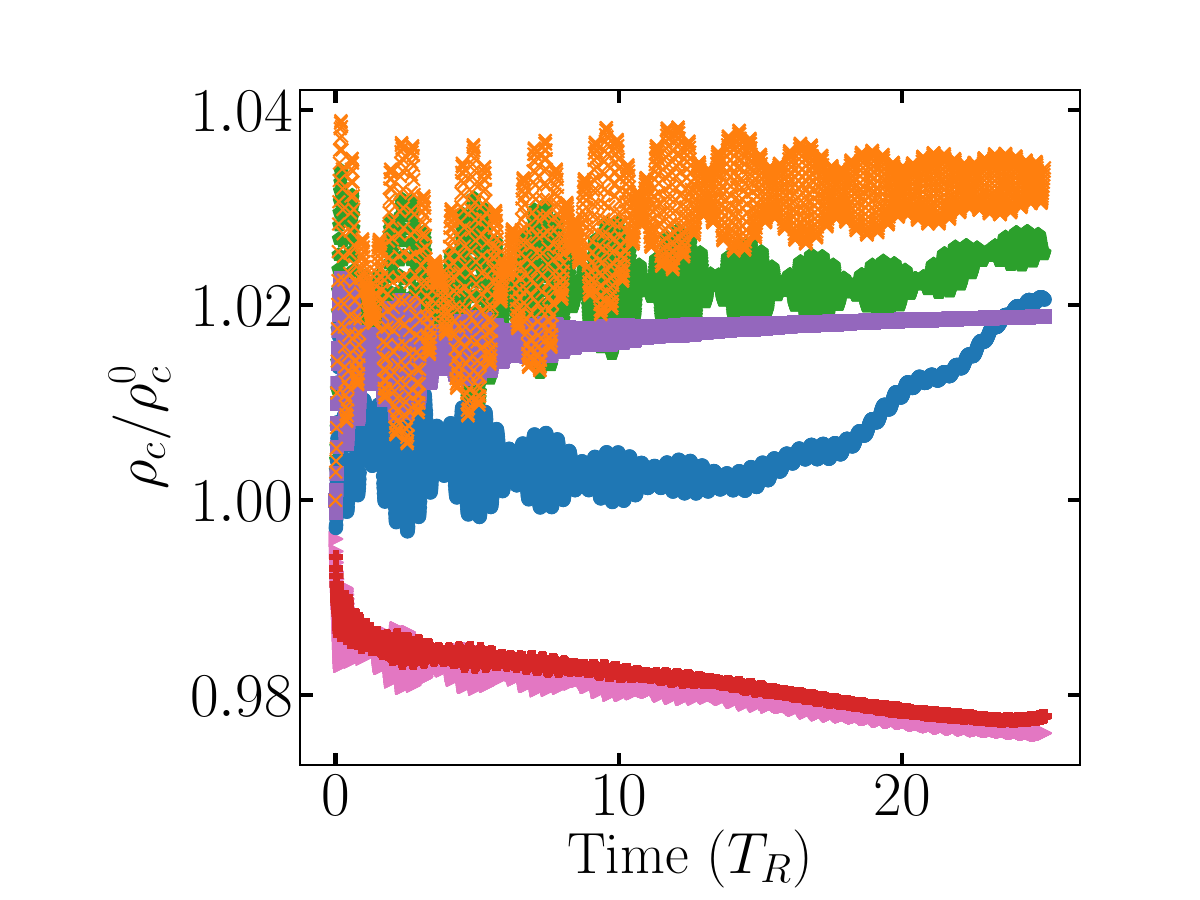}}%
    \qquad
    \subfloat[Mass in star, inertial frame]{\includegraphics[width=6.5cm]{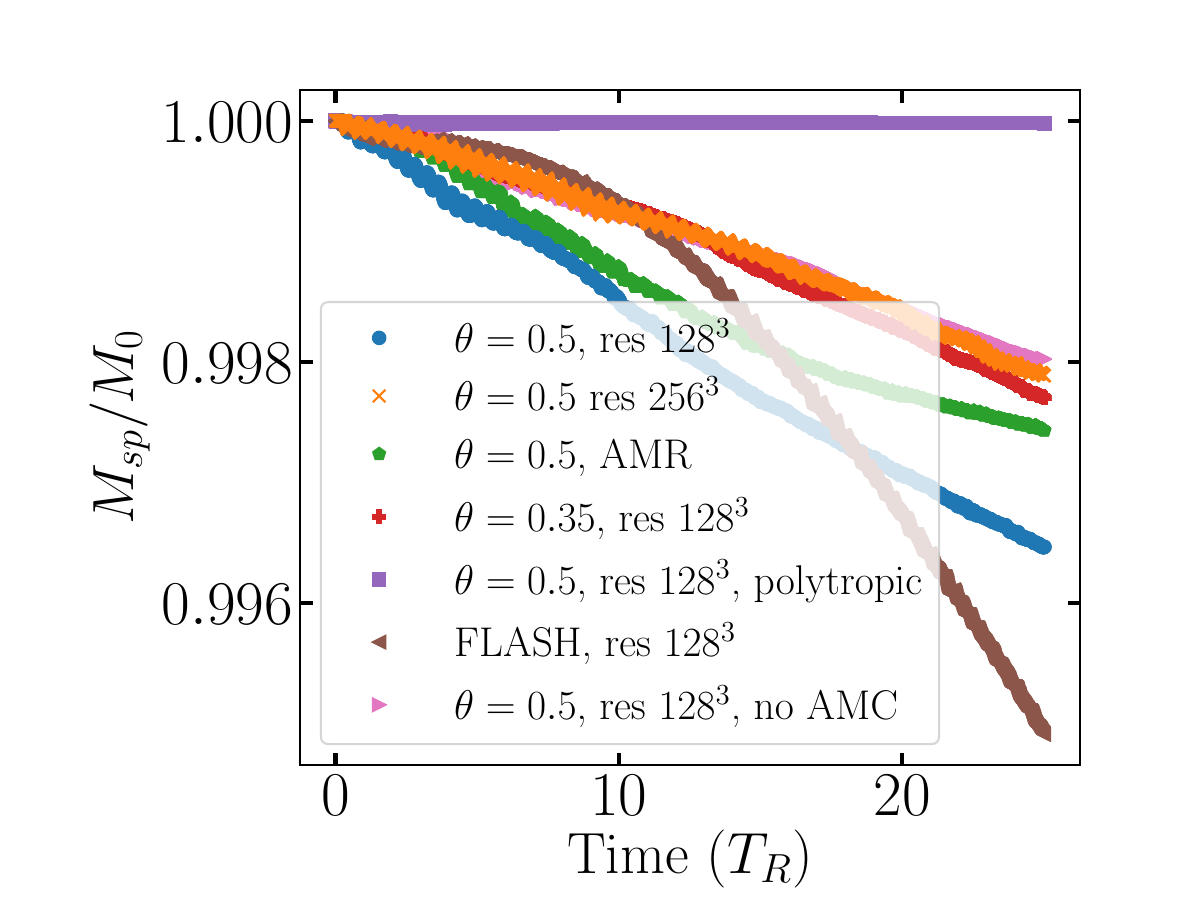}}%
    \subfloat[Mass in star, rotating frame]{\includegraphics[width=6.5cm]{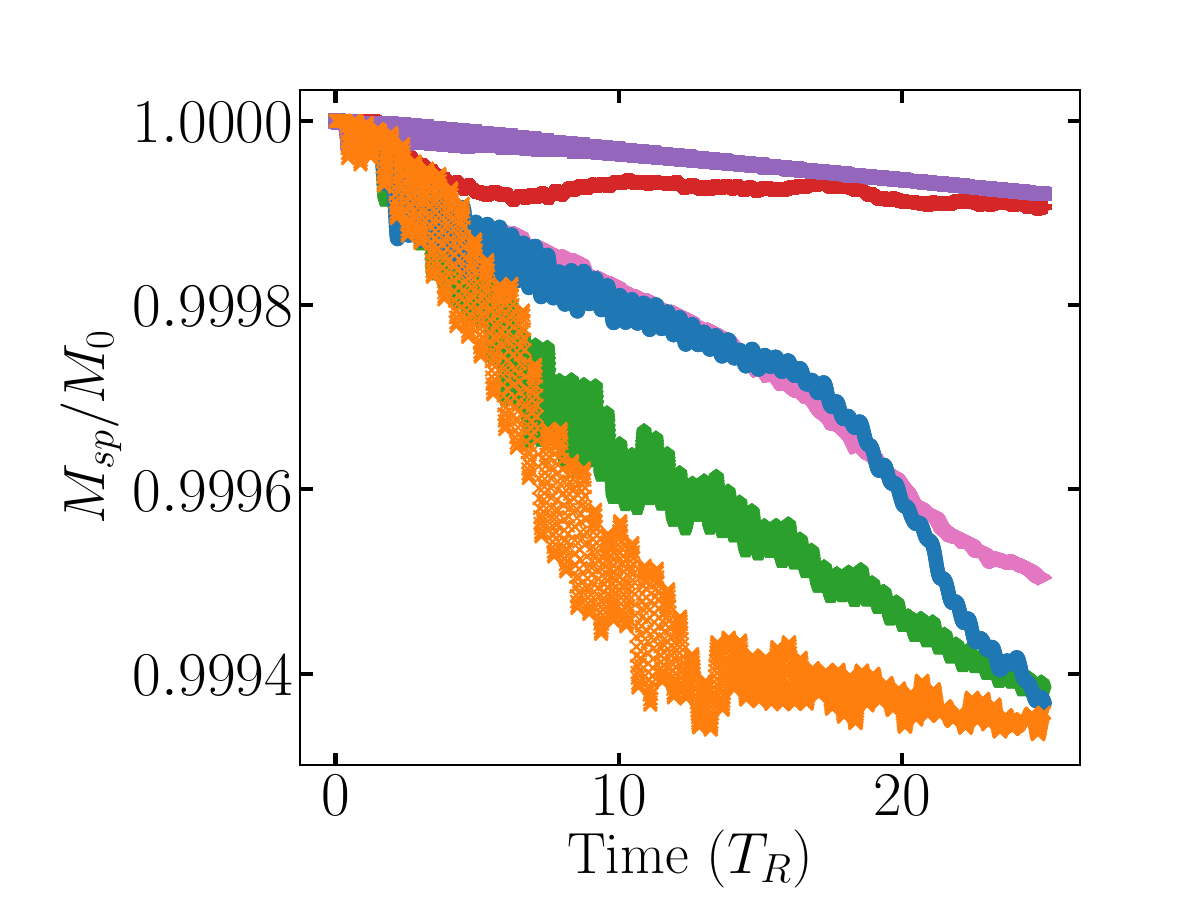}}%
    \qquad
    \subfloat[Diffusion of high density gas, inertial frame]{\includegraphics[width=6.5cm]{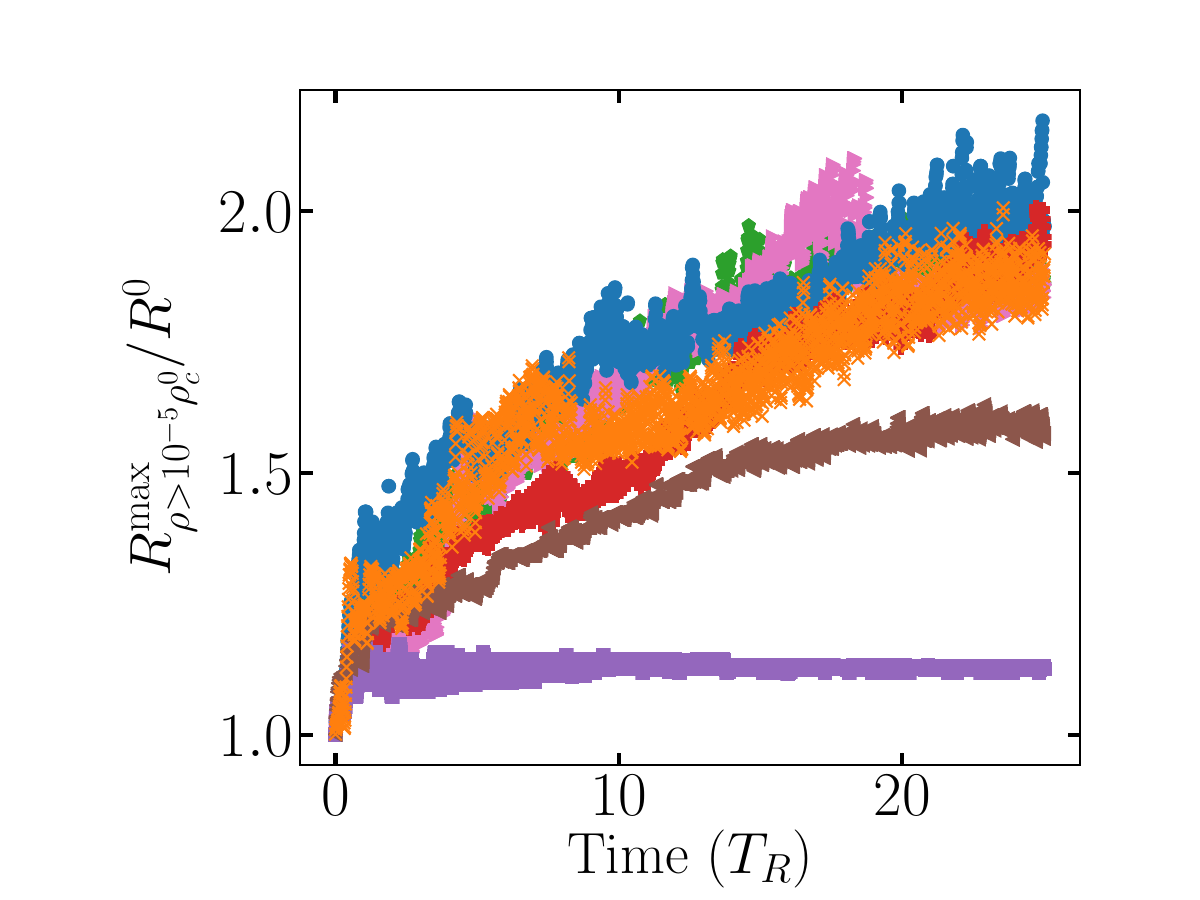}}%
    \subfloat[Diffusion of high density gas, rotating frame]{\includegraphics[width=6.5cm]{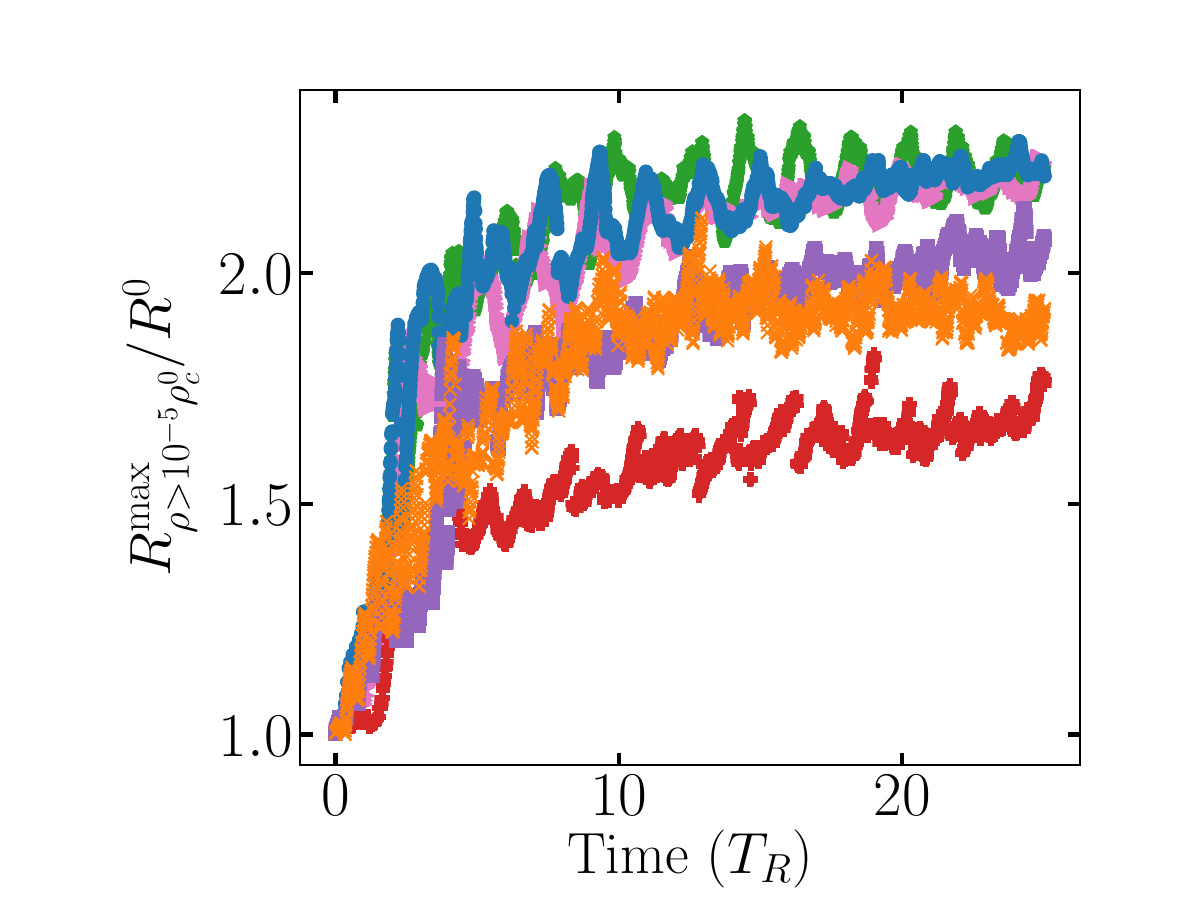}}
    \qquad
    \subfloat[Final rotation profile, inertial frame]{\includegraphics[width=6.5cm]{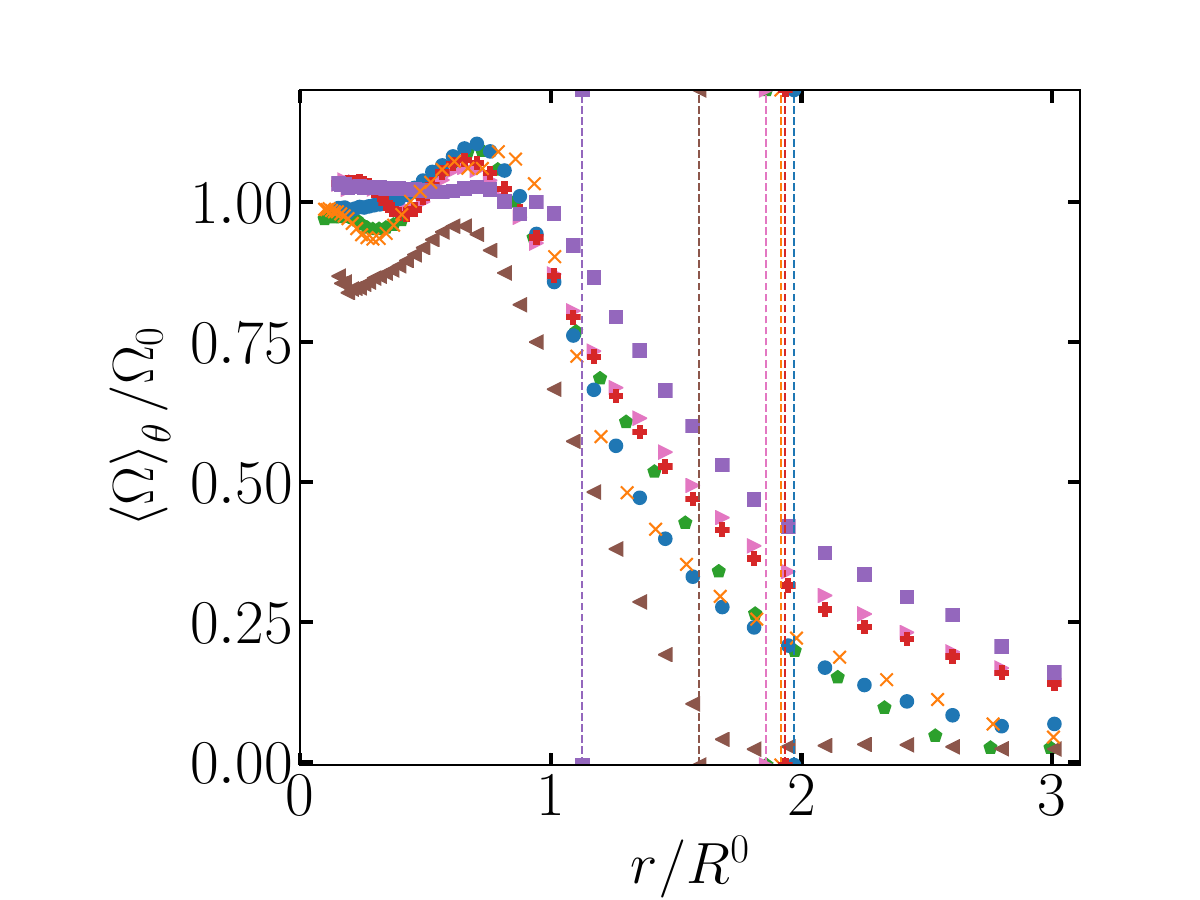}} \subfloat[Final rotation profile, rotating frame]{\includegraphics[width=6.5cm]{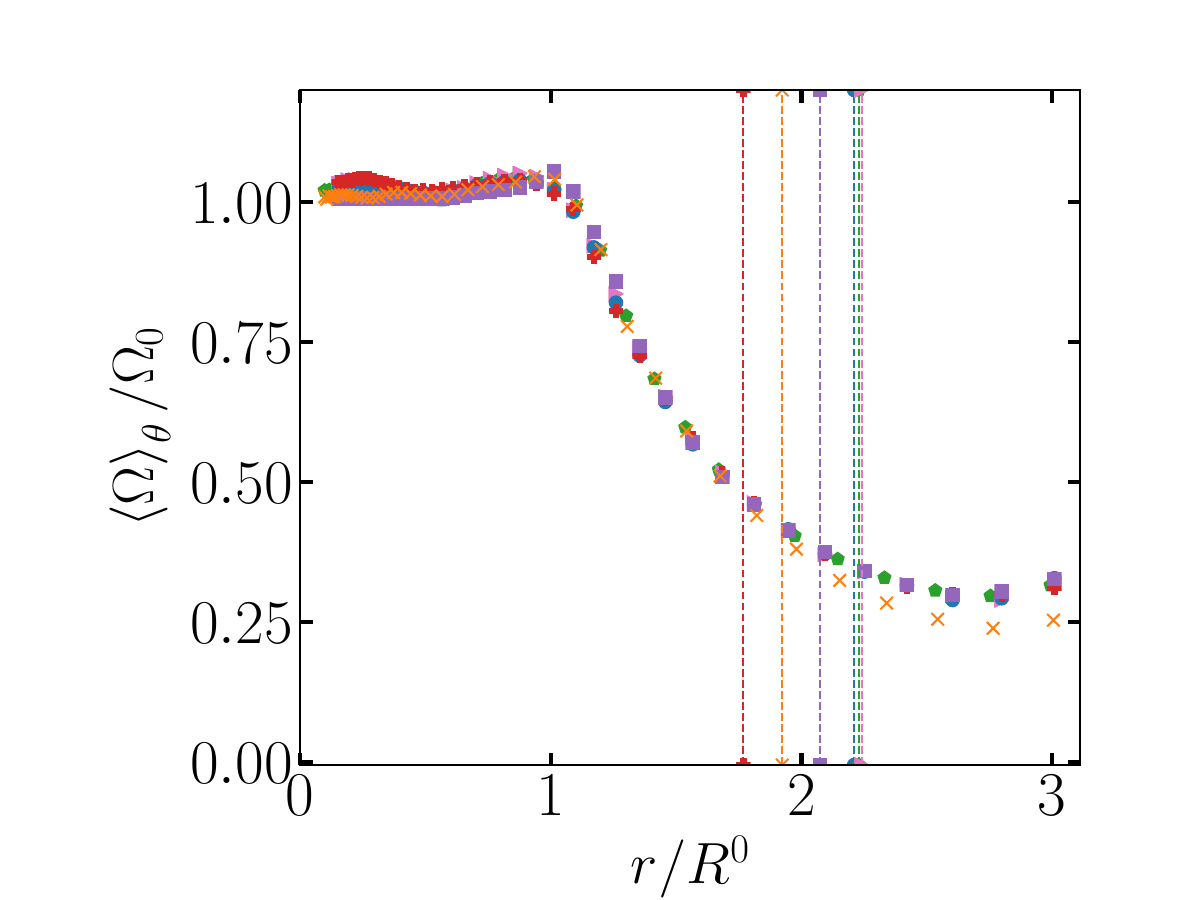}}%
        \caption{\protect\footnotesize{The rotating star test with $\Omega_0 = 0.52$ code units \revision{($0.0023$  rad~s$^{-1}$ and $0.19$ rad~s$^{-1}$ for the MS and the WD models, respectively)}. Several quantities are compared. The vertical lines in panels (g) and (h) are the maximum distance of high density gas at the end of the simulation (rightmost values of the curves in panel e and f, respectively)}}
\label{fig:rotating_star_octo193}
\end{figure*}


\begin{figure*}
\centering
    \subfloat[Total energy, inertial frame]{\includegraphics[width=6.5cm]{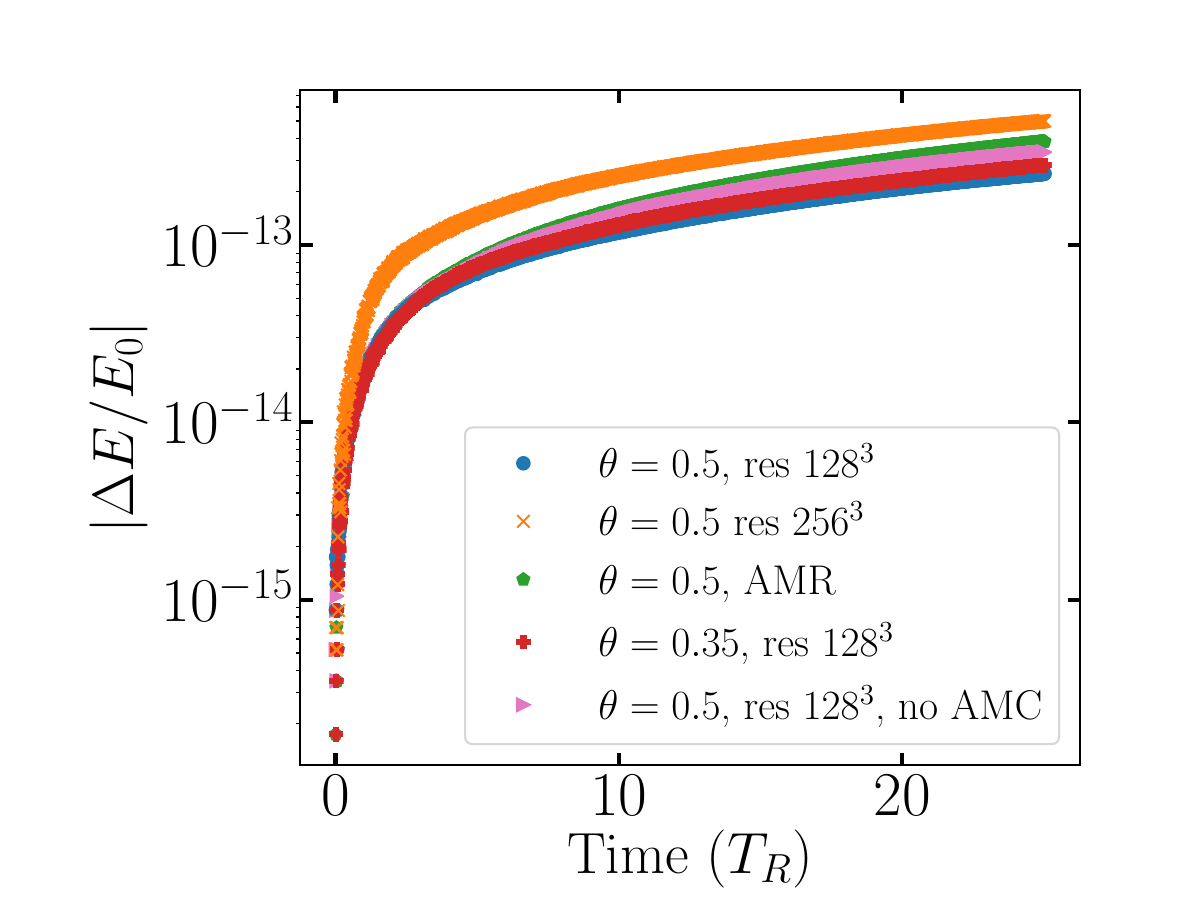}}%
    \subfloat[Total energy, rotating frame]{\includegraphics[width=6.5cm]{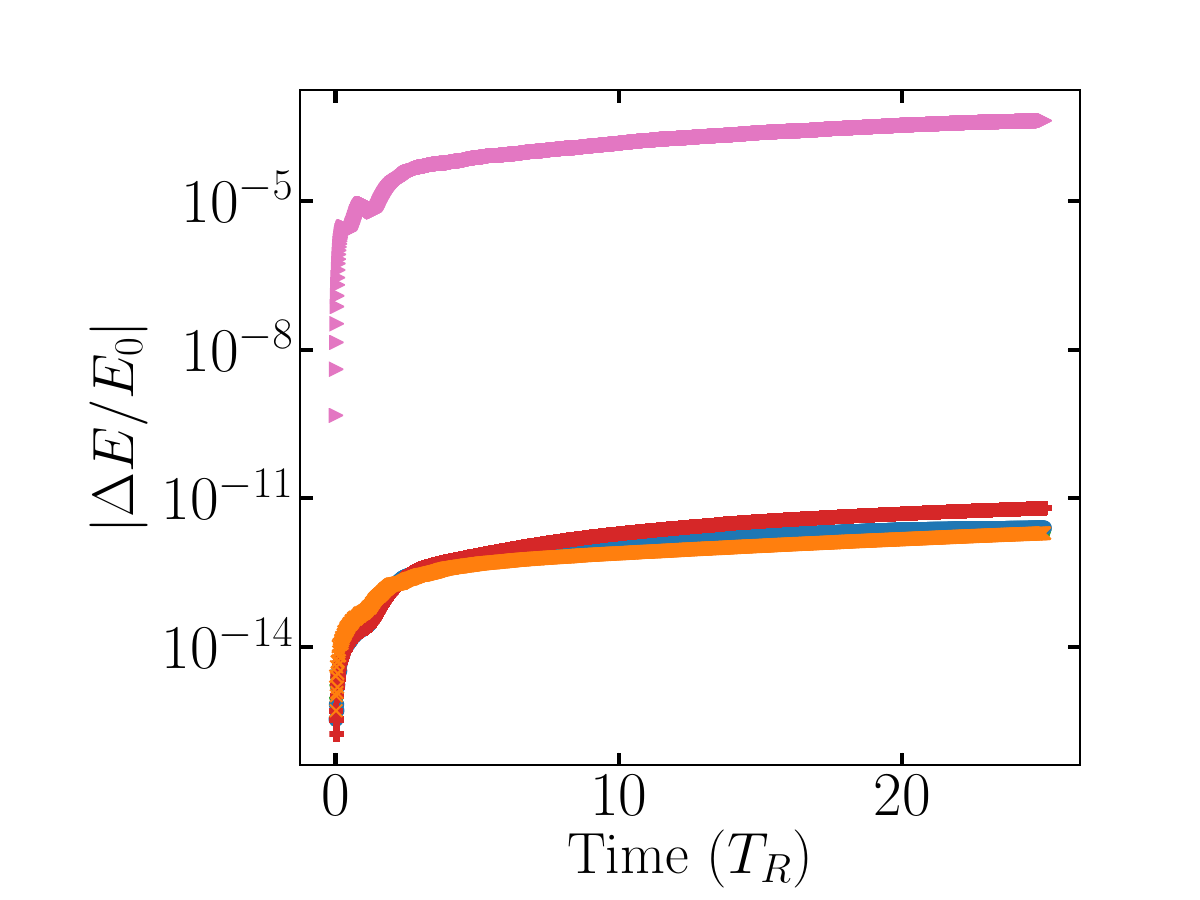}}%
    \qquad
    \subfloat[Total z-angular momentum, inertial frame]{\includegraphics[width=6.5cm]{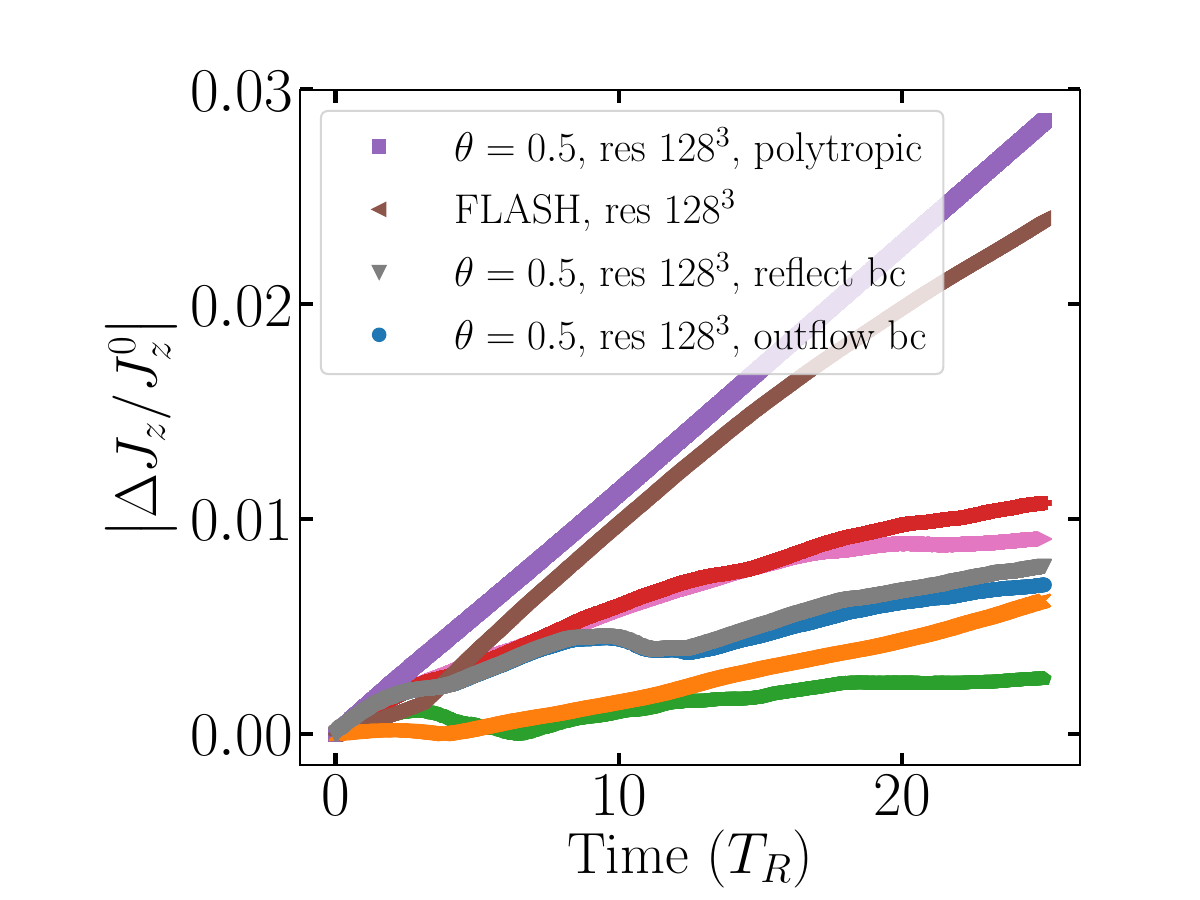}}%
    \subfloat[Total z-angular momentum, rotating frame]{\includegraphics[width=6.5cm]{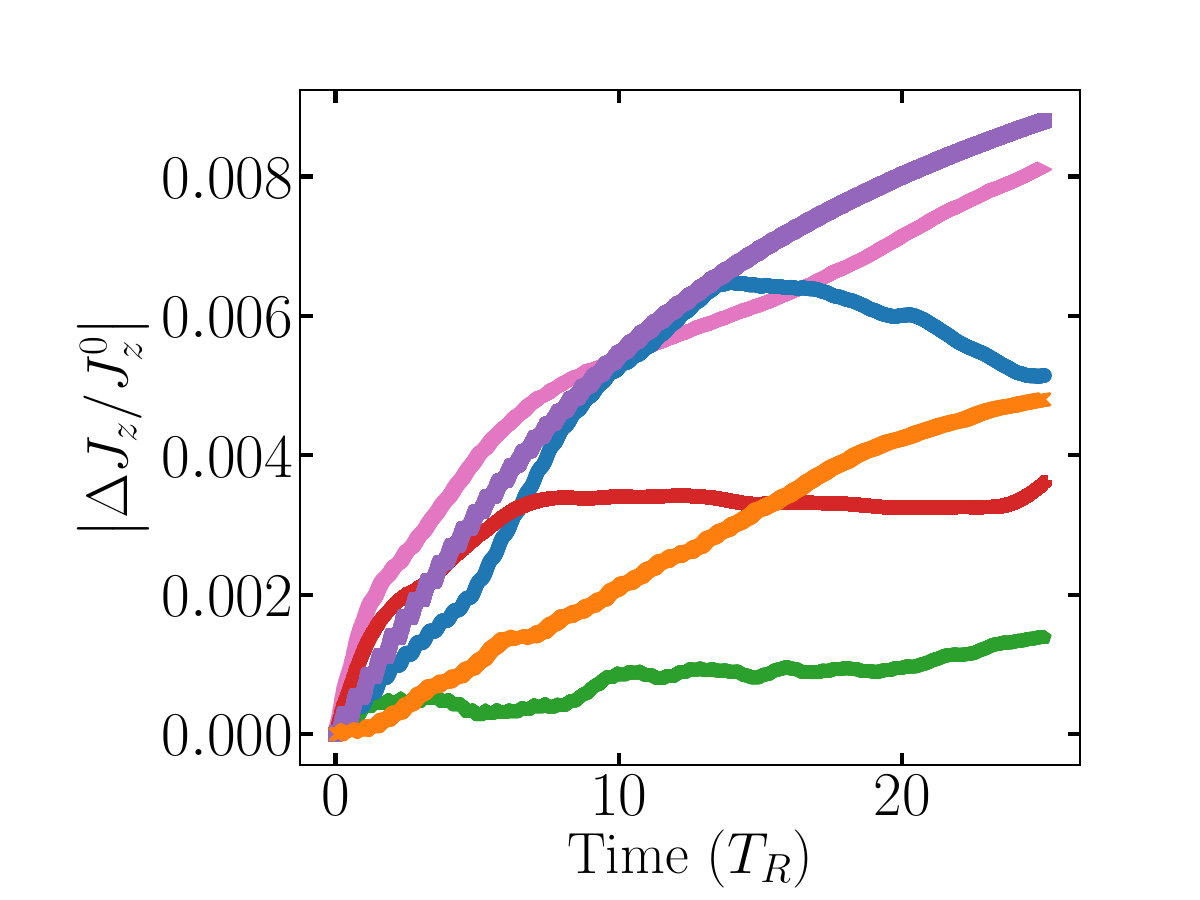}}%
     \caption{\protect\footnotesize{The rotating star test with $\Omega_0 = 0.52$ code units \revision{($0.0023$ rad/s and $0.19$ rad/s for the MS and the WD models, respectively)}. Conservation of energy and angular momentum. \revision{All simulations used an ideal gas equation of state except the simulation shown in purple squares, which used a polytropic equation of state}}}
\label{fig:rotating_star_octo193_conserved}
\end{figure*}

\subsection{Star Rotating in the Grid}
\label{ssec:rotating_star}

Here we simulate a rotating star in equilibrium with \octo\ and \flash. Its rotation profiles, $\Omega (r)$, should remain flat whether the simulation is run in the inertial frame or in a rotating frame of reference, over a certain amount of time. 

We constructed an equilibrium profile of a rotating polytrope using the Self Consistent Field method that will be described in Section~\ref{ssec:SCF}. We then interpolated the profile in the \octo\ and \flash\ grids using the same method. The fast-spinning star has an oblate shape with a ratio of 2/3 between polar and equatorial radii. The angular velocity is $0.52$ in code units ($0.0023$ rad~s$^{-1}$ and $0.19$~rad~s$^{-1}$ for the MS and the WD models, respectively), and the rotational period is $\approx 12$ code units ($45\;{\rm min}$ and $32.3\;\sec$ for the MS and the WD models, respectively).

We carry out simulations in the inertial frame (both with \octo\ and \flash) and in the rotating frame (only with \octo). For each of these frames of reference, we carry out six simulations with the same combinations of resolution, $\theta$ parameter and EoS as done for the stationary star in Section~\ref{ssec:polytrope}.

In Figures~\ref{fig:rotating_star_octo193}(a) and (b) we plot the familiar evolution of the core density scaled to the initial value. With an exception of the \flash\ simulation that has a declining trend, reaching 85 percent of its initial values over 25 rotation periods, all the \octo\ simulations retain the initial core density value within 5 percent with no noticeable difference between rotating and inertial frame. In Figure~\ref{fig:rotating_star_octo193}(c), we plot the diffusion of the stellar mass out of the original boundary in the inertial frame. Once again \flash\ performs worse with a loss of 0.5 percent of the mass over 25 rotation periods, but most of the \octo\ profiles are a close second with the polytropic EoS simulation performing distinctly better. Comparing this with panel (d) we see how all \octo\ simulations perform much better in the rotating frame, losing at most 0.05 percent of the mass. Similarly, Figure~\ref{fig:rotating_star_octo193}(e) shows that in the inertial frame, the high density inner star diffuses out moving to twice its original radius. The polytropic EoS simulation is by far the best, with \flash\ also performing well. In this case the rotating frame, panel (f), does not perform a great deal better than the inertial frame.

The acid test is the evolution of the rotation profile. All simulations in the inertial frame (Figure~\ref{fig:rotating_star_octo193}(g)) do not retain a flat profile (with the notable exception of the polytropic EoS one), while those in the rotating frame (Figure~\ref{fig:rotating_star_octo193} (h))  systematically do. However, in the inertial frame, by the end of the simulations, the spin of the gas that has diffused out of the original volume is distinctly less for all simulations than in the rotating frame. The \flash\ simulation (in the inertial frame) does not retain a flat profile, but it has the least diffusion of gas out of the boundary of the sphere of all simulations.

Finally, in Figure~\ref{fig:rotating_star_octo193_conserved} we show the degree of conservation we achieve in the \octo\ simulations. \revision{ This highlights the importance of including the angular momentum correction (AMC) of the gravity solver for the conservation of energy (upper panels) as outlined in Section~\ref{energy_equation}. In the lower panels we plot the deviation from angular momentum conservation $\Delta J_z / J_z^0$, where $\Delta J_z = J_z\left(t\right)-J_z^0$, and $J_z\left( t \right) = \int_V \left(x s_y - y s_x \right) dV$ and $J_z^0$ are the z-angular momentum inside the simulation domain, and the initial z-angular momentum, respectively, and where $s_x$ and $s_y$ are the x and y components of the inertial momenta, respectively. As we do not take into account outflows we can compare between \octo\ and \flash, showing \octo\ outperforms \flash\ in ideal gas simulations. Additionally,  by comparing between simulations with outflow (blue circles) and reflect (gray downward pointing triangles) boundary conditions, we find that outflows only very slightly affect angular momentum conservation. Overall, the AMR simulation conserves angular momentum the best, and simulations that include AMC conserve angular momentum better than the simulation without this correction. } 


In conclusion this test gives an excellent idea of the type of numerical diffusion we can expect. The rotating frame outperforms the inertial frame, and the polytropic EoS seems to be doing best in almost all cases. \flash\ performs the worst except in retaining a stable high density sphere inside the star and it can also be seen that a higher gravity accuracy ($\theta = 0.35$) improves the performance.

\noindent Click on the link\footnote{\url{https://youtu.be/7ZrxGJW2J_Y}} to view a movie of the rotating star in the rotating frame.

\section{Binary Merger Between Two Polytropes with a Mass Ratio of $0.5$}
\label{sec:binarymerger}

Here we present the results of two binary merger simulations performed in the rotating frame, starting from  identical initial conditions, but differing in the adopted EoS. The donor star is half the mass of the accretor and initially fills its Roche lobe. Both stellar spins are synchronized to the orbital frequency. In the first simulation, the two stars are constructed using the same polytropic EoS ($\epsilon_1=\epsilon_2=1$), with polytropic index $n=\frac{3}{2}$. In the second simulation, we use an ideal gas EoS (with dual energy parameters $\epsilon_1=0.001$ and $\epsilon_2=0.1$ from Equations~\ref{eq:dual_energy1} and \ref{eq:dual_energy2}).

We are motivated to model such a system for two reasons. First, by appropriate choice of units, this system approximates a binary consisting of two low mass WDs. We have set the unit conversion factors so that the accretor has a mass of $0.6$~M$_\odot$, the donor a mass of $0.3$~M$_\odot$ (1 code mass unit is 1 solar mass). The computational domain is 40 code length units on a side, or $1.7\times 10^{11}$~cm (1 length units is $4.36\times 10^9$~cm). With this scaling the separation between the two stars is $6 \times 10^{-2}$~R$_\odot$, and the period is $2.7$~minutes. The second reason is that a similar system was modelled by \citet{Motl2017} providing us with a comparison simulation as a further verification of \octo. 
 
We use 9 levels of refinement, with the highest level having a grid spacing of $4.2 \times 10^7$~cm. The accretor is $50$ cells across along its longest axis (the spinning stars have slightly larger equatorial than polar radii) and the donor is $79$ cells. 

Following \cite{Motl2017}, we initially drive the stars into deep contact by systematically removing angular momentum. This is accomplished by adding source terms to the evolution equations for the x and y components of the momentum (Equation~\ref{rhou}), 
\begin{equation}
\begin{matrix}
    S_{s_x,\mathrm{driving}} = -\frac{y}{x^2+y^2}\left(x u_y - y u_x\right) f_{\mathrm{driving}} \\ 
    S_{s_y,\mathrm{driving}} = +\frac{x}{x^2+y^2}\left(x u_y - y u_x\right) f_{\mathrm{driving}}
\end{matrix}
\end{equation}
where $x$ and $y$ are the x and y components of the position vector $\vec{x}$, $u_x$ and $u_y$ are the x and y component of the inertial frame velocity, $\vec{u}$, and $f_{\mathrm{driving}}$ is the driving rate in units of inverse time. This has the effect of reducing the z-angular momentum component of a given cell at a logarithmic rate of $f_{\mathrm{driving}}$. It does not affect the cylindrical radial or z linear momenta. Similarly to what was done by \citet{Motl2017}, we drive the system together at a rate of 1 percent per orbit for the first $2.7$ orbits.

\subsection{The Initial Stellar Model}
\label{ssec:SCF}

The initial conditions for our binary simulations were produced using our own variant of the ``Self Consistent Field" (SCF) method (\cite{Hachisu1986a,Hachisu1986b,Even2009,Kadam2016}). Given a barotropic EoS, $p=p(\rho)$, the SCF method allows one to construct an equilibrium model of a binary system with synchronously rotating stars. \octo's SCF can be used to construct systems with a polytropic or bi-polytropic EoS or a cold white dwarf EoS.  The user selects the masses of each star and the Roche lobe filling factor for the donor. When each star has a different EoS, the Roche lobe filling factor for the accretor must also be specified.  Here we document \octo's SCF as it pertains to the construction of the initial conditions for the binary models in this paper. These models use the same structural polytropic EoS for each star. One of the models is evolved with the same polytropic EoS, and the other with an ideal gas EoS. 

The effective potential is defined as
\begin{equation}
    \phi_{\mathrm{eff}}=  \phi - \frac{1}{2}\Omega_{\rm orb} r^2,
    \label{effective}
\end{equation}
where $\phi$ is the gravitational potential, $\Omega_{\rm orb}$ is the rotational frequency of the binary, and $r$ is the distance to the axis of rotation. Using the effective potential, the hydrostatic equilibrium equation in the rotating frame can be written
\begin{equation}
   \frac{1}{\rho}  \gradient{p} + \phi_{\mathrm{eff}} = 0.
   \label{hydrostatic}
\end{equation}
For the EoS we use:
\begin{equation}
p_{\mathrm{poly}} = K \rho^{1+\frac{1}{n}},
\label{polytropic}
\end{equation}
where $K$ is the polytropic constant and $n$ is the polytropic index.
Combining Equations \ref{hydrostatic} and \ref{polytropic} and integrating we arrive at:
\begin{equation}
K\left( 1 + n\right) \rho^{\frac{1}{n}} + \phi_{\mathrm{eff}} = C_{1|2},
\label{hydrostatic2}
\end{equation}
where the constant on the RHS is either $C_1$ for the primary, accreting star or $C_2$ for the secondary, donor star. While in general the polytropic constant, $K$, may be different for each star, in this paper, they are the same. 

The SCF solves for the initial conditions iteratively. The user selects the mass of each star, the polytropic index, the initial separation, and the Roche filling factor of the donor. (Note that while the user specifies the initial separation, it is the orbital angular momentum that is held constant.) The Roche filling factor is defined as 
\begin{equation}
    f = \frac{\phi_{\mathrm{edge}} - \phi_C}{\phi_{L1} - \phi_C}, 
\end{equation}
where $\phi_{\mathrm{edge}}$ is the effective potential at the edge of the donor $\phi_C$ is the  effective potential at the centre of mass of the donor, and $\phi_{L1}$ is the effective potential at $L_1$, the first Lagrange point. 

\begin{figure*}
\centering   
    \subfloat[Orbital density slice, polytropic EoS]{\includegraphics[width=6cm]{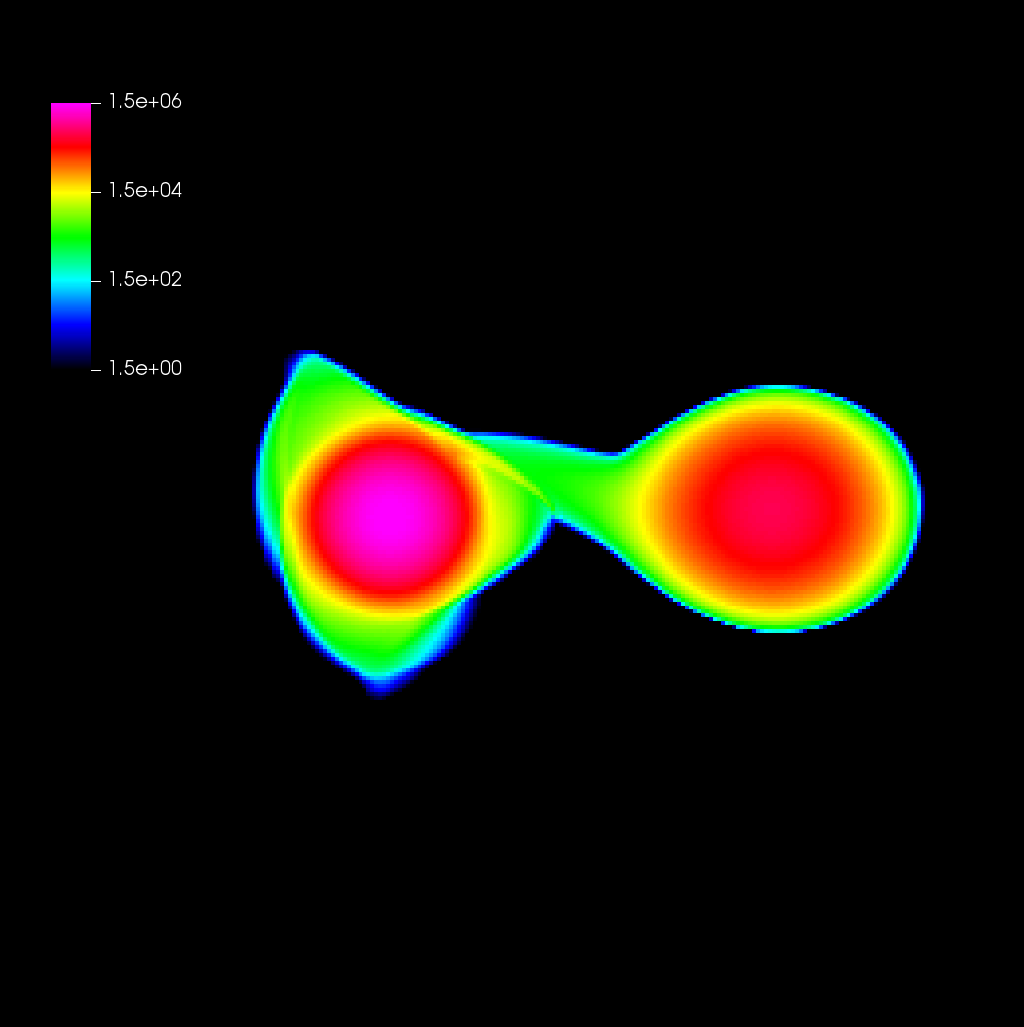}}
    \subfloat[Orbital density slice, ideal gas EoS]{\includegraphics[width=6cm]{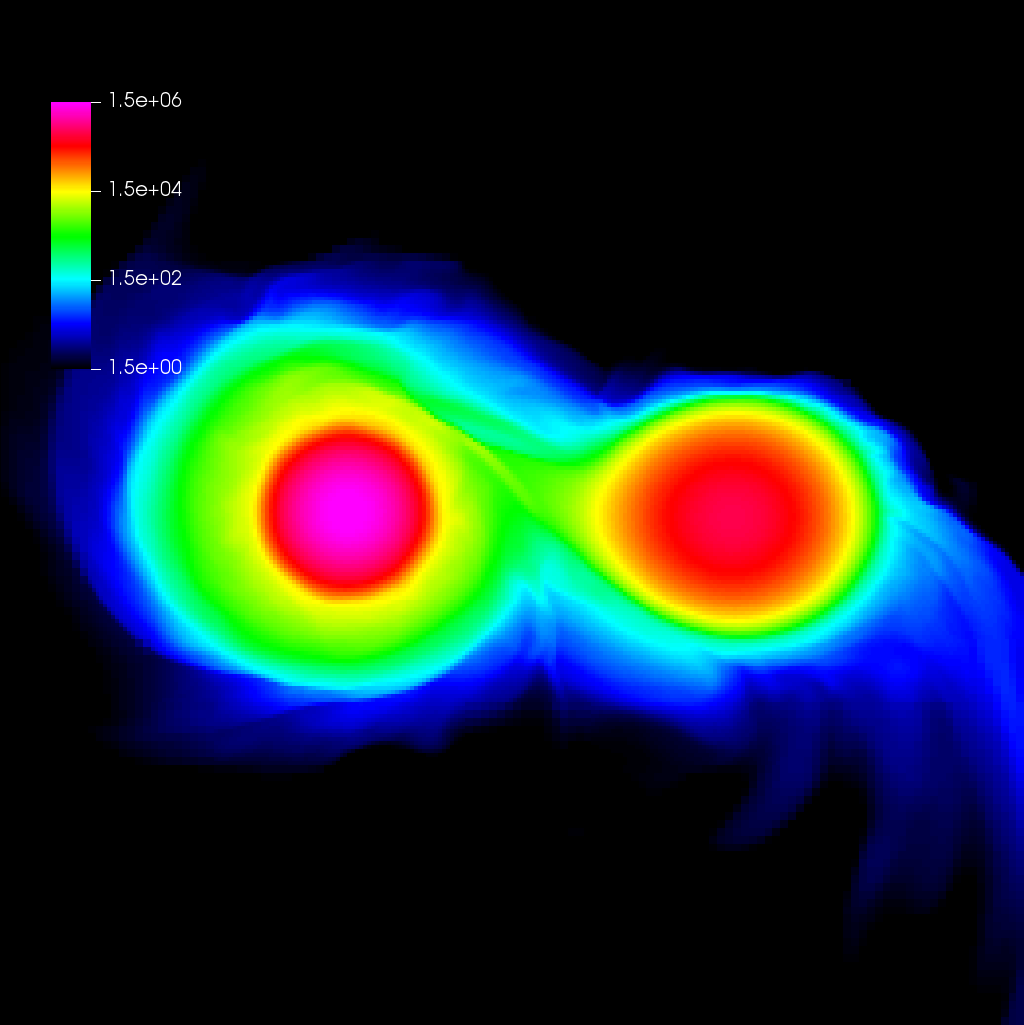}}
        \qquad
    \subfloat[Orbital donor fraction slice, polytropic EoS]{\includegraphics[width=6cm]{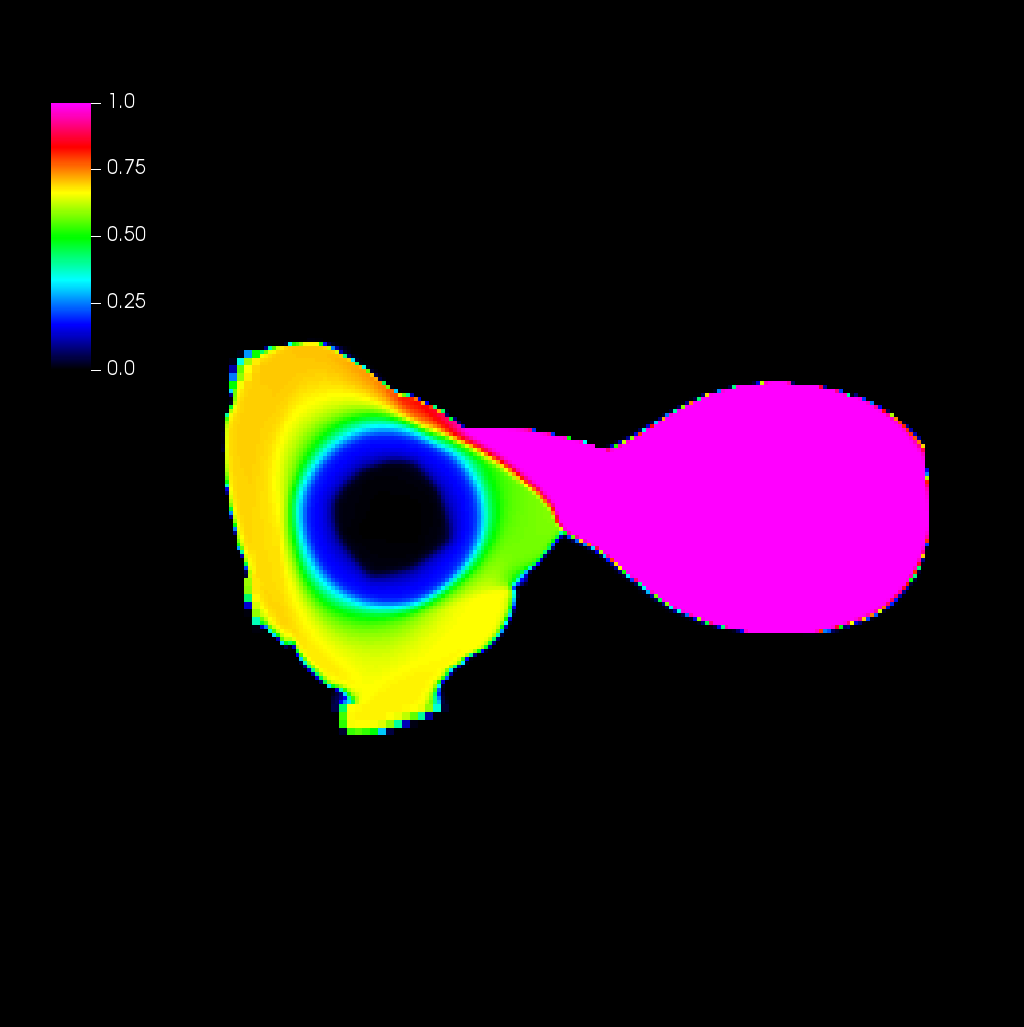}}
    \subfloat[Orbital donor fraction slice, ideal gas EoS]{\includegraphics[width=6cm]{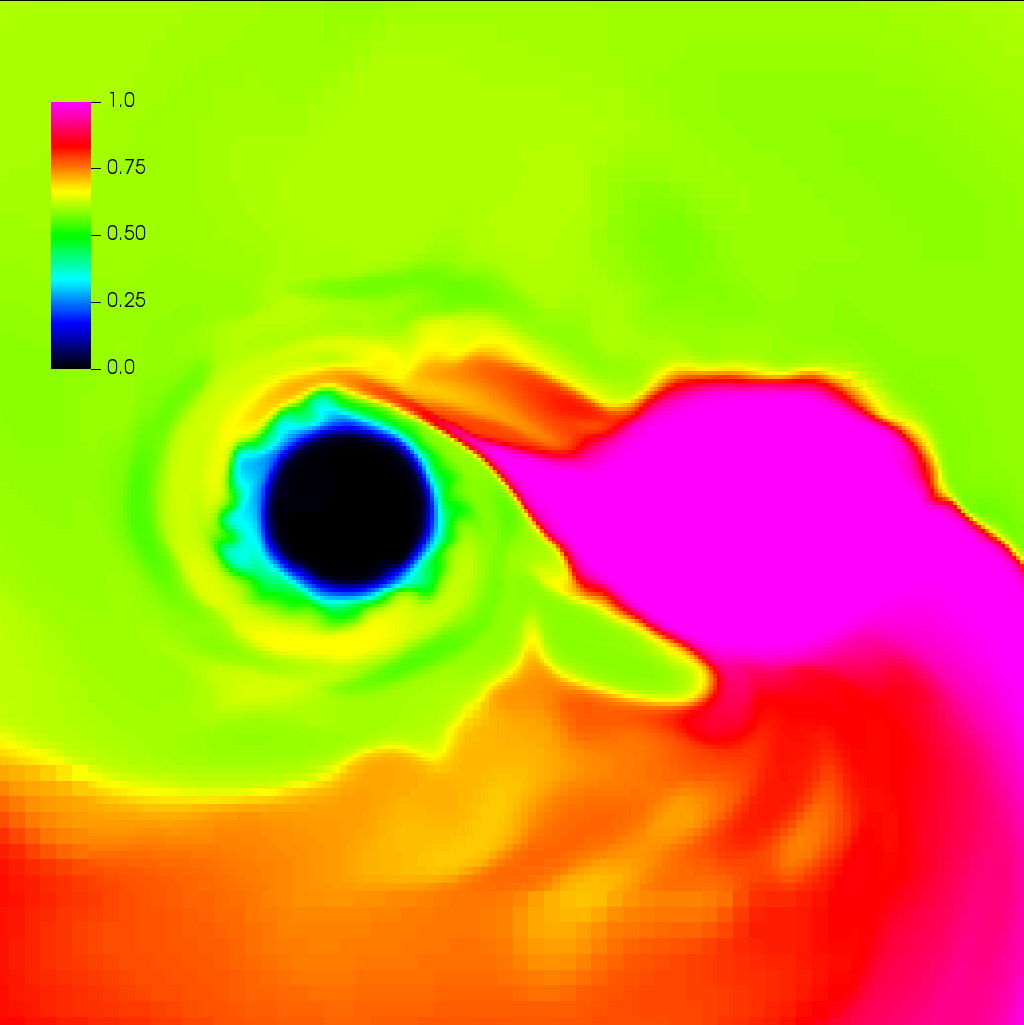}}
        \qquad
    \subfloat[Vertical density slice, polytropic EoS]{\includegraphics[width=6cm]{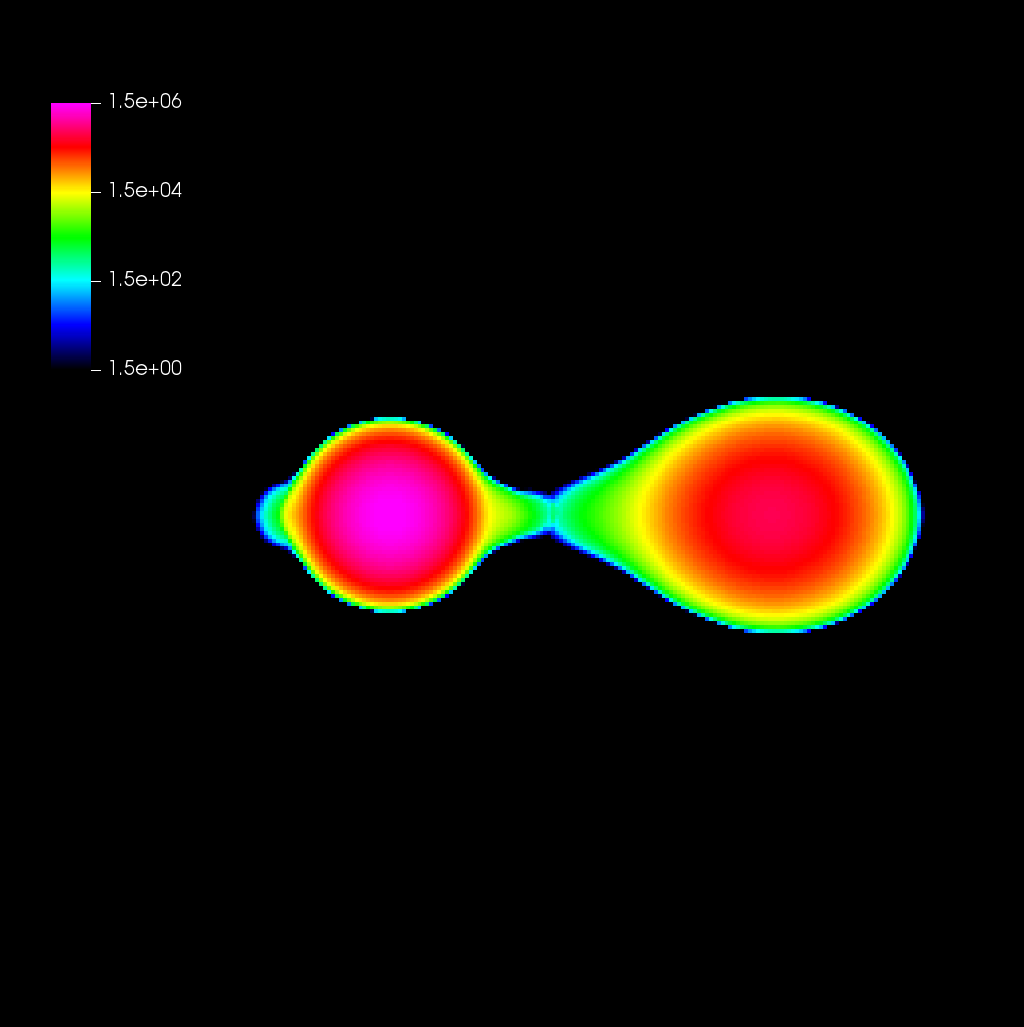}}
    \subfloat[Vertical density slice, ideal gas EoS]{\includegraphics[width=6cm]{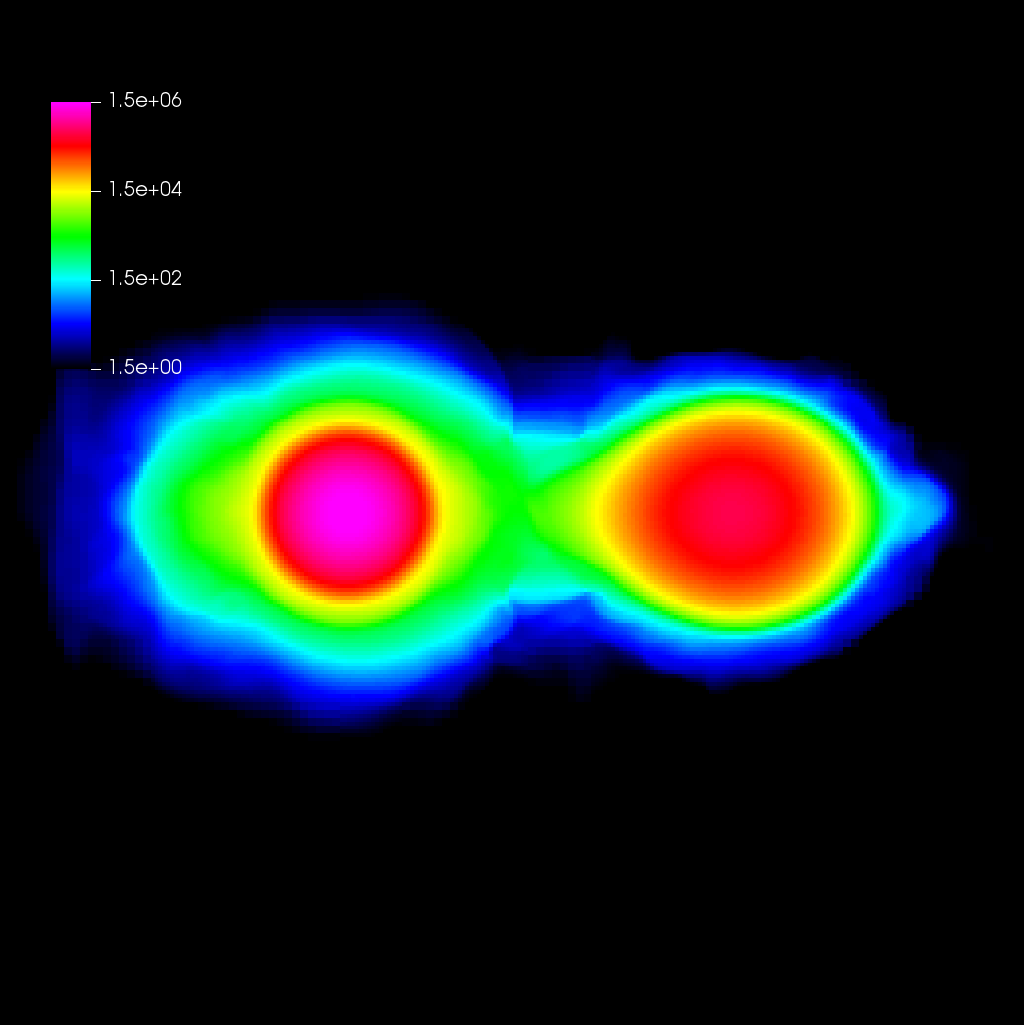}}
\caption{\protect\footnotesize{Snapshots from our $q=0.5$ simulations with a polytropic EoS (left column, at time $11.7 \times P_0$, where $P_0$ is the initial orbital period) or an ideal gas simulation (right column, at time $12.6 \times P_0$). The box size is $11 \times 10^9$~cm in all cases. The density scale is logarithmic and runs from $1.5 \ \mathrm{g}/\mathrm{cm}^3$ to  $1.5  \times 10^6  \ \mathrm{g}/\mathrm{cm}^3$} }
\label{fig:q0.5a}
\end{figure*}

For the initial iteration we place two polytropic stars in the grid at the desired separation and small enough that they are within their respective Roche lobes.  The gravity solver is also called before the iterations begin. For each iteration we: (1) multiply the densities of each star by a constant factor for each star, such that the result yields the desired mass of each star; (2) advect the entire grid such that the centre of mass lies at the centre of coordinates; (3) set the new value for $\Omega_{\rm orb}$ using
\begin{equation}
\Omega_{\rm orb} \rightarrow 
\frac{J_{0,\rm orb} (M_1+M_2)}{M_1 M_2 a^2},
\end{equation}
where $J_{0,\rm orb}$ is the  initial orbital angular momentum, $M_1$ and $M_2$ are the fixed masses of the accretor and donor, and $a$ is the orbital separation for the current iteration; (4) compute the integration constants $C_1$ and $C_2$, 
\begin{figure*}
\centering   
    \subfloat[$t=7.3 \times P_0$, polytropic EoS]{\includegraphics[width=6cm]{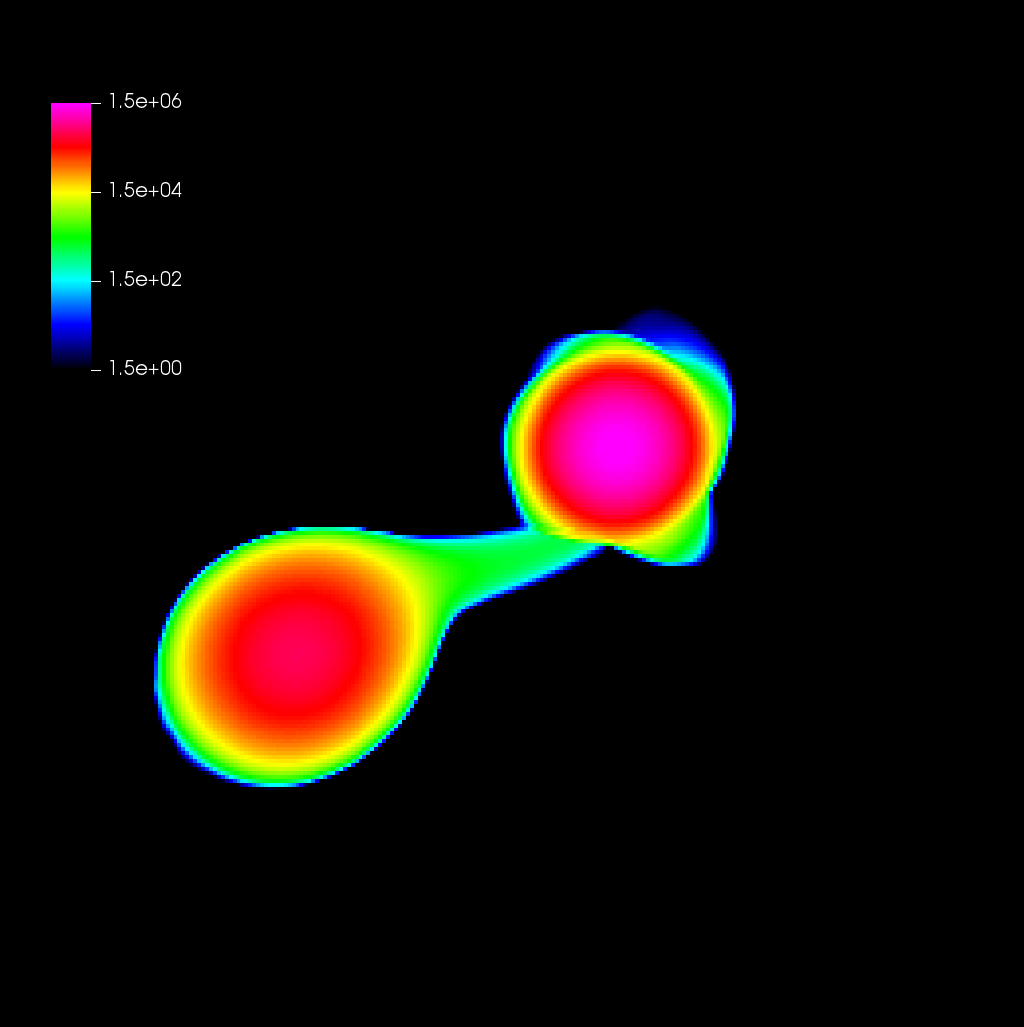}}
    \subfloat[$t=6.7 \times P_0$, ideal gas EoS]{\includegraphics[width=6cm]{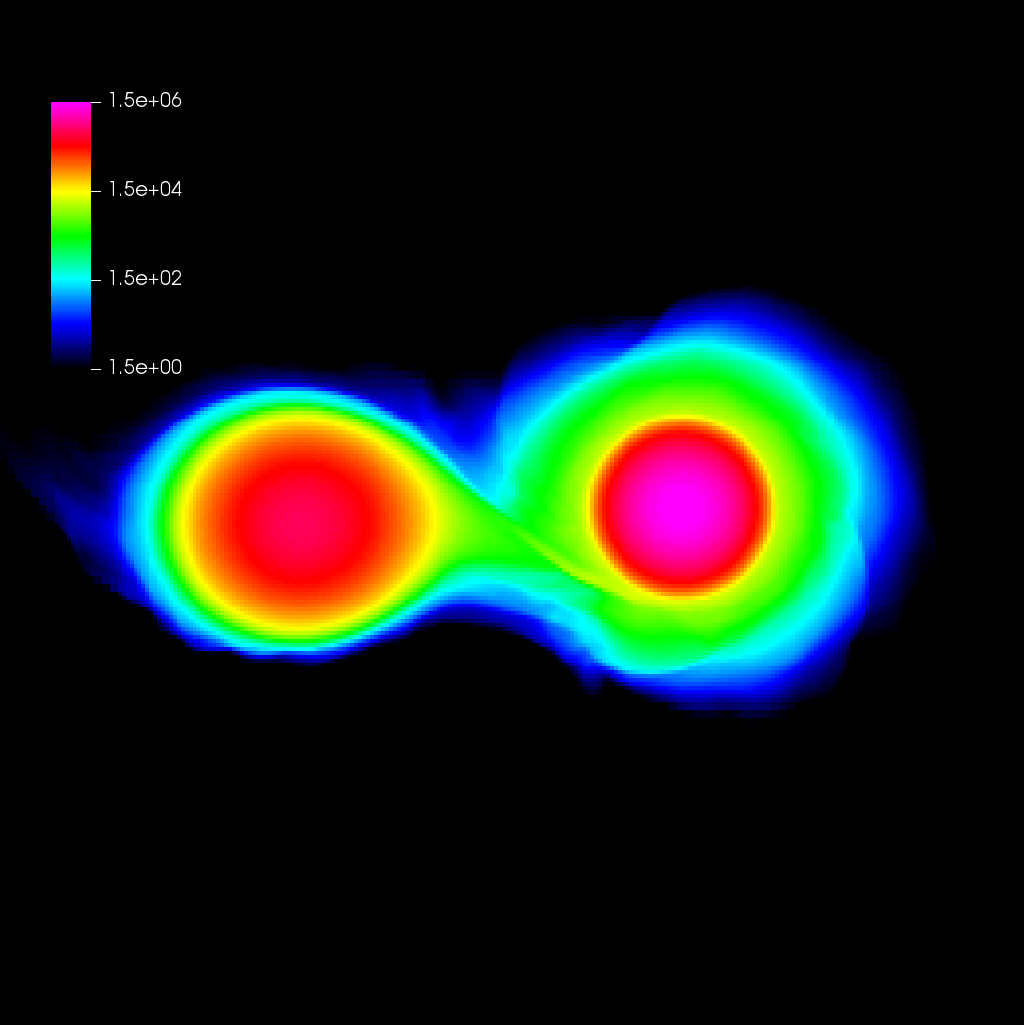}}
        \qquad
    \subfloat[$t=21.2 \times P_0$, polytropic EoS]{\includegraphics[width=6cm]{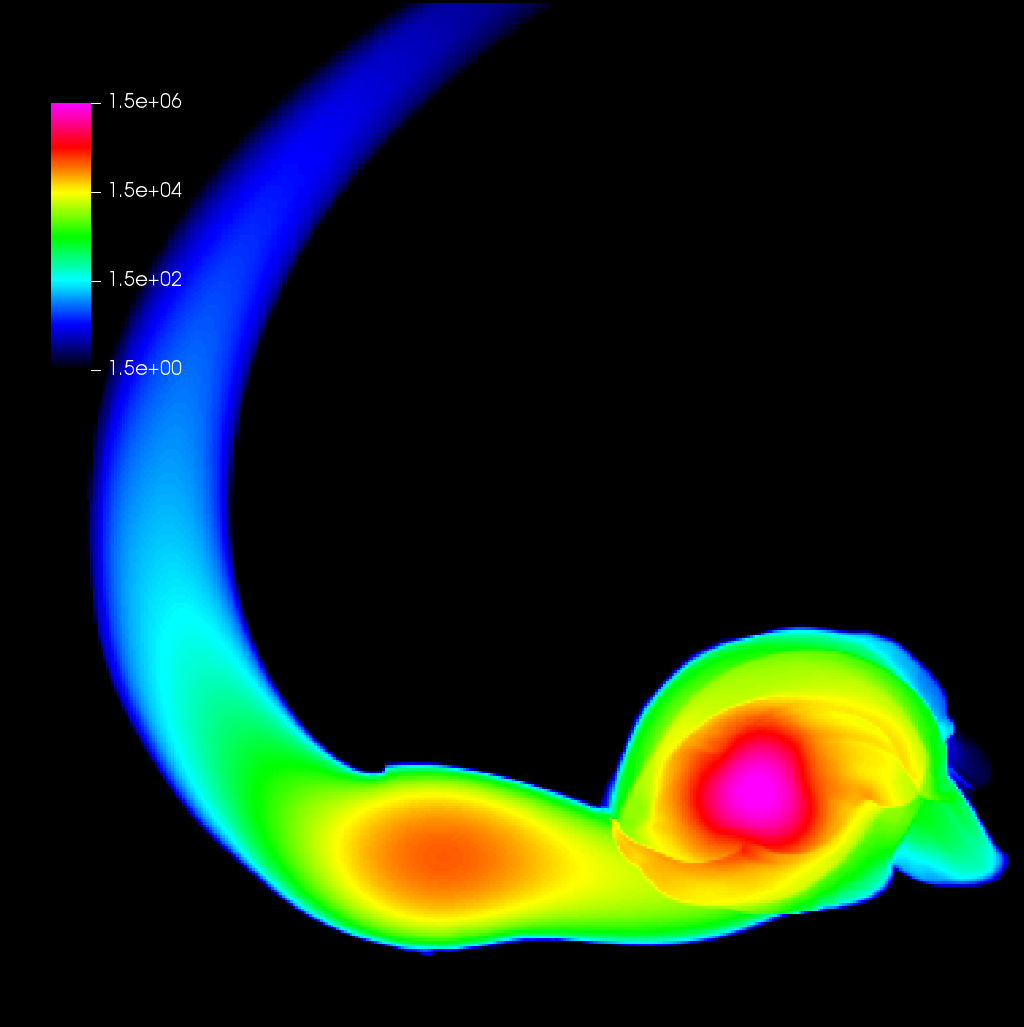}}
    \subfloat[$t=19.6 \times P_0$, ideal gas EoS]{\includegraphics[width=6cm]{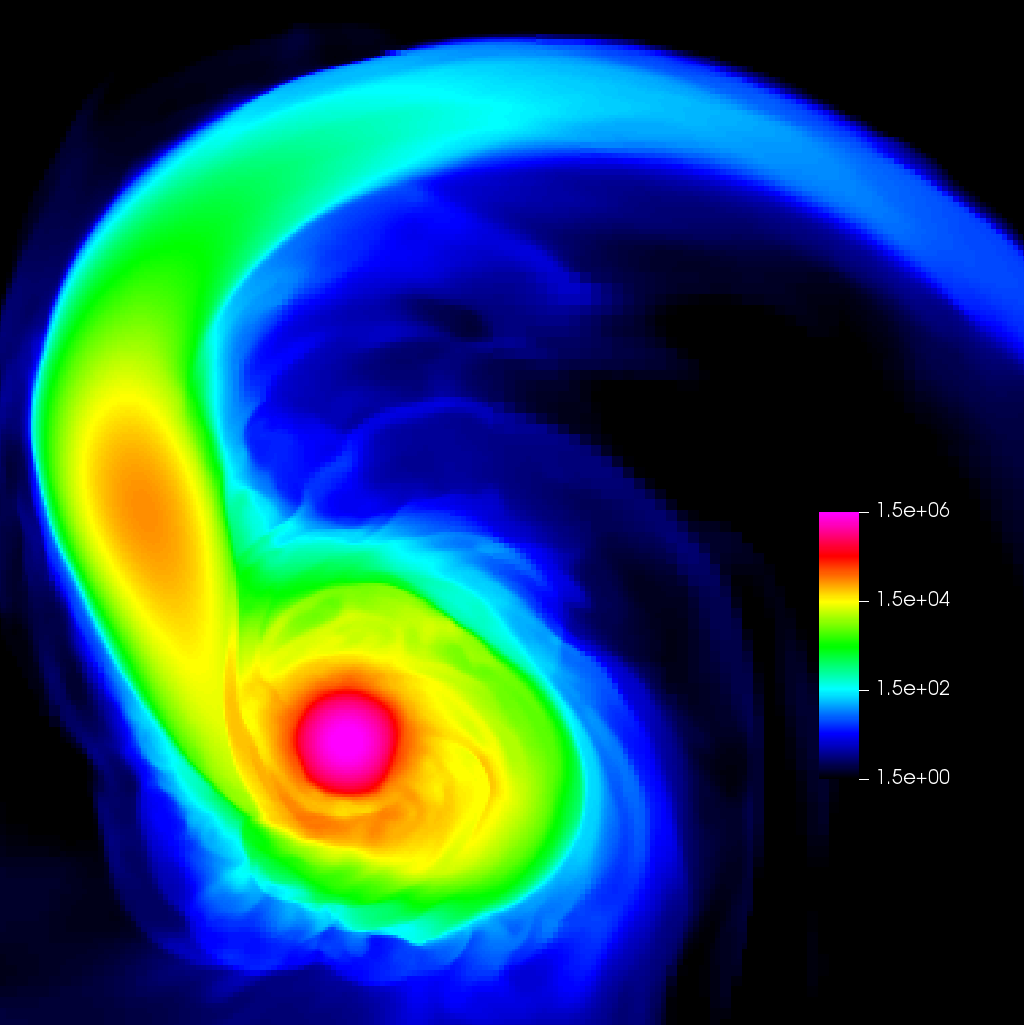}}
 
\caption{\protect\footnotesize{Density slices on the orbital plane for our $q=0.5$ simulations with a polytropic EoS (left column) or an ideal gas EoS (right column). The box size is $11 \times 10^9$~cm (top row) or $16 \times 10^9$~cm (bottom row).  The density scale is logarithmic and runs from $1.5 \ \mathrm{g}/\mathrm{cm}^3$ to  $1.5 \times 10^6 \ \mathrm{g}/\mathrm{cm}^3$. The snapshots are taken at times 7.3 (a), 6.7 (b), 21.2 (c) and $19.6 \times P_0$ (d). See Table 1 for description} }
\label{fig:q0.5b}
\end{figure*}

\begin{equation}
    C_1 = K\left( 1 + n\right) \rho_1^{\frac{1}{n}} + \phi_1,
\end{equation}
and 
\begin{equation}
    C_2 = \left(1-f\right) \phi_C + f \phi_{L1}, 
\end{equation}
where $\rho_1$ is the maximum density of the accretor and $\phi_1$ is the effective potential at the centre of the accretor; (5) compute a new value for $K$, 
\begin{equation}
K = \frac{C_2 \phi_2}{\left(n+1\right)\rho_2^{\frac{1}{n}}},
\end{equation}
where $\rho_2$ is the density at the centre of the donor, and $\phi_2$ is the effective potential at the centre of the donor ($\phi_2$ = $\phi_C$); (6) compute a new density value at each point on the grid using
\begin{equation}
    \rho = \left(\frac{\phi_{\rm eff}- C_{1|2}}{K\left(n+1\right)}\right)^n,
\end{equation}
for the accretor or donor region as appropriate; (7) solve for the potential and repeat the process from step (1).

\subsection{Simulation Results}
\label{ssec:q0.5_results}

In each model, mass transfer begins within the first couple of orbits. The transfer rate grows and the donor is tidally disrupted at approximately $21.2 P_0$ for the polytropic and $19.6 P_0$ for the ideal gas model. Here $P_0$ is the initial orbital period, which is the same for each model. The merged remnant is then evolved for a couple of additional $P_0$.

In Figure ~\ref{fig:q0.5a}, we show various images at $t=11.7 P_0$ for the polytropic simulation and $t=12.6 P_0$ for the ideal gas simulation. Panels (a) and (b) show density slices in the equatorial plane for the two EoS choices. \octo\ is able to track the material that was originally part of the donor, and the donor mass fraction is displayed in panels (c) and (d). The polytropic model clearly exhibits the same polygonal resonances at high mass transfer rates that were first reported in \cite{D_Souza_2006} and were later confirmed in the SPH simulations reported in \cite{Motl2017}. As in \cite{Motl2017}, these resonance patterns do not appear in the ideal gas model because the accretion structure is thicker due to shock heating. In Figure~\ref{fig:q0.5a}(e)--(f) we show a density cut at the same times, showing that while the ideal gas EoS simulation produces $L_2$ and $L_3$ outflow, the polytropic EoS simulation only shows a hint of $L_2$ outflow by that time. The level of detail clearly visible near $L_1$ is consistent with the Roche geometry.

 Figure~\ref{fig:q0.5b}(a)--(b) are from earlier in the evolution at approximately $t = 7 P_0$ in each model. The resonance patterns in the polytropic model progress from pentagonal shaped, to box shaped, to triangular shaped (clearly seen in in Figure~\ref{fig:q0.5b}(a)). Figure~\ref{fig:q0.5b}(c)--(d) show the system as tidal disruption is occurring. As the donor is disrupted, significant mass flows through the $L_2$ Lagrange point, resulting in the ``tails" seen in the figure. 

\begin{figure*}
\centering   
    \subfloat[Density, $t=25.0 \times P_0$, polytropic EoS]{\includegraphics[width=6cm]{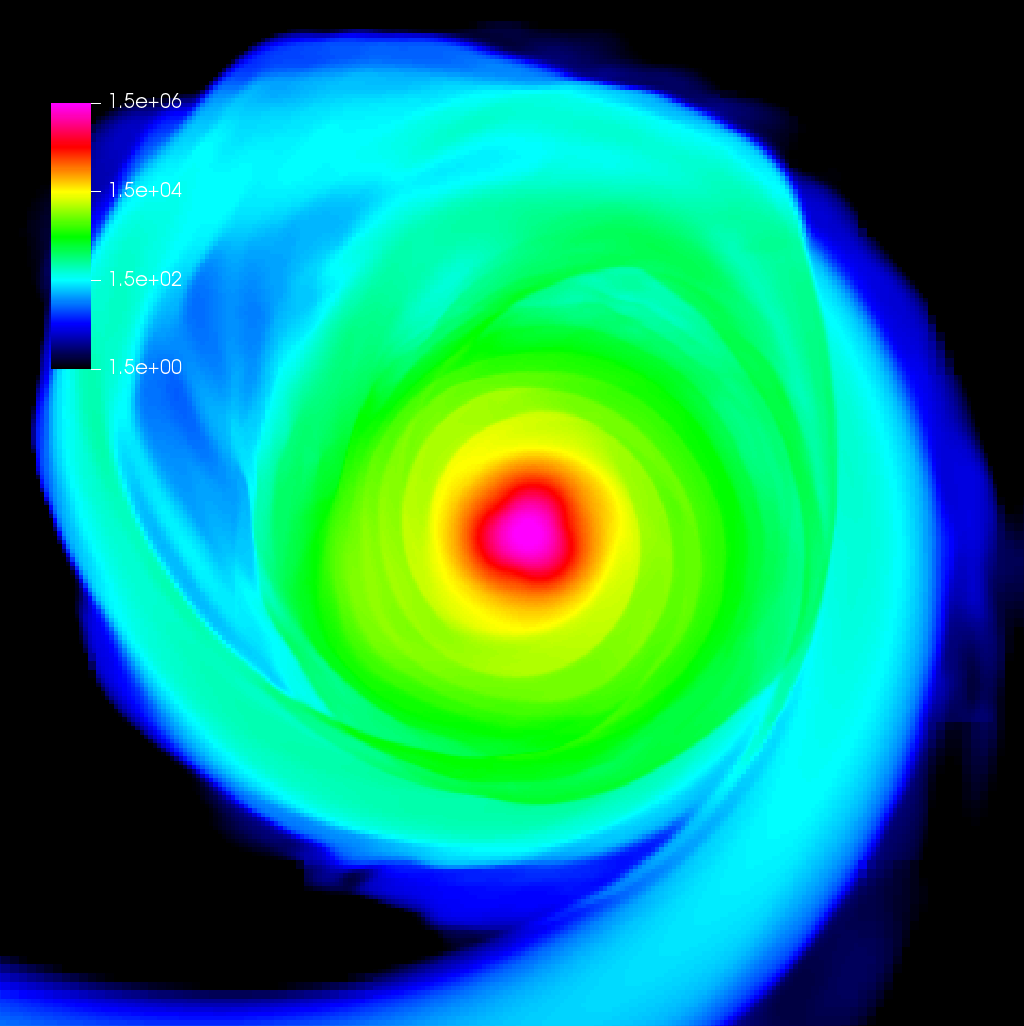}}
    \subfloat[Density, $t=23.2 \times P_0$, ideal gas EoS]{\includegraphics[width=6cm]{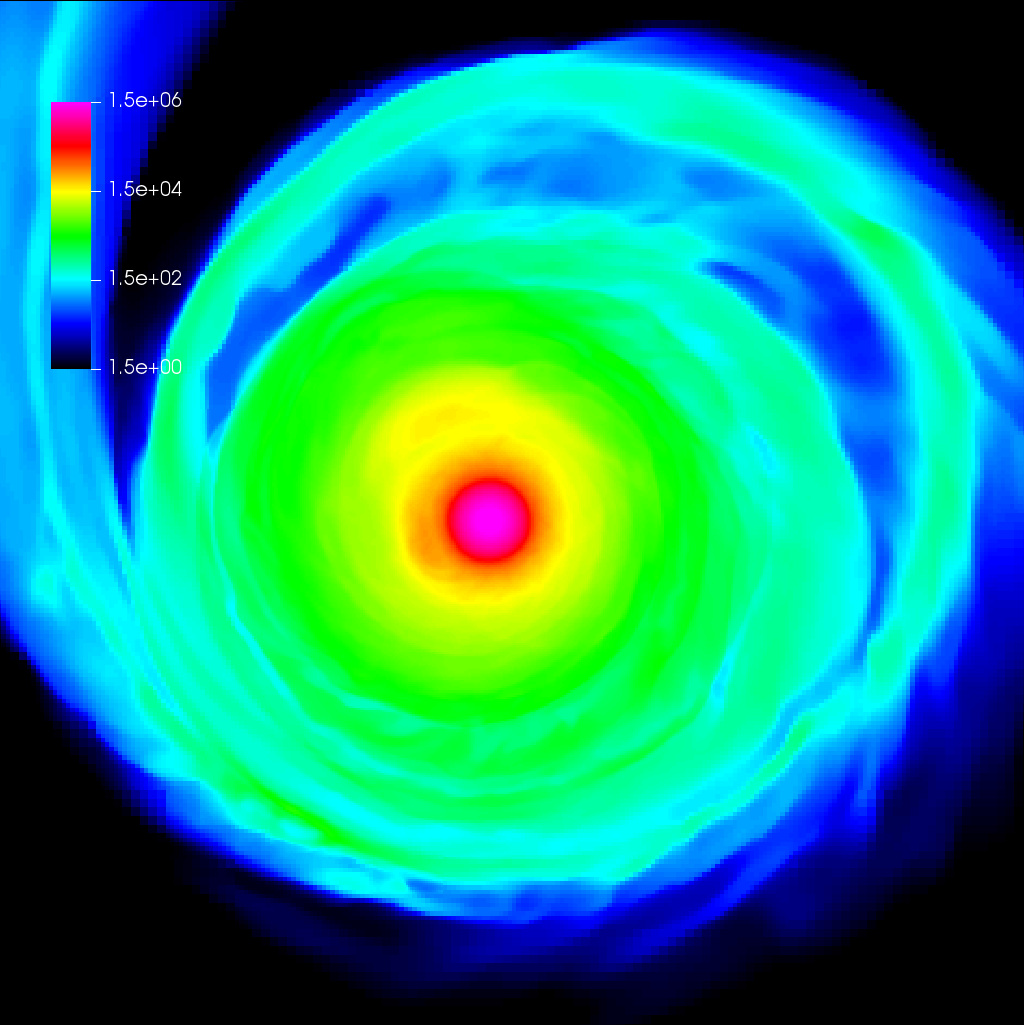}}
        \qquad
    \subfloat[Donor fraction, $t=25.0 \times P_0$, polytropic EoS]{\includegraphics[width=6cm]{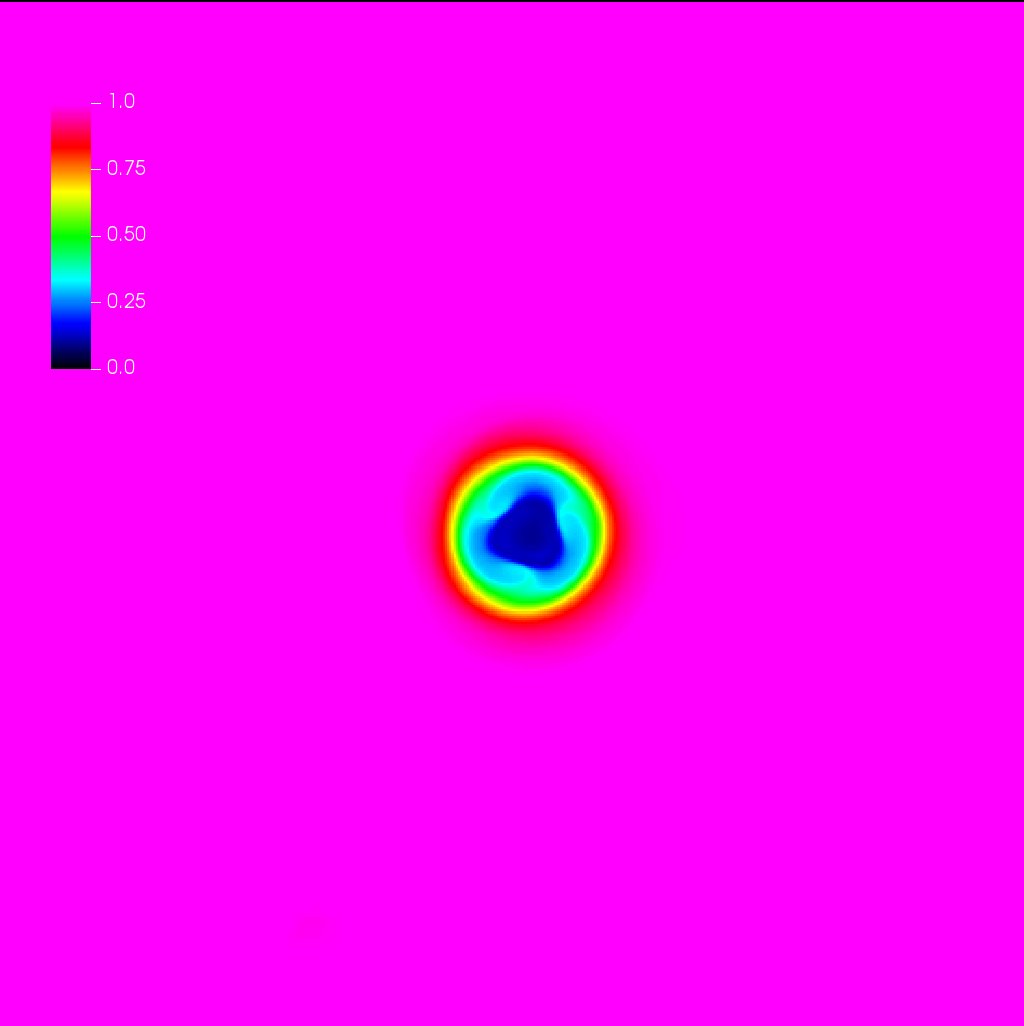}}
    \subfloat[Donor fraction, $t=23.2 \times P_0$, ideal gas EoS]{\includegraphics[width=6cm]{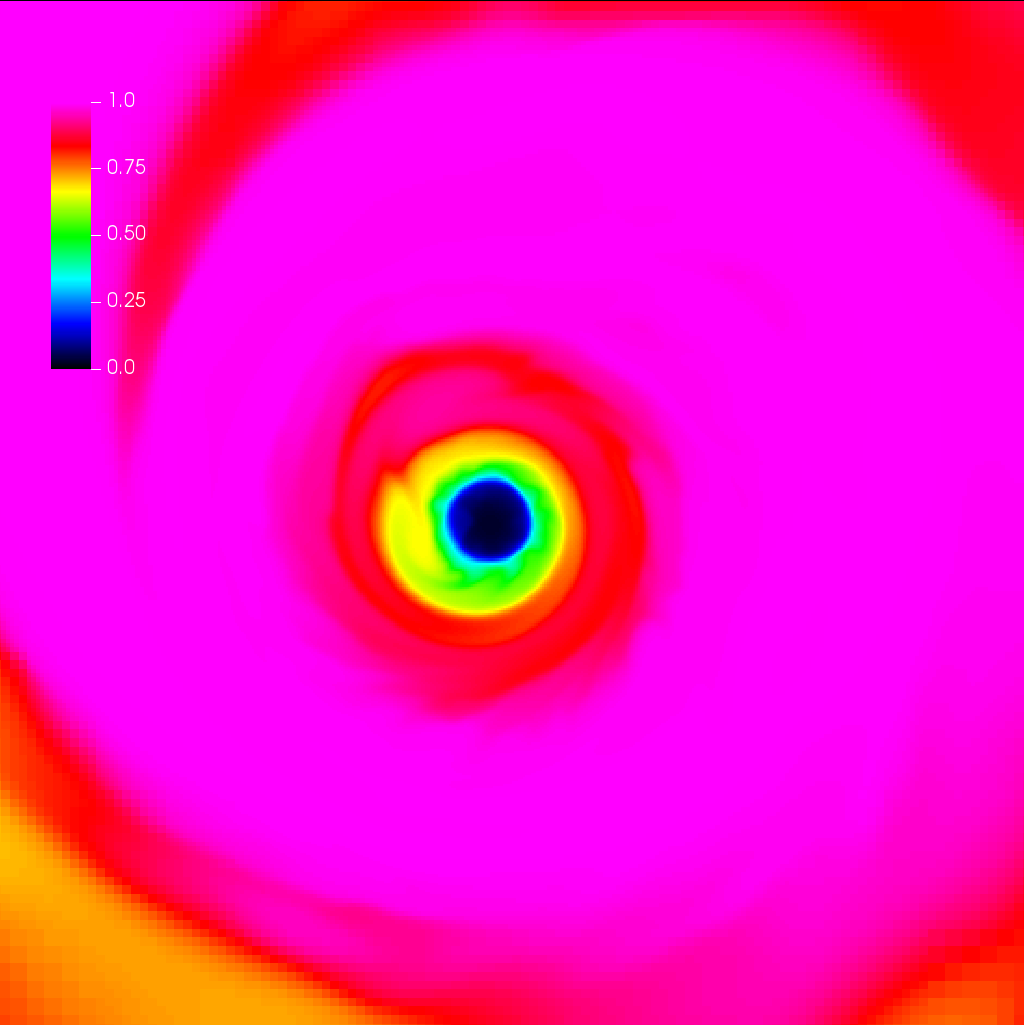}}
        \qquad
    \subfloat[Density, $t=25.0 \times P_0$, polytropic EoS]{\includegraphics[width=6cm]{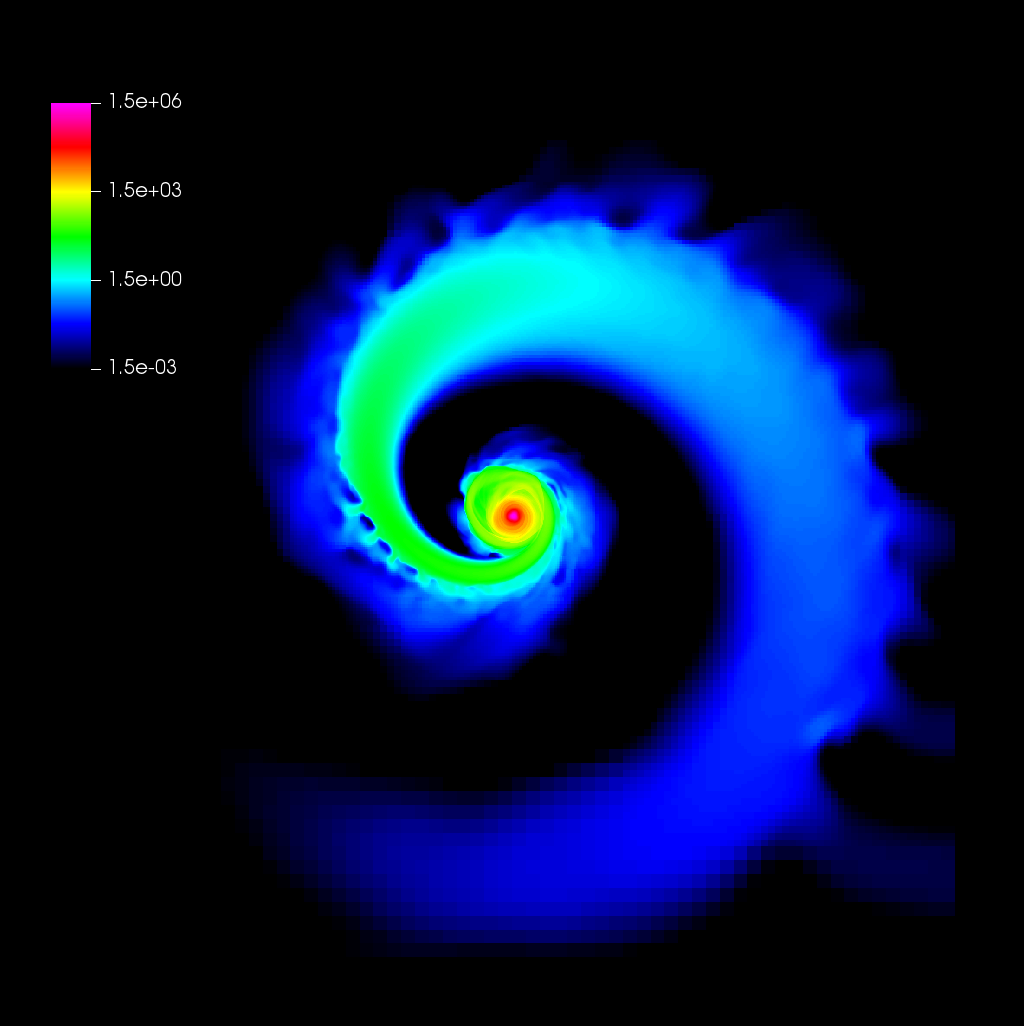}}
    \subfloat[Density, $t=23.2 \times P_0$, ideal gas EoS]{\includegraphics[width=6cm]{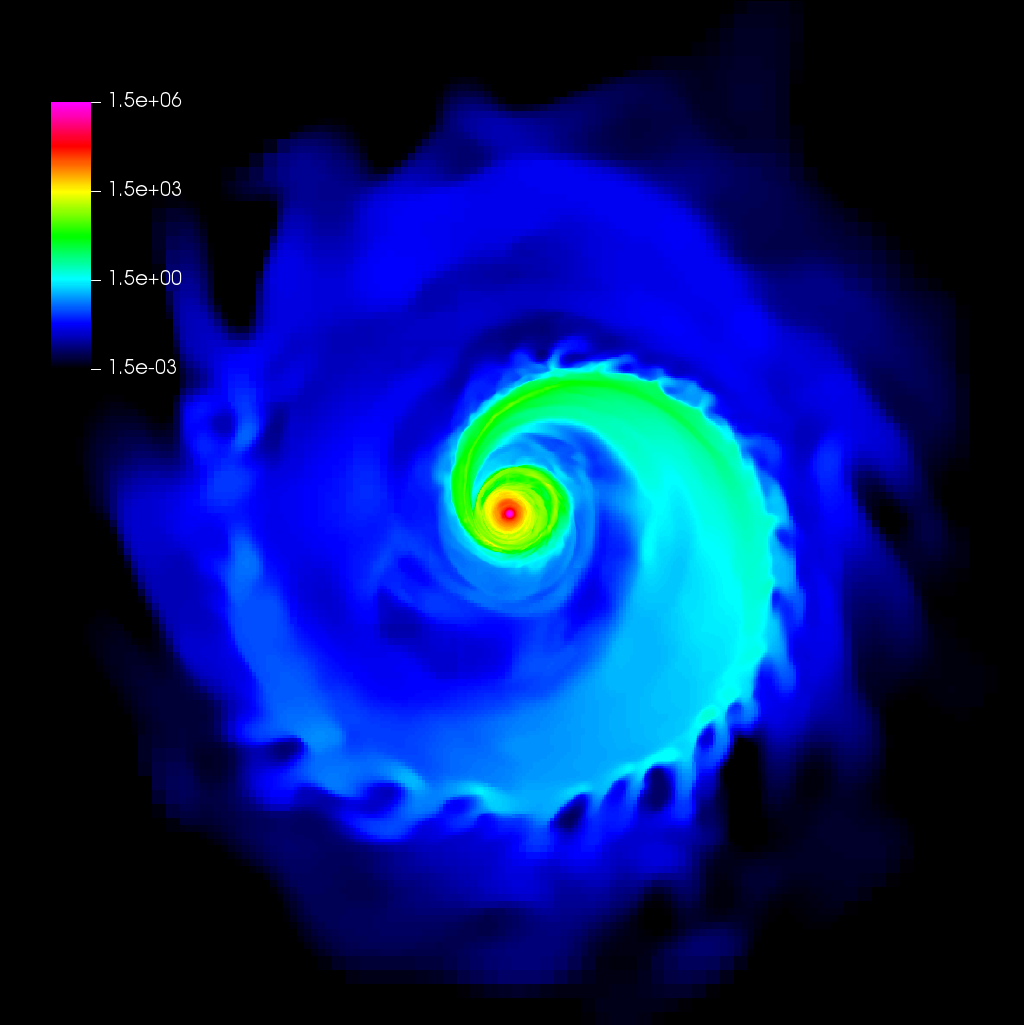}}
\caption{\protect\footnotesize{Slices on the orbital plane for our $q=0.5$ simulations with a poytropic EoS (left column) or an ideal gas EoS (right column). The box size is $20 \times 10^9$~cm (top two rows; top row: logarithmic density scale with range $1.5 - 1.6 \times 10^6$~g/cm$^3$) or $200 \times 10^9$~cm (bottom row; logarithmic density scale with range $1.5\times 10^{-3} - 1.6 \times 10^6$~g/cm$^3$). The snapshots are taken at times 25.0 (a, c, e), and $23.2 \times P_0$ (b, d, f). See Table 1 for description.} }
\label{fig:q0.5c}
\end{figure*}

The images in Figure~\ref{fig:q0.5c} are taken from two orbits after tidal disruption and merger has occurred. Panels (a) and (b) show equatorial density slices, while panels (c) and (d) show the donor fraction. The triangular shaped resonance pattern is still evident near the core in the polytropic model.  In the ideal gas model, there is still a remnant of the donor above and just to the left of centre. The bottom images are shown with a larger spatial and density range to reveal the large scale low density structures that form from the merger. 

\subsubsection{Simulation Diagnostics}
\label{ssec:simulation)diagnostics}

In order to measure the simulated properties of each star in the binary, we must first establish how to locate the stellar surface of each star. \octo\ accomplishes this iteratively. For each iteration \octo\ updates the centres of mass of each star and the orbital frequency.  Summations are done over the grid cells within the regions defined for each star, and indices $i, j,$ and $k$ referring to grid cells are implied on the RHS of the next two equations. We thus define the centres of mass of each star as
\begin{equation}
    \vec{x}_{\rm 1|2} = \frac{\sum^{1|2} \rho\ \vec{x}\  \Delta x^3 }{M_{1|2}}.
    \label{eq:xcom}
\end{equation}
The velocity of the centre of mass of each star is
\begin{equation}
    \vec{u}_{\rm 1|2} = \frac{\sum^{1|2} \rho \vec{u} \Delta x^3 }{M_{1|2}}.
\end{equation}
Given $\vec{x}_{1|2}$ and $\vec{u}_{1|2}$, we define the orbital angular frequency as
\begin{equation}
    \Omega_{\mathrm{orb}} = \frac{\left[\left(\vec{x}_1 - \vect{x}_2\right) \times \left(\vec{u}_1 - \vec{u}_2\right)\right]}{|\vec{x}_1 - \vect{x}_2|^2} \cdot \vec{e}_z,
    \label{eq:omega}
\end{equation}
where $\vec{e}_z$ is the unit vector in the z direction, out of the orbital plane.

For the first iteration, we compute the centres of mass and orbital frequency based on an initial guess for the bounding surface of each star. \octo\ evolves the original accretor and original donor mass densities as passive scalars, allowing it to track donor material within the accretor. We use the volumes defined by these densities to seed the first iteration, with cells containing majority accretor (donor) material flagged for the accretor (donor). 

\begin{figure}
\centering   
\includegraphics[width=7cm]{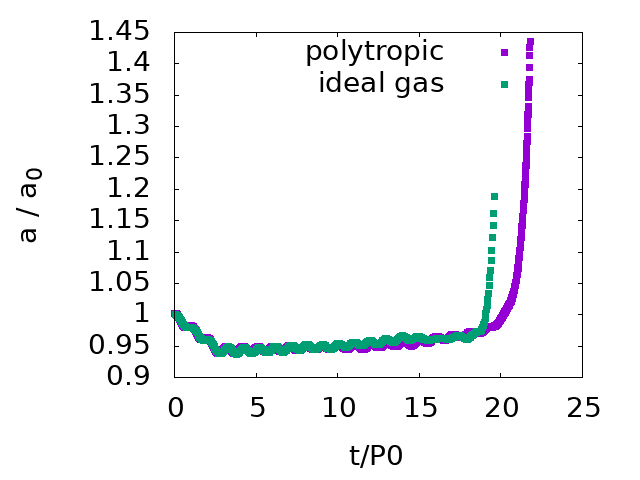}
\caption{\protect\footnotesize{Diagnostic plots from the $q=0.5$ polytropic and ideal gas simulations. Orbital separation is given as a function of the initial value.}}
\label{fig:q0.5_sep}
\end{figure}

With the stellar volumes defined, we can then calculate the centres of mass and orbital frequency. Beginning with the second iteration (out of a total of five), we use the acceleration from the effective potential to find the regions of each star. This acceleration is, \begin{equation}
        \vec{g}_{\rm eff} = \vec{g} + \vec{r} \Omega_{\mathrm{orb}}^2,
    \end{equation}
where $\vec{r}$ is the radial vector distance to the axis of rotation.
  We define the quantities
    \begin{equation}
       q_{1|2} =  \vec{g}_{\rm eff} \cdot 
       \frac{\vec{x} - \vec{x}_{1|2}}{|\vec{x} - \vec{x}_{1|2}|}.
    \end{equation}
If ${\rm min}(q_1,0) < {\rm min}(q_2,0)$, the cell belongs to the accretor. If ${\rm min}(q_1,0) > {\rm min}(q_2,0)$, the cell belongs to the donor. For the case where $q_1 \ge 0$ and  $q_2 \ge 0$, the cell is in neither star. This definition by itself may miss some cells in the centres of each star where the effective gravitational force is not guaranteed to point towards the centre of mass. For this reason, any material within a certain critical radius of the centre of mass of a star is included as part of that star regardless of the values of $q_{1|2}$. This critical radius is $\frac{1}{4} R_{L,1|2}$, where $R_{L,1|2}$ is the  Roche lobe radius of each star taken from the point mass approximation.  

\begin{figure*}
\centering   
\includegraphics[width=6cm]{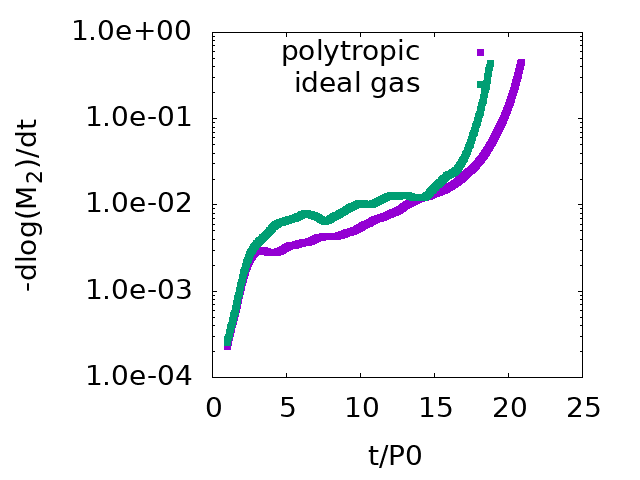}
\includegraphics[width=6cm]{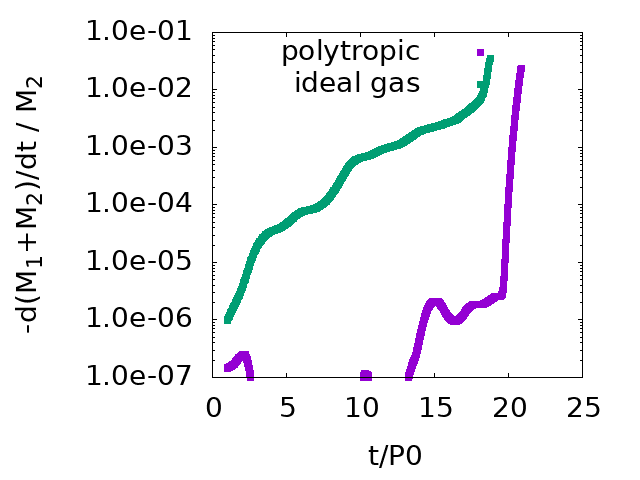}
\includegraphics[width=6cm]{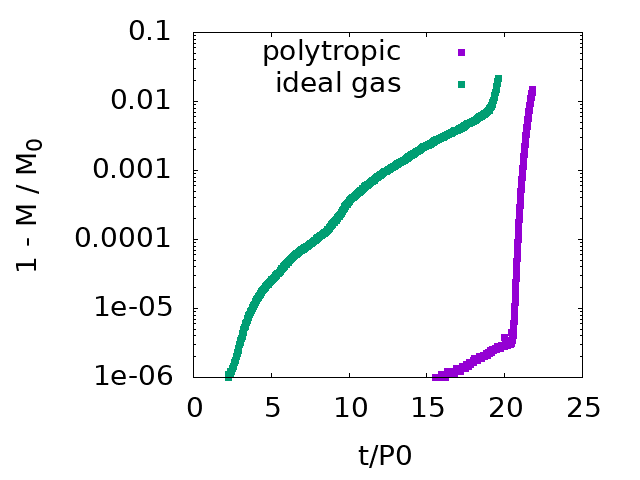}
\includegraphics[width=6cm]{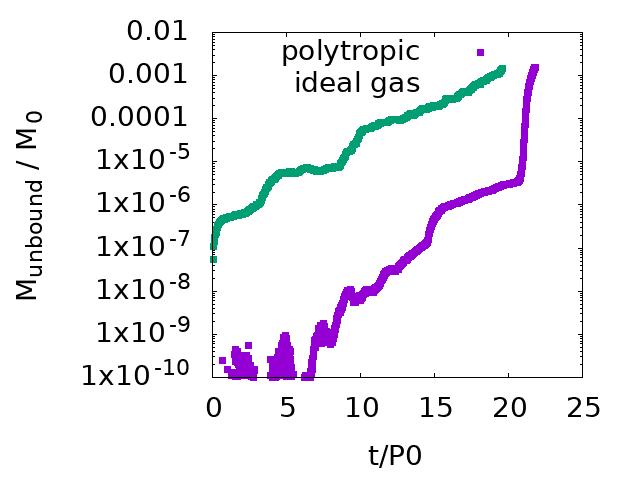}
\caption{\protect\footnotesize{Diagnostic plots from the $q=0.5$ polytropic and ideal gas simulations. Upper left: donor mass loss rate; upper right: mass loss rate from the both stars ($M=M_1+M_2$, $M_0$ is the initial mass of the system); lower left: mass outside the boundary of the two stars; lower right: mass unbound from the binary }}
\label{fig:q0.5_massloss}
\end{figure*}

With the stellar volumes defined, we can measure the properties of each star.  The mass of each star is thus defined as
\begin{equation}
    M_{1|2} = \sum^{1|2} \rho \Delta x^3.
\end{equation}
The orbital separation is, 
\begin{equation}
    a = \left|\vec{x}_2 -\vec{x}_1\right|.
\end{equation} 
We show the separation of each model in  Figure~\ref{fig:q0.5_sep}. Once the driving phase is over, the orbital separation of the two simulations increases with time up until the final merger. This shows the donor does not plunge into the accretor but instead is tidally disrupted (e.g.,  Figure~\ref{fig:q0.5b}(c)--(d)). The matter from the disrupted donor then falls onto the accretor. Our diagnostic plots that measure properties of the individual stars do not extend to this phase because there is no longer a clearly defined donor. 

Computing the time rate of change of the two stars' masses, $\dot{M}_{1|2}$, requires filtering out high frequency noise, here defined as anything greater than ($\frac{1}{2}\Omega_{\mathrm{orb}}$). The regions defining each star change with time and can vary slightly from one time step to the next. The high frequency part of this signal needs to be filtered out to measure a meaningful $\dot{M}_{1|2}$. We apply a windowed $\mathrm{sinc}$ filter using an exact Blackman window with frequencies of $\frac{1}{2} \Omega_{\mathrm{orb}, 0}$ and  $ \frac{3}{20} \Omega_{\mathrm{orb}, 0}$. We apply this filter to $M_{1|2}$ and then compute the time-derivatives with nearest neighbors by a first order discrete scheme. 

In Figure~\ref{fig:q0.5_massloss} (top left panel) we show the donor mass loss rate, normalised to the donor's mass. The time derivative is in units of initial orbital periods. The donor initially loses less than one percent per orbit, increasing to $\sim$1~percent, and then ramping up towards the time when the donor is tidally shredded. In Figure~\ref{fig:q0.5_massloss} (top right panel) we show the amount of mass that is no longer a part of either star. The two star system loses only a small fraction of the donor's mass throughout most of the evolution.  This can be mass unbound from the system or mass in a common envelope, but still bound to the system. The ideal gas model loses mass at a higher rate. This is because the accreted gas is shock heated and therefore has higher pressure. It builds up a  thicker layer around the accretor, and some of this gas is able to escape. 

In Figure ~\ref{fig:q0.5_massloss} (bottom panels) we plot the mass that is outside the surface of either stars, and the mass that is unbound from the system. Unbound mass is that for which the inertial frame kinetic energy density exceeds the inertial frame potential energy. Further from the system centre of mass, where most of the unbound material is likely to be located, this indicator is more reliable than using rotating frame quantities. Most of the mass remains bound. In the case of the ideal gas model, about 10 percent of the mass leaves the stellar boundaries, but only about 1 percent of the mass actually becomes unbound from the system. In the case of the polytropic model these figures are even lower due to lack of shock heating.  

\begin{figure*}
\centering   
\includegraphics[width=5cm]{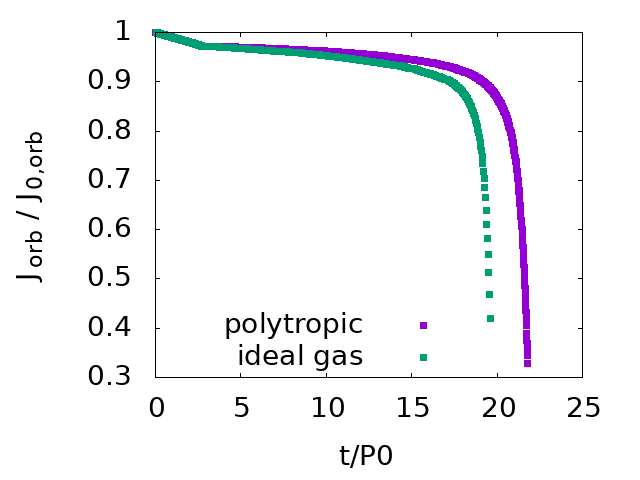}
\includegraphics[width=5cm]{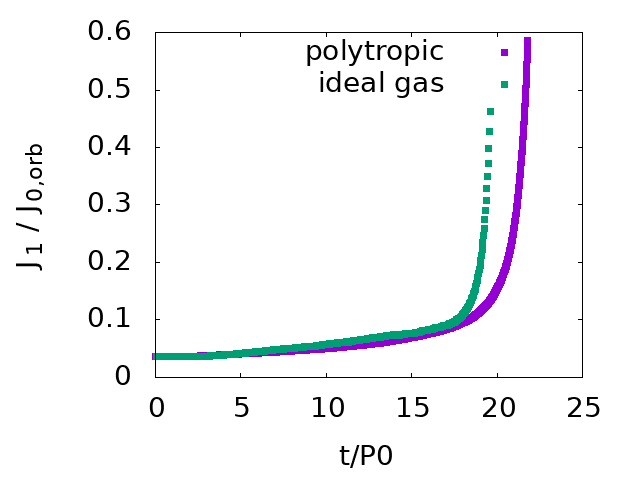}
\includegraphics[width=5cm]{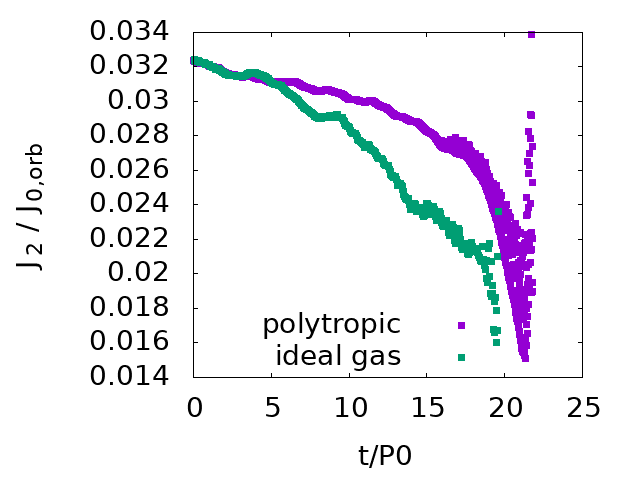}
\includegraphics[width=5cm]{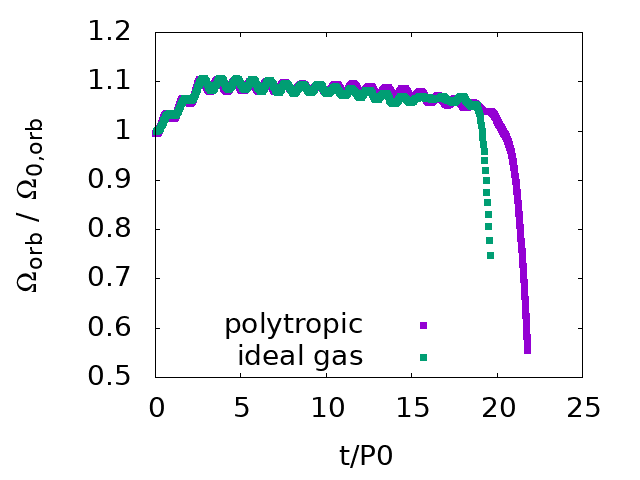}
\includegraphics[width=5cm]{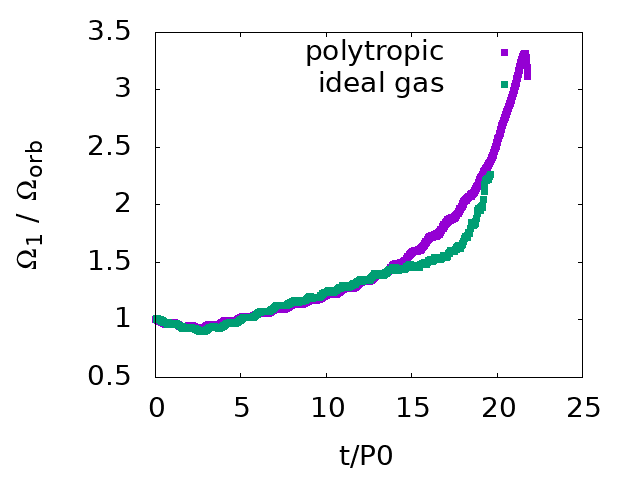}
\includegraphics[width=5cm]{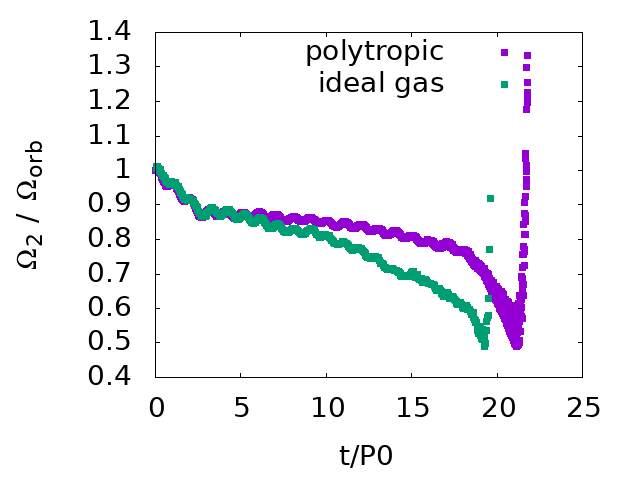}
\caption{\protect\footnotesize{Diagnostic plots from the $q=0.5$ polytropic and ideal gas simulations.  Top left: the orbital angular momentum; top middle: the accretor angular momentum; top right: the donor angular momentum. All angular momenta are normalised to the initial value of the orbital angular momentum. Bottom left: the orbital frequency; bottom middle: the spin frequency of the accretor; bottom right: the spin frequency of the donor. All frequencies are normalised to the orbital frequency at $t=0$}}
\label{fig:q0.5_rotation}
\end{figure*}

The orbital angular momentum is
\begin{equation}
    J_{\mathrm{orb}} = \frac{M_1 M_2}{M_1 + M_2} \Omega_{\rm orb} a^2.
\end{equation}
We show $J_{\mathrm{orb}}$ for both models in the top left panel of Figure~\ref{fig:q0.5_rotation} as a function of the initial orbital angular momentum ($1.5 \times 10^{51}$~g~m$^2$~s$^{-1}$). 
As expected, both simulations lose orbital angular momentum throughout their evolution. 

Each star has a spin angular momentum
\begin{equation}
    J_{1|2} = \sum^{1|2} \rho \left(\vec{x} - \vec{x}_{1|2}\right) \times \left( \vec{u} - \vec{u}_{1|2}\right) \Delta x^3,
\end{equation}
and an angular frequency assuming rigid body rotation
\begin{equation}
    \Omega_{1|2} = \frac{J_{1|2}} {\sum^{1|2}\rho \left|\left(\vec{R}-\vec{R}_{1|2}\right) \right|^2 \Delta x^3},
\end{equation}
where $\vec{R}_{1|2}$ is the projection of $\vec{x}_{1|2}$ on the xy-plane. We show $J_{1|2}$ in the middle right and lower right panels of Figure~\ref{fig:q0.5_rotation}. In each model, the accretor gains angular momentum from the donor. We show $\Omega_{1|2}$ along with $\Omega_{\mathrm{orb}}$ in Figure~\ref{fig:q0.5_rotation}.

\begin{figure*}
\centering   
\includegraphics[width=5cm]{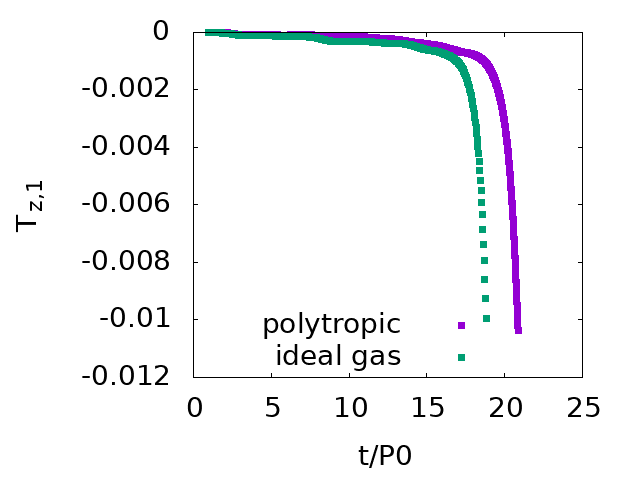}
\includegraphics[width=5cm]{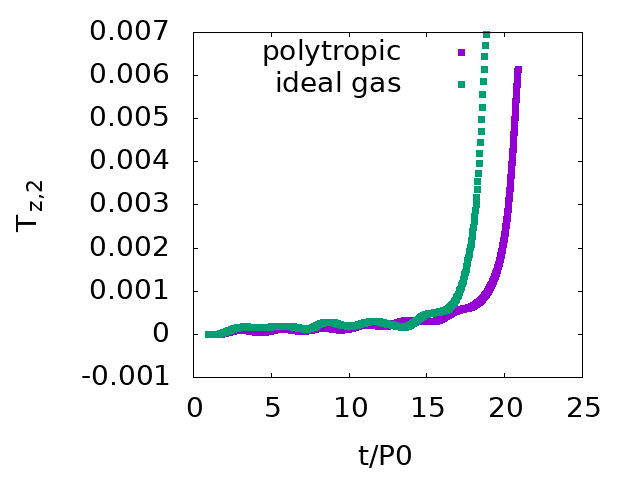} \\
\includegraphics[width=5cm]{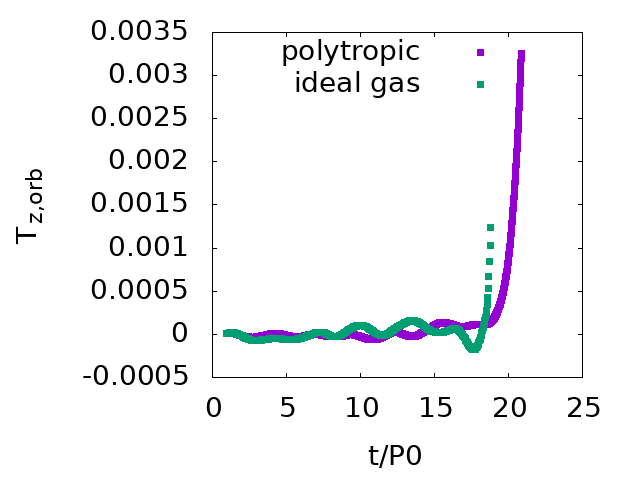}
\includegraphics[width=5cm]{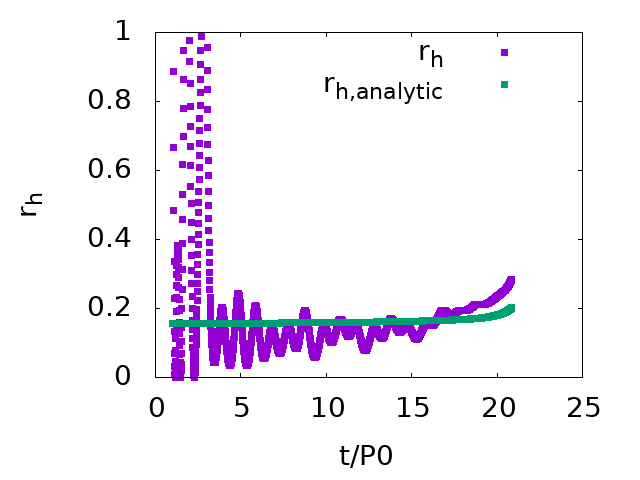}
\caption{\protect\footnotesize{Diagnostic plots from the $q=0.5$ polytropic and ideal gas simulations. Top left: the gravitational torque on the accretor; 
top right: the gravitational torque on the donor;
bottom left: the gravitational torque on the orbit; bottom right: the circularization radius for the polytropic simulation calculated using Equation~\ref{eq:dsouza} and for the analytical approximation using Equation~\ref{eq:rh}}}
\label{fig:q0.5_torque}
\end{figure*}

The gravitational torques exerted on each star are
\begin{equation}
    \vec{T}_{1|2} = \sum^{1|2} (\vec{x} - \vec{x}_{1|2}) \times \rho \vec{g}.
\end{equation}
The gravitational torque exerted on the orbit is
\begin{equation}
    \vec{T}_{\rm orb} = \vec{x}_1 \times \sum^1 \rho \vec{g} +
    \vec{x}_2 \times \sum^2 \rho \vec{g}.
\end{equation}
The z-component of $\vec{T}_{\rm orb}$ plays an important role in whether or not the binary survives mass transfer. This represents the rate at which the tidal interaction can restore the orbital angular momentum lost to mass transfer (Figure~\ref{fig:q0.5_torque} (a) to (c)). 

\cite{D_Souza_2006} gives an expression relating the time rate of change of the separation to systematic sources of orbital angular momentum, gravitational torque, and the mass transfer rate. This is
\begin{equation}
    \frac{\dot{a}}{2 a} = \left(\frac{\dot{J}}{J_{\rm orb}}\right)_{\rm sys} + \vec{T}_{z,\rm orb} - \frac{\dot{M}_2}{M_2} \left(1 - q - \sqrt{(1+q)r_h}\right).
    \label{eq:dsouza}
\end{equation}
The first term on the RHS, $\left(\frac{\dot{J}}{J_{\rm orb}}\right)_{\rm sys}$, is due to systematic changes in orbital angular momentum due to losses through $L_2$. In addition, there is a loss of orbital angular momentum during the first 2.7 orbits because of our artificial driving that needs to be accounted for when using this expression. The second term represents the contribution due to the gravitational torque on the orbit.  The final term includes, $r_h$, the effective ``circularization" radius. This is the radius from the accretor's centre of mass, which has the same specific angular momentum as the material at the $L_1$ Lagrange point, normalised to the orbital separation. While the first part of the final term, $\frac{\dot{M}_2}{M_2} \left(1 - q \right)$, accounts for the change in orbital angular momentum due to the transfer of mass from one star to the other, the second part, $\frac{\dot{M}_2}{M_2} \left( - \sqrt{(1+q)r_{h}}\right)$, accounts for the orbital angular momentum that is converted to spin angular momentum in the accretor. 

\citet{2002apa..book.....F} provide an analytical approximation to $r_h$, 
\begin{equation}
    r_{h,{\rm analytic}} = \left(1+q\right )\left(0.500-0.227 \log_{10} q\right)^4.
    \label{eq:rh}
\end{equation}
A more accurate expression that includes the torquing of the stream of accreting gas by the donor is that of \citet{Verbunt1988}. However, for the current purpouse the expression in Equation~\ref{eq:rh} is sufficiently accurate. The orbital separation, angular momentum driving rate, gravitational torque on the orbit, the mass transfer rate, and mass ratio are all known or can be calculated from the data set, allowing us to compute a value for $r_h$ using Equation~\ref{eq:dsouza}. We show this quantity compared with $r_{\rm h, analytic}$ for the polytropic model, in the bottom right panel of Figure $\ref{fig:q0.5_torque}$. In the beginning the value of  $r_{\rm h}$ varies wildly. This is during the initial period of angular momentum driving. For the majority of the run, $r_{\rm h}$ is slightly less than $r_{h,{\rm analytic}}$, indicating the spin angular momentum actually transferred to the accretor is slightly less than predicted by the analytic theory. Toward the end, the computationally-derived $r_{\rm h}$ begins to exceed the analytical value $r_{h,{\rm  analytic}}$. This is because angular momentum is now flowing through the $L_2$ point, making Equation \ref{eq:dsouza} no longer valid.

\begin{figure*}
\centering   
    \includegraphics[width=5cm]{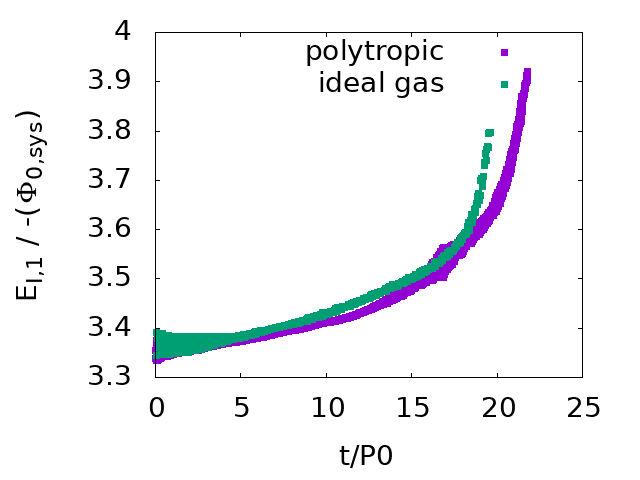}
    \includegraphics[width=5cm]{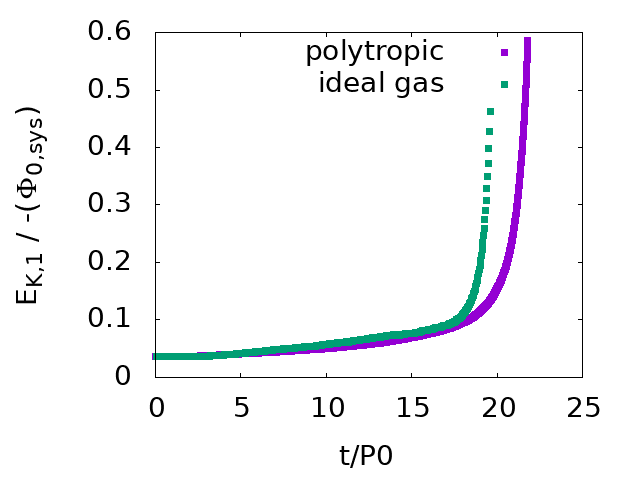}
    \includegraphics[width=5cm]{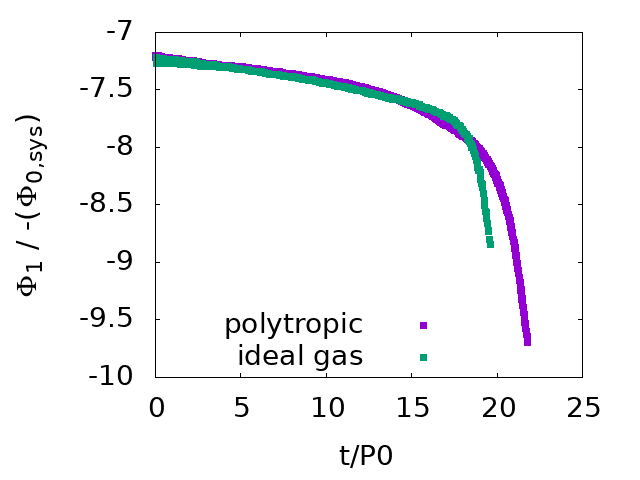}
    \\
    \includegraphics[width=5cm]{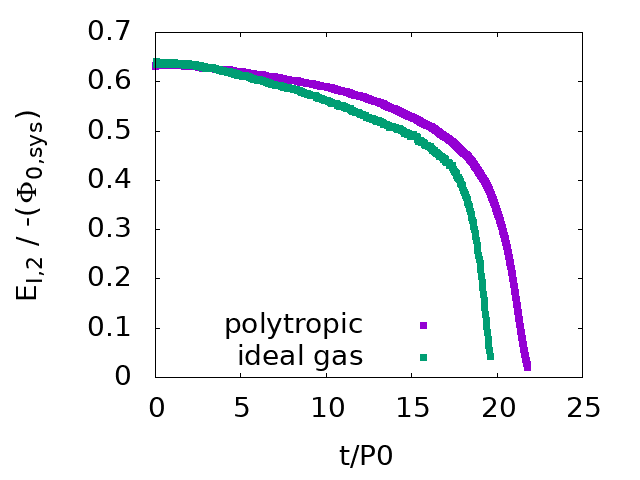}
    \includegraphics[width=5cm]{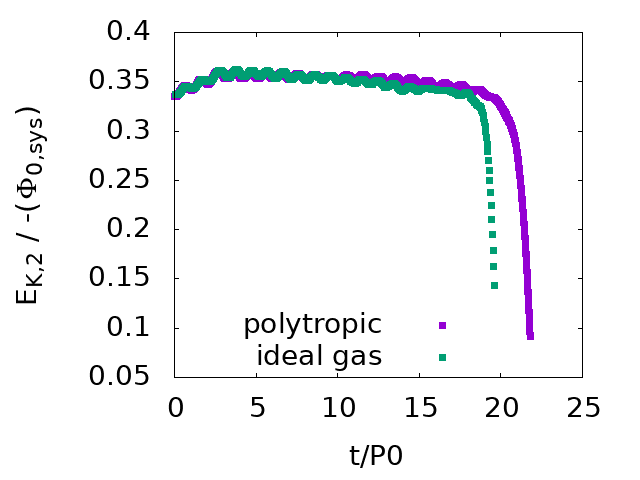}
    \includegraphics[width=5cm]{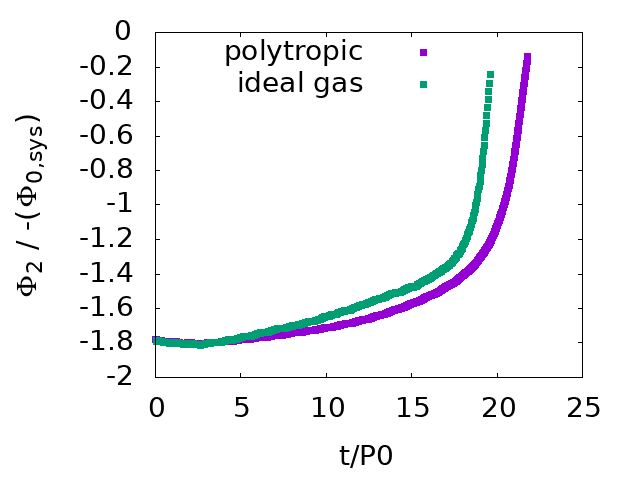}
 \caption{\protect\footnotesize{Diagnostic plots from the $q=0.5$ polytropic and ideal gas simulations. Top left: the accretor's internal energy; top middle:  the accretor's kinetic energy; top right: the accretor's potential energy. Bottom left: the donor's internal energy; bottom middle: the donor's kinetic energy; bottom right: the donor's potential energy. All energies are normalised to the absolute value of the total potential energy at $t=0$}}
\label{fig:q0.5_energies}
\end{figure*}

The internal, kinetic, and potential energies of the stars are 
\begin{equation}
    E_{I,1|2} = \sum^{1|2} \rho \varepsilon,
\end{equation}
\begin{equation}
    E_{K,1|2} = \sum^{1|2} \frac{1}{2} \rho u^2,
\end{equation}
and
\begin{equation}
    \Phi_{1|2} = \sum^{1|2} \frac{1}{2} \rho \Phi,
\end{equation}
respectively. Here the internal energy density, $\rho \varepsilon$, is obtained from the dual energy formalism. We show these energies in Figure~\ref{fig:q0.5_energies}. The donor loses internal and kinetic energy to the accretor. Although the ideal gas model includes shock heating,  since  the energy of the shock heated material around the accretor is very small compared to the total internal energy of the star, shock heating makes little apparent difference in the plots of total internal energy. As the accretor gains mass, its gravitational well deepens as its potential energy is lowered. The opposite is true of the accretor.

\begin{figure*}
\centering   
    \includegraphics[width=6cm]{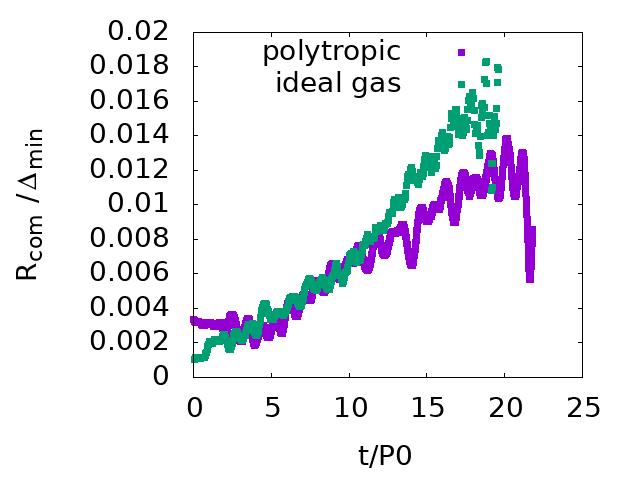}
    \includegraphics[width=6cm]{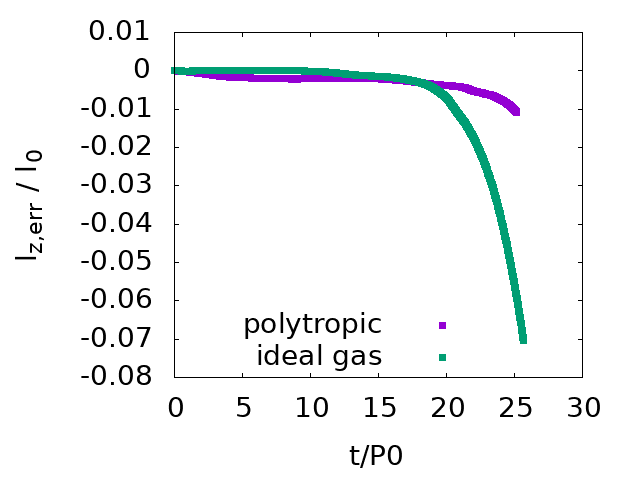}
    \includegraphics[width=6cm]{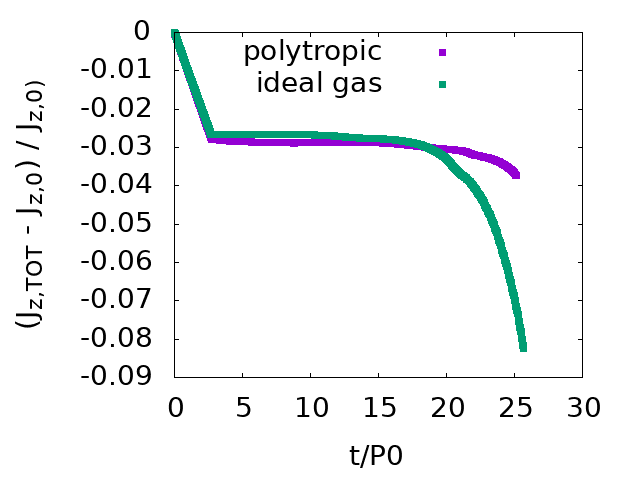}
    \includegraphics[width=5.75cm]{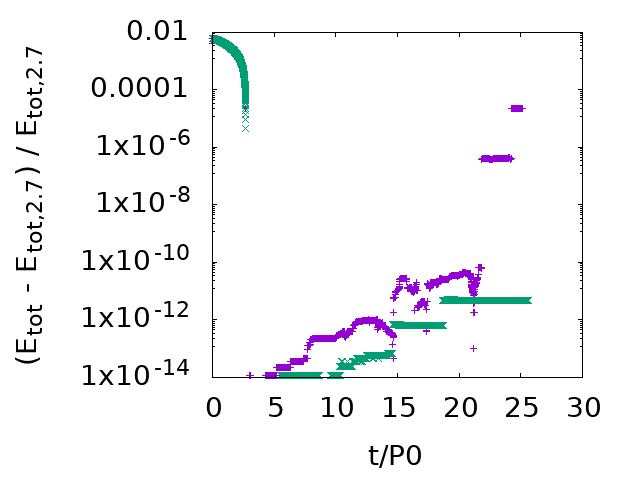}

\caption{\protect\footnotesize{Diagnostic plots from the $q=0.5$ polytropic and ideal gas simulations. Top left: the location of the system centre of mass (Equation~\ref{eq:R_com}) in units of the finest grid cell; top right: the relative angular momentum error  (Equation~\ref{eq:l_err}); bottom left: the total angular momentum loss from the computational domain (the initial decrease corresponds to the angular momentum that is artificially subtracted from the system during the driving phase); bottom right: the total energy conservation error of the system (the values before $x=2.7$ orbits are meaningless because of the driving)}}
\label{fig:q0.5_error}
\end{figure*}

Grid based codes are subject to centre of mass drift, both physical, from the flow of matter out of the grid (which, by the end of the simulations, is $2.0 \times 10^{29}$~g and $1.0\times 10^{28}$~g for the ideal gas and polytropic EoS, respectively), and numerical. The latter effect is greatly reduced by using a gravity solver that conserves linear momentum, as \octo\ does. There are still numerical viscosity effects that can push matter one way or another. In addition, the SCF code does not render initial conditions with the centre of mass perfectly at the coordinate centre. Figure~\ref{fig:q0.5_error} (top left panel) shows the magnitude of  cylindrical radial location of the system's centre of mass, 
\begin{equation}
   \left| \vect{R_{\rm com}} \right | = 
   \sqrt{x_{\rm com}^2 + y_{\rm com}^2} 
\label{eq:R_com}   
\end{equation}
in length units normalised to the finest cell width.

As discussed in Section~\ref{subsec:evolution_eq}, \octo\ evolves the angular momentum (Equation~\ref{eq:ang_mom_ev}) for the purposes of obtaining the error in the angular momentum conservation. The evolved angular momenta, $\vec{l}$, 
can be compared with the angular momenta obtained from the evolved linear momenta, $\vec{x} \times \rho \vec{u}$,  (Equation~\ref{rhou}) to obtain the error. We define the global angular momentum conservation violation error, $\vec{l}_{\rm err}$, as:
\begin{equation}
    \vec{l}_{\rm err} = \sum^{\rm V} 
   \left(\vec{l}_i - \vec{x}_i \times \rho_i \vec{u}_i\right),
\label{eq:l_err}   
\end{equation}

\noindent where the $i$ index refers to a given computational cell and the summation is over the entire computational domain. We show this value over time in Figure~\ref{fig:q0.5_error} (top right panel). In Figure ~\ref{fig:q0.5_error}, bottom left panel, we plot the total angular momentum non-conservation in the computational domain, scaled to the initial total angular momentum. Here we see that the polytropic EoS conserves momentum to approximately 1~percent over the entire simulation, while for the ideal gas EoS this is considerably worse at 7~percent (the initial drop in angular momentum in the right panel is due to the artificial driving phase and should be ignored). The reason for the difference is that the ideal gas EoS simulation expands more due to shock heating moving gas into the outskirts of the domain where lower resolution dominates with a worse performance of angular momentum conservation. Poorly resolved, sharp momentum density gradients, such as those seen in the outskirts of the simulations (see Figure~\ref{fig:q0.5c}, bottom panels, where we plot density slices) will always disfavour good conservation.


In Figure ~\ref{fig:q0.5_error}, bottom right panel, we show the energy conservation. The value in the y-axis is scaled to the energy after the driving, at 2.7 orbits. Hence data points before $x=2.7$ are not meaningful. After that x value, however, we see that the energy conservation is excellent, going from $10^{-14}$ to $10^{-11}$ for the ideal gas EoS, or to $10^{-5}$ for the polytropic EoS.


Without doubt the polytropic EoS simulation is superior in general, because the ideal gas EoS with more heating results is an inflated gas distribution, a higher rate of mass transfer from donor to accretor, and any quantities that derive from that, such as angular momentum transfer or the speed of the merger are also affected. 

These simulations were run on 400 cores over 20 nodes, of QueenBee2 (see Section~\ref{sec:scaling}) and took approximately 200 hours of wall-clock time, or just over 8 days for a total modest cost of approximately 80\,000 CPU-hours. They have between 2 and 8 million cells. If the scaling test in Section~\ref{sec:scaling} are consulted, we would conclude that a doubling of the number of cores would reduce the wall clock time between 40 and 60~percent, while any further increase of the number of cores would not be particularly advantageous. 

Naturally, this run could be repeated with substantially larger resolution (and larger number of cores), likely leading to an expansion of discovery space. However, increasing resolution with the aim of determining convergence is not straightforward because it is not clear what quantities can be reasonably expected to converge. An increase in resolution automatically lowers the mass transfer rate in the early interaction and lengthens the pre-merger time beyond what can be simulated (see for instance \citet{Motl2017} who compared a 4-million cell WD merger cell simulation to a 47-million cell one, or \citet{Reichardt2019} who compared the early mass transfer before a common envelope interaction between a $\sim$100\,000 SPH particle simulation to one with 1.3 million particles). What quantities, then, should we expect to converge to a given value with increasing resolution? It is possible that, for example, the increase in mass transfer rate over one orbital period, measured at a specific initial separation may be a quantity that is expected to converge. The determination of how to measure the accuracy of simulation results of this level of complexity is an urgent, on-going area of research.
 
\noindent Click on the link\footnote{\url{https://www.youtube.com/watch?v=0JD5E7DUImw}} to view a movie of the $q=0.5$ ideal gas simulation. 
\noindent Click on the link\footnote{\url{https://www.youtube.com/watch?v=MfArAQPPHss}} to view a movie of the $q=0.5$ polytropic EoS simulation.


\section{Code performance and scaling}
\label{sec:scaling}

In this section, we present a series of scaling tests to demonstrate the performance of \octo\, and we make comparisons with equivalent tests carried out with \flash. In order to compare the two codes, we set the  CFL coefficient to be the same in both ($\eta_{\rm CFL}=0.4$).   

We carried out two types of scaling tests. The first one was performed by executing a number of time steps $(N_{\rm steps})$ in the simulation of a stationary star. The second test included a binary system of stars and was designed to stimulate some level of regridding. Tests were performed on three supercomputers, BigRed3 (BR3), QueenBee2 (QB2) and Gadi, with variable resolutions ($128^3$, $256^3$, $512^3$, and $1024^3$) and variable values of $\theta$ (0.35 and 0.5), the gravity resolution parameter. We describe the supercomputer hardware in Table~\ref{tab:hardware}. The code and its dependencies' versions used with the three sets of scaling runs are reported in Table~\ref{tab:octo_dependencies}.

\begin{table*}
\begin{tabular}{lccccc}
\toprule
    Cluster & CPU & Memory & Interconnect & \# nodes & \# of cores \\
    \midrule
     QB2 & 2 $\times$ 10 Intel E5-2680v2 Xeon processors (+2 GPUs$^1$) & 64 GB &  56 Gb/sec (FDR) InfiniBand & 504 & 10800\\ 
     BR3 & 2 $\times$ 12 Intel Xeon Processort E5-2690 v3 & 64 GB & Cray Aries & 930 & 22\,464 \\
     Gadi & 2 $\times$ 24 core Intel Xenon Scalable `Cascade Lake' & 190 GB&
    200 Gb/sec (HDR) InfiniBand
     &  3024 & 145\,152 \\
     \bottomrule
     \multicolumn{6}{l}{$^1$ The GPUs are NVIDIA Tesla K20; NVIDIA-SMI 450.51.05, driver version: 450.51.05, CUDA version: 11.0 with 6GB of memory per device.}
\end{tabular}
\caption{Computational infrastructure used in this work. QB2 is Louisiana's Optical Network Infrastructure's QueenBee2; BR3 is the University of Indiana BigRed3, while Gadi is the Australian National Computational Infrastructure peak machine}
\label{tab:hardware}
\end{table*}

\begin{table*}
\centering
\begin{tabular}{llll}
\hline
Software & QB2 & BR3 & Gadi\\
\hline
HPX & 1.4 & 1.4 & 1.4 \\
Vc & 1.4.1 & 1.4.1 & 1.4.1 \\
Boost  & 1.69/1.68  & 1.68 & 1.72 \\
hwloc   & 1.11.1 & 1.11.12 & 1.11.12 \\ 
jemalloc & 5.1.0  &5.1.0 & 5.1.0 \\
gcc & 8.3/7.4   & 8.3 & 8.3 \\
hdf5 & 1.12/1.8   & 1.8.12 & 1.8.12 \\
silo & 4.10.2   & 4.10.2 & 4.10.2 \\
cmake & 3.13  & 3.13.2 & 3.16.2 \\
CUDA & 10.2 & - & -  \\
MPI & MVAPICH2 2.3.2/ OpenMPI 4.0 & cray-mpich 7.7.10 & intel-mpi 2019.8.254 \\
\bottomrule
\end{tabular}
  \caption{Software dependencies of \octo\ (version 0.8). Note that we used a customized version of \href{https://github.com/STEllAR-GROUP/Vc}{Vc}. Note that we used the pre-compiled MPI version on the cluster and therefore have some variation there. If the gcc compiler was recent enough to compile HPX and \octo\ we used the pre-compiled version on the cluster}. 
  \label{tab:octo_dependencies}
\end{table*}

\begin{figure*}
\centering   
    \includegraphics[width=8.5cm]{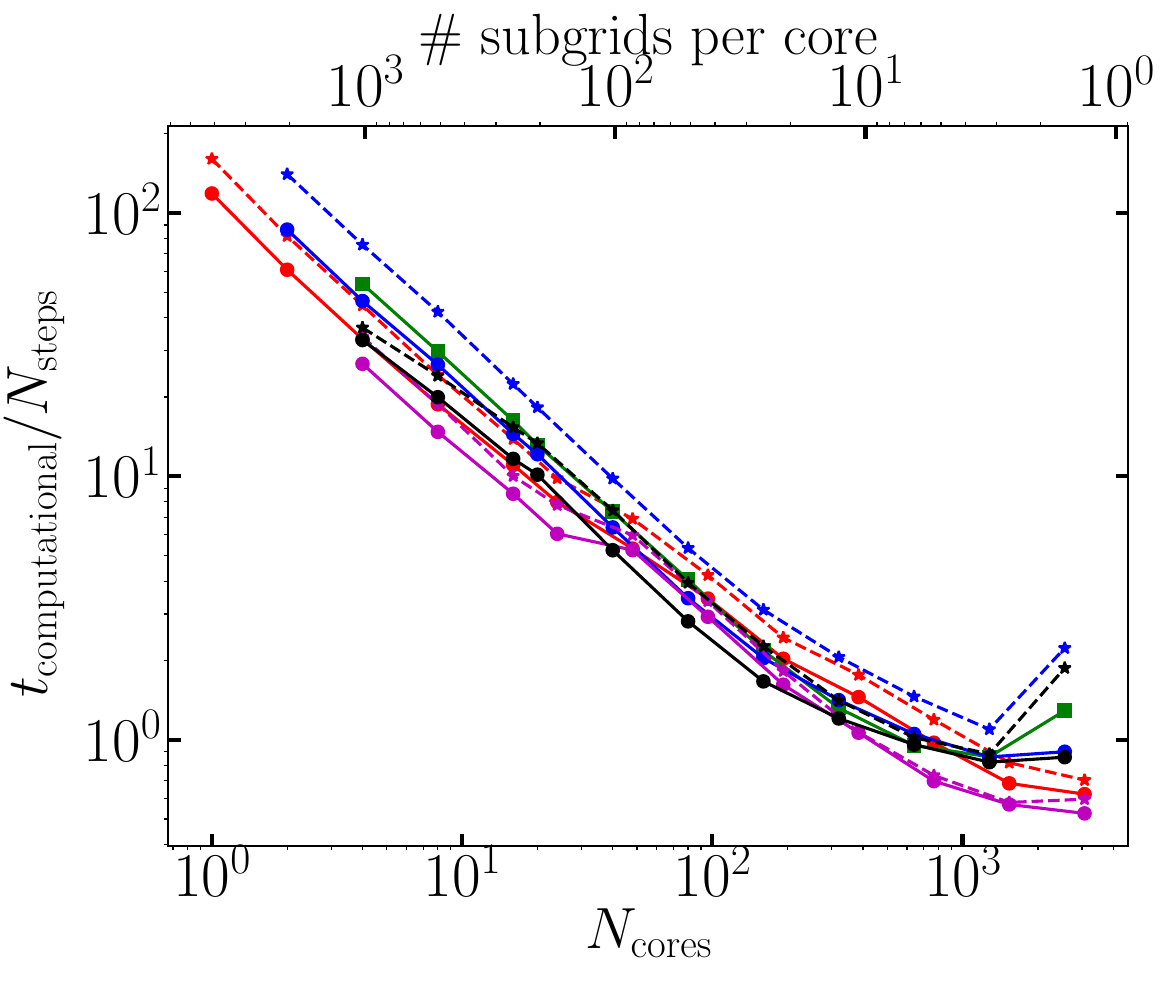}%
    \includegraphics[width=8.5cm]{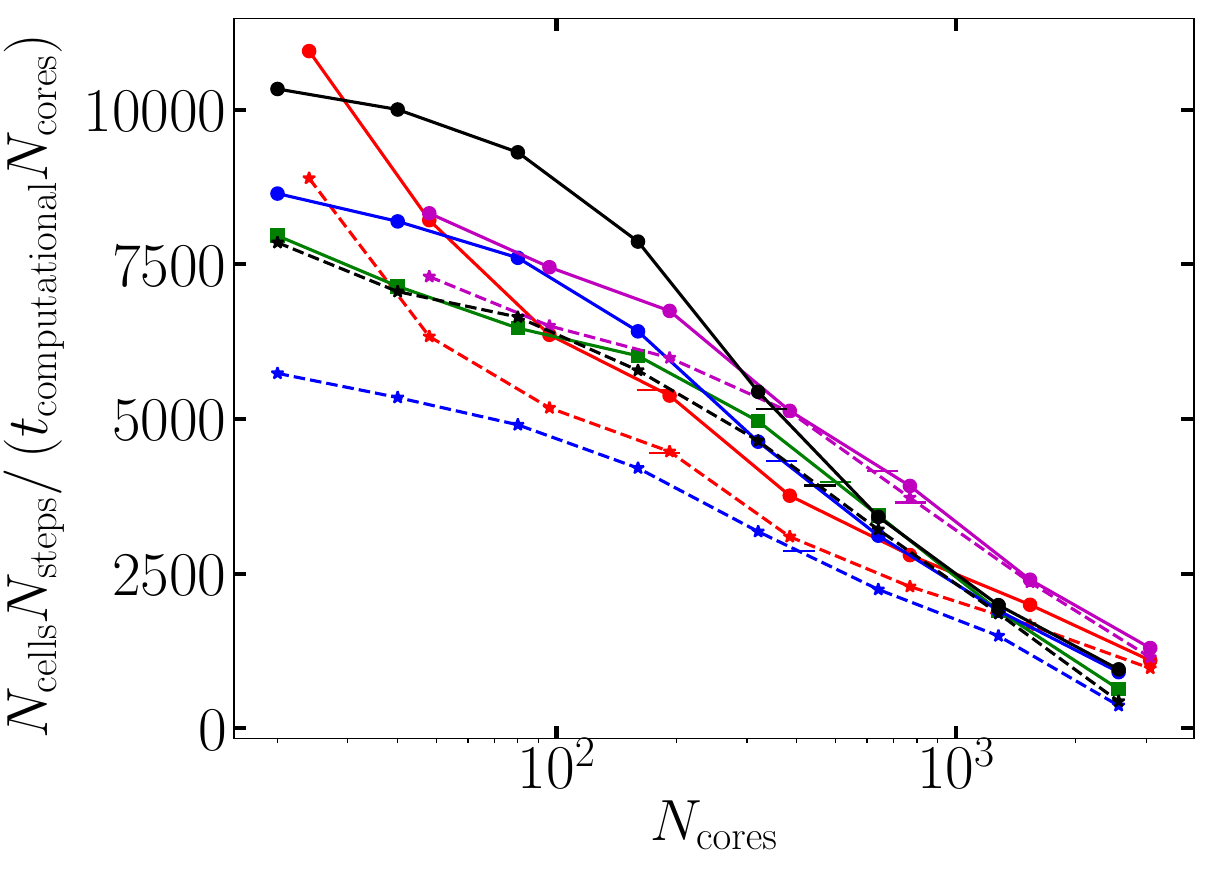}
        \qquad
    \includegraphics[width=8.5cm]{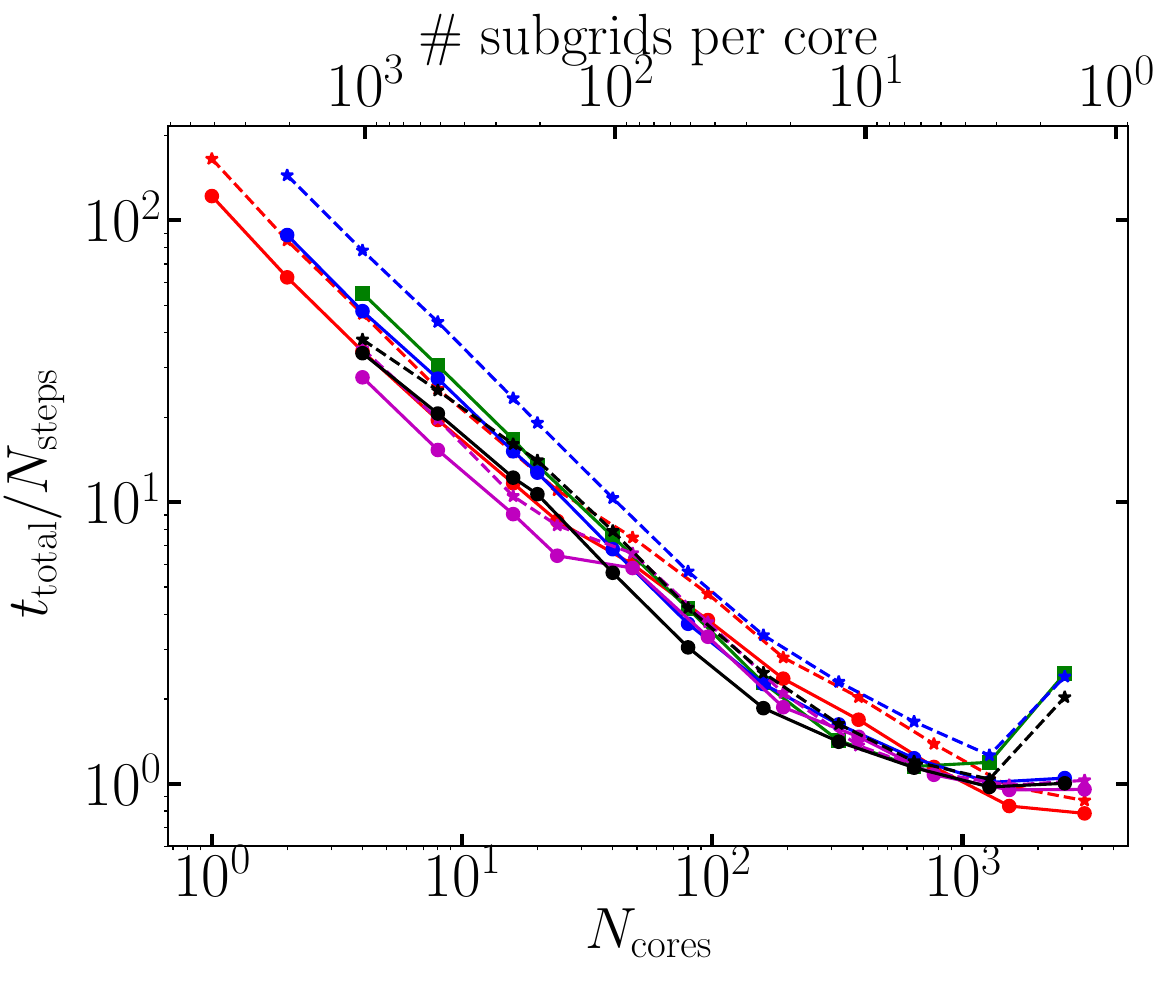} %
    \includegraphics[width=8.5cm]{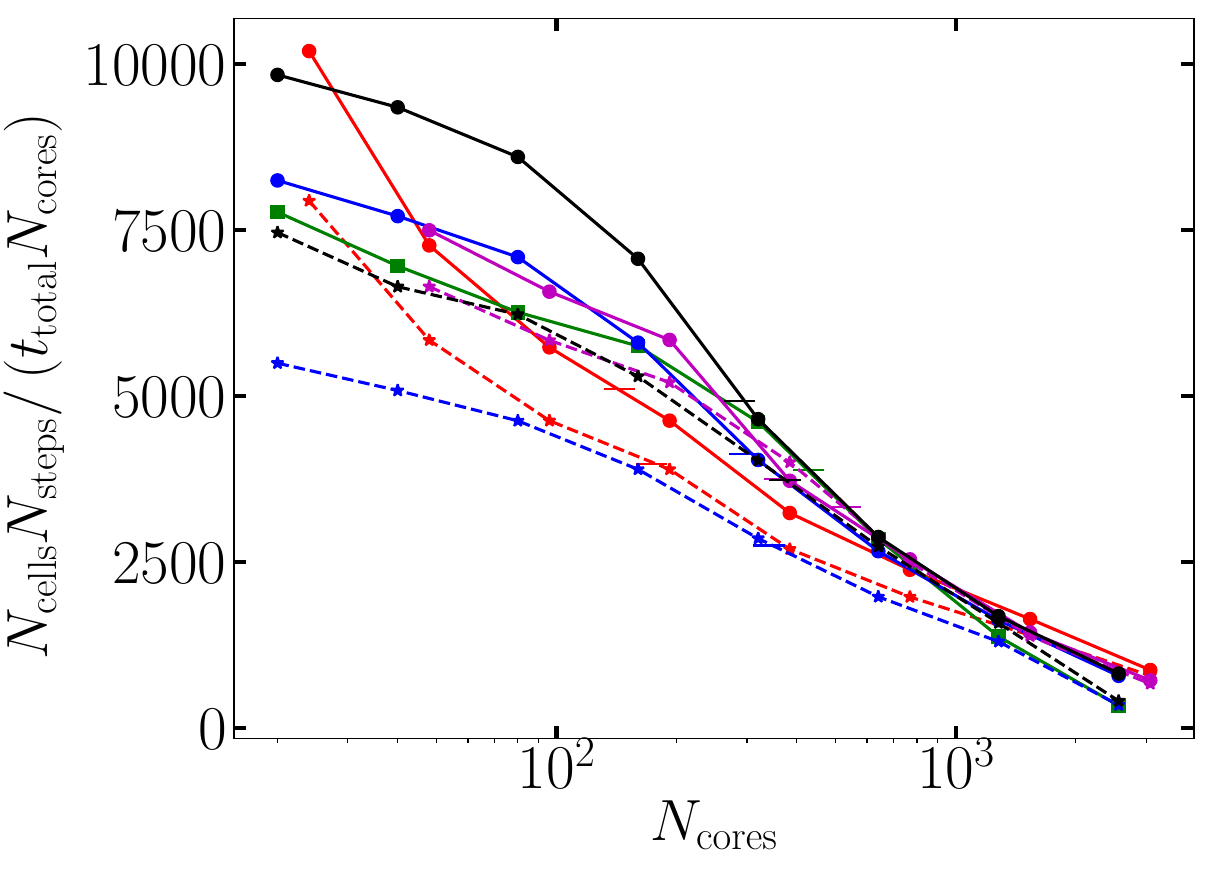} 
    
    \includegraphics[width=16cm]{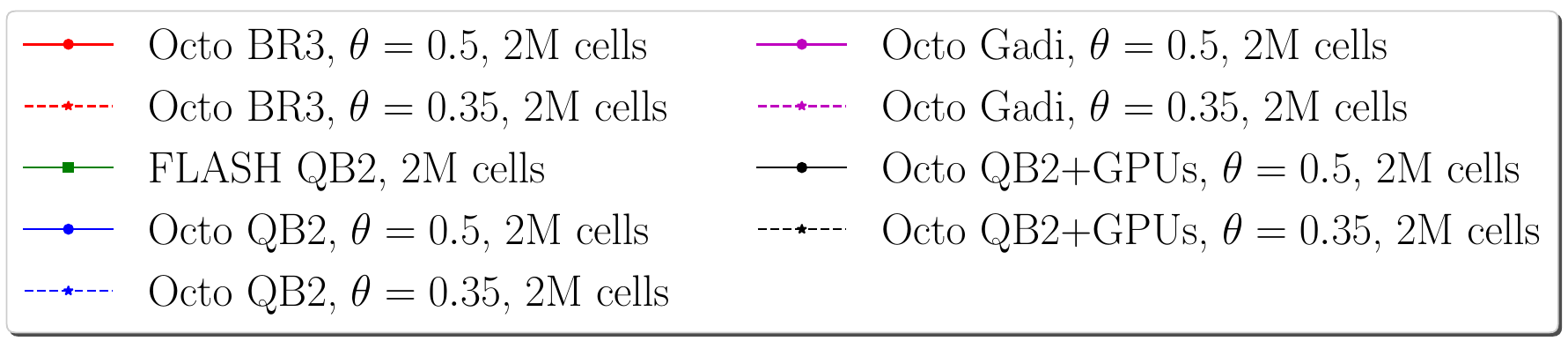}
\caption{\protect\footnotesize{Scaling test carried out on the pulsating polytrope problem of a small size. QB2 refers to QueenBee2, BR3 refers to the BigRed3. 
The simulations contain $128^3\approx2M$} cells or $16^3=4096$ subgrids.  The short horizontal segments mark the 0.5 efficiency} 
\label{fig:star_scalability_low_res}
\end{figure*}
\begin{figure*}
\centering   
\includegraphics[width=8.5cm]{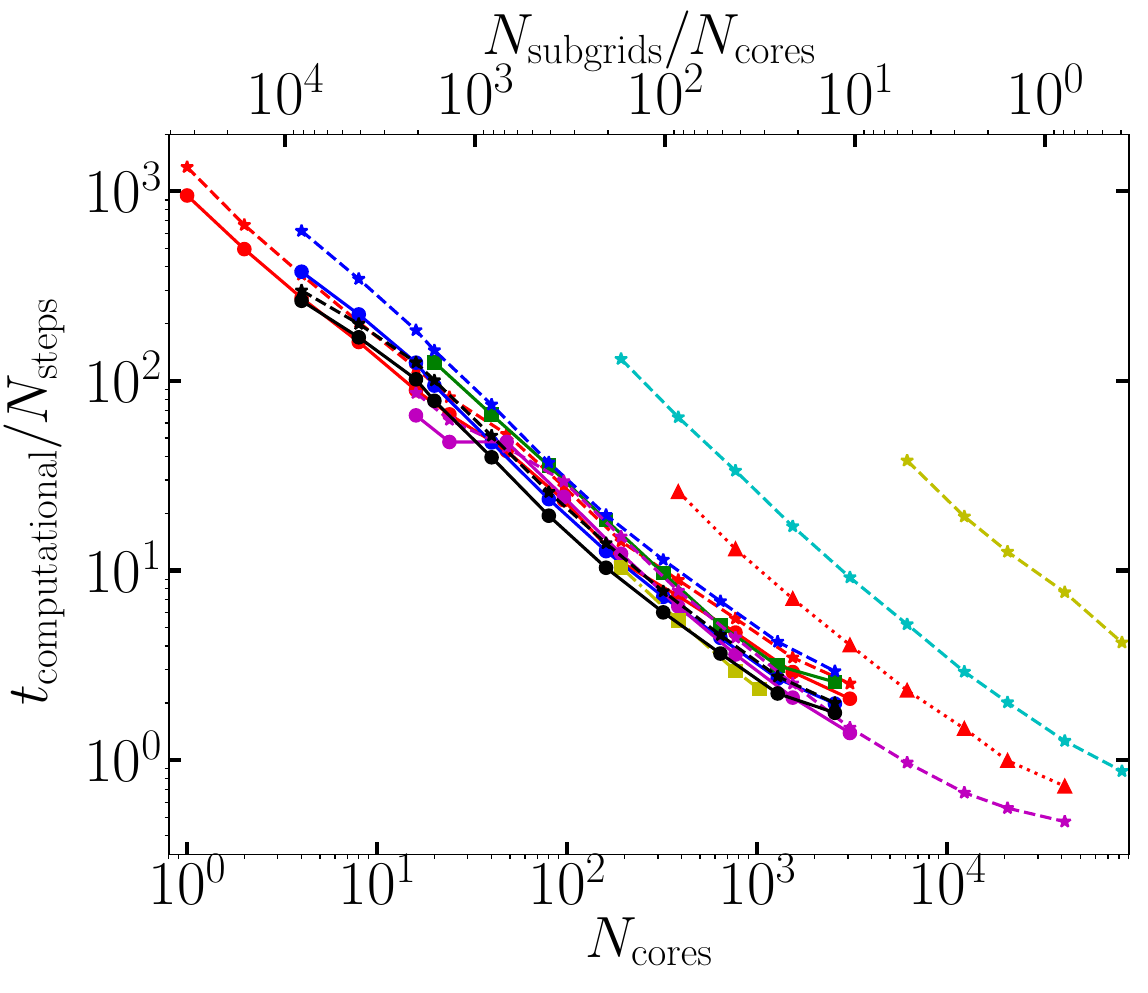}%
\includegraphics[width=8.5cm]{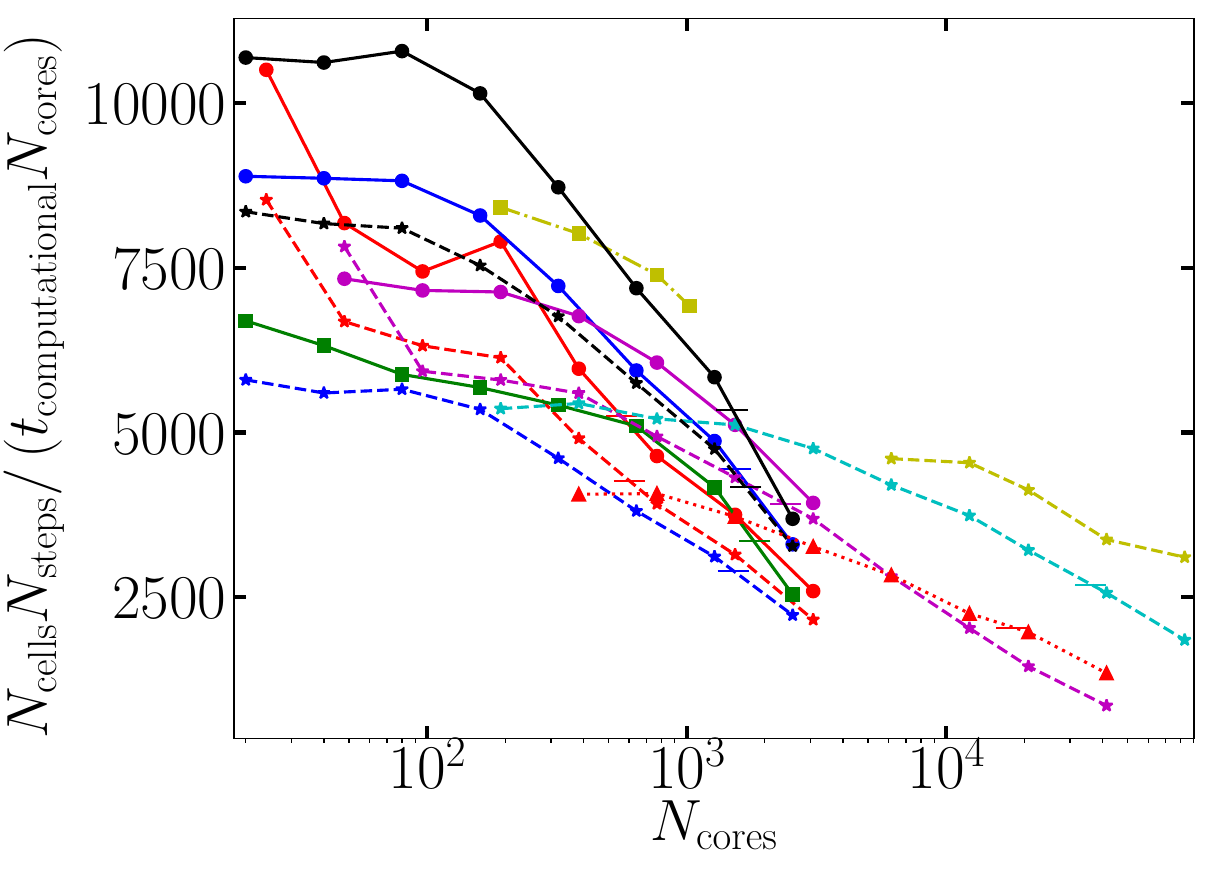}
\qquad
\includegraphics[width=8.5cm]{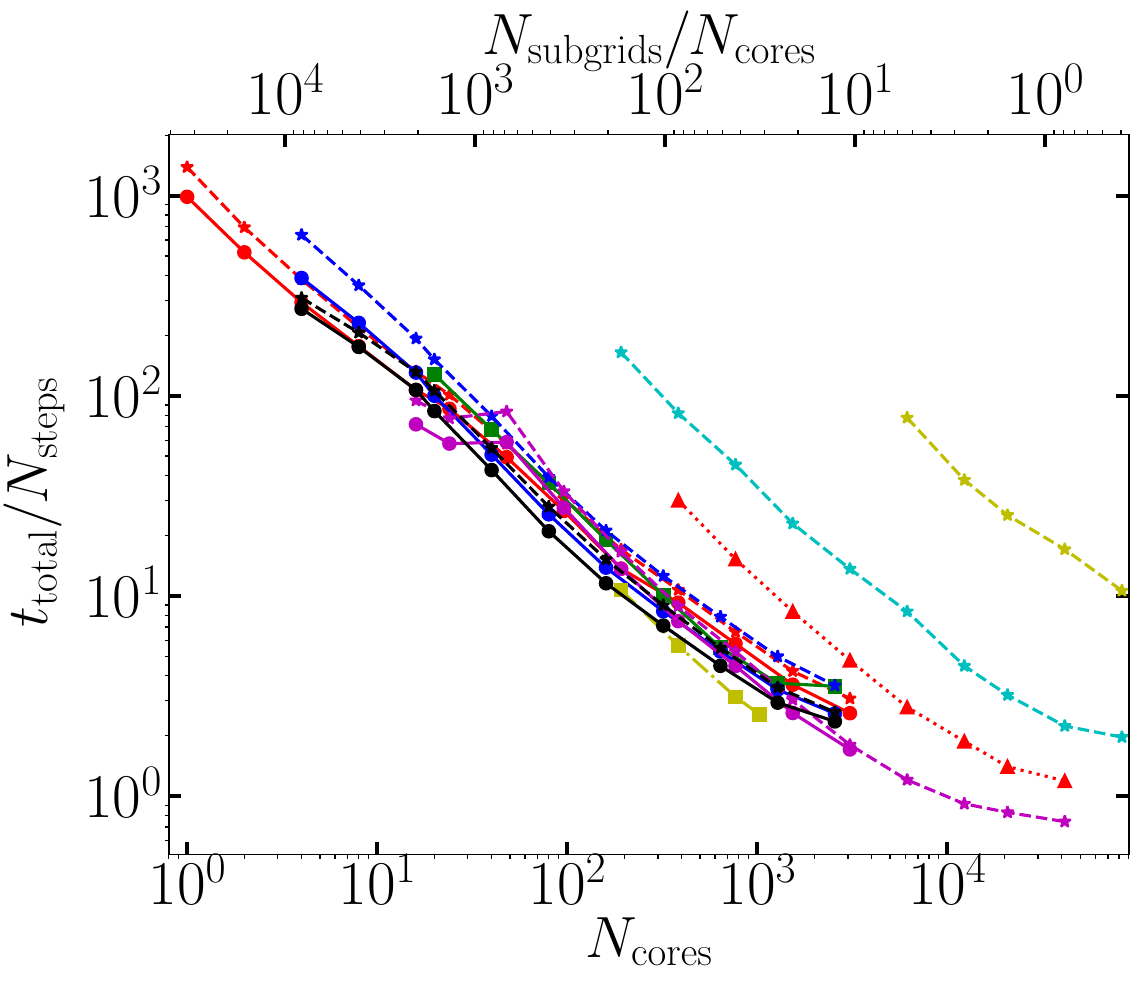} %
\includegraphics[width=8.5cm]{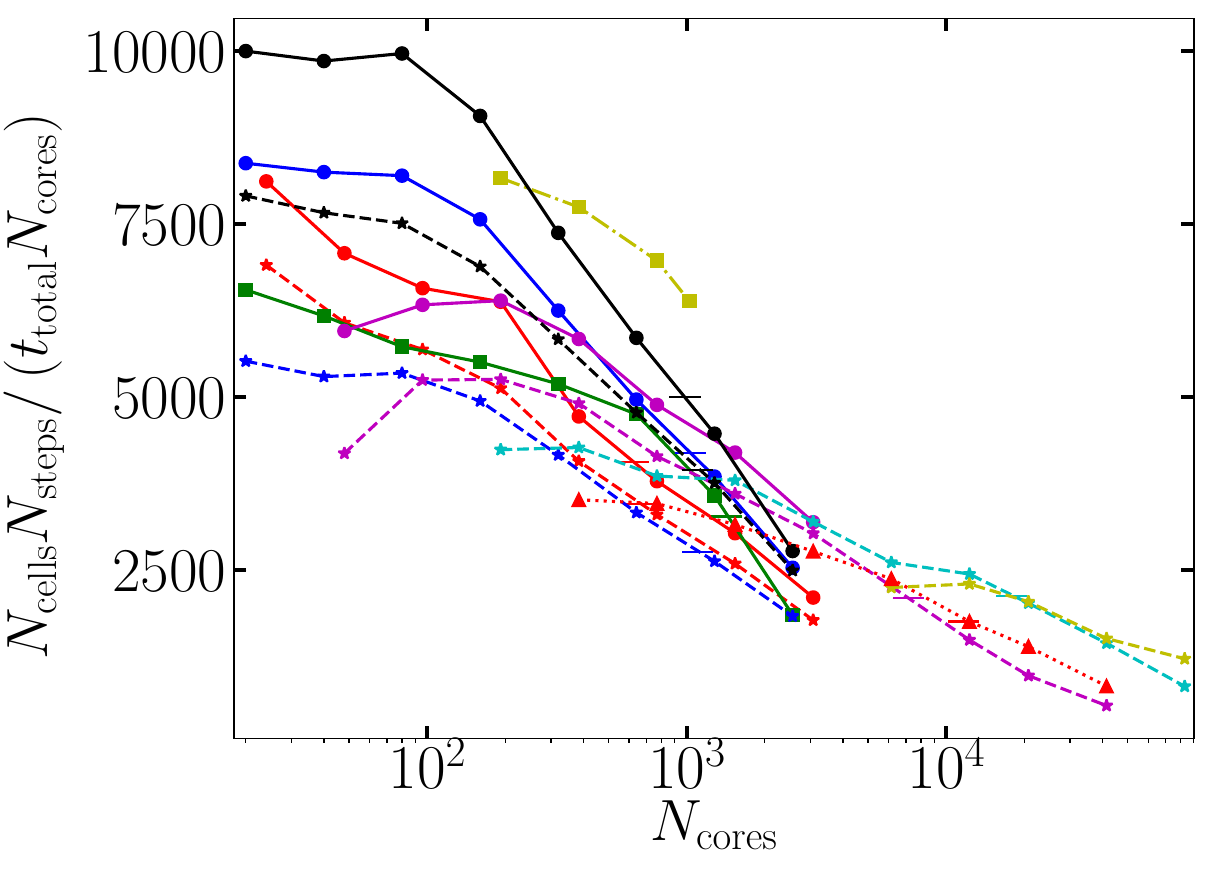} 

\includegraphics[width=16cm]{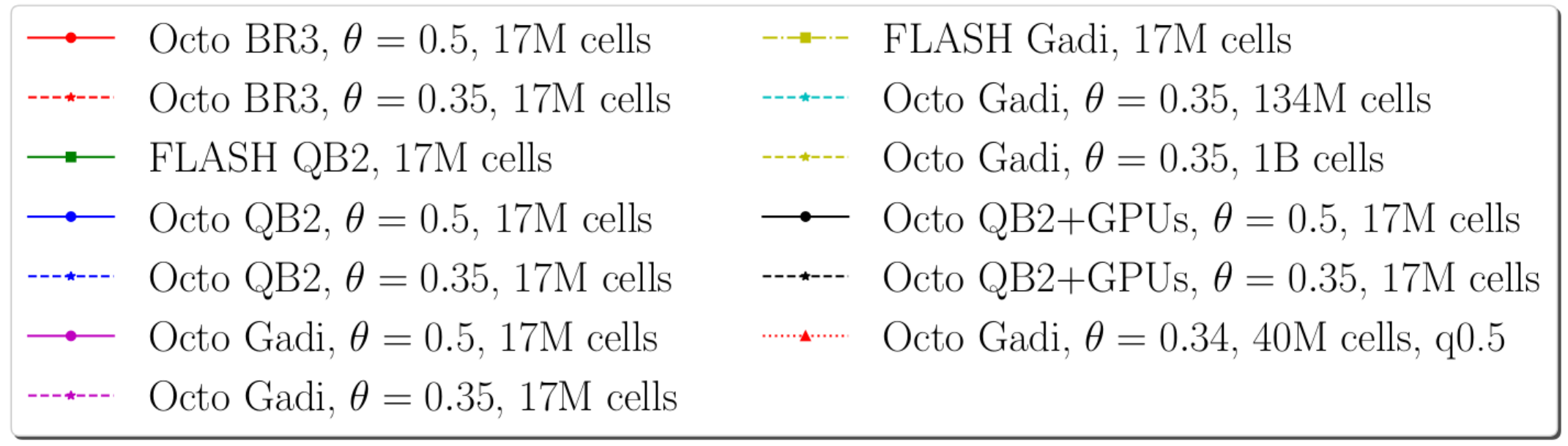}

\caption{\protect\footnotesize{Scaling test carried out on the pulsating polytrope problem of big sizes. QB2 refers to QueenBee2, BR3 refers to the BigRed3. 
The simulations contain $256^3$ (17M), $512^3$ (134M), and $1024^3$ (1B) cells or $4096,\;32\,768,\;262\,144$, subgrids, respectively. One subgrids per core of the 17M cells simulations (top axis on left panels) is 8 subgrids per core of the 134M cells, and 64 subgrids per core of the 1B cells simulations. The short horizontal segments mark the 0.5 efficiency}} 
\label{fig:star_scalability_high_res}
\end{figure*}

In Figures \ref{fig:star_scalability_low_res} and \ref{fig:star_scalability_high_res}, we present two measures of code speed: the computational time per time step and the wall-clock time per time step. We show this including and excluding the initialization time, regridding time, and the time to write out the output. While the computational time is a feature of the code, the time to write out the output is a feature of the machine and file system used and can therefore vary among otherwise similar computers. These quantities are a measure of the strong scaling properties of \octo. We also show the number of time steps $(N_{\rm steps})$ times the number of cells ($N_{\rm cells}$) processed per second, which measures the amount of work that the cores have available to do. If we had perfect scaling, these curves should be horizontal, indicating the cores have sufficient work, but they tend instead to curve downwards as the addition of more cores results in cores not having sufficient work to do. The reason why the curves tilt downwards is also the introduction of additional overheads, such as message sending, due to the addition of nodes. \revision{Note that above a certain high number of cores, each core has computational work of only several subgrids, in which case the communication between cores dominates the performance, and the scaling of the computational time becomes flat as well.}

The values used for the figures are reported in Tables~\ref{tab:Gadi_star_scaling}--\ref{tab:BR3_star_scaling}, organised by computer. In brackets, next to the values used for the plots, we calculate the ``speedup" ($ S_n = (t_N / t_N^\prime) / (N_{\rm core}^\prime / N_{\rm core})$, where $t_N$ and $t_N^\prime$ are the time step lengths for runs that use $N_{\rm core}$ and $N_{\rm core}^\prime$ cores, respectively). These efficiencies are calculated with respect to the run that uses an entire node in each machine (48 cores on Gadi, 20 cores on QB2 and 24 cores on BR3), when available or with the run with the least number of cores ($N_{\rm cores}$), when not. \revision{On the RHS panels of Figures~\ref{fig:star_scalability_low_res} and \ref{fig:star_scalability_high_res}, we mark the value of 0.5 efficiency by short horizontal segments over plotted on each curve.}

In the tables, we have reported on the {\it total time} 
also known as wall-clock time, which includes the initialization, regridding and the writing to disk part of the computation. If we had reported only the computational time 
the efficiencies would increase.


We start by comparing the code performance for different values of the gravity solver parameter $\theta$. This comparison was made with \octo\ with 2 million (solid lines) and 17 million cells (dashed lines) using BR3, QB2, QB2+GPUs, and Gadi (red, blue, black and magenta lines, respectively, Figure~\ref{fig:star_scalability_low_res}). As expected, the smaller the value of $\theta$ (the more precise the gravity solver solution) the longer the simulations take. At 2 million cells, the QB2 tests show an increase in the computational time by increasing the number of cores between 1280 and 2560, while this is less pronounced for the other two computers. This is due to a node-communication problem. The $\theta=0.35$ simulations are doing more communication between sub-grids and hence nodes. The use of CPU+GPU improves the wall clock time, but also displays the upturn. This behaviour is less pronounced for tests with more than 17 million cells  (Figure~\ref{fig:star_scalability_high_res}).

Next we compare the total time taken for the test for any code on any machine. For the 2 and 17-million cell simulations, the Gadi simulations are the fastest, although we note that for a range of core number between $\sim$30 and $\sim$300 it is the QB2 with GPU that is the fastest. This is particularly evident at 17 million cells, looking at the work per core per time plots (Figure~\ref{fig:star_scalability_high_res}, RHS panels). Whether runs with GPUs are faster depends on whether the run is dominated by computation or by communication. GPU are computationally faster, but Gadi has a 4-time faster network communication than any other network used in this paper. Further information on \octo\ runs with GPUs can be found in \citet{daiss2019piz}, and will be the subject of future papers.

We now come to the actual scaling properties. For the 2-million cell simulations, the efficiency of increasing the number of cores drops once we move past 1536 cores: on Gadi from 768 to 1546 to 3072 cores the efficiency drops from 34, to 19 to 10 percent. This is similar on QB2 and BR3 (note that on QB2 we are comparing slightly different number of cores: 640, 1280 and 2560).  At 17 million cells, the same three steps show efficiencies of 99, 86, and 72 percent on Gadi, while for the other computers, they are markedly smaller, with, for instance, 47, 37, 26 percent on BR3, and slightly better but similar values for QB2, making the runs on Gadi the fastest and most scalable.

 Using GPUs with the 2- and 17-million cell simulations reduces the time per step between a factor of 1.5 and 2;  the scalability of the hybrid CPU+GPU runs is slightly inferior to those of the pure CPU runs for the 2-million cell simulations, for which processors have less work to do, but almost identical to the non-GPU runs for the 17-million cell simulations. 

Next, we look at scalability vs. problem size by comparing runs with 2, 17, 134 million and 1 billion cells on Gadi. Looking at the efficiencies when increasing the number of cores between 1536 and 3072, we find 10, 72, and 75 percent for 2, 17, and 134 million cells, respectively which would imply that increasing the size of the problem from 17 to 134 million has not afforded us a much improved scaling. On the other hand, going from 6144 to 12\,736 to 20\,735 to 41\,472 cores and finally to 82\,944 cores, the efficiencies improve markedly upon doubling the number of cores for the 1 billion cell run compared to the 17-million cell run. In fact, it is remarkable that our 1 billion cell run is still scaling very well once we move to $\sim$80\,000 cores, giving hope that we can carry out very high resolution simulations with reasonable wall-clock times (though the actual cost of the simulation in terms of CPU-hours would remain very high).  
\begin{table*}
    \centering
    \begin{tabular}{lcccccccc}
    \hline
    Code     &  \octo\ & \octo\ & \flash\ & \octo\ &  \octo\ & \octo\ & \octo\ & \octo$^{a}$  \\
    $N_{\rm cells}$ &  2M    & 2M   & 17M & 17M   & 17M    & 134M  & 1000M  & 40M \\
    $\theta$ &  0.5   & 0.35  & -- & 0.5   & 0.35   & 0.35  & 0.35   & 0.34 \\
    \hline
    $N_{\rm cores}$ &&&&&&&&\\
    4   & 27.7 & 35.0 & -- & -- & -- & -- & -- & -- \\
    8  & 15.3 & 19.7 & -- & -- & -- & -- & -- & -- \\
    16   & 9.06 & 10.5 & -- & 72.1 & 94.7 & -- & -- & -- \\
    24  & 6.45 & 8.26 & -- & 57.7 & 77.1 & -- & -- & -- \\
    48   & 5.83 (--)  & 6.58 (--) & -- & 58.7 (--) & 83.5 (--) & -- & -- & -- \\
    96  & 3.32 (88\%) & 3.74 (88\%) & -- & 27.6 (125\%)& 33.3 (125\%)& -- & -- & -- \\
    192   & 1.87 (78\%) & 2.10 (78\%) & 10.7 (--) & 13.7 (126\%)& 16.6 (125\%)& 165 (--) & -- & --  \\
    384  & 1.47 (50\%) & 1.36 (60\%) & 5.64 (95\%) & 7.48 (117\%)& 8.90 (117\%)& 81.8 (101\%)& -- & 30.1 (--)\\
    768   & 1.08 (34\%) & 1.09 (38\%) & 3.13 (85\%) & 4.47 (99\%)& 5.27 (99\%)& 45.3 (91\%)& -- & 15.3 (98\%)\\
    1024  & -- & -- & 2.57 (78\%) & -- &  -- & -- & -- & -- \\
    1536  & 0.951 (19\%) & 0.988 (21\%) & --  & 2.60 (86\%)& 3.03 (86\%)& 23.0 (90\%)& -- & 8.37 (90\%)\\
    3072  & 0.955 (10\%) & 1.03 (10\%) & --  & 1.71 (72\%)& 1.80 (72\%)& 13.6 (75\%)& -- & 4.77 (79\%)\\
    6144  & -- & -- & -- & -- & 1.20 (55\%)& 8.36 (62\%)& 77.7 (--) & 2.78 (68\%)\\
    12\,288  & -- & -- & -- & -- & 0.915 (36\%)& 4.47 (58\%)& 38.0 (102\%)& 1.88 (50\%)\\
    20\,736  & -- & -- & -- & -- & 0.829 (23\%)& 3.20 (48\%)& 25.3 (91\%) & 1.40 (40\%) \\
    41\,472  & -- & -- & -- & -- & 0.745 (13\%) & 2.24 (34\%)& 17.1 (67\%) & 1.19 (23\%) \\
    82\,944  & -- & -- & -- & -- & -- & 1.97 (19\%)& 10.6 (54\%) & -- \\
    \hline
    \multicolumn{9}{l}{$^a$ Test using a binary simulation with $q= 0.5$.}
    \end{tabular}
    \caption{Wall clock time per time step in seconds for different scaling tests carried out on Gadi. In parenthesis we report the speedup of a given calculation with respect to the calculation that uses one node (48 cores), interpreted as the fractional reduction of the time step compared to the fractional increase in core count}
    \label{tab:Gadi_star_scaling}
\end{table*}

 

\begin{table*}
    \centering
    \begin{tabular}{lllcccccccc}
    \hline
    Code    & \flash\ & {\sc octo} & {\sc octo} & {\sc octo} & {\sc octo} & \flash\ & {\sc octo} & {\sc octo} & {\sc octo} & {\sc octo} \\
    $N_{\rm cells}$ & 2M     & 2M         & 2M         &  2M        & 2M         & 17M    & 17M        & 17M        & 17M        & 17M        \\
    $\theta$ & --     & 0.5        & 0.35       & 0.5        & 0.35       & --     & 0.5        & 0.35       & 0.5        & 0.35       \\
    GPUs      & --     & --         &  --        & \checkmark          & \checkmark           & --     & --         & --         & \checkmark           & \checkmark          \\
    \hline
    $N_{\rm cores}$ &&&&&&&&&&\\
    2  & -- & 88.7 & 144 & -- & -- & -- & -- & -- & -- & -- \\
    4  & 55.1 & 47.6 & 78.1 & 33.7 & 37.7 & -- & 388 & 637 & 272 & 308 \\
    8  & 30.5  & 27.4 & 43.6 & 20.6 & 24.9 & -- & 231  & 356 & 175 & 207 \\
    16  & 16.7  & 15.1 & 23.4  & 12.2 & 16.0 & -- & 131  & 193 & 107 & 132 \\
    20  & 13.5 (--) & 12.7(--) & 19.1 (--) & 10.7 (--) & 14.0 (--)& 128 (--) & 100 (--) & 152 (--) & 83.9 (--) & 106 (--) \\
    40  & 7.54 (90\%) & 6.80 (93\%) & 10.3 (93\%) & 5.61 (95\%) & 7.89 (89\%)& 67.9 (94\%) & 50.8 (98\%) & 79.2 (96\%) & 42.6 (99\%) & 54.7 (97\%) \\
    80  & 4.19 (81\%) & 3.70 (86\%) & 5.67 (84\%) & 3.05 (87\%) & 4.21 (83\%)& 36.6 (87\%) & 25.6 (98\%) & 39.2 (97\%) & 21.0 (100\%) & 27.9 (95\%) \\
    160  & 2.28 (74\%) & 2.26 (70\%) & 3.37 (71\%)& 1.85 (72\%) & 2.48 (71\%)& 19.0 (84\%) & 13.8 (90\%)& 21.2 (90\%) & 11.6 (91\%) & 15.2 (87\%) \\
    320  & 1.42 (59\%) & 1.62 (49\%) & 2.30 (52\%) & 1.41 (47\%) & 1.63 (54\%)& 10.1 (79\%) & 8.39 (75\%) & 12.6 (75\%) & 7.11 (74\%) & 8.98 (74\%) \\
    640 & 1.15 (37\%) & 1.23 (32\%) & 1.66 (36\%) & 1.14 (29\%) & 1.20 (37\%)& 5.51 (73\%) & 5.28 (59\%) & 7.87 (60\%) & 4.48 (59\%) & 5.49 (60\%) \\
    1280 & 1.19 (18\%) & 1.01 (20\%) & 1.26 (24\%) & 0.975 (17\%) & 1.04 (21\%)& 3.66 (55\%) & 3.40 (46\%) & 4.99 (48\%) & 2.93 (45\%) & 3.48 (48\%) \\
    2560 & 2.46 (4\%) & 1.05 (9\%) & 2.40 (6\%) & 1.00 (8\%) & 2.03 (5\%) & 3.54 (28\%) & 2.58 (30\%) & 3.57 (33\%) & 2.36 (28\%) & 2.63 (32\%)\\
    \hline
    \end{tabular}
    \caption{Wall clock time per time step in seconds for different scaling tests carried out on one stationary polytrope on QueenBee2. In parenthesis we report the speedup of a given calculation with respect to the calculation that uses one node (20 cores), interpreted as the fractional reduction of the time step compared to the fractional increase in core count }
    \label{tab:QB2_star_scaling}
\end{table*}

\begin{table*}
    \centering
    \begin{tabular}{lcccc}
    \hline
    Code     &  \octo\ & \octo\ & \octo\ & \octo\ \\
    $N_{\rm cells}$ &  2M    & 2M    & 17M   & 17M   \\
    $\theta$ &  0.5   & 0.35  &  0.5  & 0.35  \\
    \hline
    $N_{\rm cores}$ &&&&\\
    1  & 122 & 165 &987 & 1390\\
    2  & 62.7 & 84.8 &521& 694\\
    4   & 34.3 & 46.5 &2938& 382\\
    8  & 19.5 & 25.2 &177& 219\\
    16   & 11.7 & 15.0 &107&132\\
    24  & 8.57 (--) & 11.0 (--) &86.11 (--) &101 (--) \\
    48   & 6.01 (71\%) & 7.48 (74\%) &49.4 (87\%) &57.5 (88\%) \\
    96  & 3.81 (56\%) & 4.72 (58\%) &26.6 (81\%) &30.7 (82\%) \\
    192   & 2.36 (45\%) & 2.81 (49\%) & 13.8 (79\%) &17.0 (74\%) \\
    384  & 1.69 (32\%) & 2.03 (34\%) &9.25 (58\%) &10.7 (59\%) \\
    768   & 1.15 (23\%) & 1.38 (25\%) &5.77 (47\%) &6.61 (48\%) \\
    1536  & 0.834 (16\%) & 0.980 (18\%) &3.60 (37\%) &4.21 (38\%) \\
    3072  & 0.785 (9\%) & 0.872 (10\%) &2.59 (26\%) &3.07 (26\%) \\
    \hline
    \end{tabular}
    \caption{Wall clock time per time step in seconds for different scaling tests carried out on one stationary polytrope on BigRed3. In parenthesis we report the speedup  of a given calculation with respect to the calculation that uses one node (24 cores), interpreted as the fractional reduction of the time step compared to the fractional increase in core count}
    \label{tab:BR3_star_scaling}
\end{table*}

The scalability properties of \flash\ were tested on QB2 at 2- and 17-million cell resolutions. They are very comparable to those of \octo. The computational time per step is intermediate between the $\theta = 0.35$ and 0.5, but tends to be longer than even the more accurate $\theta=0.35$ \octo\ simulations when using a higher number of cores.

\begin{figure*}
\centering   
\subfloat[]{\includegraphics[width=7.5cm]{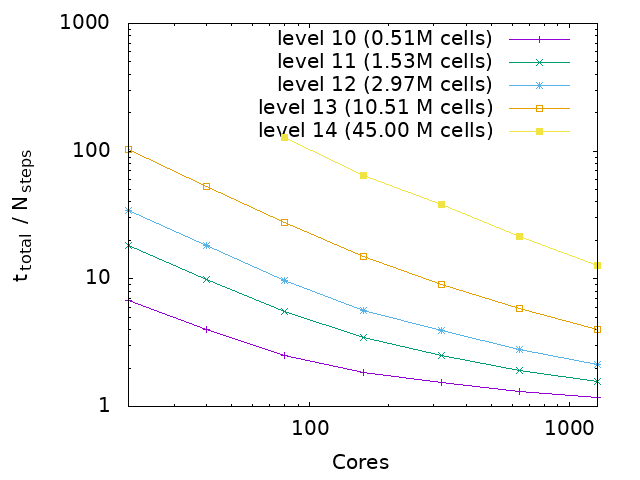}}
\subfloat[]{\includegraphics[width=7.5cm]{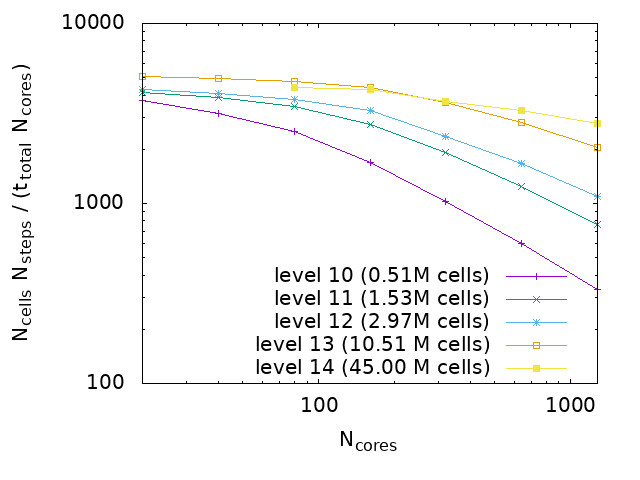}}
\caption{\protect\footnotesize{Scaling results for the first 30 time steps of a binary simulation on LONI's QB2. The left panel shows the total wall-clock time per time step as a function of core count, while the right panel shows the total throughput per core in terms of ${N_{\mathrm{cells}} N_{\mathrm{timesteps}}}/(t_{\mathrm{wallclock}}N_{\mathrm{cores}})$ }}
\label{fig:q0.5_poly_scaling}
\end{figure*}

An additional scaling test was carried out using the first 30 time steps of a binary interaction simulation similar to that of Section~\ref{sec:binarymerger}. The tests was to determine code scalability in view of AMR regridding. We performed a series of simulations on QB2, with increasing resolution by virtue of adding additional levels of refinement, see Table~\ref{tab:v1309_star_scaling} and Figure~\ref{fig:q0.5_poly_scaling}. The code performance is aligned with that we have observed for the static polytrope on the same computer cluster (Table~\ref{tab:QB2_star_scaling}), although a side by side comparison is not possible in virtue because the number of cells was not exactly the same.  

\revision{We finally carried out a last scaling test, where twelve regridding events took place in 100 time stpes, and where we started the simulation from an intermediate stage of the $q=0.5$ simulation in order to trigger substantial regridding (Table~\ref{tab:Gadi_star_scaling}, last column). This scaling runs were carried out on Gadi. Although the number of AMR boundaries varied by $\approx$ 30~percent in these runs, the scaling was similar to a uniform grid with no regridding taking place (see red-dotted triangle line in Figure~\ref{fig:star_scalability_high_res} and the last column in Table~\ref{tab:Gadi_star_scaling}). We conclude that our regridding scheme scales well to high number of cores.}


\begin{table*}
    \centering
    \begin{tabular}{lccccc}
    \hline
    Code     &  \octo\ & \octo\ & \octo\ & \octo\ & \octo \\
    $N_{\rm cells}$ &  0.51M    & 1.5M    & 3.0M   & 10.5M    & 45M  \\
    $\theta$ &  0.34   & 0.34  &  0.34  & 0.34 & 0.34  \\
    \hline
    $N_{\rm cores}$ &&&&&\\
    20   & 6.73 & 18.3 & 34.18 & 103 & --  \\
    40  & 3.98 (85\%) & 9.89 (92\%)& 18.1 (95\%) & 53.1 (97\%) & --  \\
    80   & 2.50 (67\%) & 5.53 (83\%)& 9.76 (88\%)& 27.4 (94\%)& 128  \\
    160  & 1.86 (45\%) & 3.47 (66\%)& 5.67 (76\%)& 15.0 (86\%)& 65.1 (98\%) \\
    320   & 1.53 (27\%) & 2.50 (46\%) & 3.93 (54\%)& 9.04 (71\%) & 37.9 (84\%)\\
    640   & 1.31 (16\%) & 1.91 (30\%) & 2.77 (30\%) & 5.83 (55\%) & 21.4 (75\%)\\
    1280   & 1.19 (9\%) & 1.56 (18\%) & 2.11 (25\%) & 4.02 (40\%) & 12.6 (63\%) \\
    \hline
    \end{tabular}
    \caption{Wall clock time per time step in seconds for different scaling tests carried out on QueenBee2 on a binary interaction simulation similar to the one of Section~\ref{sec:binarymerger}. In parenthesis we report the speedup of a given calculation with respect to the calculation that uses one node (20 cores), except for the $45$ million cell run, which is given with respect to the smallest number of cores tested (80 cores) 
    }
    \label{tab:v1309_star_scaling}
\end{table*}

\section{Summary and Conclusions}
\label{sec:concl}

We have presented a thorough suite of benchmark tests and a comprehensive and detailed description of the AMR hydrodynamics code \octo. Although \octo\ has been used in tandem with other codes in previous publications in astrophysics \citep{Kadam2016,Staff2018} as well as in computer science  \citep[e.g.][]{pfander2018,gregor2019,heller2019}, there has never been an actual ``method paper".

\octo\ joins a small suite of codes specifically designed to study the merger of white dwarf stars while also modelling the earlier, pre-merger phase. \octo\ is designed to be a scalable, accurate code that is able to exploit CPU and GPU computer architectures. It is currently optimised to model stellar mergers for similar-sized stars that can be modelled using a star setup via a barotropic EoS, although for the evolution one can choose among a number of analytical EoS options. It does not currently include nuclear reactions such as for example the code {\sc castro} \citep{Katz2016}. \octo's gravity accuracy outperforms \flash, with overall smaller residuals in the constant density sphere test (Section~\ref{ssec:static_sphere}), although the comparison may not be with exactly the same parameters. The static and translating polytrope tests also show \octo's superior conservation of the peak density and overall mass of the structure, as well as the centre of mass position and overall velocity (for the translating polytropes; Sections~\ref{ssec:polytrope} and \ref{ssec:polytrope_translation}). 

In the rotating polytrope test, simulations in the rotating frame distinctly outperform those calculated in the inertial frame, although even in the inertial frame the polytrope retains its mass, position and radius to excellent precision. \octo\ demonstrates a superior conservation of angular momentum and energy in this test, although the hydrodynamics is not well exercised in single grid tests where there is little gas motion (Section~\ref{ssec:rotating_star}).

We have carried out scaling tests using a static polytrope, where the entire grid is fully refined with resolutions from 2 million to 1 billion cells. We have used three different computer clusters, two university-based mid-tier computers, QueenBee2 (Louisiana State University) and BigRed3 (Indiana University) and one peak facility, Gadi, at the Australian Government's National Computational Infrastructure Centre. Strong scaling properties of the code are excellent up to $\sim82\,000$, the largest number of cores tested, but only for the largest problems of 1 billion cells. With smaller problems of 134 million cells, performance improvement is obtained up to a maximum of $\sim$20\,000 cores (after which the speedup is modest). For the 17 million cell test, similar to the number used in the binary simulation in Section~\ref{sec:binarymerger}, the optimum number of cores is $\sim$6000.

We have also carried out somewhat limited scaling tests of a binary problem using QueenBee2 and Gadi to measure the performance when regridding takes place and found that the scaling properties of the code are similar to those assessed using the stationary polytrope with no AMR regridding.

While the emphasis in this paper is not \octo's ability to use mixed CPU-GPU architectures, we have carried out a small test that showed that the use of 2 GPU units per node improves the step time by a factor of 1.5-2. Given the increased emphasis on the use of GPUs for this type of computation (see references in Section~\ref{ssec:HPX}) and we will further address this capability in future work.

Finally, we have used \octo\ to carry out intermediate resolution WD merger simulations with an ideal gas and a polytropic EoS and compared the results to a similar simulations carried out by \citet{Motl2017} using two different codes, the grid code \flower, and the SPH code {\sc snsph} \citep{Fryer2006}. The total run time of this 2-8 million cell simulation was approximately 8 days and it used 80\,000 CPU-hours on 400 cores of the computer cluster QueenBee2. By comparison the high resolution cylindrical simulation in \citet{Motl2017}, with an approximately similar resolution, ran for over a year. Because of the relatively small wall-clock times achieved by \octo\ thanks to its excellent scalability, higher resolution runs are possible.

Our improvements to \octo\ include (i) applying high resolution only to the binary through adaptive mesh refinement (ii) exploiting the more favorable Courant condition from Cartesian coordinates while maintaining acceptable conservation and (iii) leveraging a parallel framework that enables simulations on a very large number of computing elements. We have effectively reduced the wall clock time for a merger simulation to complete from over a year to essentially one day. These developments enable a wide variety of numerical experiments that would not be possible or conclusive if run only at lower resolution.

An interesting question is how to best exploit resolution improvements. We know from \citet{Motl2017} that simply increasing the resolution will lower the initial mass-transfer rate and greatly increase the early, pre-merger phase of the simulation. This would extend enormously the pre-merger time, making the simulation impossibly long. Under such circumstances we would be forced to increase the artificial angular momentum extraction (the driving) to speed up the simulation to a time when the mass-transfer rate is high enough that the merger takes place. As such, the mass transfer rate and the pre-merger timescale cannot be measured with this type of simulation. On the other hand, there are other parameters of interest that would benefit from higher resolution. An example is the actual structure of the flow during the low mass-transfer rate time, or the details of the flow at or after the time of merger, including the determination of dredged-up elements. 

In future papers we will implement a number of additions. First we will experiment with increasing the photospheric resolution, while at the same time implementing a radiation transport algorithm. With this the immediate aim is to model the light signature, something that would greatly benefit from sufficient resolution at the photosphere as thoroughly explained by \citet{Galaviz2017}. After that, the following improvement will be to use model stars imported from 1D stellar structure simulations, instead of being constructed using barotropic solutions, and that can use a tabulated EoS. A follow-up study is already underway to simulate a $q=0.1$ binary which mimics the V1309 Sco system - this would be only the second simulation of this system, after the 100\,000 SPH particle hydrodynamic simulation of \citet{Nandez2014}, and a $q=0.7$ WD binary thought to be a typical progenitor system for the R Coronae Borealis stars \citep[e.g.,][]{Clayton2011}.

\section*{Acknowlegements}
This work was supported by National Science Foundation Award 1814967.
The numerical work was carried out using the
computational resources (QueenBee2) of the Louisiana Optical Network
Initiative (LONI) and Louisiana State University’s High Performance Computing (LSU HPC). Our use of BigRed3 at Indiana University was supported by Lilly Endowment, Inc., through its support for the Indiana University Pervasive Technology Institute.
This research was undertaken with the assistance of resources (Gadi) from the National Computational Infrastructure (NCI Australia), an NCRIS enabled capability supported by the Australian Government. We wish to thank Dale Roberts at NCI and Richard Miller at Macquarie University for helping the team to run scaling tests on an exorbitant number of Gadi cores. Finally we thank Macquarie University for funding the Gadi computer time from the university allocation. 

\section*{Data Availability}
The source code of \octo\ \citep{dominic_marcello_2021_4432574} is available on GitHub\footnote{\url{https://github.com/STEllAR-GROUP/octotiger}} released under the Boost Software License Version 1. The build scripts to build \octo and its dependencies are available on GitHub\footnote{\url{https://github.com/diehlpk/PowerTiger}} as well. The input files to run the \octo\ simulations are available on Zenodo \citep{dominic_c_marcello_2020_4393374}. Table~\ref{tab:octo_dependencies} lists the software version used on the different clusters. Note that we used the pre-compiled MPI version on the cluster and therefore have some variation there. If the gcc compiler was recent enough to compile HPX and \octo\ we used the pre-compiled version on the cluster.

For movies of the simulations and download of selected simulation data see our YouTube channel\footnote{\url{https://youtube.com/playlist?list=PL7vEgTL3FalaLHjQQH5UtqEVR9-WkFeZp}}.

%
%


\clearpage
\bibliographystyle{mnras}
\bibliography{bibliography}

\begin{thebibliography}{}
\makeatletter
\relax
\def\mn@urlcharsother{\let\do\@makeother \do\$\do\&\do\#\do\^\do\_\do\%\do\~}
\def\mn@doi{\begingroup\mn@urlcharsother \@ifnextchar [ {\mn@doi@}
  {\mn@doi@[]}}
\def\mn@doi@[#1]#2{\def\@tempa{#1}\ifx\@tempa\@empty \href
  {http://dx.doi.org/#2} {doi:#2}\else \href {http://dx.doi.org/#2} {#1}\fi
  \endgroup}
\def\mn@eprint#1#2{\mn@eprint@#1:#2::\@nil}
\def\mn@eprint@arXiv#1{\href {http://arxiv.org/abs/#1} {{\tt arXiv:#1}}}
\def\mn@eprint@dblp#1{\href {http://dblp.uni-trier.de/rec/bibtex/#1.xml}
  {dblp:#1}}
\def\mn@eprint@#1:#2:#3:#4\@nil{\def\@tempa {#1}\def\@tempb {#2}\def\@tempc
  {#3}\ifx \@tempc \@empty \let \@tempc \@tempb \let \@tempb \@tempa \fi \ifx
  \@tempb \@empty \def\@tempb {arXiv}\fi \@ifundefined
  {mn@eprint@\@tempb}{\@tempb:\@tempc}{\expandafter \expandafter \csname
  mn@eprint@\@tempb\endcsname \expandafter{\@tempc}}}

\bibitem[\protect\citeauthoryear{Amini \& Kaiser}{Amini \&
  Kaiser}{2019}]{amini2019assessing}
Amini P.,  Kaiser H.,  2019, in 2019 IEEE/ACM Third Annual Workshop on Emerging
  Parallel and Distributed Runtime Systems and Middleware (IPDRM). pp 26--33

\bibitem[\protect\citeauthoryear{Bordner \& Norman}{Bordner \&
  Norman}{2012}]{10.5555/2462077.2462081}
Bordner J.,  Norman M.~L.,  2012, in Proceedings of the Extreme Scaling
  Workshop. BW-XSEDE '12.
University of Illinois at Urbana-Champaign, USA

\bibitem[\protect\citeauthoryear{{Bryan}, {Norman}, {Stone}, {Cen}  \&
  {Ostriker}}{{Bryan} et~al.}{1995}]{Bryanetal1995}
{Bryan} G.~L.,  {Norman} M.~L.,  {Stone} J.~M.,  {Cen} R.,   {Ostriker} J.~P.,
  1995, \mn@doi [Computer Physics Communications]
  {10.1016/0010-4655(94)00191-4}, \href
  {https://ui.adsabs.harvard.edu/abs/1995CoPhC..89..149B} {89, 149}

\bibitem[\protect\citeauthoryear{{Bulla}, {Sim}, {Pakmor}, {Kromer},
  {Taubenberger}, {R{\"o}pke}, {Hillebrandt}  \& {Seitenzahl}}{{Bulla}
  et~al.}{2016}]{Bulla2016}
{Bulla} M.,  {Sim} S.~A.,  {Pakmor} R.,  {Kromer} M.,  {Taubenberger} S.,
  {R{\"o}pke} F.~K.,  {Hillebrandt} W.,   {Seitenzahl} I.~R.,  2016, \mn@doi
  [\mnras] {10.1093/mnras/stv2402}, \href
  {https://ui.adsabs.harvard.edu/abs/2016MNRAS.455.1060B} {455, 1060}

\bibitem[\protect\citeauthoryear{{Burdge} et~al.,}{{Burdge}
  et~al.}{2020}]{Burdge20}
{Burdge} K.~B.,  et~al., 2020, \mn@doi [\apjl] {10.3847/2041-8213/abca91},
  \href {https://ui.adsabs.harvard.edu/abs/2020ApJ...905L...7B} {905, L7}

\bibitem[\protect\citeauthoryear{{Clayton}}{{Clayton}}{2012}]{clayton_rcb}
{Clayton} G.~C.,  2012, Journal of the American Association of Variable Star
  Observers (JAAVSO), \href
  {https://ui.adsabs.harvard.edu/abs/2012JAVSO..40..539C} {40, 539}

\bibitem[\protect\citeauthoryear{{Clayton} et~al.,}{{Clayton}
  et~al.}{2011}]{Clayton2011}
{Clayton} G.~C.,  et~al., 2011, \mn@doi [\apj] {10.1088/0004-637X/743/1/44},
  \href {https://ui.adsabs.harvard.edu/abs/2011ApJ...743...44C} {743, 44}

\bibitem[\protect\citeauthoryear{Colella \& Woodward}{Colella \&
  Woodward}{1984}]{COLELLA1984174}
Colella P.,  Woodward P.~R.,  1984, \mn@doi [Journal of Computational Physics]
  {https://doi.org/10.1016/0021-9991(84)90143-8}, 54, 174

\bibitem[\protect\citeauthoryear{Copik \& Kaiser}{Copik \&
  Kaiser}{2017}]{copik2017using}
Copik M.,  Kaiser H.,  2017, in Proceedings of the 5th International Workshop
  on OpenCL. pp~1--7

\bibitem[\protect\citeauthoryear{D'Souza, Motl, Tohline  \& Frank}{D'Souza
  et~al.}{2006}]{D_Souza_2006}
D'Souza M. C.~R.,  Motl P.~M.,  Tohline J.~E.,   Frank J.,  2006, \mn@doi [The
  Astrophysical Journal] {10.1086/500384}, 643, 381

\bibitem[\protect\citeauthoryear{Dai{\ss} et~al.,}{Dai{\ss}
  et~al.}{2019}]{daiss2019piz}
Dai{\ss} G.,  et~al., 2019, in Proceedings of the International Conference for
  High Performance Computing, Networking, Storage and Analysis. pp 1--37

\bibitem[\protect\citeauthoryear{Daiß et~al.,}{Daiß
  et~al.}{2019}]{gregor2019}
Daiß G.,  et~al., 2019, From Piz Daint to the Stars: Simulation of Stellar
  Mergers using High-Level Abstractions (\mn@eprint {arXiv} {1908.03121})

\bibitem[\protect\citeauthoryear{{De Marco}, {Long}, {Jacoby}, {Hillwig},
  {Kronberger}, {Howell}, {Reindl}  \& {Margheim}}{{De Marco}
  et~al.}{2015}]{DeMarco2015}
{De Marco} O.,  {Long} J.,  {Jacoby} G.~H.,  {Hillwig} T.,  {Kronberger} M.,
  {Howell} S.~B.,  {Reindl} N.,   {Margheim} S.,  2015, \mn@doi [\mnras]
  {10.1093/mnras/stv249}, \href
  {https://ui.adsabs.harvard.edu/abs/2015MNRAS.448.3587D} {448, 3587}

\bibitem[\protect\citeauthoryear{{Dehnen}}{{Dehnen}}{2000}]{Dehnen2000}
{Dehnen} W.,  2000, \mn@doi [\apjl] {10.1086/312724}, \href
  {https://ui.adsabs.harvard.edu/abs/2000ApJ...536L..39D} {536, L39}

\bibitem[\protect\citeauthoryear{Després \& Labourasse}{Després \&
  Labourasse}{2015}]{DESPRES201528}
Després B.,  Labourasse E.,  2015, \mn@doi [Journal of Computational Physics]
  {https://doi.org/10.1016/j.jcp.2015.02.032}, 290, 28

\bibitem[\protect\citeauthoryear{{Diehl}, {Seshadri}, {Heller}  \&
  {Kaiser}}{{Diehl} et~al.}{2018}]{8638479}
{Diehl} P.,  {Seshadri} M.,  {Heller} T.,   {Kaiser} H.,  2018, in 2018
  IEEE/ACM 4th International Workshop on Extreme Scale Programming Models and
  Middleware (ESPM2). pp 19--28, \mn@doi{10.1109/ESPM2.2018.00006}

\bibitem[\protect\citeauthoryear{{Duquennoy} \& {Mayor}}{{Duquennoy} \&
  {Mayor}}{1991}]{duquennoy}
{Duquennoy} A.,  {Mayor} M.,  1991, Astronomy and Astrophysics, \href
  {http://adsabs.harvard.edu/abs/1991A%26A...248..485D} {248, 485}

\bibitem[\protect\citeauthoryear{Even \& Tohline}{Even \&
  Tohline}{2009}]{Even2009}
Even W.,  Tohline J.~E.,  2009, \mn@doi [Astrophys. J. Suppl.]
  {10.1088/0067-0049/184/2/248}, 184, 248

\bibitem[\protect\citeauthoryear{{Frank}, {King}  \& {Raine}}{{Frank}
  et~al.}{2002}]{2002apa..book.....F}
{Frank} J.,  {King} A.,   {Raine} D.~J.,  2002, {Accretion Power in
  Astrophysics: Third Edition}

\bibitem[\protect\citeauthoryear{{Fryer}, {Rockefeller}  \& {Warren}}{{Fryer}
  et~al.}{2006}]{Fryer2006}
{Fryer} C.~L.,  {Rockefeller} G.,   {Warren} M.~S.,  2006, \mn@doi [\apj]
  {10.1086/501493}, \href {http://adsabs.harvard.edu/abs/2006ApJ...643..292F}
  {643, 292}

\bibitem[\protect\citeauthoryear{{Fryxell} et~al.,}{{Fryxell}
  et~al.}{2000}]{Fryxel2000}
{Fryxell} B.,  et~al., 2000, \mn@doi [\apjs] {10.1086/317361}, \href
  {https://ui.adsabs.harvard.edu/abs/2000ApJS..131..273F} {131, 273}

\bibitem[\protect\citeauthoryear{{Galaviz}, {De Marco}, {Passy}, {Staff}  \&
  {Iaconi}}{{Galaviz} et~al.}{2017}]{Galaviz2017}
{Galaviz} P.,  {De Marco} O.,  {Passy} J.-C.,  {Staff} J.~E.,   {Iaconi} R.,
  2017, \mn@doi [\apjs] {10.3847/1538-4365/aa64e1}, \href
  {http://adsabs.harvard.edu/abs/2017ApJS..229...36G} {229, 36}

\bibitem[\protect\citeauthoryear{{Hachisu}}{{Hachisu}}{1986a}]{Hachisu1986a}
{Hachisu} I.,  1986a, \mn@doi [\apjs] {10.1086/191121}, \href
  {https://ui.adsabs.harvard.edu/abs/1986ApJS...61..479H} {61, 479}

\bibitem[\protect\citeauthoryear{{Hachisu}}{{Hachisu}}{1986b}]{Hachisu1986b}
{Hachisu} I.,  1986b, \mn@doi [\apjs] {10.1086/191148}, \href
  {https://ui.adsabs.harvard.edu/abs/1986ApJS...62..461H} {62, 461}

\bibitem[\protect\citeauthoryear{Heller et~al.,}{Heller
  et~al.}{2019a}]{BILLIONS}
Heller T.,  et~al., 2019a, \mn@doi [The International Journal of High
  Performance Computing Applications] {10.1177/1094342018819744}, 33, 699

\bibitem[\protect\citeauthoryear{Heller et~al.,}{Heller
  et~al.}{2019b}]{heller2019}
Heller T.,  et~al., 2019b, \mn@doi [The International Journal of High
  Performance Computing Applications] {10.1177/1094342018819744}, 33, 699

\bibitem[\protect\citeauthoryear{Hillebrandt \& Niemeyer}{Hillebrandt \&
  Niemeyer}{2000}]{typeia}
Hillebrandt W.,  Niemeyer J.~C.,  2000, \mn@doi [Annual Review of Astronomy and
  Astrophysics] {10.1146/annurev.astro.38.1.191}, 38, 191

\bibitem[\protect\citeauthoryear{{Hurley}, {Roberts}  \& {Wright}}{{Hurley}
  et~al.}{1966}]{Hurley1966}
{Hurley} M.,  {Roberts} P.~H.,   {Wright} K.,  1966, \mn@doi [\apj]
  {10.1086/148532}, \href
  {https://ui.adsabs.harvard.edu/abs/1966ApJ...143..535H} {143, 535}

\bibitem[\protect\citeauthoryear{{Ivanova} et~al.,}{{Ivanova}
  et~al.}{2013}]{Ivanovaetal2013}
{Ivanova} N.,  et~al., 2013, \mn@doi [\aapr] {10.1007/s00159-013-0059-2}, \href
  {https://ui.adsabs.harvard.edu/abs/2013A&ARv..21...59I} {21, 59}

\bibitem[\protect\citeauthoryear{{Ivezi{\'c}} et~al.,}{{Ivezi{\'c}}
  et~al.}{2007}]{Ivezic2007}
{Ivezi{\'c}} {\v Z}.,  et~al., 2007, in {Valsecchi} G.~B.,  {Vokrouhlick{\'y}}
  D.,   {Milani} A.,  eds,  IAU Symposium Vol. 236, Near Earth Objects, our
  Celestial Neighbors: Opportunity and Risk. pp 353--362 (\mn@eprint {}
  {astro-ph/0701506}), \mn@doi{10.1017/S1743921307003420}

\bibitem[\protect\citeauthoryear{Jetley, Gioachin, Mendes, Kale  \&
  Quinn}{Jetley et~al.}{2008}]{jetley2008massively}
Jetley P.,  Gioachin F.,  Mendes C.,  Kale L.~V.,   Quinn T.,  2008, in 2008
  IEEE International Symposium on Parallel and Distributed Processing. pp 1--12

\bibitem[\protect\citeauthoryear{{Jha}}{{Jha}}{2017}]{typeiax}
{Jha} S.~W.,  2017, {Type Iax Supernovae}.
p.~375, \mn@doi{10.1007/978-3-319-21846-5_42}

\bibitem[\protect\citeauthoryear{Kadam, Motl, Frank, Clayton  \&
  Marcello}{Kadam et~al.}{2016}]{Kadam2016}
Kadam K.,  Motl P.,  Frank J.,  Clayton G.,   Marcello D.,  2016, \mn@doi
  [Monthly Notices of the Royal Astronomical Society] {10.1093/mnras/stw1814},
  462, stw1814

\bibitem[\protect\citeauthoryear{{Kadam}, {Motl}, {Marcello}, {Frank}  \&
  {Clayton}}{{Kadam} et~al.}{2018}]{Kadam2018}
{Kadam} K.,  {Motl} P.~M.,  {Marcello} D.~C.,  {Frank} J.,   {Clayton} G.~C.,
  2018, \mn@doi [\mnras] {10.1093/mnras/sty2540}, \href
  {https://ui.adsabs.harvard.edu/abs/2018MNRAS.481.3683K} {481, 3683}

\bibitem[\protect\citeauthoryear{Kaiser, Brodowicz  \& Sterling}{Kaiser
  et~al.}{2009}]{kaiser2009parallex}
Kaiser H.,  Brodowicz M.,   Sterling T.,  2009, in 2009 International
  Conference on Parallel Processing Workshops. pp 394--401

\bibitem[\protect\citeauthoryear{Kaiser, Heller, Adelstein-Lelbach, Serio  \&
  Fey}{Kaiser et~al.}{2014}]{kaiser2014hpx}
Kaiser H.,  Heller T.,  Adelstein-Lelbach B.,  Serio A.,   Fey D.,  2014, in
  Proceedings of the 8th International Conference on Partitioned Global Address
  Space Programming Models. pp 1--11

\bibitem[\protect\citeauthoryear{Kaiser et~al.,}{Kaiser
  et~al.}{2020}]{Kaiser2020}
Kaiser H.,  et~al., 2020, \mn@doi [Journal of Open Source Software]
  {10.21105/joss.02352}, 5, 2352

\bibitem[\protect\citeauthoryear{Kale \& Krishnan}{Kale \&
  Krishnan}{1993}]{10.5555/871085}
Kale L.~V.,  Krishnan S.,  1993, Technical report, CHARM++: A Portable
  Concurrent Object Oriented System Based on C++.
USA

\bibitem[\protect\citeauthoryear{{Kashyap}, {Haque}, {Lor{\'e}n-Aguilar},
  {Garc{\'\i}a-Berro}  \& {Fisher}}{{Kashyap} et~al.}{2018}]{Kashyap2018}
{Kashyap} R.,  {Haque} T.,  {Lor{\'e}n-Aguilar} P.,  {Garc{\'\i}a-Berro} E.,
  {Fisher} R.,  2018, \mn@doi [\apj] {10.3847/1538-4357/aaedb7}, \href
  {https://ui.adsabs.harvard.edu/abs/2018ApJ...869..140K} {869, 140}

\bibitem[\protect\citeauthoryear{{Katz}, {Zingale}, {Calder}, {Swesty},
  {Almgren}  \& {Zhang}}{{Katz} et~al.}{2016}]{Katz2016}
{Katz} M.~P.,  {Zingale} M.,  {Calder} A.~C.,  {Swesty} F.~D.,  {Almgren}
  A.~S.,   {Zhang} W.,  2016, \mn@doi [\apj] {10.3847/0004-637X/819/2/94},
  \href {https://ui.adsabs.harvard.edu/abs/2016ApJ...819...94K} {819, 94}

\bibitem[\protect\citeauthoryear{{Kippenhahn}, {Kohl}  \&
  {Weigert}}{{Kippenhahn} et~al.}{1967}]{HeWD1}
{Kippenhahn} R.,  {Kohl} K.,   {Weigert} A.,  1967, Zeitschrift f\"{u}r
  Astrophysik, \href {http://adsabs.harvard.edu/abs/1967ZA.....66...58K} {66,
  58}

\bibitem[\protect\citeauthoryear{{Kippenhahn}, {Thomas}  \&
  {Weigert}}{{Kippenhahn} et~al.}{1968}]{HeWD2}
{Kippenhahn} R.,  {Thomas} H.-C.,   {Weigert} A.,  1968, Zeitschrift f\"{u}r
  Astrophysik, \href {http://adsabs.harvard.edu/abs/1968ZA.....69..265K} {69,
  265}

\bibitem[\protect\citeauthoryear{Kurganov, Noelle  \& Petrova}{Kurganov
  et~al.}{2000}]{KURGANOV00}
Kurganov A.,  Noelle S.,   Petrova G.,  2000, SIAM J. Sci. Comput, 23, 707

\bibitem[\protect\citeauthoryear{{MacLeod} \& {Loeb}}{{MacLeod} \&
  {Loeb}}{2020}]{MacLeod2020}
{MacLeod} M.,  {Loeb} A.,  2020, \mn@doi [\apj] {10.3847/1538-4357/ab822e},
  \href {https://ui.adsabs.harvard.edu/abs/2020ApJ...893..106M} {893, 106}

\bibitem[\protect\citeauthoryear{{Marcello} \& {Tohline}}{{Marcello} \&
  {Tohline}}{2012}]{Marcello2012}
{Marcello} D.~C.,  {Tohline} J.~E.,  2012, \mn@doi [\apjs]
  {10.1088/0067-0049/199/2/35}, \href
  {https://ui.adsabs.harvard.edu/abs/2012ApJS..199...35M} {199, 35}

\bibitem[\protect\citeauthoryear{{Marcello}, {Kadam}, {Clayton}, {Frank},
  {Kaisar}  \& {Motl}}{{Marcello} et~al.}{2016}]{Marcello2016}
{Marcello} D.,  {Kadam} K.,  {Clayton} G.~C.,  {Frank} J.,  {Kaisar} H.,
  {Motl} P.,  2016, in Accretion Processes in Cosmic Sources. p.~55

\bibitem[\protect\citeauthoryear{Marcello, Shiber, Marco, Frank, Clayton, Motl,
  Diehl  \& Kaiser}{Marcello et~al.}{2020}]{dominic_c_marcello_2020_4393374}
Marcello D.~C.,  Shiber S.,  Marco O.~D.,  Frank J.,  Clayton G.~C.,  Motl
  P.~M.,  Diehl P.,   Kaiser H.,  2020, {Files for reproducing results in
  Octo-Tiger: a new, 3D hydrodynamic code for stellar mergers that uses HPX
  parallelisation}, \mn@doi{10.5281/zenodo.4393374}, \url
  {https://doi.org/10.5281/zenodo.4393374}

\bibitem[\protect\citeauthoryear{Marcello et~al.,}{Marcello
  et~al.}{2021}]{dominic_marcello_2021_4432574}
Marcello D.,  et~al., 2021, STEllAR-GROUP/octotiger: Benchmark paper,
  \mn@doi{10.5281/zenodo.4432574}, \url
  {https://doi.org/10.5281/zenodo.4432574}

\bibitem[\protect\citeauthoryear{{Mason}, {Diaz}, {Williams}, {Preston}  \&
  {Bensby}}{{Mason} et~al.}{2010}]{Masonetal2010}
{Mason} E.,  {Diaz} M.,  {Williams} R.~E.,  {Preston} G.,   {Bensby} T.,  2010,
  \mn@doi [\aap] {10.1051/0004-6361/200913610}, \href
  {https://ui.adsabs.harvard.edu/abs/2010A&A...516A.108M} {516, A108}

\bibitem[\protect\citeauthoryear{{Meyer} \& {Meyer-Hofmeister}}{{Meyer} \&
  {Meyer-Hofmeister}}{1979}]{MeyerMHofmeister1979}
{Meyer} F.,  {Meyer-Hofmeister} E.,  1979, \aap, \href
  {https://ui.adsabs.harvard.edu/abs/1979A&A....78..167M} {78, 167}

\bibitem[\protect\citeauthoryear{Motl, Tohline  \& Frank}{Motl
  et~al.}{2002}]{Motl_2002}
Motl P.~M.,  Tohline J.~E.,   Frank J.,  2002, \mn@doi [The Astrophysical
  Journal Supplement Series] {10.1086/324159}, 138, 121

\bibitem[\protect\citeauthoryear{{Motl}, {Frank}, {Tohline}  \&
  {D'Souza}}{{Motl} et~al.}{2007}]{Motl2007}
{Motl} P.~M.,  {Frank} J.,  {Tohline} J.~E.,   {D'Souza} M. C.~R.,  2007,
  \mn@doi [\apj] {10.1086/522076}, \href
  {https://ui.adsabs.harvard.edu/abs/2007ApJ...670.1314M} {670, 1314}

\bibitem[\protect\citeauthoryear{{Motl}, {Frank}, {Staff}, {Clayton}, {Fryer},
  {Even}, {Diehl}  \& {Tohline}}{{Motl} et~al.}{2017}]{Motl2017}
{Motl} P.~M.,  {Frank} J.,  {Staff} J.,  {Clayton} G.~C.,  {Fryer} C.~L.,
  {Even} W.,  {Diehl} S.,   {Tohline} J.~E.,  2017, \mn@doi [\apjs]
  {10.3847/1538-4365/aa5bde}, \href
  {https://ui.adsabs.harvard.edu/abs/2017ApJS..229...27M} {229, 27}

\bibitem[\protect\citeauthoryear{{Nandez}, {Ivanova}  \& {Lombardi}}{{Nandez}
  et~al.}{2014}]{Nandez2014}
{Nandez} J.~L.~A.,  {Ivanova} N.,   {Lombardi} Jr. J.~C.,  2014, \mn@doi [\apj]
  {10.1088/0004-637X/786/1/39}, \href
  {http://adsabs.harvard.edu/abs/2014ApJ...786...39N} {786, 39}

\bibitem[\protect\citeauthoryear{{Pakmor}, {Kromer}, {Taubenberger}, {Sim},
  {R{\"o}pke}  \& {Hillebrandt}}{{Pakmor} et~al.}{2012}]{Pakmor2012}
{Pakmor} R.,  {Kromer} M.,  {Taubenberger} S.,  {Sim} S.~A.,  {R{\"o}pke}
  F.~K.,   {Hillebrandt} W.,  2012, \mn@doi [\apjl]
  {10.1088/2041-8205/747/1/L10}, \href
  {https://ui.adsabs.harvard.edu/abs/2012ApJ...747L..10P} {747, L10}

\bibitem[\protect\citeauthoryear{{Pejcha}, {Metzger}  \& {Tomida}}{{Pejcha}
  et~al.}{2016}]{Pejchaetal2016b}
{Pejcha} O.,  {Metzger} B.~D.,   {Tomida} K.,  2016, \mn@doi [\mnras]
  {10.1093/mnras/stw1481}, \href
  {https://ui.adsabs.harvard.edu/abs/2016MNRAS.461.2527P} {461, 2527}

\bibitem[\protect\citeauthoryear{Pfander, Dai\ss, Marcello, Kaiser  \&
  Pfl\"{u}ger}{Pfander et~al.}{2018}]{pfander2018}
Pfander D.,  Dai\ss G.,  Marcello D.,  Kaiser H.,   Pfl\"{u}ger D.,  2018, in
  Proceedings of the International Workshop on OpenCL. IWOCL '18.
ACM, New York, NY, USA, pp 19:1--19:8, \mn@doi{10.1145/3204919.3204938}, \url
  {http://doi.acm.org/10.1145/3204919.3204938}

\bibitem[\protect\citeauthoryear{{Reichardt}, {De Marco}, {Iaconi}, {Tout}  \&
  {Price}}{{Reichardt} et~al.}{2019}]{Reichardt2019}
{Reichardt} T.~A.,  {De Marco} O.,  {Iaconi} R.,  {Tout} C.~A.,   {Price}
  D.~J.,  2019, \mn@doi [\mnras] {10.1093/mnras/sty3485}, \href
  {https://ui.adsabs.harvard.edu/abs/2019MNRAS.484..631R} {484, 631}

\bibitem[\protect\citeauthoryear{{Ricker}, {Timmes}, {Taam}  \&
  {Webbink}}{{Ricker} et~al.}{2019}]{ricker2019}
{Ricker} P.~M.,  {Timmes} F.~X.,  {Taam} R.~E.,   {Webbink} R.~F.,  2019,
  \mn@doi [IAU Symposium] {10.1017/S1743921318007433}, \href
  {https://ui.adsabs.harvard.edu/abs/2019IAUS..346..449R} {346, 449}

\bibitem[\protect\citeauthoryear{{Rucinski}}{{Rucinski}}{2010}]{2010AIPC.1314...29R}
{Rucinski} S.,  2010, in {Kalogera} V.,  {van der Sluys} M.,  eds, ~ Vol. 1314,
  American Institute of Physics Conference Series. pp 29--36,
  \mn@doi{10.1063/1.3536391}

\bibitem[\protect\citeauthoryear{{Schneider}, {Ohlmann}, {Podsiadlowski},
  {R{\"o}pke}, {Balbus}, {Pakmor}  \& {Springel}}{{Schneider}
  et~al.}{2019}]{Schneider2019}
{Schneider} F. R.~N.,  {Ohlmann} S.~T.,  {Podsiadlowski} P.,  {R{\"o}pke}
  F.~K.,  {Balbus} S.~A.,  {Pakmor} R.,   {Springel} V.,  2019, \mn@doi [\nat]
  {10.1038/s41586-019-1621-5}, \href
  {https://ui.adsabs.harvard.edu/abs/2019Natur.574..211S} {574, 211}

\bibitem[\protect\citeauthoryear{{Sedov}}{{Sedov}}{1946}]{Sedov1946}
{Sedov} L.~I.,  1946, Journal of Applied Mathematics and Mechanics, 10, 241

\bibitem[\protect\citeauthoryear{Shu \& Osher}{Shu \& Osher}{1989}]{SHU1989}
Shu C.-W.,  Osher S.,  1989, \mn@doi [Journal of Computational Physics]
  {https://doi.org/10.1016/0021-9991(89)90222-2}, 83, 32

\bibitem[\protect\citeauthoryear{{Smith}}{{Smith}}{1984}]{1984QJRAS..25..405S}
{Smith} R.~C.,  1984, Quarterly Journal of the Royal Astronomical Society,
  \href {http://adsabs.harvard.edu/abs/1984QJRAS..25..405S} {25, 405}

\bibitem[\protect\citeauthoryear{{Sod}}{{Sod}}{1978}]{Sod1978}
{Sod} G.~A.,  1978, \mn@doi [Journal of Computational Physics]
  {10.1016/0021-9991(78)90023-2}, \href
  {https://ui.adsabs.harvard.edu/abs/1978JCoPh..27....1S} {27, 1}

\bibitem[\protect\citeauthoryear{Solheim}{Solheim}{2010}]{amcvn}
Solheim J.-E.,  2010, \mn@doi [Publications of the Astronomical Society of the
  Pacific] {10.1086/656680}, 122, 1133

\bibitem[\protect\citeauthoryear{{Staff} et~al.,}{{Staff}
  et~al.}{2018}]{Staff2018}
{Staff} J.~E.,  et~al., 2018, \mn@doi [\apj] {10.3847/1538-4357/aaca3d}, \href
  {https://ui.adsabs.harvard.edu/abs/2018ApJ...862...74S} {862, 74}

\bibitem[\protect\citeauthoryear{{Tauris} \& {van den Heuvel}}{{Tauris} \& {van
  den Heuvel}}{2014}]{Tauris2014}
{Tauris} T.~M.,  {van den Heuvel} E.~P.~J.,  2014, \mn@doi [\apjl]
  {10.1088/2041-8205/781/1/L13}, \href
  {https://ui.adsabs.harvard.edu/abs/2014ApJ...781L..13T} {781, L13}

\bibitem[\protect\citeauthoryear{{The C++ Standards Committee}}{{The C++
  Standards Committee}}{2017}]{cxx17_standard}
{The C++ Standards Committee} 2017, Technical report, {ISO International
  Standard ISO/IEC 14882:2017, Programming Language C++}.
{Geneva, Switzerland: International Organization for Standardization (ISO).}

\bibitem[\protect\citeauthoryear{{The C++ Standards Committee}}{{The C++
  Standards Committee}}{2020}]{cxx20_standard}
{The C++ Standards Committee} 2020, Technical report, {ISO International
  Standard ISO/IEC 14882:2020, Programming Language C++}.
{Geneva, Switzerland: International Organization for Standardization (ISO).}

\bibitem[\protect\citeauthoryear{Thoman et~al.,}{Thoman
  et~al.}{2018}]{thoman2018taxonomy}
Thoman P.,  et~al., 2018, The Journal of Supercomputing, 74, 1422

\bibitem[\protect\citeauthoryear{{Tylenda} et~al.,}{{Tylenda}
  et~al.}{2011}]{tylenda2011}
{Tylenda} R.,  et~al., 2011, \mn@doi [\aap] {10.1051/0004-6361/201016221},
  \href {http://adsabs.harvard.edu/abs/2011A%26A...528A.114T} {528, A114}

\bibitem[\protect\citeauthoryear{{Verbunt} \& {Rappaport}}{{Verbunt} \&
  {Rappaport}}{1988}]{Verbunt1988}
{Verbunt} F.,  {Rappaport} S.,  1988, \mn@doi [\apj] {10.1086/166645}, \href
  {https://ui.adsabs.harvard.edu/abs/1988ApJ...332..193V} {332, 193}

\bibitem[\protect\citeauthoryear{{Warner}}{{Warner}}{2003}]{warner03}
{Warner} B.,  2003, {Cataclysmic Variable Stars},
  \mn@doi{10.1017/CBO9780511586491.
}

\bibitem[\protect\citeauthoryear{{Webbink}}{{Webbink}}{1992}]{Webbink1992}
{Webbink} R.~F.,  1992, in {van den Heuvel} E. P.~J.,  {Rappaport} S.~A.,  eds,
   NATO Advanced Study Institute (ASI) Series C Vol. 377, X-Ray Binaries and
  Recycled Pulsars. pp 269--280

\makeatother
\end{thebibliography}
\bsp

\end{document}